# Multidimensional Fluorescence Imaging and Super-resolution Exploiting Ultrafast Laser and Supercontinuum Technology

EGIDIJUS AUKSORIUS

Department of Physics

## Imperial College London

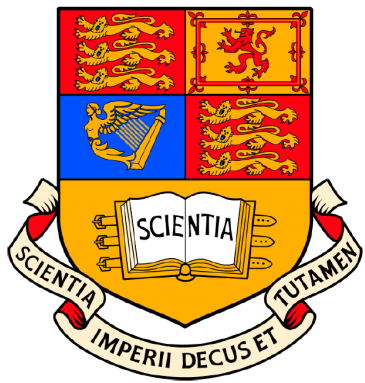

Thesis submitted in partial fulfilment of the requirements for the degree
*of*
Doctor of Philosophy of the University of London

December, 2008

*Skiriu tėvams*

# Abstract

This thesis centres on the development of multidimensional fluorescence imaging tools, with a particular emphasis on fluorescence lifetime imaging (FLIM) microscopy for application to biological research. The key aspects of this thesis are the development and application of tunable supercontinuum excitation sources based on supercontinuum generation in microstructured optical fibres and the development of stimulated emission depletion (STED) microscope capable of fluorescence lifetime imaging beyond the diffraction limit. The utility of FLIM for biological research is illustrated by examples of experimental studies of the molecular structure of sarcomeres in muscle fibres and of signalling at the immune synapse. The application of microstructured optical fibre to provide tunable supercontinuum excitation source for a range of FLIM microscopes is presented, including wide-field, Nipkow disk confocal and hyper-spectral line scanning FLIM microscopes. For the latter, a detailed description is provided of the supercontinuum source and semi-confocal line-scanning microscope configuration that realised multidimensional fluorescence imaging, resolving fluorescence images with respect to excitation and emission wavelength, fluorescence lifetime and three spatial dimensions. This included the first biological application of a fibre laser-pumped supercontinuum exploiting a tapered microstructured optical fibre that was able to generate a spectrally broad output extending to $\sim 350$ nm in the ultraviolet. The application of supercontinuum generation to the first super-resolved FLIM microscope is then described. This novel microscope exploited the concept of STED with a femtosecond mode-locked Ti:Sapphire laser providing a tunable excitation beam by pumping a microstructured optical fibre for supercontinuum generation and directly providing the (longer wavelength) STED beam. This STED microscope was implemented in a commercial scanning confocal microscope to provide compatibility with standard biological imaging, and exploited digital holography using a spatial light modulator (SLM) to provide the appropriate phase manipulation for shaping the STED beam profile and to compensate for aberrations. The STED microscope was shown to be capable of recording superresolution images in both the lateral and axial planes, according to the settings of the SLM.

# Acknowledgments

I would like to express my thanks to all those that helped me during my research and thesis writing. I am indebted to my supervisors, Paul French and Mark Neil for all the help, supervision and trusting me with this challenging but very interesting project. Thanks to Paul for his guidance, drive and enthusiasm and also for his support and patients when things did not go according to the plan or materialise 'before Christmas'. Thanks to Mark for his skilful advice on all things regarding microscopy and spatial light modulators. I would also like to thank Chris Dunsby who taught me everything in the lab from laser physics to microscopy. Thanks to Peter Lanigan who introduced me to all the particularities of the Leica SP2 + B&H microscope system and to James McGinty who has been ordering numerous items when I urgently needed them and for being useful with the whereabouts of things in the Lab. Further thanks to Bosanta Boruah and Dylan Owen who contributed more directly to the 'physics' results in this thesis. I would also like to acknowledge help from Roy Taylor and Sergei Popov for their advice on fibre laser and supercontinuum generation and in particular thanks to Andrei Rulkov for his numerous efforts to keep my fibre laser running and to John Travers for all things supercontinuum. Also thanks to Gordon Kennedy for helping me with the 'big' lasers on the STED setup. Everyone in the optics workshop was always helpful and some of the setups would have not existed without their help, therefore thanks to Martin Kehoe, Simon Johnson, Martin Dowman, James Stone and Paul Brown. Thanks again to Chris and Valerie for proofreading this thesis. Special thanks to Vincent for being a great mate both in the office and out. I am grateful to all other present and past member in the group whose help was sometimes so important: Dan, Jose, Richard, Raul, Bebhinn, Delisa, Damien, Pieter, Khadija, Ian, Cliff, Ewan, David, Sunil, Hugh, Stephane, Tom, Anca, Ali and anyone else I may have forgotten. I also enjoyed your company in the H-bar and other 'great' places in South Ken! Many thanks to my friends whom I met here in London that made my life much more interesting outside the lab, especially thanks those from Lithuanian community, *Ceilidh* and *Jonkelis* dancing clubs. In particular, thanks to Agnė, Sebastian, Natalia and Anne for their friendship over those years. Also, I am grateful to Michael for his help and support in the final stages of the writing-up. And finally, but not least, thanks to my whole family and friends back in Lithuania who haven't seen me much during this time but always supported and believed in me.

# Acronyms

*described in:*

| | | |
|---|---|---|
| 2 D | | Two-Dimensional |
| 3 D | | Three-Dimensional |
| B&H | 3.3.2 | Becker and Hickl GmbH (*company*) |
| CCD | 3.2.1 | Charge Coupled Device |
| CW | | Continuous Wave |
| EEM | 4.3.3 | Excitation Emission Matrix |
| FLIM | 3 | Fluorescence Lifetime Imaging |
| FRET | 2.2.8 | Fluorescence Resonance Energy Transfer |
| FWHM | | Full Width Half Maximum |
| GOI | 3.3.1 | Gated Optical Intensifier |
| GVD | 4.2.2 | Group Velocity Dispersion |
| $I^5M$ | 2.4.3 | Incoherent Interference Illumination Image Interference Micr. |
| IRF | 3.2.3 | Instrument Response Function |
| MOF | 4.2.2 | Microstructured Optical Fibre |
| NA | 2.3.1 | Numerical Aperture |
| OTF | 2.3.1 | Optical Transfer Function |
| PMT | 3.2.1 | Photo-Multiplier Tube |
| PSF | 2.3.1 | Point Spread Function |
| SLM | 5.2.7 | Spatial Light Modulator |
| SMF | 4.2.2 | Single Mode Fibre |
| STED | 2.6.1 | Stimulated Emission Depletion |
| $\tau$ | 2.2.7 | Fluorescence lifetime |
| TCSPC | 3.2.3 | Time Correlated Single Photon Counting |
| Ti:Sapphire | | Titanium Sapphire |

# Contents

# 1.   Introduction

This thesis concerns the development and biological application of multidimensional fluorescence microscopy, including fluorescence lifetime imaging (FLIM) and its extension to super-resolved far field imaging using stimulated emission depletion (STED) microscopy [1].

Progress in microscopy has been a key element to success in areas such as biology for which it has been used since the end of the 16[th] century when the first microscope was invented. Optical (ultraviolet-visible-near infrared) radiation is well suited to the study of biological samples since the photon energies correspond to electronic energy transitions in many organic molecules and the consequent light-matter interactions can be used to study and contrast many different species. Fluorescence microscopy is the most sensitive of the available optical contrast enhancing techniques (dark field, phase contrast etc) due to the ability to separate weak fluorescence signal from the strong excitation light. It allows the observation of 'live' biochemistry on a microscopic level, with the advantage of preserving the cellular context of biochemical connectivity, compartmentalization and spatial organization [2]. Moreover its application to biology has been particularly stimulated by development of new fluorescent probes that enable highly biochemically specific labelling and especially by advances in fluorescence labelling of proteins by gene transfer [3, 4]. In addition, new techniques have made it possible to label proteins with small organic fluorophores, nanocrystals [4, 5]. A new type of labels – fluorescing nanodiamonds colour centres [6, 7] are being developed that do not blink and barely photobleach [8], therefore holds a great promise for future imaging. Various photoswitchable fluorophores [9-11] has recently been developed to enable super-resolution imaging using various fluorescence microscopy techniques [12]. Thus fluorescence microscopy has become the most popular imaging tool in cell biology with a huge variety of fluorescence imaging and labelling techniques available [13].

The instrumentation associated with optical microscopy (light sources, detectors, electronics, computers etc) has been developed and refined to provide exquisite sensitivity and molecular contrast [14]. Lasers are used as excitation sources in many types of microscope, and particularly in confocal scanning microscopes because of their ability to provide bright, spatially coherent and collimated radiation that allows light to be focused to a diffraction limited spot. Therefore a laser is an ideal light source for confocal microscopy and is one of the factors behind its widespread use. With the developments in laser technology, many new types of microscopy have been developed, for example, the introduction of ultrashort pulse lasers led to many nonlinear optical microscopies and their extensive application to biology [15]. Time domain fluorescence lifetime imaging (FLIM) has also developed with the advent of ultrashort pulse laser technology. Spectral tunability that provides excitation of many different fluorophores is another desirable property of lasers. As a result, supercontinua generated by propagating ultrashort pulses through microstructured optical fibres (MOF) have recently become a popular excitation source for microscopy. These sources can cover the spectral region from the ultraviolet to the infrared and have many qualities typically associated with lasers, including ultrashort pulses and spatial coherence [16, 17], effectively making them 'white' lasers. The work reported in this thesis includes the application of supercontinuum sources to various types of multidimensional fluorescence imaging, including FLIM and STED microscopy.

At the same time, increasingly sophisticated approaches have been developed to analyse fluorescence signals [18]. The fluorescence spectrum (emission and excitation) is perhaps the most important parameter after the intensity and can provide important information about a fluorescing sample, for example to contrast different molecular species or different local molecular environments. It can also be used to measure the distance between two specific fluorophores by measuring the Förster resonance energy transfer (FRET) efficiency. Fluorescence lifetime is perhaps the next most important parameter and can provide additional information to the fluorescence intensity and spectrum [19]. If, for example, spectral discrimination cannot distinguish two different fluorophores, then lifetime can be used. Moreover fluorescence lifetime imaging can provide a more reliable method to image energy transfer processes. In this thesis FLIM was used to report FRET in protein interactions and to provide enhanced contrast in actomyosin images of various states. FLIM was also combined with imaging of fluorescence excitation-emission matrices to further increase the contrast. This thesis also contains the first demonstration of FLIM beyond the diffraction limit using STED microscopy.

Together the wide range of fluorescence imaging techniques can be applied to study almost every biological process but, until recently, the spatial resolution has been fundamentally limited to a fraction of the wavelength of the optical radiation, due to diffraction. Diffraction, which occurs due to the wave-like nature of light, prevents the light being focussed to an infinitesimally small spot, which therefore limits the resolution of a microscope. For visible radiation this corresponds to ~ 200 nm which is not sufficient to resolve some biological structures. A straightforward approach is to reduce the radiation wavelength, as in X-ray microscopy, or abandon focusing light at all, as in scanning near field microscopy. Although these and other techniques (such as electron microscopy and scanning probe microscopies) offer superior resolution and allow single molecule visualisation, these techniques require significant sample preparation, such that in vivo imaging is usually not possible. Scanning probe microscopy techniques are limited to surface imaging and electron and X-ray microscopy are damaging to biological samples. Thus, optical microscopy has been a tool of choice to study biological processes but the diffraction limit to spatial resolution has precluded the direct imaging of individual molecules. There is therefore great interest in developing optical techniques that are capable of visualising molecular processes below the diffraction limit while providing the rich molecular information associated with fluorescence imaging. Recently a number of 'super-resolution' fluorescence microscopy techniques have been developed [20]. This thesis used STED microscopy to improve spatial resolution.

**Chapter 2** of this thesis provides a general review of fluorescence spectroscopy and fluorescence microscopy, and discusses various super-resolution fluorescence techniques.

Then **Chapter 3** describes the specific FLIM techniques used for the work reported in this thesis, which are time-correlated single photon counting (TCSPC) and wide-field time-gated FLIM detection. These time domain techniques require pulsed excitation that was provided by either a mode-locked Ti:Sapphire laser system or by an ultrafast supercontinuum source pumped by an ultrafast laser – the latter approach being novel at the beginning of this thesis research. After reviewing FLIM in some detail, its application to biological studies is illustrated by descriptions of experimental studies of FLIM-FRET imaging of protein interactions at the immunological synapse between white blood cells and target cells and the application of FLIM to detect actomyosin states in mammalian muscle sarcomeres.

**Chapter 4** then discusses the application of supercontinuum excitation sources to fluorescence microscopy, beginning with a review of the underlying physics and outlining the different techniques for controlling the spectral properties of the output and the range of applications. Experimental characterisations of a home-built Ti:Sapphire laser pumped supercontinuum source and a fibre-laser pumped supercontinuum source extending down to the ultraviolet spectral region is then presented. The end of the Chapter presents application of supercontinuum excitation sources to wide-field and Nipkow confocal FLIM microscopy and to hyperspectral FLIM implemented in a line-scanning confocal microscope. The latter is an extension of FLIM microscopy to multidimensional fluorescence imaging that provides 6 D resolved imaging – acquiring the excitation-emission-lifetime matrix for each pixel in a 3-D optically sectioned image.

**Chapters 5** and **6** concern the development of a super-resolving STED microscope. **Chapter 5** describes the programmable manipulation of the spatial properties of light using a spatial light modulator (SLM). Implementing laser beams with specific spatially varying intensity and phase profiles is essential for STED microscopy and this was realised for the first time using programmable digital holograms that provided unprecedented flexibility as well as the capability to compensate for aberrations in the optical system. **Chapter 6** then describes the first application of supercontinuum generation to super-resolved fluorescence imaging using STED microscopy, including a detailed discussion of the experimental system, which was implemented in a commercial laser scanning confocal microscope, and the first demonstration of STED FLIM microscopy. **Chapter 7** concludes the thesis, reviewing progress and presenting suggestions for future work.

*Author declaration*

Work presented in the beginning of this thesis – **Chapters 3** and **4** are a result of collaboration with different people at Imperial College as appropriately acknowledged in the respective Chapters. Author's main work is associated with developing stimulated emission depletion (STED) microscope, which is presented in **Chapters 5** and **6**. The developed setups and the experimental results presented therein are the author's own work unless otherwise stated.

# 2. Fluorescence Microscopy and Super-resolution

## 2.1 Introduction

Fluorescence is the most sensitive optical technique to probe molecules because it can be efficiently separated from excitation light due to the Stokes shift. Therefore, fluorescence microscopy is preferred imaging modality to study cell biology, where subtle changes in the signal have to be imaged. Fluorescence microscopy is a mature field but still keeps developing towards higher sensitivity, versatility, and temporal, spectral, and spatial resolution [21]. The field now extends to single molecule imaging with the time resolution, which, for example, allows the observation of single molecule conformation changes [22, 23]. This shows that optical microscopy has evolved from being a purely intensity imaging techniques to nano-spectroscopy, where many different fluorescence parameters can be recorded at the same time for small volumes of sample.

A lot of attention is currently drawn to address fundamental drawback of fluorescence microscopy – the limited spatial resolution, caused by the diffraction limit. Contrary to the well-known Abbe's resolution limit, fluorescence microscopy is in principle able to achieve unlimited resolution. One way to break the resolution limit is by using structured illuminating light and the nonlinear dependence of fluorescence on the illumination intensity. Structured illumination is able to improve the resolution by a factor of two on its own, but in combination with nonlinearity between the illumination intensity and fluorescence, the resolution improvement becomes, in principle, unlimited. Scanning microscopes (like confocal or STED microscopes) can also be thought of as structured illumination microscopes but with structured illumination brought to an extreme – a focal point. Therefore the same logic applies. Other, recently emerged, super-resolution techniques are based on sequential localisation of individual fluorescent molecules and currently are able to achieve similar resolution to structured illumination microscopy (~ 20 nm).

In this Chapter fluorescence phenomenon, fluorescence microscopy and various fluorescence microscopes capable of achieving super-resolution are reviewed.

## 2.2     Fluorescence

### 2.2.1     Jablonski diagram

Molecules consist of atoms that are held together by sharing electrons residing in the outer electronic orbital of each atom. A molecule can absorb energy and store it in a form of complicated motion of nuclei and electrons. Due to quantisation a molecule has a discrete energy level structure that can be broadly separated into the *electronic*, *vibrational* and *rotational* levels as shown in Jablonski diagram in Figure 1. The smallest amount of energy that can be absorbed is associated with molecular rotations and is called rotational energy. More energy can be deposited in vibrational energy form that is in vibrations of molecule's nuclei. The largest portion of energy, called electronic energy, can be absorbed by promoting outer molecular electron to a higher molecular orbital that is on the average further away from the centre of a molecule.

### 2.2.2     Franck-Condon principle

When a molecule absorbs a photon with energy high enough to transition the molecule from one electronic state to another, the molecule is then said to go from the ground state $S_0$ to the excited state $S_1$ (or in general $S_n$). This transition typically involves a change in vibrational and rotational energy as well. The most likely energy transitions, however, are the ones, which involve minimal nuclear distance change when the molecule goes from $S_0$ to $S_1$. This is because the electronic transition, being very fast (~ 1 fs), occurs faster than the nuclei can significantly change their position (Franck-Condon principle) due to huge electron and nuclei mass difference. Therefore, those vibronic levels in the excited state are favoured that have the most similar nuclei configuration to that in the ground stage. In the quantum mechanical formulation a molecule has the higher the probability to absorb a photon the more the vibrational wavefunctions of the levels associated with the transition overlap (Franck-Condon factor). That does not, however, mean that the most probable electronic transitions will be those that involve the same vibronic levels (for example $v_0 \rightarrow v_0$ or $v_1 \rightarrow v_1$) as would be the case if the equilibrium position between the nuclei would not change after excitation.

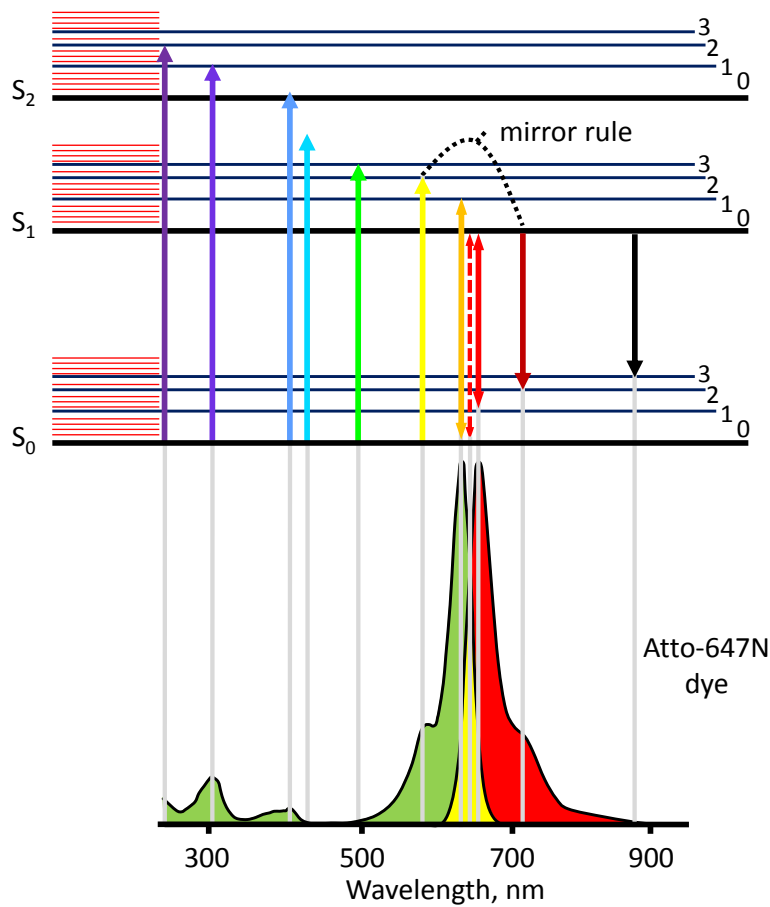

Figure 1. Jablonski diagram with fluorescence excitation and emission spectra (below) of the Atto-647n dye. Jablonski diagram shows molecular energy structure composed of rotational (red horizontal lines) vibronic (dark blue) and electronic (black) levels. Various hypothetical transitions are drawn for excitation and emission with arrow colour representing photon wavelength. Green area represents absorption, red represent emission and yellow is an overlap area that marks a region where photon absorption and photon emission is possible.

Normally, the equilibrium position does change due to the change of electron configuration in the excited state, which increases repulsion between nuclei. Most of the molecules at room temperature in the ground state occupy the zeroth vibronic level (probability of a molecule finding in the higher vibronic levels follows Boltzmann distribution) and, therefore, most of the transitions start from there ($v = 0$). However, there are many different vibronic levels in $S_0$ to choose from, which allows different transition to occur, but with different strengths depending on the Franck-Condon factor. The multiple transitions form the fluorescence excitation spectrum of a molecule, as illustrated in Figure 1.

### 2.2.3    Vibrational relaxation and Kasha's rule

When a molecule is excited to a high vibronic level of the $S_1$ state, it experiences fast relaxation ($\sim$ ps) to the zeroth vibronic level of $S_1$ due to the so called *vibrational relaxation*, which results from the interactions of the excited molecule with its environment. The interaction slows down the molecular rotations and vibrations and thus the energy of the molecule is dissipated into the environment radiationlessly – without emitting any photon. Similarly, if a molecule is excited to a higher electronic level ($S_0 \rightarrow S_2$ in Figure 2), it usually undergoes another type of radiationless relaxation – *internal conversion* (from $S_2$ to $S_1$). However, the conversion is sometimes possible for $S_1 \rightarrow S_0$ and even $S_2 \rightarrow S_0$, as illustrated in Figure 2. However, the latter is extremely rare to occur and is known as Kasha's rule –'fluorescence cannot happen directly from $S_2$'. Internal conversion happens because of vibro-rotational levels existing between $S_2$ and $S_1$ and sometimes $S_0$, so that molecule can climb down that ladder through relaxation. In most molecules there is a gap between $S_1$ and $S_2$ where vibro-rotational level do not exist and, therefore, excitation at particular wavelength is not possible (aqua transition arrow in Figure 1). When in $S_1$, a molecule can go either to $S_0$ or it could be further excited to $S_2$ (Figure 2), which is less likely process to occur.

### 2.2.4    Fluorescence emission

After vibrational relaxation, the molecule typically spends a comparatively long time ($\sim$ ns) before spontaneously emitting a photon and returning to the ground state – fluorescing. In a similar way to excitation, the emission can happen to various vibronic levels of the ground state but with different probabilities, which forms an emission spectrum of a molecule (right spectrum in Figure 1). The emitted photons are typically of lower energy because energy other then electronic (vibrational and rotational) is lost during interactions with the molecular environment; the shift in energy and, therefore, in wavelength is called Stokes shift. However, the emitted photon can have shorter wavelength than the absorbed one if, for example, a molecule is not in the zeroth vibronic level of the ground state (usually happens at higher temperatures), when it absorbs a photon and emission transition happens to lower vibronic level than the molecule was excited from. This can happen anywhere in the overlap area between the excitation and emission spectra (yellow area in Figure 1). The overlap area indicates a spectral region where a molecule is, in principle, capable of absorbing photons and of emitting them. On the left hand side of this area, the molecule is more likely to absorb a photon than to emit (doubled arrow orange transition line in Figure 1), whereas on the right hand side a molecule is more likely to emit a photon than to absorb (doubled arrow red transition line). The point where a molecule has the same probability of absorbing and of emitting is $S_1 v_0 \leftrightarrow S_0 v_0$ transition (doubled arrow red dashed transition line). A molecule has also got a defined limit of the longest wavelength (lowest energy) photon that it can emit, which is between the lowest vibronic level of the excited state, $S_1 v_0$ and the highest vibronic level of the ground state, $S_0 v_{highest}$ (black transition line in Figure 1). A reverse process: $S_0 v_{highest} \rightarrow S_1 v_0$ is hardly possible since there are virtually no molecules occupying the $S_0 v_{highest}$, therefore, the excitation spectrum intensity is practically zero there. If $S_0$ and $S_1$ electronic states have similar vibronic level structure or, more precisely, similar shapes of the potential curves, then, for example, transitions $S_0 v_0 \rightarrow S_1 v_2$ and $S_1 v_0 \rightarrow S_0 v_2$ (see the corresponding yellow and dark red transition lines in Figure 1) will have the same probability – Franck-Condon factor. When that is true for many transitions, the absorption and excitation spectra have mirror symmetry – as can be seen around the overlap area in Figure 1. In general, vibrational level structure of ground and excited states are different, with the excitation spectrum shape reflecting the distribution of excited state vibrational levels and the emission spectrum – distribution of ground state energy levels. Individual transitions can only be seen at low temperatures since at room temperatures the transition experience homogenous broadening and fluorescence spectrum losses details.

### 2.2.5    Phosphorescence

Most commonly electrons in the excited state of a molecule have its magnetic spins oriented antiparallel with respect to each other and, therefore, a molecule does not form an overall magnetic moment.  However, if one of the electrons flips its spin then the two parallel spins form the magnetic moment, which can have three different orientations with respect to the applied magnetic field: parallel, antiparallel or perpendicular. The flip of the spin is very unlikely process and is called *intersystem crossing*. The three different orientations have different energies in magnetic field and, therefore, a molecule have three different energy levels. Because of that an energy level of a molecule that has antiparallel spins is called triplet state (*T*) (as opposed to the singlet state (*S*) where spins are antiparallel and do not form a magnetic moment). A molecule in a triplet state (*T₁*) has lower energy then in a corresponding singlet state (*S₁*), as illustrated in Figure 2. A molecule from the lowest triplet state, *T₁* can be excited to the higher triplet state, *T₂*, which is known to lead to photobleaching. From *T₁* a molecule can emit a photon and return to the ground state. However, this involves the electron changing its spin one more time. As changing spin is an unlikely process, an associated emission of photon is also unlikely and therefore happens on a slow time scale (~ μs). This is called phosphorescence.

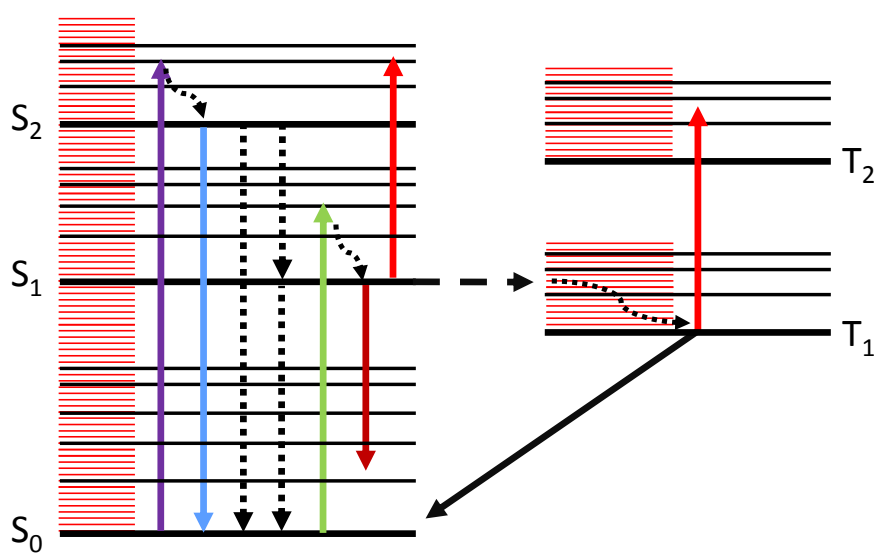

Figure 2. Jablonski diagram for various possible electronic transitions. Excited molecule can go to *S₁* (green arrow) or *S₂* (purple arrow) where it experiences fast vibrational relaxation to the ground vibronic level of *S₁* or *S₂* respectively. From there it could experience various internal conversions (black dashed arrows). $S_2 \rightarrow S_0$ is rather unlikely for internal conversion and especially for fluorescence (aqua arrow). From *S₁* molecule can be either excited to *S₂* (red arrow) or undergo intersystem crossing to *T₁* (black dashed horizontal arrow) but most probably fluoresce and reach ground state (dark red curve). The molecule in *T₁* can transition to *S₀* (black arrow) or be further excited to *T₂* (red arrow).

In general, the highest molecular orbitals in the molecule determine the properties of fluorescence, like absorption and emission wavelength, quantum yield etc. [24].

### 2.2.6    Fluorescence quantum yield

The fluorescence of a single molecule is random but a characteristic fluorescence rate constant, $k_{fluor}$ (photons / second) can be defined to describe behaviour of many molecules. Some molecules, however, do not emit a photon through fluorescence, as discussed, since they can return to a normal state through other processes like internal conversion, phosphorescence or energy transfer, which will be discussed later. These processes can be accounted by a one non-radiative rate constant, $k_{nr}$. Non-radiative decay reduces the quantum yield, $Q$ – a factor that defines the efficiency of a molecule to produce a number of emitted photons for a number of absorbed photons. Quantum yield accounts for non-radiatively lost photons:

$$Q = k_{fluor} / (k_{fluor} + k_{nr})$$

eq. 1

### 2.2.7    Fluorescence lifetime

Due to spontaneous nature of fluorescence, a molecule, in principle, can take any time before it emits a photon ($0 - \infty$). However, as discussed above, for an ensemble of molecules a fluorescence rate constant, $k_{fluor}$ can be defined. In general, a sum of fluorescence and non-radiative rate constants ($k_{fluor} + k_{nr}$) describe how fast the excited entity of molecules returns to the ground level, therefore, an inverse sum of the rate constants characterises an average fluorescence lifetime, $\tau$:

$$\tau = 1 / (k_{fluor} + k_{nr})$$

eq. 2

The change of a population, $n$ in the excited level follows exponential decay with time and can be expressed through $\tau$ as:

$$n = n_0 e^{-t / \tau}$$

eq. 3

Where $n_0$ is the initial population of the excited state and $t$ is the time after the excitation. Physically eq. 3 says that $1 / e$ ($\sim 37$ %) of the excited molecules will leave the state within the time window of $\tau$. The fluorescence dynamics will be governed by eq. 3. Fluorescence lifetime (and therefore quantum yield) can give information about surrounding media of molecules since $k_{nr}$ can change due to the molecule's local environment.

### 2.2.8    Fluorescence energy resonance transfer (FRET)

A molecule in the excited state can interact with its surroundings and therefore may lose its vibrational / rotational energy by transferring it to the neighbouring molecules. In aqueous solutions water is the most likely recipient of energy. If an interacting molecule is itself a fluorophore, then the excitation energy in the excited molecule can be transferred to the neighbouring molecule and excite it. This is referred to as resonant energy transfer (RET), because no energy is lost through the transfer. The phenomenon is also widely known as Förster resonance energy transfer (FRET) since it was Förster who first described the process [25]. Often word Fluorescence is added in front of RET but that is misleading because no actual fluorescence happens here. Förster resonance energy transfer happens between molecules that are only a few nanometers away through a non-radiative dipole-dipole interaction. For this to happen, the excited molecule (donor) has to have its emission spectrum overlaping with the absorption spectrum of the recipient molecule (acceptor). FRET can be seen as one of the competing effects that can de-excite molecule with rate constant, $k_{fret}$ defined as:

$$k_{fret} = 1 / \tau \times (R_0 / r)^6$$

eq. 4

Where $R_0$ is the so-called Förster distance, at which the energy transfer of 50 % happens; $r$ is the distance between the centres of the donor and acceptor molecules. It is evident from the equation that $k_{fret}$ is inversely proportional to the fluorescence lifetime but has normalised distance coefficient ($R_0 / r$) that scales with the power of 6, which mean that FRET quickly dies away when the distance is increased. The power of 6 in eq. 4 comes from the properties of the electrical field near the dipole, which varies as a function of $1/r^3$, therefore an interaction between the two dipoles is described as a function of $1/r^6$. Similarly to fluorescence quantum yield, FRET efficiency can be defined as:

$$E = k_{fret} / ( k_{fret} + k_{nr} + k_{fluor})$$

eq. 5

By substituting $k_{fret}$ in eq. 5 by eq. 4 and using eq. 2, one obtains:

$$E = 1 / (1 + (R_0 / r)^6)$$

eq. 6

The equation above shows that FRET can be used to measure nanometre distances ($r$) by measuring its efficiency [26].

## 2.3      Fluorescence microscopy

Fluorescence microscopy is based on optical microscopy principles, which will therefore be discussed first. This will be followed by a description of other optical contrast enhancement techniques and, finally, various fluorescence microscopy techniques that can improve spatial resolution will be reviewed.

### 2.3.1    Optical microscopy

Optical or light microscopy involves passing visible light, transmitted or reflected from the sample, through a single or multiple lenses to allow a magnified view of the sample [27]. The first microscopes were built in 16[th] century and relied on a single lens. Microscopes with two lenses were later constructed, but it was only in 19[th] century that they outperformed the single lens microscopes when the optics behind image formation was understood and manufacturing of optical elements like the objective lenses were perfected. Sample illumination in those early microscopes relied on the so-called critical illumination, where focusing of illumination source onto the sample was performed. Images acquired with critical illumination are a result of the multiplication between the actual structure of the sample and the structure of the illumination source. This causes a problem if the illumination source is not homogeneous. Therefore diffusers had to be used or light sources were used slightly defocused. That changed with the introduction of *Köhler illumination* [28], which enabled formation of an uniform sample illumination with incoherent large area light source. In the simplest form of Köhler illumination the light source is put in the back focal plane of the condenser lens so that it would image the source to infinity. If the sample is put in the way of the beam a homogeneous illumination across the sample is thus created. The objective on the other side of the sample can gather transmitted light and form an image, which can be detected with an imaging detector (like a CCD) or by raster scanning with a point detector (like photomultiplier or avalanche photo diode behind a small pinhole). Köhler illumination normally involves additional lenses that enable control of the illumination field and aperture, as is shown in Figure 4 and Figure 5, and explained in 2.3.3.

### *Spatial resolution*

An optical microscope can collect a restricted amount of information from an object due to the diffraction limit. This has an effect on the optical resolution of a microscope. Optical resolution was investigated independently by Rayleigh and Abbe in the 19[th] century. They both came up with similar conclusions but had different approaches. Abbe used the following description to explain the performance of a microscope (non-fluorescence). Optically every object can be considered as a composition of various gratings with different periodicity and orientation. Each grating produces zeroth (undiffracted), ±1 and higher diffraction orders. A lens can collect a limited number of diffraction orders because of its limited size or more precisely its numerical aperture (NA), as illustrated in Figure 3. If the lens captures the zeroth and the first diffraction orders then an image of the grid is reproduced on the other side of the lens. The image however will not be a true representation of the grating since this would need the infinite number of the diffraction orders to be collected. The two diffraction orders creates sinusoidal pattern in the image plain (not shown in the figure). Nevertheless, the individual stripes in the grating can be discerned in the formed image. In the case when all the diffraction orders miss the lens (except zeroth) the stripes in the grating cannot be distinguished and an image of grating is not formed. The Abbe's criterion then states that the smallest period of a grating (Δx or Δy) that can still be imaged is:

$$\Delta x_{Abbe}, \Delta y_{Abbe} = 0.5\lambda \, / \, (n \cdot sin\alpha)$$

eq. 7

Where $n \cdot sin\alpha = NA$ – numerical aperture of an objective lens. $\lambda$ – illumination wavelength.

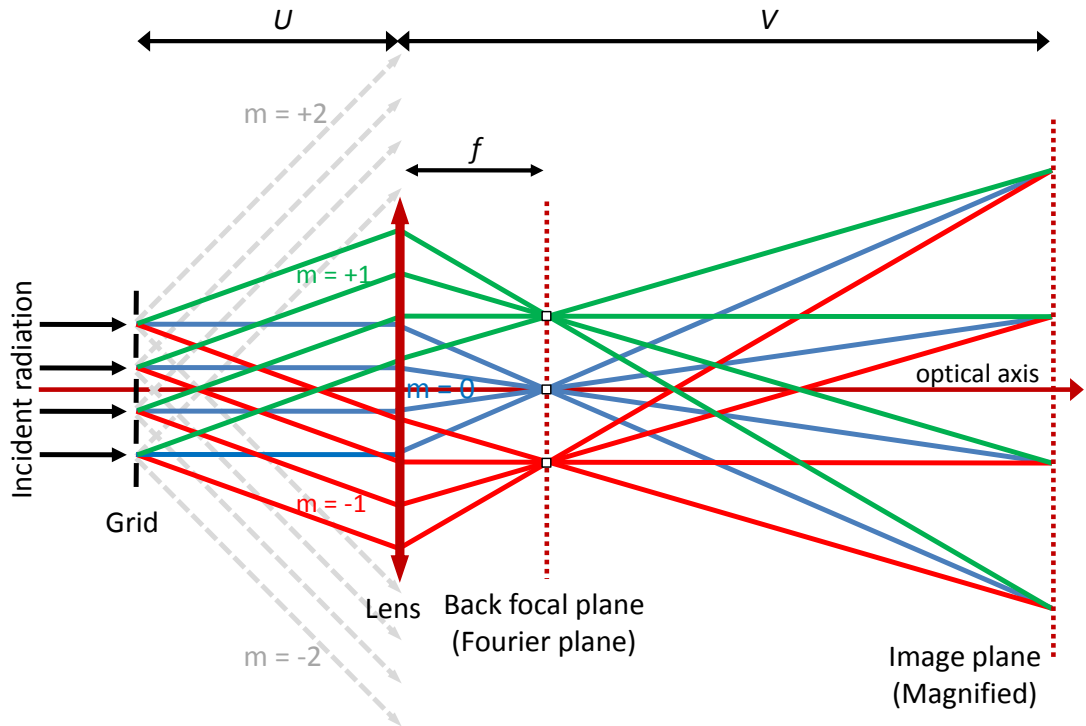

Figure 3. Illustration of Abbe's image formation. A grating is illuminated with coherent light source, which diffracts the light into the different diffraction orders. 0 (blue) and ±1 (green and red) orders are collected by the lens and ±2 (gray) misses it. The collected diffraction orders are spatially separated in the back focal plane (Fourier plane) of the lens according to the spatial frequencies they represent: zero spatial frequency appears on a lens optical axis whereas higher frequencies – further away from it. In the image plane the orders interfere to form the magnified image (by *U / V*) of the grating. *f* – focal length of the lens; *m* – diffraction order.

Rayleigh, on the other hand, used the following reasoning to define the resolution. A diffraction pattern of a point source formed in the image plane (a plane orthogonal to the optical axis) by a lens can be described mathematically by the Airy function [29]. The Airy function can be expressed as:

$$I(v) = I_0 \left(2J_1(v) / v\right)^2$$

eq. 8

Where $J_1(v)$ is the first-order Bessel function of the first kind, $v = 2\pi r n \sin\alpha/\lambda$ – a dimensionless optical coordinate and $r$ – radial coordinate. The Airy function has series of defined zeros – dark circles around a bright focal spot. The first dark circle is located at $v_0 = 1.22\ \pi$ in the optical coordinates. In the real coordinates the diameter, $d_{Airy}$ of the bright focal spot (that is surrounded by the first dark circle) can be expressed as:

$$d_{Airy} = 1.22\lambda / (n \cdot sin\alpha)$$

eq. 9

This is often called the Airy disc. In terms of the Rayleigh criterion, two points can be resolved if the maximum of the Airy function of one point is located in the first minimum of the other. This distance is equal to the radius of the first dark circle − $d_{Airy} / 2$. Therefore, the lateral resolution, $\Delta r$ can be defined as:

$$\Delta r_{Rayleigh} = 0.61\lambda / (n \cdot sin\alpha)$$

eq. 10

In other words the two point sources will be resolved laterally if they are not closer than by $\Delta r$. A diffraction pattern of a point source formed by a lens in the axial plane (parallel to the optical axis) is defined by a different function [29]. Nevertheless a formula for the axial Rayleigh criterion can be derived using similar reasoning, since the axial diffraction pattern also features a bright focal spot and a region of a minimum intensity. Therefore, the axial resolution, $\Delta z$ can be defined as:

$$\Delta z = 2\lambda / (n \cdot sin^2\alpha)$$

eq. 11

The resolution achieved with a wide field fluorescence microscope is limited to ~ 200 nm laterally and ~ 500 nm axially, because it is either physically difficult ($\alpha >$ 68°, n > 1.5) or biologically incompatible ($\lambda < 350$ nm) to improve any of the factors or parameters in the Abbe's or Rayleigh. In fluorescence imaging, the diffraction pattern of the spot is also referred to as the point spread function (PSF). Its Fourier transform is called the optical transfer function (OTF) and describes the spatial frequencies that can be transmitted by a microscope.

### 2.3.2    Other contrast enhancing microscopy techniques

In microscopy one aims at enhancing contrast and structural details of a studied object. Bright field microscopy is the simplest and the oldest way to provide contrast, which measures scattered or transmitted (non-absorbed) light that gives rise to changes in brightness or colour of a sample. However, the transmitted signal is observed against the background of light source radiation so that a detector with a very wide dynamic range is required to detect it [30]. Therefore, the bright field microscopy lacks contrast and, most importantly, cannot provide optically sectioned images. This is different in the dark field (as opposed to the bright field) microscopy, which is an optical microscopy contrasting technique that works by illuminating the sample with a light cone of higher numerical aperture (NA) than the objective lens that captures the light through the sample can collect. The objective, therefore, does not collect any light unless the sample scatters it. A sample thus appears as bright objects on a dark background. Many biological samples are transparent and some of them are hardly visible because of that, but they may exhibit a variation in refractive index that would make light to acquire different phase due to different optical path lengths across the sample if light passes through it. Unfortunately, our eyes are not able to detect changes in phase and so the optical path difference has to be somehow converted into the intensity variations. This can be done through interference, as is implemented in the phase contrast and differential interference contrast microscopes [27]. Other types of microscopy that can provide contrast include polarisation microscopy, which can sense polarisation properties of a sample. Thus, the microscope can, for example, image birefringence of a sample that generally comes from highly aligned structures, such as fibres, crystals etc. Regardless of their ability to enhance contrast through variations in the birefringence, refractive index or thickness of a sample, these microscopes cannot be used to indentify individual objects other than by their shape (cell organelles, for example) and cannot detect objects smaller than the diffraction limit. Although various staining can be used to enrich the contrast, these do not have a well-understood interaction with the sample. The advantage of fluorescence microscopy is that it blocks the excitation light and weak fluorescence signals can be recorded coming from specific sites that have been labelled with appropriate fluorophores, therefore it is highly sensitive and specific technique.

### 2.3.3    Wide-field fluorescence microscopy

#### Epi-fluorescence

The first fluorescence microscopes functioned in the transmission (or diascopic) mode, which made it difficult to separate the excitation light from the fluorescence of the sample since the substantial part of the excitation light could go through the sample (and reach the detector). Nevertheless, to get rid of the excitation light, the dark field imaging arrangement can be used, which ensures that only fluorescence and scattered excitation light go through, whereas the direct excitation light is not collected by the objective lens. The scattered light can now be easily blocked with a filter since it is now much weaker.

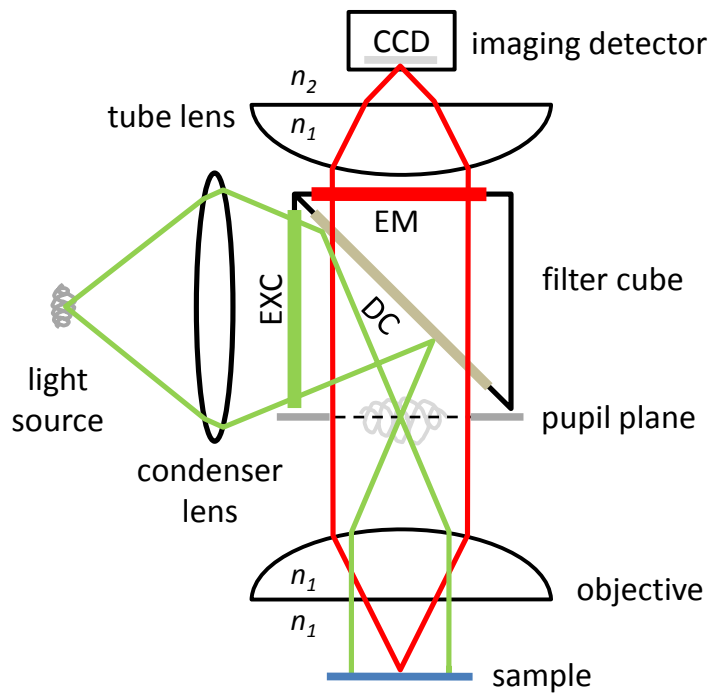

Figure 4. Basic optical scheme of a epi-fluorescence wide-field infinity-corrected (oil immersion) microscope with Köhler illumination. Illumination and image formation paths are shown in green and red curves, respectively. Köhler illumination creates homogeneous illumination at the sample with an incoherent large area light source. Fluorescence goes through the filter cube and is imaged onto the sensitive are of the imaging detector (CCD). EXC – excitation filter; EM – emission filter; DC – dichroic filter.

However, such a way to discriminate fluorescence from excitation limits the simultaneous use of, for example, fluorescence and phase microscopy, and also restricts the use of full potential of the high numerical aperture objectives for fluorescence collection. Instead, the fact that excitation light is less backscattered (reflected) than transmitted can be exploited. To reduce the excitation signal, the fluorescence can be collected in the reflection or epi-illumination mode, as shown in Figure 4. Here, the objective lens is used for both, the illumination of the sample and for the collection of fluorescence. Köhler illumination is made to work in the reflection mode (epi-illumination). A condenser lens is used here to produce a magnified image of the light source on the back focal plane (pupil plane) of the objective. The objective then images the conjugate image of the light source to infinity, thus creating a homogeneous illumination across a sample. Figure 4 shows a simplified optical set up of a microscope that, in addition to Köhler illumination in epi-illumination mode, uses oil immersion and infinity corrected objective lens.

### *Filter cube*

A dichromatic beam splitting mirror, referred to as a dichroic (DC), is normally used to spatially separate excitation and fluorescence beams. The dichroic filter is designed to reflect the 'blue' part of the spectrum and to transmit the 'red'. This allows to spatially separate excitation and fluorescence signals with good efficiency. However, the *out of band* rejection of a dichroic filter is normally not very high allowing some excitation light from a broadband light source to go through together with fluorescence and, therefore, an additional emission filter is normally used in front of the detector. In addition to that an excitation filter is usually used in front of the broad band excitation source to select an appropriate excitation spectral band. The three elements – excitation filter, dichroic and emission filter together can be combined into one element – a *filter cube*, as shown in Figure 4. This allows changing of all three elements rapidly if one needs to image different dye with different spectral characteristics.

### *Infinity corrected optics*

Infinity corrected objective images a sample, which is placed in front focal plane, to infinity. This helps to avoid certain optical aberrations and allows one to insert different optical elements in the optical path after the objective lens (or in some other part of the illumination where the light is collimated), without causing any unwanted distortions across the beam that would appear if the light were impinging on the element at some angle. For example, a quarter wave plate would convert a divergent or convergent beam across it into elliptically polarised light with various degree of ellipticity across the wavefront rather than homogenously circularly polarised. For high resolution / magnification, an oil immersion objective is used so that the light sees the area between the objective and a sample as one optical media, which helps to minimise spherical aberrations.

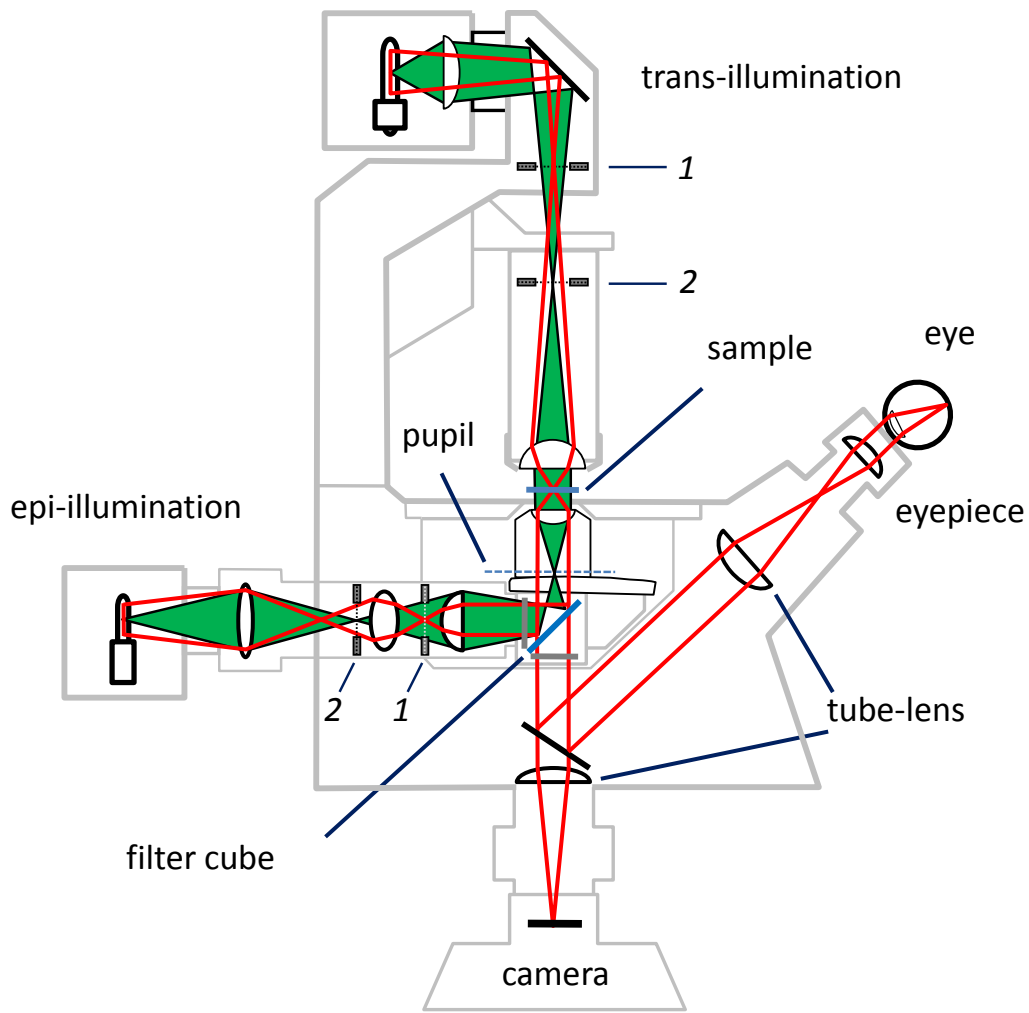

Figure 5. Anatomy of a modern inverted wide-field microscope equipped for transmitted and epi-illumination. Köhler Illumination paths for trans / epi-illumination are shown in filled green colour and image formation path is shown in red. Fluorescence can be detected with the camera or viewed through the eyepiece. The filter cube is removed (rotated out) for trans-illumination imaging. The transmitted image can be viewed in the same way as the fluorescence image (camera or eyepiece). *1* – Field diaphragm; *2* – Aperture diaphragm.

Since the microscope ensures Köhler illumination by imaging the light source onto the pupil plane of the objective, this can mean that the pupil plane will not be completely filled with light because of the structure of the light source, therefore potentially limiting the resolution of such a microscope (by limiting angles that can impinge on the sample). However, illumination is not crucial to resolution since the most important fact in the wide-field microscopy is high aperture collection of fluorescence. Large area, non-coherent light sources are routinely used for illumination, although expanded lasers can also be used but their spatial coherence has to be destroyed to avoid speckled illumination. The fluorescence image can be detected with an imaging detector (for example CCD camera) or viewed through an eyepiece (not shown in Figure 4). A typical modern wide-field epi-fluorescence microscope, also equipped with the illuminator (lamp housing, lenses and diaphragms) for transmitted light imaging, is presented in Figure 5. Köhler illumination is set for both the transmission (trans-illumination) and fluorescence (epi-illumination) imaging. There are two diaphragms in the illumination path in both arms that control the field of view and aperture of the illumination independently. This enables control over the size of the illumination field (Field diaphragm in Figure 5) and over the angle of illumination (Aperture diaphragm), respectively. Diaphragms also help with the alignment of the Köhler illumination. This microscope is therefore able to record transmission, reflection (bright field) and epi-fluorescence images. Moreover, it can be equipped with various contrast-enhancing techniques as discussed above. Furthermore, it is possible to attach a beam scanning (or stage scanning) unit and implement pinhole detection for confocal imaging, therefore both wide-field and confocal imaging are possible within a single microscope body. It is appropriate to also mention here that the microscope can function as a platform for many types of microscopy, including FLIM and STED microscopy, which is central theme of this thesis and will be discussed later.

### 2.3.4    *Multidimensional fluorescence microscopy*

A number of fluorophores that can be told apart is limited to two or three because of the broad fluorescence spectrum. As discussed above the fluorescence spectrum in room temperature has a broad bandwidth, which makes it difficult to separate two different fluorophores with similar spectral characteristics by way of band pass interference filters. Therefore, the detected signal if two filters were used would contain fluorescence of both fluorophores. However, if the full fluorescence spectrum is recorded then a contribution of each fluorophore can calculated by matrix inversion if the spectra of each fluorophore is known [31]. There are many different approaches of spectral imaging. Other parameters, like fluorescence polarisation or lifetime can also be recorded [18].

## 2.4    Fluorescence microscopy beyond the diffraction limit

In the past decade there has been substantial progress in improving the resolution of fluorescence microscopy [20]. Various optical techniques have been invented to improve the resolution with most of them using conventional optics. The first efforts to improve spatial resolution by modifying the pupil plane of the microscope were made by G. Toraldo di Francia in 1952. However, the attempts were not very successfully because the PSF of the microscope had high secondary lobes, which were difficult to eliminate. Later it was discovered that the resolution barrier of a microscope could be shifted by increasing the OTF support region by up to two times using structured illumination [32, 33]. However, it was found that to actually break the resolution barrier, some kind of nonlinear transition in fluorophore has to be employed [34] The latter will be discussed in more details in Section 2.6.

### 2.4.1    Wide-field fluorescence deconvolution microscopy

As discussed earlier, an image recorded in a wide-field microscope can be blurred and some information coming from the fluorescent object may be compromised by out of focus light. This is the fluorescence image formed in an optical microscope is a result of the convolution between an object and the microscope's PSF. Mathematically, the recorded image, *I* can be written as a convolution between the real object, *R* and the microscope's PSF:

$$I(x, y, z) = R(x, y, z) \otimes PSF(x, y, z)$$

eq. 12

Deconvolution microscopy aims to computationally reverse this process and restore the real object, *R*. The first efforts to reassign out of focus light back to the original plane were carried out in the 1980's [35, 36]. It is most effective if the PSF is experimentally measured, by, for example, imaging a fluorescent bead with the diameter smaller than the diffraction limit. A set of images is recorded by progressively defocusing the fluorescent bead, to get a three dimensional PSF of a subdiffraction bead. If we Fourier transform eq. 12, the operation of convolution can be changed to multiplication, which will simplify the deconvolution procedure. In the Fourier space the deconvolution operation can be written as:

$$R\tilde{}(k_x, k_y, k_z) = \Gamma(k_x, k_y, k_z) / PSF\tilde{}(k_x, k_y, k_z)$$

eq. 13

Where $R\tilde{}$, $\Gamma$ and $PSF\tilde{}$ are the Fourier transforms of $R$, $I$ and $PSF$, respectively, and $k_x$, $k_y$, $k_z$ – the spatial frequencies. To find the real object, $R$, an inverse Fourier transform of the solution, $R\tilde{}$, in eq. 13 has to be taken. However, it is difficult to collect a large enough number of photons from a sub-diffraction size fluorescent bead to accurately measure its PSF. This leads to noisy PSF and furthermore noise will be introduced and amplified in the solution, $R\tilde{}$, because $PSF\tilde{}$ is a denominator in eq. 13. Noise tends to affect the high spatial frequencies ($k_{xyz}$), therefore objects like edges in the initial image, $I$ will be distorted after the deconvolution. To address this problem iterative algorithms are used, with non negative constraints imposed on the solution, and smoothing of the PSF to avoid noise amplification [37]. Following the improvements in deconvolution algorithms and with the increases in computing power have allowed deconvolved images with resolution beyond the diffraction limit to be obtained [38].

### 2.4.2 Confocal microscopy

Confocal microscopy was the first kind of optical fluorescence microscope that was able to increase resolution and also to provide optical sectioning. Confocal microscopy was invented by Minsky in 1957 [39] to study thick brain slices [40]. His microscope was able to produce optically sectioned images of the specimen by using a pinhole to block out-of-focus light that came from scatter in a specimen, as shown in Figure 6. Minsky used another lens (condenser) on the other side of the specimen to image a point source on a specimen obtained from a pinhole placed in front of a light source. The out-of-focus light cannot get through the pinhole since its image plane does not lie in the plane of the pinhole. Thus by raster scanning the specimen, an image with optical sectioning better than that of a wide-field microscope is obtained. However, the technique did not become widespread immediately and it only gained momentum in the 1980s [41, 42] when the lasers [43] computers and microelectronics [44], necessary to successfully operate the system became available.

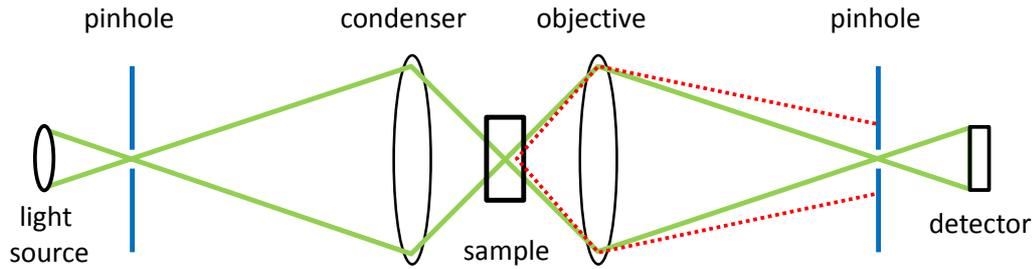

Figure 6. Schematics of the transmission confocal microscope showing principles of confocal imaging and optical sectioning. A point source is projected demagnified with a lens (condenser) on a specimen and signal is detected with another lens (objective) in transmission mode. Out of focus light (doted red curve) is rejected by another pinhole in front of detector.

Confocal fluorescence microscopy, similarly to wide-field fluorescence microscopy, is normally carried out in the reflection or epi-fluorescence mode as shown in Figure 7.

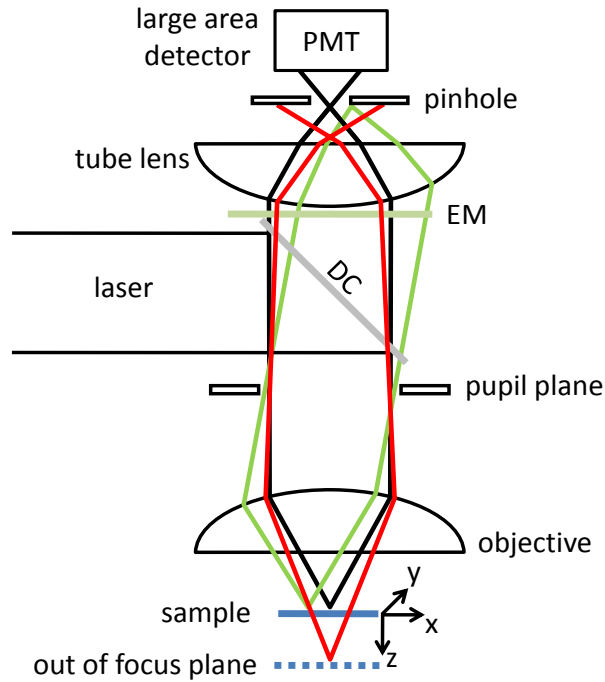

Figure 7. Basic optical scheme of epi-fluorescing laser scanned confocal microscope with infinity-corrected optics and oil immersion objective. Collimated laser beam is focused by the objective to a spot which is scanned across the sample with the sample scanning stage ($x$, $y$, $z$). Fluorescence is imaged onto the point detector (PMT) through a pinhole. Fluorescence originating from out of focus planes or from adjacent point is blocked by the pinhole (red and green curves, respectively).

By decreasing the pinhole size, more out-of-focus light can be rejected to improve sectioning, but at the expense of detected fluorescence intensity. In a severe case when the pinhole is almost closed the fluorescence signal is strongly dominated by noise.

The optimum pinhole size is the size of an Airy disc as defined by eq. 9. The pinhole set to the size of the Airy disc (often called 1 Airy) lets through a significant amount of signal but also effectively rejects out of focus light. The Airy disc in a sample plane of a microscope is magnified by the objective and tube lens system (as illustrated in Figure 8) and therefore, the size of the pinhole in the image plane has to be selected to match that of the Airy disc.

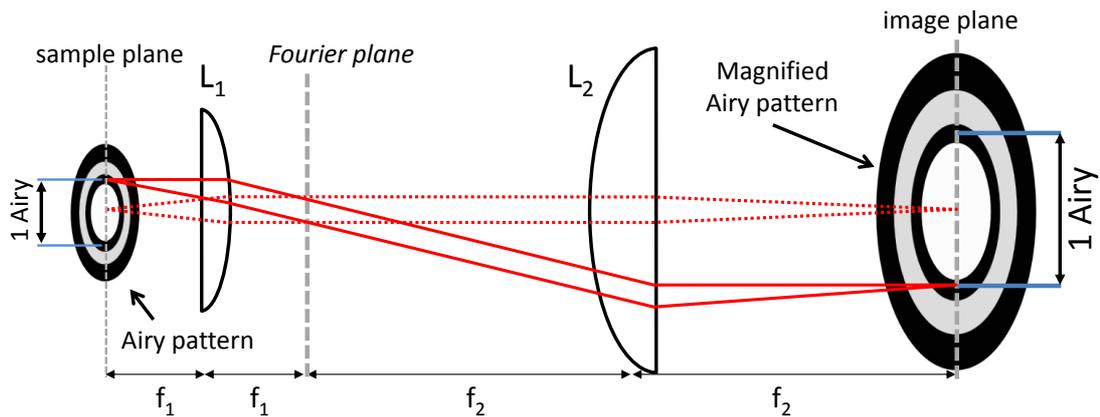

Figure 8. Illustration of the magnification of the Airy disk pattern in a microscope. An Airy pattern formed in the sample plane by the laser illumination (not shown in this picture), is magnified with the objective (L1) and the tube (L2) lenses, arranged in the 4-f system, by a factor of $M = f2 / f1$, where $f1$ and $f2$ are focal length of the objective and the tube lenses, respectively. Therefore, if the pinhole were to be put in the image plane in order to discriminate the out of focus light, its size should be that of the Airy disk in the sample plane multiplied by the factor $M$.

The decrease in pinhole size has an effect on lateral resolution too, since signal coming from off-axis is rejected, as shown in Figure 7. The PSF of the confocal microscope can be shown to be an autocorrelation of the PSF of a standard fluorescence microscope, and is therefore narrower by a factor of $\sqrt{2}$. In practice, however, the resolution improvement is compromised by the signal loss and therefore is seldom realised [30]. An infinity corrected oil immersion objective lenses are routinely used to increase resolution and reduce aberrations. Confocal microscope has two disadvantages due to its scanning mode of operation. One of them is sample photobleaching, since the excitation light is focused to a tight spot to record fluorescence from each pixel. To get a good signal a high excitation applied to compensate signal loses experienced by rejecting out-of-focus light. The other disadvantage is image recording time because scanning in three dimensions takes comparatively long time. This can be critical when imaging live objects, which can move during the acquisition process.

### *Fast confocal microscopy*

Most confocal microscopes that use galvonometric scanning can acquire images in 0.1-1 sec. The scanning speed might be too slow if some dynamic processes are observed especially when 3 D images have to be acquired. There are a few ways to increase the scanning speed [45]. One of them is to use faster scanners, which oscillate at their resonant frequency [46] and can be driven at 8 kHz rate. Microscopes with such scanners are now commercially available [47]. The down side of these scanners are nonlinear photobleaching occurring due to non-homogeneous scanning speed across the sample. The other ways to increase scanning speed include use of high-speed rotating polygonal mirror [48], spatial light modulator (SLM, described in Section 5.2.7) [49] or acousto-optical modulator (AOM). In the latter case the fast beam steering is performed by rapidly changing frequency of the sound wave. The problem with AOM, however, is that the deflection is wavelength dependent and, therefore, the broadband light is dispersed (ultrashort radiation, for instance). To reduce the acquisition time some parallelization in fluorescence acquisition can also be introduced. The first such kind of microscope was called a tandem microscope [50] and its operation principle was based on the Nipkow disk. This type of microscope was still a confocal microscope but instead of one pinhole it employed many of them in a form of the spinning Nipkow disk. It worked together (in tandem) with an identical disc that created an illumination pattern identical to the detection pinholes. As both discs rotate in unison, multiple beams are scanned across the sample and fluorescence is collected from the multiple points. With this kind of microscope a real-time confocal image could be observed through eyepiece. The disadvantage of the Nipkow disk microscope is that only a small fraction (~ 1 %) of the illuminating light makes it through the pinholes to the specimen. Therefore this microscope is often combined with an array of microlenses to increase throughput efficiency. There is a trade-off between the number of pinholes and sectioning strength, because, if the spacing between holes is too small nearest neighbours begin to pollute one another. This problem can be circumvented by using an array of closely packed apertures modulated in time such that they open and close in a completely uncorrelated fashion [51]. The result is the sum of a conventional and a confocal image. A separate conventional image must be taken and subtracted in order to obtain the desired confocal image. An alternative to rotating disc is to employ a SLM to generate a mask of pixel-sized pinholes as in programmable array microscopy (PAM) to eliminate mechanical moving parts [52]. Instead of having to project and detect through pinholes one can use slit illumination and detection. This dramatically increases scanning speed and light throughput, however, it works at the expense of decreased optical sectioning [53, 54]. Another way to speed up acquisition is line focus scanning where imaging is performed by scanning line of illumination along the sample [55]. The fluorescence, in such a configuration, can be collected through a slit rather than a pinhole to discriminate the out-of-focus light.

### 2.4.3    Other axial resolution improvement techniques

A fluorescence microscopy has an axial resolution that is lower than the lateral resolution due to the fluorescence excitation and collection angle configuration [29]. The asymmetry between lateral and axial resolution stems from the fact that the numerical aperture of the objective used does not cover a full sphere – i.e. a solid $4\pi$. One approach to increase axial (and lateral) resolution would be to not use far field optics at all, since it is diffraction that limits the resolution. Scanning probe microscopy techniques, as discussed in section 2.7, demonstrate superior axial resolution, although they are constrained to imaging only the surface of the specimen – one optical plane and therefore optical sectioning is not possible. Similarly total internal reflection fluorescence (TIRF) microscopy [56, 57] cannot perform 3 D imaging but it can image surface of the specimen with a superior axial resolution. In this microscopy the sample is illuminated by the evanescent wave that has an exponentially decaying intensity profile that penetrates along optical axis ~ 100 nm and excites fluorophores. Far-field optics is of greater interest since it can provide optical sectioning and is compatible with live cell imaging. One way to improve resolution is by gathering light over a larger set of angles around the sample using opposite objectives that enables increasing the overall NA of the imaging system [58]. Second way is by using the structured illumination where higher sample spatial frequencies that cannot be directly imaged are captured by the lens in an indirect manner. Various configurations have been developed for exciting / collecting light over the wider range of angles [32, 59].

### *Standing wave microscopy*

One of the first configurations was standing wave microscopy [60], where a mirror was placed under the sample to reflect light not collected by the objective. The interference between reflected and incoming light creates a flat standing wave of fluorescence excitation. However, its PSF exhibits high side lobes close to the main maximum, which made the method suitable only for thin objects (thinner than the one period of the standing wave).

### *4pi microscopy*

A PSF with smaller side lobes can be created by the technique called 4pi microscopy, where wavefronts produced by two opposite facing objectives are coherently added [61]. The resulting spot is raster scanned across a specimen and fluorescence can be detected through both objectives. Three variants of the technique exists, described as A [61], B [62] and C. Type A uses both objectives to excite the sample with counter propagating beams interfering in the sample, and fluorescence is collect through one of them. In type B, one objective is used for excitation and both objectives are used to collect fluorescence light from both sides of the specimen, which then propagates through paths of equal optical length to the detector and interferes there. In type C, both methods are combined, which gives interfering excitation and detection. Type C is used most often. The central focal spot produced by the two objectives extends in $\Delta z \approx \lambda / 3n$. However, the illumination still does not cover the complete $4\pi$ solid angle ($\alpha \approx 68^o < 90^o$) and the PSF features sidelobes in the axial direction. To reduce these lobes, squaring of PSF is often implemented by two-photon excitation [61, 63]. Deconvolution can also be used to get rid of the lobes and further improve axial resolution [59]. The system is thus able to increase the axial resolution by up to seven times ($\sim 80$ nm resolution) in live cells [64]. Lateral resolution is only minimally increased compared to a standard confocal setup. By using the latest high NA objectives ($\alpha = 74^o$), single photon 4pi microscopy is possible since the side lobes fall below 50 % [65]. The 4pi microscope has now been commercialized (Leica TCS 4PI).

The 4pi concept, as will be discussed in Section 5.2.5, has been also implemented in STED microscopy [66] and multiphoton multifocal microscopy [67].

### *Theta microscopy*

A similar technique called theta microscopy was developed, where the light source illuminates the sample with a second objective that is perpendicular to the observation axis [68]. This configuration makes the PSF almost spherical. However, only objectives with small NA can be used because of practical constraints (pairs of high NA objectives cannot get close enough together due to their short working distances). This method is therefore mostly used for imaging large samples in 3 D. A wide field version of the theta microscope involves illuminating an entire plane with a light sheet [69] and is called Single Plane Illumination Microscopy (SPIM). Both microscopes claim axial resolution better than of confocal microscopy [70]. SPIM is mostly used for imaging large sample, for example, embryos [71].

### *$I^nM$*

Wide field variants of the 4pi microscope also exist, known under the general name of $I^nM$, which also use two objectives. When the sample is excited through one of the objectives and fluorescence images, from the same focal plane, are collected and interfered on a CCD, the method is called Image Interference Microscopy, $I^2M$. When the sample is illuminated from both objectives with incoherent light and fluorescence is collected through one of them, the method is called $I^3M$ - incoherent, interference, illumination microscopy. If both are combined ($I^2M + I^3M$), the technique is called $I^5M$, where interference occurs at the sample and at the CCD [72-74]. The difference between standing wave microscopy and $I^5M$ is that former generates structured illumination with coherent and the latter with incoherent illumination respectively. PSF side lobes are effectively separated and minimized in the sequence: *Standing wave microscopy→$I^nM$→4piA→4piC.*

### *Structural illumination*

To provide optical sectioning in a widefield microscope, the fact that non-zero spatial frequency undergoes attenuation with defocus can be exploited. Figure 9 shows images of a progressively defocused grid recorded in a widefield microscope where one can see that higher frequencies die away quickly with defocus but that the zero (background) stays.

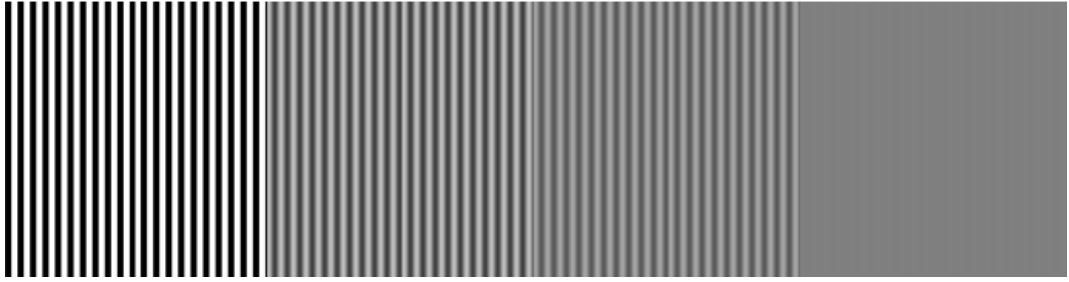

Figure 9. Images of a fluorescent grid as recorded by progressively defocusing it (from left to right). Figures obtained from T. Wilson's Lab, Oxford.

If a grid is projected on a sample and excites fluorescence, only the fluorescence signal coming from the focal plane will be modulated whereas the out-of-focus background will look more homogeneous. To get rid of the background and the modulation on the image, two other images are also recorded with a $2\pi/3$ and $4\pi/3$ phase shifts of the grid from which an un-modulated, sectioned image can be calculated [75]. The illumination pattern on the sample can be also formed by interfering two laser beams [76]. The principle has been implemented commercially by Zeiss (Apotome, Zeiss, Goettingen, Germany) and by Optigrid (Qioptiq Imaging Solutions, Rochester, New York, USA).

### 2.4.4 Structured illumination for lateral resolution improvement

Use of structured illumination in order to increase lateral resolution was first proposed in 1963 [77]. The first demonstration of this technique, however, was performed only in 1998 [78]. This principle is know in the literature under different names, such as harmonic excitation light microscopy, laterally modulated excitation microscopy and patterned excitation microscopy. Since optical microscopy has a classical resolution limit that does not allow to observe sample structures with details spaced closer than $\Delta x$, spatial frequency components higher than $k_0 = 1/\Delta x = 2NA/\lambda$ are therefore not transmitted through the microscope. To overcome this limit, a spatial frequency mixing approach, similar to demodulation in radio electronics, can be used. The idea behind it is to multiply the sample structure with a known spatial frequency that contains high spatial frequencies in order to generate sum and difference frequencies. The easiest way to implement this is to project a grid onto a sample with high spatial frequency, $k_{grid}$ that is, on the other hand, sufficiently low to be transmitted through the microscope. The resulting sample fluorescence will now include down-shifted spatial frequencies, in a manner analogous to *Moiré* fringes (shown in Figure 10).

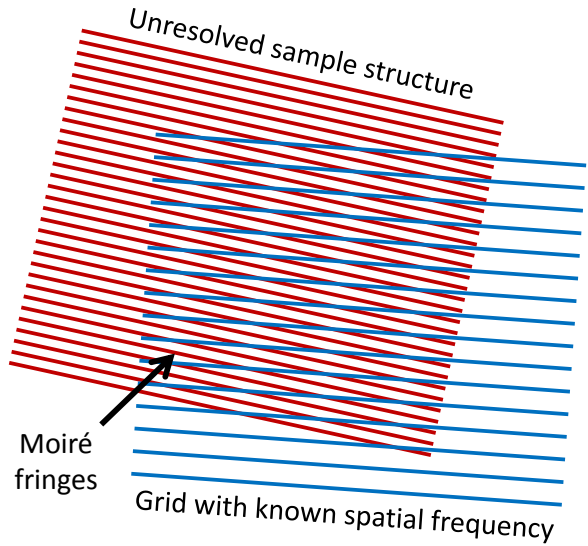

Figure 10. Forming of Moirés fringes.

The Moiré pattern has new components with a smaller spatial frequency, $k_{moiré}$, which can be resolved with a microscope since it falls within its pass band:

$$k_{moiré} = k_{sample} - k_{grid}$$

eq. 14

The maximum detectable spatial frequency can be increased with this method by $k_{grid}$, i.e. from $k_{sample} = k_0$ to $k_0 + k_{grid}$. The $k_{grid}$ itself cannot be made larger than $k_0$ - the resolution limit, so, the limiting spatial frequency that could be detected is increased to the twice the initial resolution: $k_{sample} = 2k_0$. It is possible to remove the grid pattern in the recorded fluorescence image by some mathematical post-processing and additional image recording. A minimum of three images with different phases of the grid pattern are required to eliminate the grid pattern from the image. The obtained image will possess improved resolution in one direction (perpendicular to the grid stripes projected). To get improvement in other directions, the same phase shifting procedure should be repeated for them [33]. Therefore, the orientation of the grid should be changed, together with the phase, to get full field resolution improvement. Instead of projecting a grid on a sample, two laser beams can be used to form an interference pattern. Using four laser beams, a 2-dimensional interference pattern can be formed, which could reduce the number images required to get a whole-field image resolution improvement [79, 80]. Now, five images, with different phases with respect to both lateral axes, are sufficient to reconstruct the original image instead of the required 9, in the case of projecting grid or interfering two laser beams. A new type of wide-field light microscopy, $I^5S$, was demonstrated that combines the lateral performance of structural illumination microscopy with the axial performance of $I^5M$, resulting in a spatial resolution of ~ 100 nm in all three dimensions [74]. It employs two opposing objectives to generate a complex three-dimensional interference pattern of multiple beams. A modified form of structured illumination microscopy that provides true three-dimensional imaging without missing-cone problem, with twice the spatial resolution of the conventional microscope in both the axial and lateral dimensions has also been recently demonstrated [81]. It significantly reduces the complexity of the system while still reaching an axial resolution of approximately 280 nm. Multicolour 3 D structured illumination microscopy was recently performed in this way [82].

## 2.5     Non-linear optical microscopy

The nonlinear response of certain molecules can be used to contrast different molecules and regions in biological specimens. This nonlinearity in the specimen can also be exploited to provide high resolution, optically sectioned imaging [83, 84]. Nonlinear processes involve multiple photons interacting simultaneously with the sample and, therefore, they inherently provide optical sectioning since the signal typically comes from very confined focal volumes where there is a reasonable probability that the required number of photons arrive together. Depth penetration is also increased in the nonlinear microscopy since usually red or infrared light is used to which biological object is less scattering compared to the visible light. Typically nonlinear optical processes do not increase the resolution because the resolution improvement brought in by $n$ photons is counteracted by the fact that the corresponding photons are $n$ times longer wavelength and therefore forms $n$ times large focal spot. Therefore, the techniques discussed here are introduced under the context of providing image contrast rather than for breaking the diffraction resolution limit, although it does provide better resolution than wide field microscopy due to inherent optical sectioning. Various nonlinear processes (generally called multiphoton fluorescence and multiphoton scattering) can be employed, such as two photon absorption (2PA), second harmonic generation (SHG), third harmonic generation (THG) and coherent anti-stokes Raman scattering (CARS), principles of which are shown in Figure 11.

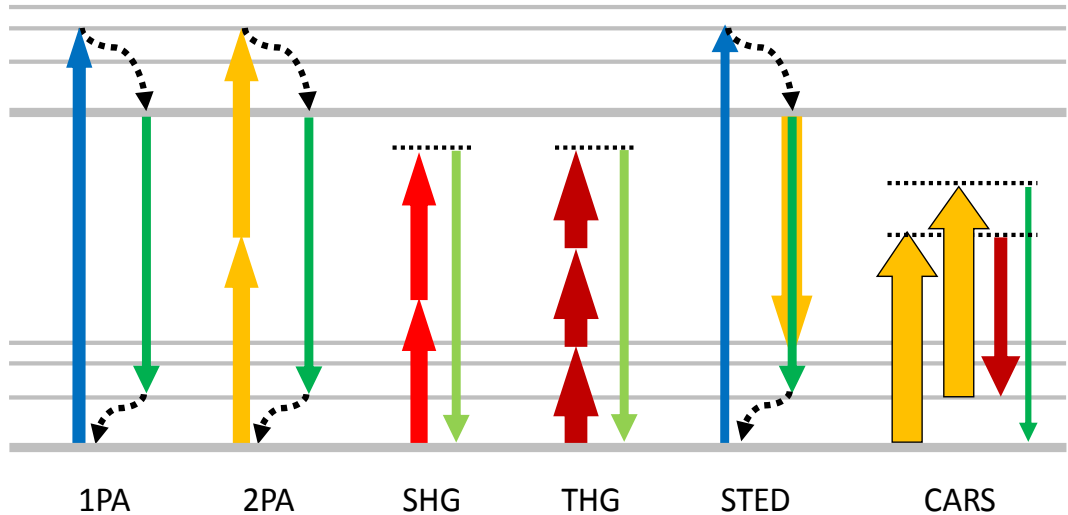

Figure 11. Various nonlinear processes. 1PA – one photon absorption (linear). 2PA – two-photon absorption: yellow – excitation beam, dark green – fluorescence. SHG – second harmonic generation: red – excitation beam, green – SHG signal. THG – third harmonic generation: dark red excitation beam, green – fluorescence. CARS: yellow – pump beams at $\omega_p$, dark red – Stokes beam at $\omega_s$, green – anti-Stokes signal at $\omega_{as} = 2\omega_p - \omega_s$.

The nonlinear response in the molecules comes from the induced polarisation vector $P$ that responds nonlinearly with applied electric field, $E$:

$$P(E) = \varepsilon_0 \left( \chi^{(1)} E + \chi^{(2)} E^2 + \chi^{(3)} E^3 + \ldots + \chi^{(i)} E^i \right)$$

eq. 15

Where $\chi^{(1)} = n^2 - 1$ is the linear and $\chi^{(2)}$ and $\chi^{(3)}$ are nonlinear susceptibilities. The $\chi^{(i)}$ represents different optical effects:

   $i = 1$ – First order process: absorption / reflection.

   $i = 2$ – Second order processes (e.g. second harmonic generation).

   $i = 3$ – Third order processes: (e.g. third harmonic generation, multiphoton absorption, coherent anti-stokes Raman scattering, self-phase modulation).

Usually the third order susceptibility is much weaker than the first and the second; therefore a much higher intensity is needed to observe processes like third harmonic generation. However not all of biological specimens exhibit a non zero second order susceptibility since this requires non centrosymmetric media. Therefore, the second harmonic generation can only be observed in some cases.

The history of nonlinear optical microscopy starts in the 1970s, when a proof of principle of second harmonic generation microscopy was demonstrated in crystals [85, 86]. In 1982 Coherent anti-Stokes Raman Scattering (CARS) microscopy was demonstrated [87] but it was not until the 1990s that the nonlinear microscopy became popular with availability of convenient ultrafast lasers. To invoke multiphoton effects one would normally require high excitation power, which may be incompatible with the power tolerance level of biological objects. Therefore it is usually the lower order nonlinear effects that are used (two photon absorption and second harmonic generation, for example).

### 2.5.1  Multiphoton microscopy

Two-photon-excited fluorescence microscopy was first demonstrated in the 1990 [88] and three-photon-excited fluorescence microscopy in 1997 [89]. These two microscopes are generally called *multiphoton* microscopes and are the most often used variants. The multiphoton microscope setup is similar to a confocal microscopy but the detection path is different and, in addition, ultrashort laser pulses are used. The fluorescence signal comes only from focal volume, but when imaging deep into tissue it can be highly scattered on its way out through the tissue. This necessitates collection from as wide area as possible in order to collect enough photons by imaging the pupil plane of the collecting objective lens on a large area detector located close to the objective. This is called a non-descanned detection. If otherwise a standard descanned path would be used then the scattered fluorescence originating deep from tissue would be clipped by optical elements (various apertures and pinhole, for example) and therefore lost [90]. Multiphoton absorption features small absorption cross-section, therefore pulsed lasers are often used because of the high peak powers that they can achieved. For laser pulses of width *t* and repetition rate *f*, the signal, compared to that of *cw* regime, is increased by a factor of:

$$k = (t \times f)^{n-1}$$

eq. 16

Where *n* is the number of photons involved in the nonlinear process. For example, for a single photon excitation ($n = 1$) $\rightarrow k = 1$, therefore no increase in signal is achieved if pulsed radiation is used instead of *cw* with the same average power. In case of two-photon excitation ($n = 2$) $\rightarrow k = 1 / (t \times f)$. This shows that signal increase scales linearly with pulse width when the average power is kept constant showing that pulsed laser sources are more suitable. Nevertheless multiphoton excitation microscopy has also been demonstrated with *cw* lasers [91-93]. Multiphoton absorption microscope image acquisition can be made faster, in a fashion similar to that used with confocal microscope, by using rotating disc. The so called Multifocal Multiphoton Microscope [94] uses a rotating disk of microlenses to quickly acquire sectioned fluorescence images with parallel beams. Another way of increasing acquisition speed through parallelisation is to use a diffractive optical element [95] or a beamsplitter [96] to produce multiple excitation beams which are then scanned simultaneously in the object plane. A commercial version of the latter technique is now available (TriMScope, LaVision). The instrument is also used extensively in our group [97] to achieve fast image acquisition with superior resolution and that can also enable fast 3-D fluorescence lifetime imaging [98].

### 2.5.2    *Second and third harmonic generation microscopy*

Second harmonic generation microscopy was demonstrated some 30 years ago using crystals [85]. Second harmonic generation images of biological samples were recorded later [99] but this only recently became popular as an imaging technique [100]. It is especially useful to provide complementary information to two photon microscopy [84] as illustrated in Figure 12 (a). In general, harmonic generation does not involve absorption if the pump wavelength is chosen away from any molecular absorption bands and so no energy should be dissipated in the sample. This is not true when the harmonic generation is enhanced near an electronic resonance, where absorption can also occur, which could lead to photodamage if high intensities are used [101]. Second harmonic generation occurs in non-centrosymmetric media such as an interface (membranes etc.) or electric field induced asymmetric environments. The second harmonic generation signal can be readily separated from the incident radiation since it is twice the frequency, and it is usually easily separated from autofluorescence as well. The technique can be useful to probe membrane structure and can measure membrane potential with single molecule sensitivity [102]. Third harmonic generation microscopy is also now widely used. It was first demonstrated in 1997 [103] and applied to a living system a year later [104, 105].

### 2.5.3    *Coherent anti-stokes Raman scattering microscopy*

Molecules in a biological specimen can also be contrasted by their vibrational spectra. However, conventional methods used to record vibrational spectra, like infrared microscopy exhibits relatively low resolution due to the use of long wavelength light [106]. Raman microscopy can also be used to directly sense biological molecules [107] but conventional Raman microscopy typically requires long integration times and high intensity lasers due to the weak nonlinearity. Nevertheless live cell imaging was recently demonstrated [108] using slit scanning Raman microscopy [109]. Stronger vibrational signals from Raman-active vibrational modes can be obtained using coherent anti-stokes Raman scattering (CARS) [87]. There two laser beams: a pump beam at frequency $\omega_p$ and a Stokes beam at $\omega_s$ are used. When the difference frequency $\omega_p - \omega_s$ matches any Raman active vibrational frequencies their resonance occurs, resulting in a strong anti-Stokes signal at $\omega_{as} = 2\omega_p - \omega_s$. It, for example, has been used for cellular [110] and tissue imaging [111]. This is also a multiphoton process that therefore provides optical sectioning and improved lateral resolution.

### 2.5.4    *Multimodal microscopy*

Second harmonic generation microscopy can give complementary information to two-photon microscopy since it reports on different properties of a biological specimen. Coherent anti-stokes Raman scattering can provide different contrast mechanism. The techniques can be combined together; for instance, collagen can be detected using second harmonic generation whereas elastin can be detected using fluorescence generated by two-photon excitation [84]. Two-photon microscopy can also be combined with coherent anti-stokes Raman scattering [111]. Second and third harmonic generation microscopies was recently demonstrated with two-photon microscopy [112] and coherent anti-stokes Raman scattering [113]. An example of images obtained by combining the three different techniques is shown in Figure 12.

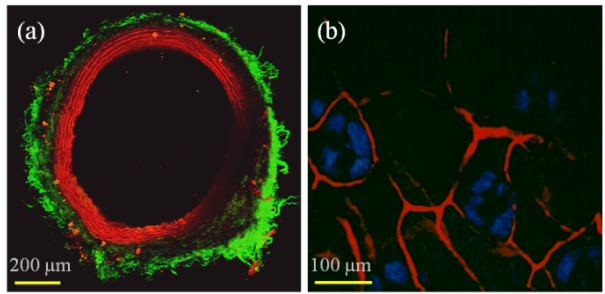

Figure 12. Multimodal imaging. (a) Merged image of two-photon and second harmonic generation (SHG) images (3 D reconstruction of a rat carotid section; green – SHG, red – two-photon fluorescence). Figure obtained from E. Beaurepaire, Ecole Polytechnique, France. (b) Merged image of two-photon and coherent anti-stokes Raman scattering (CARS) images (image of mouse skin; blue – CARS, red – two-photon fluorescence). Figure taken from [111].

## 2.6    Fluorescence microscopy with unlimited resolution

The first fluorescence microscopy with in principle unlimited resolution was demonstrated using stimulated emission depletion (STED) microscopy [1, 114]. Similar techniques, in terms of technical implementation, demonstrated later relied on ground state depletion [115, 116] and fluorophore photoswitching [117, 118]. Saturated structured illumination microscopy [119, 120] was successfully demonstrated later and is probably the most straightforward technique to implement technically. However, none of these techniques could outperform STED microscopy in terms of achievable resolution and compatibility with cell imaging. Recently demonstrated localisation-based super-resolution techniques [121, 122] can achieve similar to STED microscopy performance, and with relatively straightforward implementation, however, at the cost of a longer acquisition time. All those microscopy technique are discussed in more details below.

### 2.6.1    Stimulated emission depletion microscopy

Stimulated emission depletion (STED) microscopy, a concept first proposed by S. Hell and J. Wichmann in 1994 [1], is one of the most promising techniques for improving the resolution of far-field optical microscopy beyond the diffraction limit. In STED microscopy, fluorescence emanating from the periphery of the focused excitation beam is suppressed by a second beam (STED) that depletes the excited state population through stimulated emission (Figure 13).

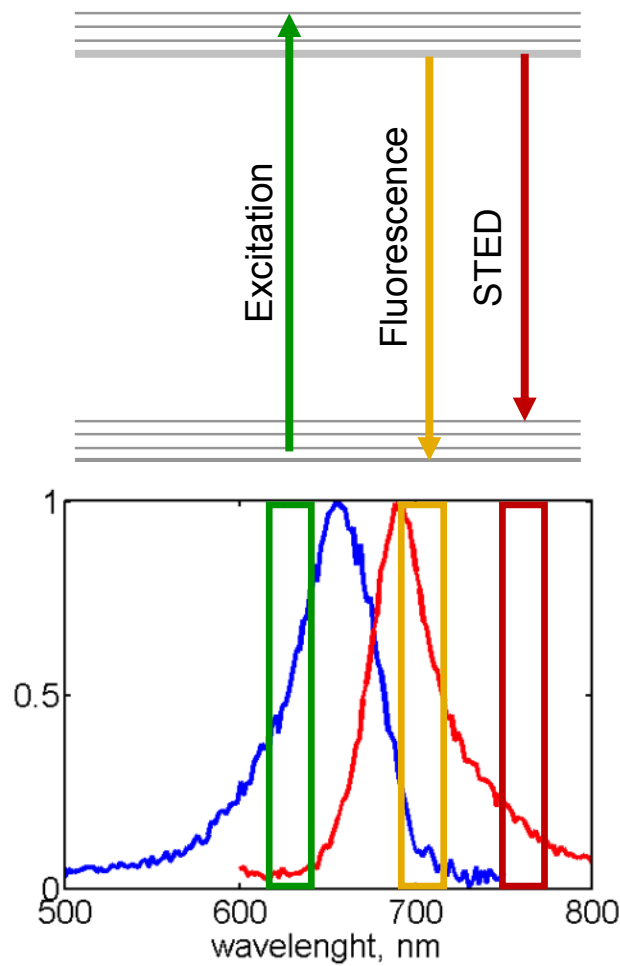

Figure 13. Principles of STED. (top) Jablonski diagram. (bottom) Corresponding spectral windows for excitation, STED and detection. The fluorescence excitation (blue curve) and emission (red curve) spectra are for the 'Dark Red' fluorescing beads from Molecular Probes used in this thesis for STED experiments. A molecule can be excited by the excitation beam but before it fluoresces, a powerful, spectrally red-shifted STED beam can be used to de-excite it.

This can effectively narrow the PSF of the microscope to permit super-resolved image to be acquired. This can happen if STED beam is spatially modulated in such a way that, upon focusing, it results in a PSF with one or more regions where the light intensity is zero and there is no stimulated emission. The STED beam should possess a high spatial gradient to make the non-depleted area as small as possible. The most optimum form of the STED beam PSF, to increase resolution laterally, is doughnut shaped, as discussed in Chapter 5. Molecules that are in the region of zero STED intensity and also in an excited state are left to decay by spontaneous emission.

Fluorescence coming from this region can be confined in an area that could be smaller than the diffraction limit. This forms the basis for super-resolution. Both beams are focused by a high NA objective, and fluorescence is recorded by a detector as shown in Figure 14.

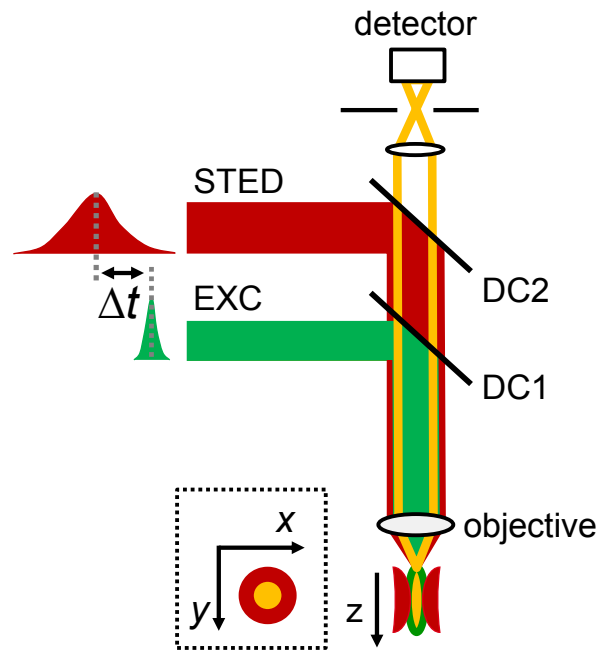

Figure 14. STED microscopy schematics. Excitation and STED pulses are overlapped by means of two dichroic mirrors and then focused with the objective onto the sample. If STED beam has its wavefront engineered so that upon focusing it forms, for example, a doughnut shaped point spread function (red drawing), then it quenches the fluorescence from the rim of the excitation point spread function (green drawing). The remaining fluorescence therefore comes from the smaller region, which can be smaller than the diffraction limit (yellow). Scanning such a pair of beams over the sample builds up a super-resolved image.

Inherently this is a scanning microscope, since only one point is interrogated at a time, and the focal point needs to be scanned all over the sample. However, there could be some degree of parallelization with multiple scanning points separated by more than the resolution limit. Today, resolutions down to $15 - 20$ nm are typically achieved [123] and significant biological applications have already been demonstrated [124]. The latest results show that PSF as narrow as ~ 5-6 nm (FWHM) can be generated in bulk diamonds with fluorescent nitrogen vacancies [125].

There is a close, nonlinear relation between STED intensity and fluorescence signal from the ultra-sharp un-depleted region: the larger the intensity of the STED beam, the smaller the fluorescence signal coming from the region. Therefore a new resolution formula can be derived that describes achieved resolution as a function of power. A new parameter - saturated intensity, $I_{sat}$ can be defined as the STED intensity at which half of the fluorescence is suppressed [117, 126]. It is the ratio $I_{max} / I_{sat}$ (where $I_{max}$ is the maximum intensity in the spatially modulated STED beam), that will uniquely determine the PSF size in the ultrasharp fluorescence region. It can be shown that at fixed $I_{sat}$ the PSF size varies inversely as the square root of the STED intensity. Consider a standing wave intensity modulation:

$$I(r) = sin^2(2\pi n r / \lambda)$$

eq. 17

Where $r$ is an optical coordinate. A fragment of the function is shown in Figure 15. If we know that $I_{sat}$ is the STED intensity at which half of the fluorescence is suppressed and $I_{max}$ is a maximum STED intensity, then there is an area $\delta x$ formed, where fluorescence goes from half of its intensity to full (where no STED is applied) and drops back to half.

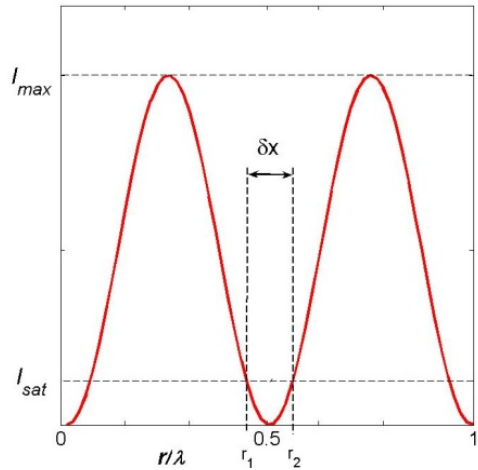

Figure 15. Standing wave.

This area therefore corresponds to the full-width half-maximum (FWHM) of the PSF. In order to find that area we can express $I_{sat}$ as a function of coordinate:

$$I_{sat} = I_{max} sin^2(2\pi n r / \lambda)$$

eq. 18

and therefore the area in the spot where the molecules has retained at least 50 % of the fluorescence can be found to be of:

$$\delta x = r_2 - r_1 = (\lambda / \pi n) arcsin(I_{sat} / Imax)^{1/2} \approx \lambda / \pi n \, (I_{sat} / I_{max})^{1/2}$$
eq. 19

The resolution of the microscope can then be written as:

$$\Delta x = \lambda / \pi n \varsigma^{1/2}$$
eq. 20

Where $\varsigma = I_{max} / I_{sat}$. In microscopy, the spatial distribution of the STED beam can be produced by the objective lens itself. If it is produced through the finite aperture of the objective lens, then the smallest possible spot is co-determined by semiaperture angle:

$$\Delta x = \lambda / \pi n sin(\alpha)\varsigma^{1/2}$$
eq. 21

This equation allows diffraction unlimited spatial resolution. According to the formula, when $\varsigma \rightarrow \infty$, the resolution limit can be squeezed almost without limit, in contrast to Abbe's resolution formula. Theoretically the PSF can be squeezed to the size of a fluorescence molecule. To increase $\varsigma$, either $I_{sat}$ can be made smaller or $I_{max}$ larger. The maximum value of the latter depends on fluorophore photophysical properties and cannot be very large for fluorescent organic molecules because of photobleaching at high intensities. On the other hand $I_{sat}$ cannot be improved much either since it depends on the fluorophore cross-section for stimulated emission, which is typically less than or comparable with that of a laser medium.

### 2.6.2    Ground state depletion microscopy

Another way to achieve super-resolution is to exploit other kind of transitions in fluorophores that can be saturated. For example, one can shelve fluorophores to the triplet state [115, 116] in photostable fluorophores. In its first experimental implementation [116], an extensive pumping for 0.5 ms with a doughnut or line shaped laser beam was used to shelve the most of the fluorophores to the triplet state, and thus to deplete the ground state in all molecules, except in those located in the centre of the shaped laser beam. The depletion was then followed by a weaker probe beam that excited the remaining molecules, residing in the ground state, to the first excited stated. Fluorescence from there was subsequently recorded and another

~50 ms was allowed for triplet state to relax ($\tau_T$ = 10 ms) to the ground state before moving to the adjacent pixel. This implementation allowed achieving resolution of ~ 50 nm. Instead of waiting for the triplet state relaxation, both pump and probe beams can be used at the same time, if the probe is modulated and fluorescence is detected with the lock-in detection, as demonstrated in Ref. [127]. This resulted in the reduction of image acquisition time and also allowed achieving resolution of ~ 7 nm, partly thanks to the very photostable bulk diamond crystal with nitrogen vacancies, used as an imaging sample.

### 2.6.3    *Photoswitching beyond the diffraction limit*

Another property of fluorophores that can be used to achieve subdiffraction imaging is photoswitching between fluorescent and dark states by radiating fluorophores with light at different wavelengths [10]. It has been demonstrated that, for example, asFP595 fluorescing protein can be switched on with yellow and off with blue light [128, 129]. Switching off is in principle possible with a wide range of wavelengths but the achievable depletion efficiency is comprised with longer wavelength since the probability for a molecule to be excited increases with wavelength. The important property of the switching 'off' is the possibility to saturate it (prerequisite to break the resolution barrier) at very low intensities. The principle allowed achieving resolution of ~ 40-100 nm in both scanning and wide field implementation [118, 130, 131]. The use of photoswitchable proteins has its limitation due to the comparatively long switch 'on' time, which puts limitation on the imaging speed. The 'on' time can be reduced by increasing the 'yellow' radiation intensity, however, at the expense of achievable depletion efficiency. There have been recently many proteins developed for super-resolution imaging [10], however, other types of labels, like photochromic synthetic compounds, were also successfully used to achieve similar results [132]. The disadvantage with photoswitching is that each molecule has to undergo at least a few cycles of photoswitching when a diffraction-limited beams scans an image, which can lead to photobleaching of molecules before an image is acquired. The problem is absent in the localisation based techniques where a molecule is activated just once and fluorescence is collected before it photobleaches or is deliberately switched-off, as explained in the next Section.

### 2.6.4    Localisation beyond the diffraction limit

Rather than modifying the excitation light pattern to yield a smaller PSF, as in STED and saturated structured illumination microscopy, image resolution below the diffraction limit may be achieved by precisely determining the positions of the fluorophores labelling the sample. The nanometre distances between different objects separated by more than the diffraction-limited resolution can be measured by 'centroiding' methods regardless of the diffraction limit. A few objects can be localised within PSF provided that the objects have some distinct characteristics. The precision of this localization process can be given approximately by

$$\Delta x = s \ / \ (N)^{1/2}$$

eq. 22

Where $\Delta x$ is the error in localization, $s$ is the standard deviation of the PSF and N is the number of photons detected [133]. This concept has been used to track small particles with nanometre-scale accuracy [134]. Recently it has been shown that, even when the emitter is a single fluorescent dye molecule, its position can be determined with a precision as high as ~ 1 nm [135], for example, different absorption [136], emission [137] or even fluorescence lifetime [138] properties has been used as a mean to differentiate individual fluorophores beyond the diffraction limit. In addition, molecule photobleaching [139] or quantum dot blinking statistics [140] has been used. A similar approach is to use different points in time to record the position of moving subresolved objects [135]. However, these techniques are only able to distinguish 2-5 fluorophores in the diffraction-limited volume. New principles have been recently developed based on localisation which can provide, in principle, unlimited resolution (reviewed in [141]). This concept, independently developed by three research groups, has been given three names: switching off individual fluorophores – stochastic optical reconstruction microscopy (STORM) [142], based on serial activation and subsequent photobleaching – Photoactivated Localization Microscopy (PALM) [122], or Fluorescence Photo-activation Localization Microscopy (FPALM) [121]. The resolution of the final image is not limited by diffraction, but by the precision of each localization. These results demonstrated a resolution improvement of an order of magnitude over conventional imaging, but require no specialised setups. So far, multicolour imaging [143], 3 D imaging [144] and live cell imaging [145] have been demonstrated.

### 2.6.5    Saturated structured illumination microscopy

Conventional fluorescence microscopy normally operates at low excitation intensities – in the linear regime where fluorescence intensity is proportional to the excitation. If the excitation intensity is increased, a population of fluorescence molecules in the excited state shows saturation because the molecules have a nonzero excitation lifetime and the number of molecules in the focal volume is limited. This broadens PSF that leads to the decreased resolution. However, saturated PSF contains higher spatial frequency components that can contain information about smaller structures than the diffraction limited microscope could achieve. Resolution improvement by a factor of two using structured illumination is, as recently shown, not the ultimate limit. Further improvement is possible if nonlinear intensity distortions can be induced on a sample with a structural illumination pattern. If the fluorescence pattern no longer matches the illumination pattern, then higher pattern modes must be present to describe the spatial fluorescence structure formed.

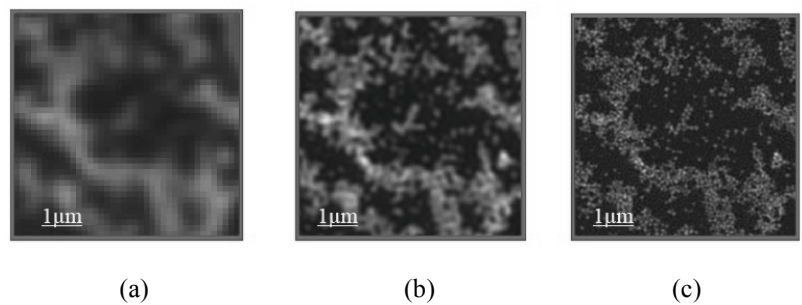

(a)                          (b)                          (c)

Figure 16. Resolution improvement using linear and nonlinear structured illumination techniques. (a) conventional microscopy, (b) linear structured illumination, and (c) saturated structured illumination microscopy. The measured mean FWHM of isolated beads in (c) is 59 nm. The illumination wavelength was 533 nm. Beads size – 50 nm. Figures taken from [120].

It is as if, together with the projected grid, higher pattern modes are present that are finer than is physically possible to project because of limitations set by microscope resolution. This can be achieved through saturating the fluorescence signal by intense illumination [119]. This technique is known under different names, such as: saturated / nonlinear structured / patterned illumination microscopy [119, 120]. This work demonstrated a resolution of 46 nm [120], as shown in Figure 16 (c), which far exceeded that achievable with linear methods, shown in Figure 16 (a,b). However, the intense illumination needed to achieve saturation of the fluorescence can lead to photobleaching of biological samples.

In the related method, a temporal dynamics of saturated $S_0 \rightarrow S_1$ transition can be used to retrieve the super-resolved information [146]. Rather than choosing to decode the super-resolved information spatially, a temporally modulated excitation can be used and fluorescence signal of higher harmonics, carrying super-resolved information, can be extracted from saturated fluorescence signal [147]. This method was recently successfully demonstrated on biological specimens [148].

## 2.7     Other super-resolution techniques

In order to place this review of fluorescence microscopy beyond the diffraction limit in context, it is instructive to review other available techniques that allow the study of biological objects beyond far field resolution limit. This helps to highlight the advantages of optical microscopy.

### 2.7.1     Electron, X-ray and near-field microscopy

In order to resolve smaller features, other kinds of microscopes have been developed that probe specimens with shorter wavelength radiation. Electron microscopy, developed in the 1930's, exploits the wave nature of electrons to achieve unsurpassed resolution and is able to visualise single molecules and even single atoms. Much current research aims to image the 3-D structure of various proteins with sub-nanometre resolution by using cryo-electron microscopy [149]. Unfortunately, EM can only be used with fixed specimens, requiring samples to be prepared with a metal coating and imaged under vacuum. In the case of cryo-EM, biological samples are rapidly frozen. X-ray microscopy stands between EM and optical microscopy in terms of resolution and can produce images of individual molecules – as in X-ray crystallography – with a resolution of less than 20 nm [150]. Although, biological samples still need freezing, X-ray microscopy does not require such significant preparation as electron microscopy. X-ray radiation is, however, severely phototoxic, making it also challenging to image biological systems.

One approach to overcoming diffraction when using optical radiation is to use a very small aperture such that light passing through it has not had a sufficient distance to diffract before impinging on the object as well as going back through it. This was be shown in 1984 by scanning a small (sub-diffraction limit) aperture sufficiently close to the sample [151]. The spatial resolution is determined by the size of the aperture, which can be less than 12 nm [152]. This approach is called scanning near-field optical microscopy and can be used to image, spectroscopically probe and modify surfaces [153]. It can visualise single molecules [154] since only those molecules that are directly under the aperture will be excited and so the lateral resolution is defined by the size of aperture but is limited to surface imaging. The axial resolution also depends on the aperture size, but the intensity of light in the axial direction decreases exponentially in the near field and these effects together result in axial resolutions of $\sim 10 - 50$ nm. The aperture can be provided by a metal-coated tapered fibre tip and control of its exact positioning over the object can be implemented using piezo actuators. It is difficult to maintain the appropriate distance between the tip and the object when imaging cells in a buffer solution and so cells are usually dried out before imaging. However, scanning near-field optical microscopy has been successfully used in biology as reviewed in [155], for example, to study DNA molecules with fluorescence energy resonance transfer (FRET) [22], co-localisation of proteins in erythrocytes infected by malaria [156] and receptor clustering [157].

### 2.7.2    *Scanning probe microscopy*

SNOM is a member of a broader class of microscopes generally called scanning probe microscopy. It generally refers to measuring a physical parameter while scanning a probe with high precision over the sample. Some scanning probe microscopy techniques do not use light to probe the sample but rather some other physical parameter such as electrical attraction force, magnetic force, current etc. The first such kind of microscope constructed was scanning tunnelling microscope (STM) developed in 1982 [158-160], which probes materials using the electron tunnelling current as the probe is brought in proximity to the sample. While lateral and axial resolutions of 0.1 nm and 0.01 nm respectively are achievable, the application to biology is limited by the requirements that both the sample and the probe need to be conductive, under vacuum and extremely clean. Atomic force microscopy (AFM), developed in 1986, helped to avoid some of these limitations [161]. AFM uses an atomically sharp probe attached to a cantilever that interacts mechanically with the sample surface. It can image atoms and single molecules [162] and was rapidly applied to biology [163]. Its main disadvantages for biological research are associated with the unavoidable interaction of tip with the sample and the limitation to surface imaging. Therefore optical microscopy is preferred to study biological processes.

## 2.8     Summary and outlook

Fluorescence microscopy is the most used technique to study cell biology. Other optical contrast enhancing techniques like phase contrast or dark field are also routinely used in biology, however they cannot indentify subcellular components other than by their shape and do not have the sensitivity to detect single molecules. Nonlinear contrast enhancing techniques like second harmonic generation or scanning probe microscopy also suffer from similar disadvantages. For example second harmonic generation is useful to contrast well-ordered protein assemblies (like collagen fibres) but lacks sensitivity and specificity for other components. Scanning near field optical microscopy is able to use spectroscopic information from fluorophores and provide high resolution (~ 10 nm), however, it is a surface imaging technique as are other scanning probe microscopes (like scanning tunnelling and atomic force microscopes). Electron microscope provides the highest resolution of all existing techniques, however, the sample needs elaborate preparation and live cell imaging is not possible. Similarly X-ray imaging is also not suitable for live cell biology experiments. Standard fluorescence microscopy can provide high sensitivity for single molecule detection, as well as provide various spectroscopic information and image in 3 D. However, its resolution until recently was not good enough to image single molecules. Resolution can be increased by collecting light over a larger aperture as in $I^2M$ and *4pi-B* microscopies or by using spatially structured illumination that frequency-mix-in high resolution information into the pass-band of the microscope, to either increase resolution axially, by using axially structured illumination, as in standing wave microscopy, $I^3M$ and *4pi-A*, or increase resolution laterally by using laterally structured illumination. Axial resolution can also be increased by using laterally structured illumination. Axial resolution was for a long time much poorer than in the lateral direction because of fluorescence excitation and collection geometries. These have been addressed in a series of new developed techniques like $I^5M$ and 4pi. It is interesting to note that spectral properties of the fluorescence markers, in particular their molecular states, may not only be used to generate signal, but also to dramatically increase the spatial resolution as is the case in STED microscopy or localisation beyond diffraction limit. For many years it was believed that optical microscope resolution was limited to ~ 200 nm. However in last 20 years it was shown that through spatially modulated illumination the resolution limit can be improved by a factor of two. However, the nonlinear relation between excitation intensity and emission can improve it further. Stimulated emission, predicted by A. Einstein in 1917, not only prepared the ground for the invention of the laser, but also for the first far-field fluorescence microscope with diffraction-unlimited resolution. A method called stimulated emission depletion emerged in 1990's claiming to beat resolution limit imposed by diffraction. Recently a resolution of ~ 20 nm was achieved with the technique and theoretically it is only limited to the size of a fluorescent molecule. Concepts of another resolution-breaking technique based on saturable reversible optical transitions and structured illumination, called saturable structured illumination microscopy, has been recently introduced and demonstrated on biological objects. An entirely different approach has been recently demonstrated that achieves super-resolution by individually localising fluorescent molecules. Most promising of the latter are the techniques based on serial photoactivation – stochastic optical reconstruction and photo-activated localisation microscopies.

# 3.   Fluorescence Lifetime Imaging

## 3.1     Introduction

Fluorescence microscopy is routinely used in biology. Apart from fluorescence intensity imaging, many of other fluorescence parameters like excitation / emission spectrum, polarisation and lifetime can be used to better characterise a specimen [18, 164]. A particular advantage of fluorescence lifetime is the fact that the lifetime measurements do not depend on fluorophore concentration. Fluorescence lifetime imaging (FLIM) in its simplest form can be used as a contrast enhancing technique to differentiate various fluorescing species or to sense local fluorophore environment as well as interactions with other molecules [19]. However, probably the most common FLIM application is to report Förster resonance energy transfer (FRET) between two molecules.

This Chapter reviews various fluorescence lifetime measurement techniques with a focus on imaging (FLIM), particularly implemented with gated optical intensifiers (GOI) and time correlated single photon counting (TCSPC), which were both used in various FLIM experiments described in this thesis. FLIM with a GOI was used to demonstrate application of supercontinuum as an excitation source in wide-field, Nipkow and line scanning FLIM microscopes as explained in Chapter 4. TCSPC was used in conjunction with confocal microscope for the biological applications described in the end of this Chapter and in conjunction with STED microscopy, as described in Chapter 6.

## 3.2     Instrumentation for time-domain fluorescence lifetime measurements

Fluorescence lifetime measurement techniques can be classified as time-domain [165, 166] or frequency domain [167, 168] approaches. Both are related through the Fourier transform and therefore potentially equivalent in terms of the information they can provide. However, technical implementation can be very different and therefore one technique can have advantages depending on the application. The time-domain techniques record fluorescence decay profiles by directly measuring fluorescence intensity as a function of time whereas frequency-domain techniques measure phase and amplitude differences as a function of frequency. Both of them can be implemented in confocal or multiphoton laser scanning microscopes. Frequency domain technique can use a sinusoidaly modulated *cw* light sources with modulation frequencies, *f* usually over MHz rates, so that $f \approx 1/\tau$ ($\tau$ – fluorescence lifetime as defined in page 19). The phase delay of the modulated fluorescence is then measured with respect to the excitation light. The longer the lifetime, the bigger is the phase delay of the fluorescence. Frequency-domain techniques were not used in this thesis and will not be further discussed. For time-domain techniques a laser with ultrashort pulses is used with a pulse width normally by an order of magnitude smaller ($\sim 0.1$ ns) than the lifetime of fluorescence being measured and the repetition rate of pulses low enough (typically a few tens of MHz) to allow fluorescence to decay before a new pulse arrives. Time-domain techniques used for fluorescence lifetime determination can be classified into two broad classes: *analogue* (usually time gated detection) and *photon counting* (TCSPC for example). Before reviewing these techniques it is instructive to first discuss the detection systems used for time domain fluorescence lifetime measurements.

### 3.2.1    *Detectors for fluorescence microscopy*

There are several different detector types employed for time-domain fluorescence lifetime measurements and in scanning microscopy [169, 170]. The most widespread method to detect photons is to use a photomultiplier tube (PMT). This is the oldest and still the most popular way to detect light. A photon impinging on a photocathode of the PMT generates electrons (photoelectrons) that are multiplied with a number (10 – 12) of dynodes aligned one after another before hitting an anode [171]. The dynodes together generate an electrical pulse which is of $\sim 10^6$-$10^8$ times stronger than the initial photoelectron. The width of the electrical pulse, called the single electron response (SER) is generally a couple of ns and therefore is not suitable for direct fluorescence lifetime measurements because the resolution of the detector should ideally be at least an order or two smaller than the fluorescence lifetime being measured in order to get an accurate lifetime estimate without requiring deconvolution techniques. However, the factor that limits temporal resolution of a PMT is its transit time spread (TTS), which is a measure of distribution of transit time through the PMT. The transit time spread of PMT arises from the different trajectories and velocities that photoelectrons can initially take in the PMT and is usually $\sim 300$ ps. Special photocathode and anode geometries are used to minimise the transit time spread. There are two types of PMTs: side-on and head-on [169]. Side-on PMTs exhibit better quantum efficiency but poorer transit time spread than head-on ones. A PMT exhibits dark noise, which comes from the thermal electron emission of dynodes and cathode. Leakage currents and radioactivity in the glass also contribute to the dark noise. Other important characteristics of PMT temporal response are pre-pulsing and after-pulsing. Pre-pulsing usually is caused by photoelectron emission from the first dynode whereas after-pulsing is believed to come from ion feedback and luminescence of the dynode material and the glass of the tube. A faster version of PMT exists called a micro channel (MC) PMT, where the number of different trajectories that an electron can take is reduced by using many small diameter (3-15 μm) channel-like dynodes aligned in parallel [169]. This microchannel plate (MCP) is followed by a cathode. The configuration reduces the single electron response from a few *ns* to around 300 ps and most importantly – the transit time spread down to 30 ps. To obtain higher gain, two or three microchannel plates can be arranged one after another. A special form of PMT is available that can spatially discriminate arriving photons by using fine mesh dynodes with an array of anodes. The highest direct temporal resolution is achieved with a streak camera (can be < 1 ps) and the full fluorescence decay can be recorded in a 'single-shot' acquisition. Other detectors that exhibit good temporal resolution and sensitivity are avalanche photo diodes (APD) and single photon avalanche photo diodes (SPAD); however, their small detection area (compared to PMT's) limits their use in fluorescence microscopy [172]. Another type of a detector commonly used in fluorescence microscopy is charge-coupled device (CCD) [173]. This is a wide field detector and, therefore, it can acquire image in single shot. Images can be acquired as fast as ~1 million frames per second with the newest cameras [174], however at the expense of reduced resolution and high power illumination. The detector (working at slower rates) is normally used in the wide-field fluorescence imaging but can also be used in scanning such as Nipkow disc, slit or even single point scanning microscopy for recording spectrally resolved fluorescence signal, for example. CCD is also an integral part of a wide-field fluorescence lifetime imager – gated optical intensifier as explained in Section 3.3.1

### 3.2.2   Analogue time-domain techniques

The analogue techniques, unlike photon counting techniques are usually used when fluorescence signal is relatively strong. Whole fluorescence decay profiles can be recorded in one shot provided the temporal resolution of the detector is sufficient, for example, with streak cameras [175, 176]. Averaging can be done over many pulse periods in case of lower fluorescence signals, so that fluorescence decay waveform with high enough $S / N$ can be built over the time. Alternatively the decay can be sampled at progressively increasing delays with respect to excitation pulse. Thus a full decay can be recorded if the sampling is scanned over the decay [166, 177, 178] as shown in Figure 17.

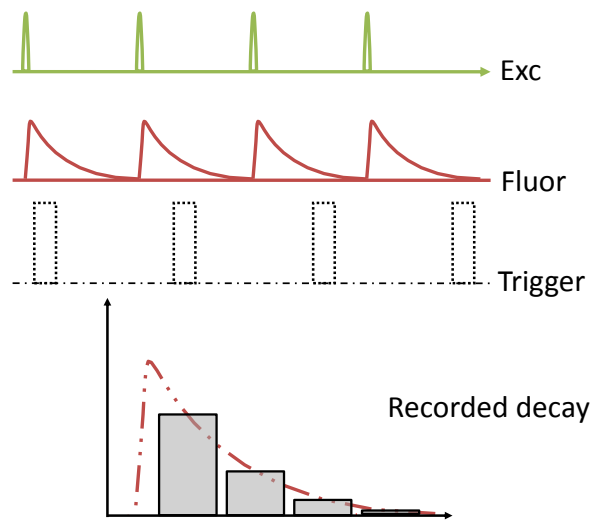

Figure 17. Time-domain recording in analogue mode. Fluorescence decay curve is sampled at progressively increasing delays by triggering some ultrafast shutter or intensifier.

The scan can be carried out by either using an ultrafast shutter that gates fluorescence at specified times or by using an ultrafast intensifier that intensifies fluorescence at specific gates. The latter is usually used in wide field imaging where intensifiers based on micro channel plates, called gated optical intensifiers (GOI) are used [179]. This principle is used in this thesis and will be described in Section 3.3. The fluorescence emission can be sequentially detected in at least two gates that are delayed with respect to the excitation pulse by different delay time. If only two gates are used then fluorescence lifetime can be relatively easily determined analytically from the following equation:

$$\tau = \Delta t \: / \: ln(I_2 \: / \: I_1)$$

eq. 23

Where $\Delta t$ is the time difference between the two gates, and $I_1$ and $I_2$ are the integrated fluorescence intensities in each gate (of equal width). The technique is also called rapid lifetime determination (RLD). The equation returns the correct lifetime only for single exponential decay profiles. In the case of multi-exponential decay profiles, it calculates the average lifetime. To find the individual lifetimes of the multi-exponential decay profiles, the number of gates has to be increased [180]. The multi-exponential decay can be described by:

$$I(t) = \sum \alpha_i \: exp(-t \: / \: \tau_i)$$

eq. 24

Where $\alpha_i$ and $\tau_i$ are amplitude (right after the excitation, $t = 0$) and fluorescence lifetime, respectively of individual component $i$. Both $\alpha_i$ and $\tau_i$ can be found by, for example, fitting multiple exponential decay profiles using maximum likelihood or nonlinear least square algorithms to the acquired data [181, 182]. The method where only one gate is used per excitation and the full fluorescence decay profile is sampled over multiple periods is not photon efficient. Detection schemes where all photons are collected with gates opened sequentially after each and every excitation pulse can be implemented by using separate counters for each gate [183, 184]. Each gated counter is enabled with preset delay relative to the excitation pulse for a specific time. This therefore enables faster FLIM image acquisition, but to date has only been implemented in single channel scanning FLIM systems.

### 3.2.3    *Photon counting time-domain techniques*

When the signal coming to the detector is so low that individual photons can be detected, then so called 'single photon counting' can be used. One of the main advantages of single photon counting is noise discrimination. The signal coming from a single photon counting detector is composed of the individual electronic pulses that represent individual photons and noise as illustrated in Figure 18 (a). Noise usually has lower amplitude and therefore can be efficiently discriminated from the signal by setting an intensity threshold. However, the signal has an amplitude distribution that overlaps with a noise amplitude distribution, which does not allow loss-less noise removal, and of course the detected signal exhibits shot noise.

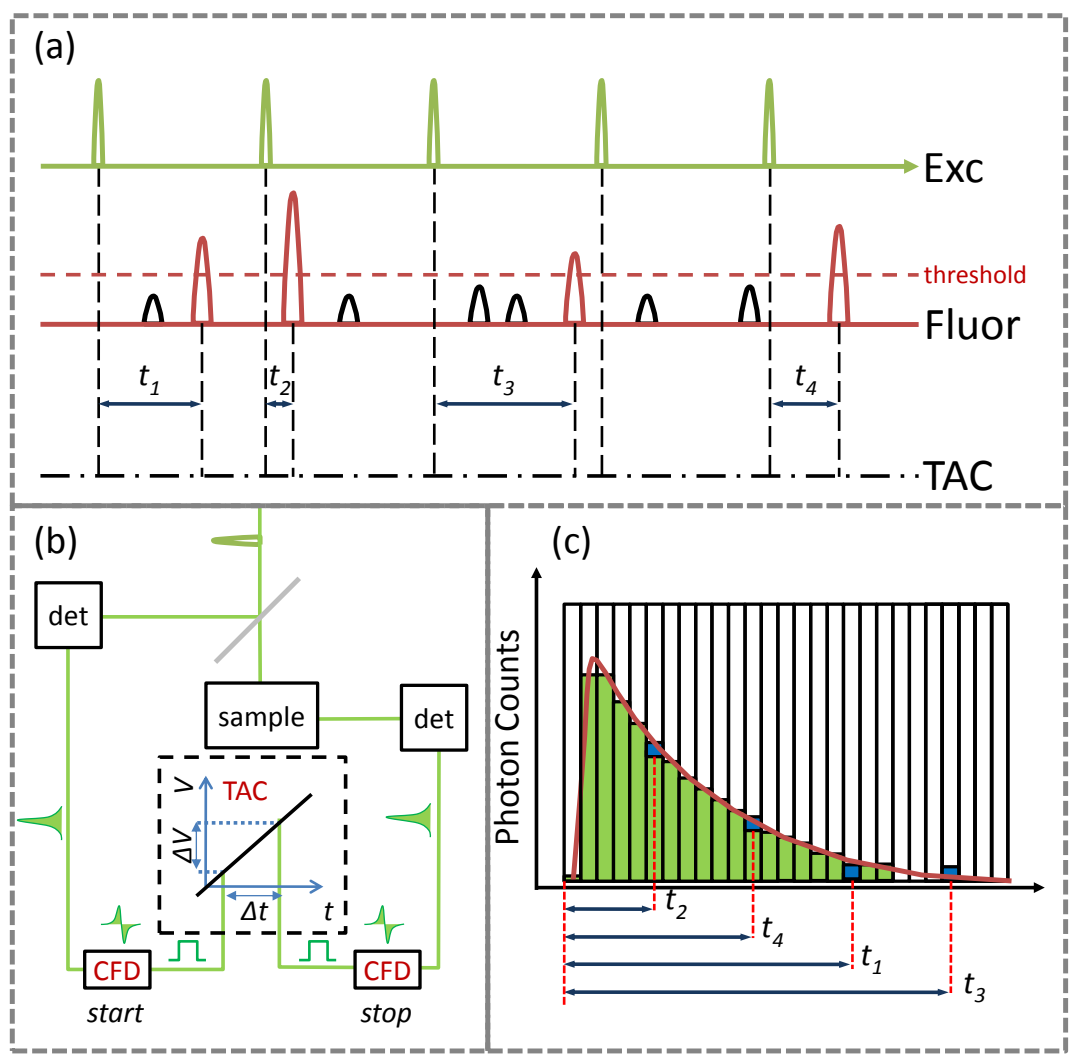

Figure 18. Time correlated single photon counting (TCSPC). (a) Periodical laser pulses (green) excite molecules that later emits single photons. Photons are converted into electrical signal by a detector. The signal is thresholded to discriminate real signal (red) from noise (black) – this is a basis of single photon counting. Constant fraction discriminator (CFD) is used to determine an exact arrival time of excitation and the emission photons, and time-to-amplitude converter (TAC) is used to accurately measure the time difference between the two – this forms a basis of TCSPC. (b) TCSPC principle (time-measurement block). (c) A photon count (shown as red square) is added to the histogram with amplitude-to-digital converter (ADC) and the fluorescence decay is thus recorded when many photons are collected.

A broad signal amplitude distribution makes it difficult to determine the exact pulse arrival time if pulses are other than square shaped. This result in timing jitter if a leading edge discriminator is used, that is, if the timing of pulse is measured by using a particular threshold. However, if half of the electric signal is delayed by a certain amount, inverted and then added with the other half, the zero intensity point of the resulting curve appears at the same point in time, independently of pulse intensity. This method, called constant fraction discriminator (CFD), is therefore immune to signal amplitude variation introduces a jitter of only ~ 50 ps compared to that of 1 ns if leading edge discrimination is used. For fluorescence lifetime determination, CFD is used together with a time-to-amplitude converter (TAC) that measures the exact time delay between the excitation and emission pulses. This is called time correlated single photon counting (TCSPC) [185]. The general principle of TCSPC is more explicitly shown in Figure 18 (b). A laser pulse triggers the time-to-amplitude converter that ramps a voltage on a capacitor linearly with time. A photon signal arriving from the PMT stops the increase of the voltage and the voltage on capacitor is read by the amplitude-to-digital converter. Its output prescribes a photon to its correct time bin in the fluorescence arrival time histogram, from which the decay profile can be determined. When the time differences for many events are measured, a probabilistic histogram of photon arrival time with respect to the excitation is built, as shown in Figure 18 (c). Time resolution of such a system is described by instrument response function (IRF), which is a response of the instrument to a zero lifetime sample. The TCSPC module works correctly if there is only one photon detected per period. When more than one photon is detected, the system starts to be saturated and *pile-up* is observed. Described 'start' and 'stop' scheme is repeated with the frequency of pulsed laser repetition rate. This makes the capacitor in the TAC module to be charged and discharged at, for example 80 MHz, if Ti:Sapphire laser is used. However, after the discharge the system cannot detect another photon for 125 ns (in modern photon counting boards like SPC-830, B&H that was used in this thesis) because of the electronics in the TCSPC module. The time that the system is busy is called the *dead time, $t_d$* of an instrument. Assuming a continuous supply of photons (in case of continuous wave excitation) at count rate of $f_{ph}$, the fraction of detected photons, *d*, equals to:

$$d = 1 / (1 + f_{ph} \times t_d)$$

eq. 25

With the excitation rate equal to the count rate, i.e. $f_{ph}$ = 1 / 125 ns = 8 MHz, the fraction of detected photons is 1 / 2 and $f_{ph}$ = 80 MHz it is 1 / 11. Therefore the dead time is a problem at high photon count rates (when, for example high intensity excitation is used) since many of photons will not be detected because they '*pile-up*' but there is room for only one to be detected. Earlier arriving photons will then be more likely to be detected than the ones coming later, therefore the system will effectively record a shorter life time than is actually the case. The problem can be addressed by the reversed 'start'-'stop' regime where detected photons would start the TAC and laser pulses would stop it, rather than the other way around. This works well when photons are detected at lower rate than the repetition rate of a laser used, which normally is the case in the photon counting regime and can be imposed by decreasing the excitation power. This limitation, however, combined with the sequential pixel scanning, makes confocal or multiphoton TCSPC a relatively slow FLIM technique.

## 3.3    Instrumentation for time domain fluorescence lifetime imaging microscopy

Here the gated optical intensifier (GOI) and TCSPC as used for FLIM experiments in this thesis are explained in more details.

### 3.3.1    Time-gated wide-field FLIM with gated optical intensifier

The gated optical intensifier was used in various FLIM microscopes setups in this thesis including the wide-field, Nipkow and line scanning FLIM microscopes to demonstrate application of supercontinuum as an excitation source as described in Chapter 4. The principle of fluorescence lifetime imaging with a gated optical intensifier is shown in Figure 19. At the heart of the gated optical intensifier is the multichannel plate, which has been already described in Section 3.2.1, in the context of multichannel plate photomultiplier. Individual channels of the multichannel plate are used to accelerate electrons created by the photons that were emitted from different sample areas. Therefore, photon emitted from a sample are sampled in parallel but the crosstalk between neighbouring channels, however, limits the ultimate spatial resolution that multichannel plate can achieve.

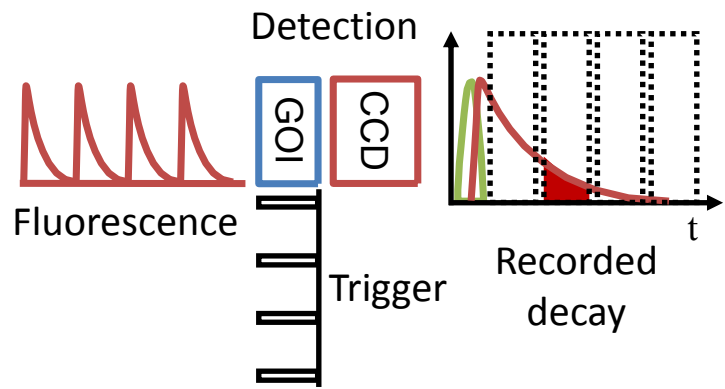

Figure 19. Principles of fluorescence lifetime imaging with the gated optical intensifier (GOI).

An image coming from the sample is intensified by the gated optical intensifier at a specific time gate delayed with respect to the excitation pulse. This gated image is read out with a CCD camera. The time delay is progressively increased until all fluorescence decay profile is probed, as illustrated in Figure 17. The gated optical intensifier therefore simultaneously samples the fluorescence decay profiles for each pixel in the image. The gain of the gated optical intensifier can be switched on the time-scale of ~ 100 ps picoseconds. A fast-gated optical intensifier (model HRI, Kentech Instruments Ltd, Didcot, UK) was used in this thesis, a description of which is available in earlier group's publications [186, 187]. Briefly, this instrument had a gate width of ~ 200 ps and could operate at 80 MHz rate. A gated optical intensifier with a better temporal resolution is also available (~ 80 ps) and was used in the earlier group's publications [180]. However it operates at ~ 10 kHz rate and requires kHz pulsed laser systems. In general such lower repetition rate systems provide inferior S / N compared to 80 MHz systems. In order to image as fast as possible, a segmented gated optical intensifier was developed to acquire FLIM image in a single shot, as described in another group's publication [188]. This particular gated optical intensifier has 4 segments and each of them is used to record time gated image but at different delays in one shot. Video rate FLIM imaging is thus achieved 29 frames per second (fps).

### 3.3.2    Laser scanning FLIM with time correlated single photon counting

The time measurement principles of TCSPC were explained in Section 3.2.3 and Figure 18 represents classic TCSPC. Modern TCSPC systems are now able to perform multidimensional measurements and besides the time measurement channel, can have a scanning interface, a detector channel register, and a large histogram memory [189], as shown in Figure 20 for the SPC-730 / 830 TSCPC module (B&H, Berlin, Germany). An example of another commercially available TCSPC module is the HydraHarp 400 (PicoQuant, Germany). Both modules can be installed into a PC. TCSPC is a particularly attractive add-on to a multiphoton microscope since the ultrashort laser required for TCSPC is already available in the multiphoton microscopy setup.

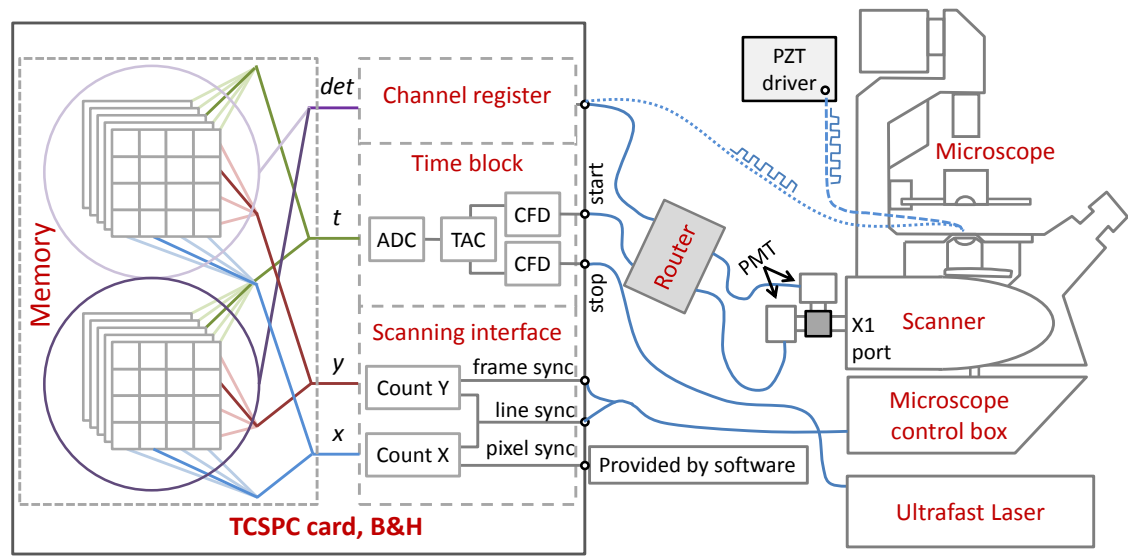

Figure 20. Schematic of the multidimensional TCSPC system in conjunction with the confocal microscope and ultrashort pulse laser. A router was used with two PMTs detecting in different spectral regions. The detectors are mounted on scanner's descanned X1 port. When only one PMT was used, a router was removed. The TCSPC card was also synchronised with muscle stretching (as carried out with piezo stretcher) by connecting mechanical response signal (blue round dotted curve) of the muscle fibre to the channel register. The detected photons were recorded in a multidimensional memory corresponding to it physical coordinates $x$ and $y$, arrival time, $t$ and the detector, $det$ that it was detected with. Pixel sync signal is provided by software since the microscope does not generate it. Shutter system used to protect detectors is not shown.

Figure 20 shows the experimental setup that was used throughout this thesis for various TCSPC FLIM experiments. The FLIM part of the system was developed by other past members of the group. Besides the multidimensional TCSPC module (SPC-730 / 830, B&H, Berlin, Germany), it consists of a commercial laser scanning confocal microscope (TCS SP2, Leica, Manheim, Germany) and an ultrashort pulse Ti:Sapphire laser (Tsunami, Spectra-Physics). Fluorescence was detected with one or more PMT (PMH-100, B&H, Berlin, Germany) through the descanned port (X1) situated on the scanner unit. A router (HRT-41, B&H, Berlin, Germany) was used to combine signals coming from multiple detectors when more than one was used. The system also included a detector gain and shutter control card (DCC-100, B&H, Berlin, Germany) but it is not shown in the Figure 20. The PMT was modified to produce an overload signal that would trigger a shutter in front of the detector via the shutter control card if the count rate exceeded $\sim 3 \times 10^6$ counts / sec, as explained in Peter Lanigan's thesis [190]. The system was used for some biological applications with frequency doubled ultrashort pulses as discussed in the next section. Frequency doubled femtosecond pulses were stretched to picosecond in order to reduce sample photobleaching. The temporal stretching was performed in a glass (F2 glass, Schott glass UK) block that is described in more details in Ref. [190]. The sides of the block were polished [190] to allow the total internal reflection of the light inside the block, as shown in Figure 21, in order to increase path length. The light in the block (block dimensions: $8 \times 1.1 \times 2.5$ cm), forming $3 \times 2$ structure (like in Figure 21), would travel $\sim 65$ cm within the block in total and therefore should allow stretching femtosecond pulses (at 470 nm) to $\sim$7.3 ps [190].

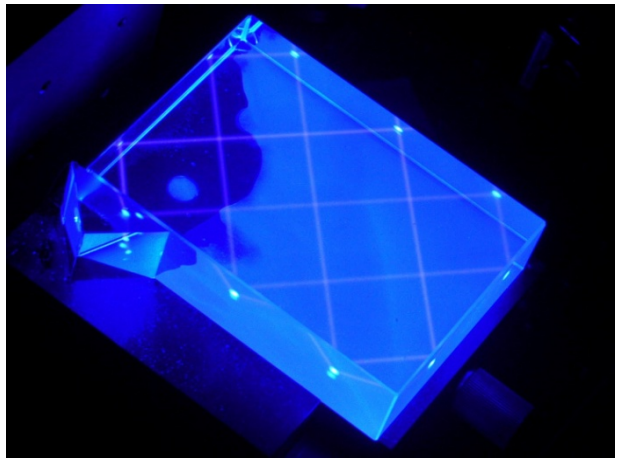

Figure 21. Femtosecond pulse stretching in a glass bock. The total internal reflection configuration was used in order to increase a path length.

The time measurement 'block' of TCSPC module contains the typical TCSPC block in the reversed start-stop configuration. The block determines the detection time $t$ with respect to the next laser pulse for each photon. The scanning interface receives the *scan* signals from the galvanometer scanners (through the microscope's control box) that indicate the beginning of the frame (*frame sync*) and the end of line (*line sync*). There is no pixel clock is available with the Leica SP2 microscope and therefore an internal pixel clock has to be used. The scanning interface determines the physical location ($x$ and $y$) of the pixel for each photon. The channel register receives a signal from the router if more than one detector is used, or it can be used with any other external synchronisation device. The channel register creates new opportunities for multidimensional imaging; together with the ($x, y, t$) detection of a photon, the detector channel number $n$ for the current photon is also read into the detector channel register. The register number $n$, for example, could represent the wavelength of the detected photon if light is split into different wavelength intervals in front of the detectors [191, 192]. If multiple lasers are used, an interleaved excitation can be set up, where $n$ would represent the excitation wavelength in the channel register of each laser. A combination of excitation at multiple wavelengths and detection with multiple detectors can also be implemented. This means that the detected photon can be assigned to a laser that excited the molecule and the detector that detected emitted photon from the excited molecule. This system would thus be able to record fluorescence excitation-emission-lifetime data at each pixel [193]. It can be used useful in FRET experiments where alternating excitation of the donor and of the acceptor individually allows identifying cases where the acceptor is not present [194] resulting in so called zero FRET efficiency [195]. The channel register can, in principle, be used to register any physical state of an experiment. For example, as it will also be shown later in this Chapter, it can be used in muscle fibre experiments to report if the detected photon came from the fibre while it was being stretched by piezo actuators or was still. This microscope system was later modified to serve as STED microscope, as explained in Chapter 6.

## 3.4    Fluorescence lifetime imaging applications in biology

### 3.4.1    Introduction

Fluorescence lifetime imaging is a powerful technique for cellular imaging that allows to follow biochemical reactions in cells on microscopic scale [196]. Fluorescence lifetime can be used to contrast different fluorescing species, with lifetime offering additional opportunities for contrast compared to intensity and spectra. For example, cell autofluorescence involves many different fluorescing species (NADH, Elastin etc) that have similar spectra but different lifetimes [197]. Some of these autofluorophores may change their lifetimes upon binding to proteins or other substances (typically ~ 20 %). For example, NADH changes its lifetime upon binding to malate dehydrogenase [198]. In general, fluorophores can change their lifetime when binding to proteins, lipids, DNA [199] etc. Thus fluorescence lifetime measurements can be used to sense molecular binding. Some fluorophores can exist in protonated and deprotonated form, the equilibrium of which is pH dependant, and therefore if the two forms have different lifetimes they can be used as pH sensors [200, 201]. The presence of some molecules can be imaged through collision induced lifetime changes. Oxygen is one of such molecule [202]. FLIM can also be used to image ion concentrations [203], refractive index change [204], local viscosity [205] and some other properties [24, 206]. In many of these applications, the fact that fluorescence lifetime does not depend on fluorophore concentration is critical. As a result, fluorescence lifetime is widely used to map out different fluorescing molecules, their specific properties and changes in their environment. Another important area where FLIM is widely used is FRET [207]. As described in Section 2.2.8, FRET is a process in which energy is transferred non-radiatively from one fluorophore to another chromophore and, if combined with imaging, it can be used in co-localisation studies on a nanometre scale [26], and to map protein-protein interactions [208] or protein conformational changes [209]. FRET efficiency is measured in those experiments to either find the distance between the two molecules or to prove that they are physically linked. There are a variety of other methods to measure FRET efficiency [207] and one of the most reliable is fluorescence lifetime imaging. If one defines $\tau_D = 1 / (k_{nr} + k_{fluor})$, i.e. the fluorescence lifetime of a donor, and $\tau_{DA} = 1 / (k_{fret} + k_{nr} + k_{fluor})$, the fluorescence lifetime of a donor in the presence of an acceptor, then the eq. 5 can yield:

$$E = 1 - \tau_{DA} / \tau_D$$

eq. 26

From eq. 6 and eq. 26 it is clear that, by measuring the fluorescence lifetime of the donor in the presence and absence of the acceptor molecule, one can learn about the relationship between the molecules. In principle one can calculate the intermolecular distance but only under the assumption that the molecules have their dipole moments in parallel, otherwise some orientation factor needs to be included. FRET is more commonly used to get qualitative information (of, for example, whether or not two molecules are close) rather than for quantitative measurements (to get an exact distance between fluorophore). FRET imaging was first investigated by using donor [210] or acceptor [211] photobleaching. Both the $\tau_D$ and $\tau_{DA}$ can be determined from the same image if regions with and without the acceptor can be recorded. This can also be done by first recording the image with the acceptor molecule and acquiring an image with the acceptor photobleached out [196]. An example of fluorescence lifetime imaging used to report FRET by means of acceptor photobleaching is shown in the next section, which describes the application of FLIM-FRET to image protein phosphorylation inside the cell.

### 3.4.2    *Application of FLIM to measure epidermal growth factor receptor phosphorylation*

The purpose of the following example of FLIM application to biology is to solely exemplify how FLIM can be used to measure FRET in cells using acceptor photobleaching. The microscope described above (Figure 20) was used for imaging. Data acquisition and analyses was performed by the author.

The activity of epidermal growth factor (EGF) and its receptor (EGFR) have been identified as key drivers in the process of cell growth and replication. Heightened activity at the EGF receptor is implicated in many cancers [212]. Activation of EGFR by epidermal growth factor (EGF) and other ligands, which bind to its extracellular domain, is the first step in a series of complex signalling pathways. To monitor the receptor activation, its phosphorylation by epidermal growth factor can be observed through FRET between donor – green fluorescent protein (GFP) labelled EGFR and receptor – Cy3 labelled anti-phosphotyrosine specific molecular antibody (mAb), as illustrated in Figure 22. The following example illustrates the use of FLIM and acceptor photobleaching to indicate FRET between the donor and the acceptor pairs in MCF7 cells. FRET between the GFP and Cy3 can be interpreted as a direct interaction between the phosphorylated receptor and the antibody [213]. Figure 23 highlights GFP expressing cells (not all cells are expressing).

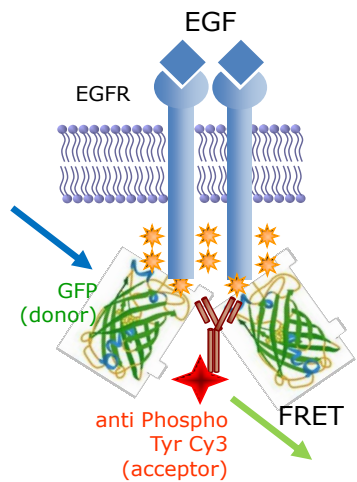

Figure 22. Activation of epidermal growth factor receptor (EGFR) with epidermal growth factor (EGF). The activation is detected with FRET between GPR and Cy3.

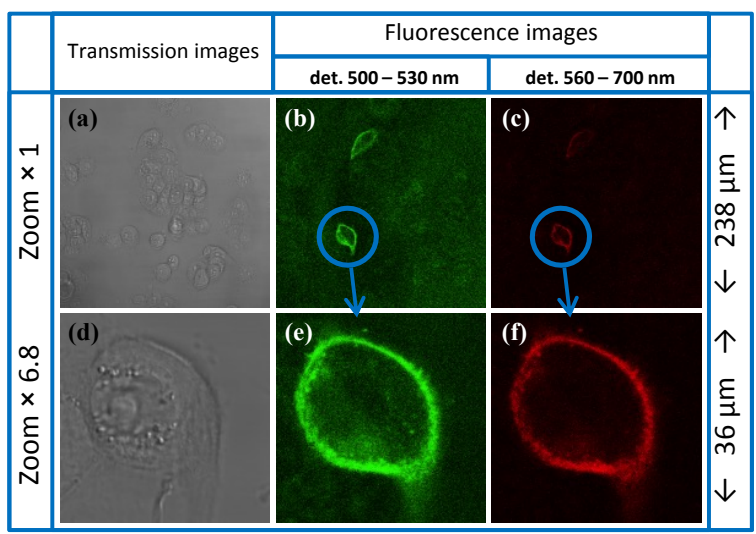

Figure 23. Transmission (a and d) and fluorescence intensity images of the cells recorded with 488 nm excitation. Fluorescence images recorded in the Green detection (b and e) and the Red (c and f) detection bands with the detection pinhole set to the size of 1 Airy disc (see Figure 8 for explanation). Water immersion objective with NA = 1.2 and ×63 magnification was used.

The Green and Red spectral detection bands were selected to optimally separate GFP and Cy3 signals, respectively. 488 nm excitation was exciting both, GFP and Cy3 molecules, therefore, to photobleach Cy3 a different laser had to be used. Acceptor molecules (Cy3) were photo-bleached in a part of the cell with a laser operating at 532 nm (5 mW, frequency doubled Nd:YAG). Photobleaching the acceptor was necessary in order to measure lifetime of the donor without presence of the acceptor, as explained above, in Section 3.4.1. To check if the acceptor molecules were actually photobleached, they were imaged using the same laser before and after photobleaching, as shown in Figure 24 (a) and (c).

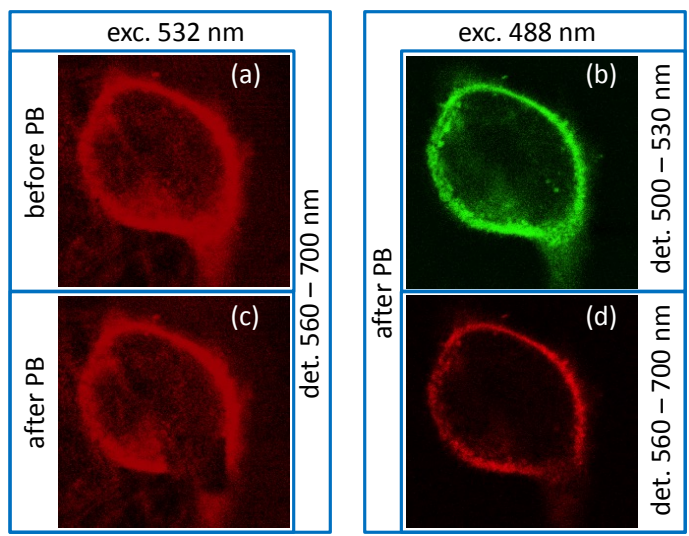

Figure 24. Photobleaching a square in the part of the cell. Excitation with 532 nm reveals that Cy3 molecules were photobleached whereas excitation with 488 nm shows that GFP are still present after photobleaching.

Clear photobleached square indicates that Cy3 molecules were completely photobleached. Additional set of images were acquired with 488 nm excitation to mainly check the status of the GFP molecules after photobleaching of Cy3 molecules. Images acquired with 488 nm excitation, shown in Figure 24 (d), after photobleaching proves that GFP was not photobleached since the square pattern, photobleached on Cy3 molecules, cannot be seen in the Green channel (a detection window for GFP). These images also indicate that no FRET is occurring in this cell, since no increase in intensity is observed (in the square region) for GFP. The Red channel shows blurred square which is indicative of GFP signal leaking into the red channel. Images of fluorescence lifetime were recorded using pulsed 470 nm excitation (frequency doubled Ti:Sapphire laser) before and after photobleaching the acceptor as shown, in Figure 25.

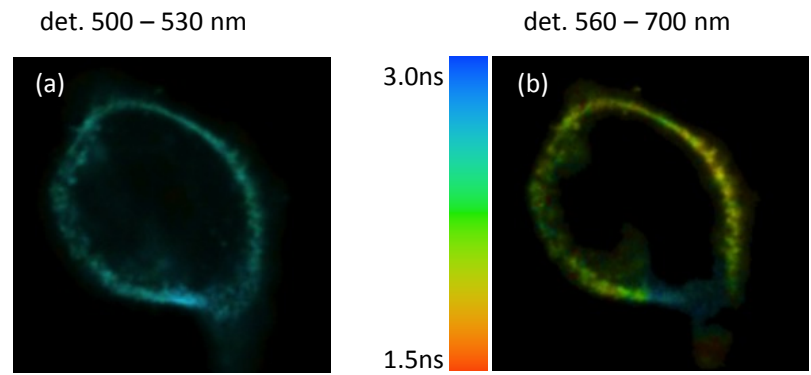

Figure 25. Merged intensity FLIM images in two spectral channels of the cell after photobleaching. No FRET is observed in this cell.

Again, the Green channel (500-530 nm) does not show any traces of the photobleached square, whereas the Red channel shows an increase in average fluorescence lifetime, which indicates that Cy3 was photobleached effectively and the longer lifetime is because of GFP signal leaking into the Red channel (560-700 nm).

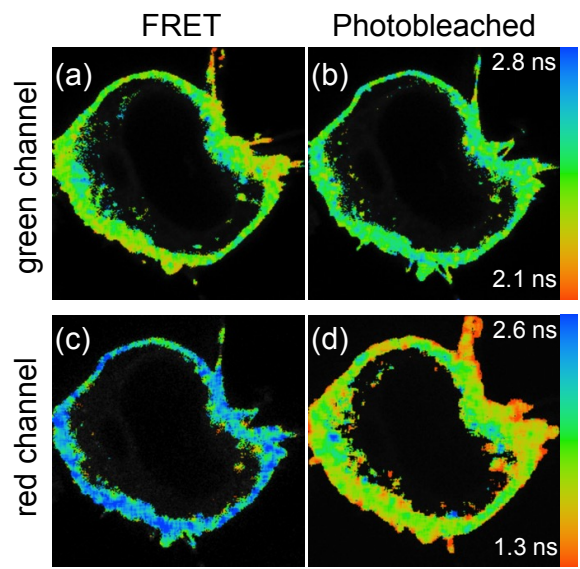

Figure 26. FLIM images in different spectral windows before and after photobleaching. A decrease in fluorescence lifetime of the GFP was observed in the presence of Cy3 (a) and could be reversed by photobleaching the Cy3 label for couple of minutes (b). Acceptor (c) is totally photobleached since only GFP fluorescence is seen (d). Image size – 36 μm across.

This correlates with no lifetime recovery in donor being observed in the 'Green' channel after photobleaching (data not shown) indicating that no FRET was occurring in this particular cell. A cell that does show FRET is shown in Figure 26, for which a clear lifetime change in both channels is seen. The GFP image in Figure 26 (a) shows FRET occurring at some parts of the membrane, as indicated by lifetime shortening.

After photobleaching, the donor lifetime recovery is observed and a more homogenous lifetime image is seen, as shown in Figure 26 (b). Figure 27 (a) shows that average donor fluorescence increases by ~ 50 ps after photobleaching. Some areas in images in Figure 26 demonstrate recovery by up to ~ 200 ps suggesting the FRET efficiency of ~ 10 %, following from eq. 26. Figure 27 (b) indicates that the acceptor is totally photobleached since the lifetime in red channel after photobleaching becomes similar to that of GFP (~ 2.3 ns) and so the remaining fluorescence is likely to be GFP cross-talk.

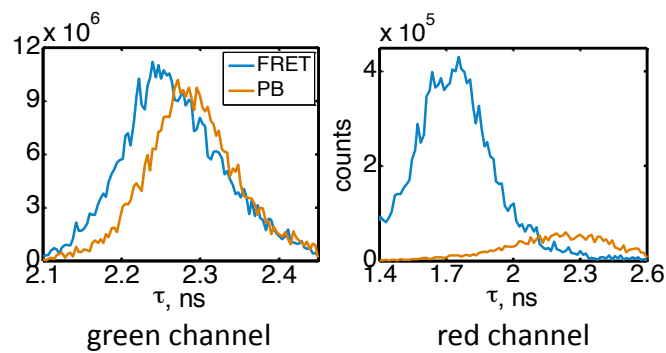

Figure 27. Fluorescence lifetime histograms of FLIM images in Figure 26.

### 3.4.3    Application of FLIM to study cell signalling at immune synapses

In the work described below the author's input was developing and setting up some of the experiments. Sample preparation and data acquisition were performed by Bebhinn Treanor and others. The work was summarised in Ref. [214], which reported supramolecular organisation of killer immunoglobulin (Ig)–like receptor (KIR) phosphorylation at natural killer cell immunological synapses. The reported results suggest that it would be interesting to use STED microscopy to resolve structure of the supramolecular organisation.

Natural killer (NK) cells perform surveillance of other cells to look for infection or other type of malfunction [215]. Natural killer cells 'probe' target cells using killer Ig-like receptor, which recognises major histocompatibility complex (MHC) class I proteins expressed on the target cells [216]. If the target cell do not express or down regulate the major histocompatibility complex protein then the natural killer cells carry out their cytolysis. If natural killer cells recognise the major histocompatibility complex then the killer Ig-like receptor is phosphorylated. FLIM used to report FRET can, in general, be employed to image protein phosphorylation of any GFP labelled receptor at an intercellular contact. Here, killer Ig-like receptor phosphorylation was imaged at the natural killer cell immune synapse measuring FRET between GFP labelled KIR2DL1 and a Cy3 labelled anti-phosphotyrosine monoclonal antibody (mAb), (see Figure 28).

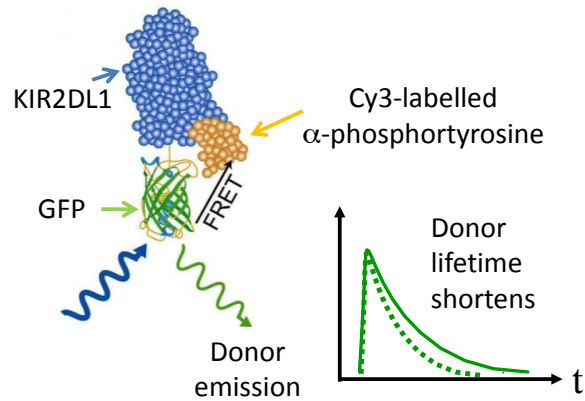

Figure 28. Phosphorylation of killer like receptor (KIR) can be detected by measuring FRET signal between Cy3 labelled α-phosphortyrosine monoclonal antibody and GFP labelled KIR2DL1. FRET distribution in cell can be measured as a lifetime shortening of donor (GFP) using FLIM.

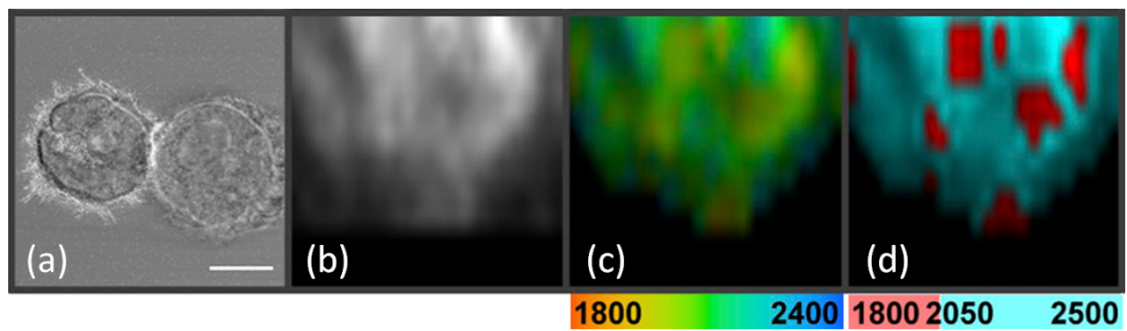

Figure 29. Observing killer immunoglobulin-like receptor phosphorylasion at the immune synapse between the natural killer and target cell conjugate using FLIM to image FRET. 3 D imaging of FRET was obtained by FLIM images being acquired every 0.5 μm throughout the conjugate. (a) The conjugate intensity image. En face reconstructions are shown of KIR2DL1-GFP intensity (b) and lifetime (in ps) plotted on a continuous scale (1,800–2,400 ps) (c) and a discrete scale (d) with red colour denoting lifetime values lower than 2050 ps and cyan colour higher than 2050 ps. 2050 ps value corresponds to 5 % of FRET efficiency (as defined by eq. 26). Scale bar – 8 μm. Figure taken from [214].

FLIM-FRET was used to investigate spatio-temporal micro-organisation of KIR in immune synapse formation [214]. Visualization of killer Ig-like receptor phosphorylation in natural killer cells contacting target cells expressing major histocompatibility complex class I proteins revealed that killer Ig-like receptor signalling is spatially restricted to the intercellular contact. Another important conclusion is that contrary to the expected homogeneous distribution of killer Ig-like receptor signalling across the intercellular contact, phosphorylated of the receptor was confined to microclusters within the larger immune synapse region as illustrated in Figure 29. It would be interesting to be able to resolve those clusters using STED microscopy as will be discussed later in this thesis.

### 3.4.4    Application of FLIM to study molecular motors

In the work described below the author's input was developing and setting up some of the experiments. Sample preparation and data acquisition were mainly performed by Delisa Ibanez Garcia. The work was summarised in Ref. [217], where FLIM was shown to be a useful tool to detect actomyosin states in mammalian muscle sarcomeres. The contraction of muscles is driven by the adenosine tri-phosphate (ATP) dependent interaction of actin and myosin filaments (Figure 30). These molecular events are responsible for force generation and the movement of molecular motors can be followed by imaging the changes in fluorescent properties of suitably placed fluorophores when muscle fibre contraction is perturbed.

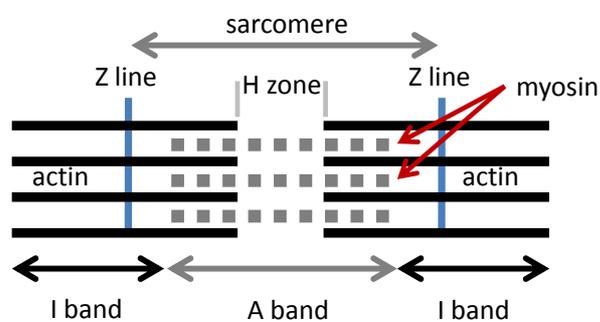

Figure 30. Structure of sarcomere.

Fluorescence lifetime imaging microscopy (FLIM) of a probe attached to the muscle motor protein myosin may provide information about the changes in its conformation upon ATP hydrolysis, product release and the steps involved in force generation and shortening. Here FLIM was applied to study actomyosin states in skeletal muscle fibres. It was used to understand changes in the conformational state of ATP bound to myosin. ATP-myosin binding was monitored using an ATP-analogue labelled with a Coumarin-based fluorophore (DEAC) for different sarcomere lengths (look Figure 30), as explained in more detail in Ref. [217]. Figure 31 shows that FLIM provides contrast between *I-band* and *A-band* (non-overlap and overlap regions) of the muscle fibre sarcomeres attributed to the binding of the ATP-DEAC to the myosin heads and hydrophobic reactions with actin respectively.

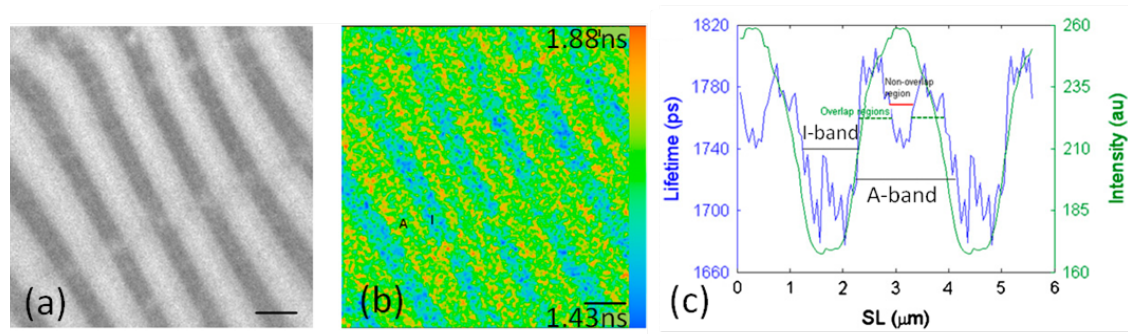

Figure 31. Fluorescence lifetime image of sarcomeres in muscle fibre. (a) Intensity image. (b) FLIM image. The *I band* has an average τ of 1.71 ns, overlap region (green line) in the A-band of 1.78 ns and nonoverlap region (red line) – 1.75 ns. (c) Lifetime (blue curve) and intensity (green) profiles over one sarcomere length (but displayed for two). Scale bar - 2 μm. Figures taken from [217].

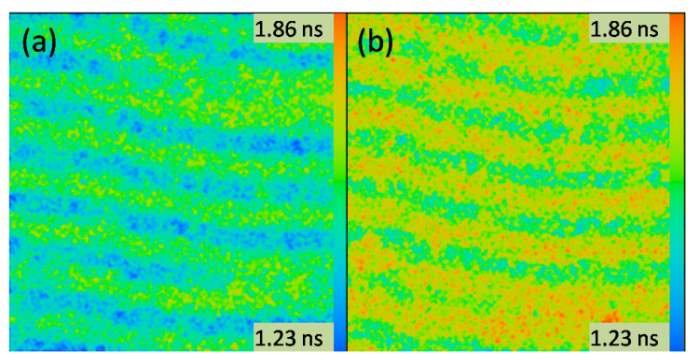

Figure 32. FLIM of muscle fibre recorded during stretching (b) and release (a). The images demonstrate an increase in the average lifetime when strain is applied to the fibre.

The demonstrated sensitivity of the ATP analogue to changes in the microenvironment illustrates the potential of FLIM to study changes in the actomyosin complex during muscle contraction. To this end FLIM acquisition was synchronised with muscle stretches performed with a specially designed piezo stage that could stretch a muscle fibre with adjustable amplitude and frequency. The image acquisition was synchronised through the channel register of the TCSPC module, as shown in Figure 20 (Section 3.3.2), to acquire one image while the muscle fibre was being stretched and another while it was being released. The resulting images, shown in Figure 32, indicate that there is an increase in the average lifetime when the strain is applied to the fibre.

## 3.5     Summary and Outlook

This Chapter demonstrated that fluorescence lifetime imaging can provide additional insight in cellular imaging. Time correlated single photon counting (TCSPC) and time-gated fluorescence lifetime imaging (FLIM) techniques and instrumentation, used throughout this thesis, were described and their applications to image biological systems were briefly illustrated. FLIM application to biology was illustrated with the TCSPC system working in conjunction with the confocal microscope and ultrashort pulse laser. Specifically, FLIM was employed to read out epidermal growth factor receptor phosphorylation following epidermal growth factor stimulation in cells by detecting FRET between green fluorescent protein-labelled epidermal growth factor receptor and Cy3 labelled anti-phosphotyrosine molecular antibodies. The decrease in green fluorescent protein lifetime due to FRET was shown to be reversible by photobleaching the donor Cy3 molecules. In another experiment the phosphorylation of killer Ig-like receptor was measured by detecting FRET between green fluorescent protein-labelled killer immunoglobulin-like receptor unit and a Cy3 labelled anti-phosphotyrosine molecular antibodies. It revealed that killer immunoglobulin-like receptor signalling was spatially restricted to the intercellular contact and occurred in clusters, therefore, it would be interesting to look at it (clusters) with STED microscopy, described in Chapter 6. FLIM was also applied to study actomyosin states in skeletal muscle fibres.

# 4. Supercontinuum Application to Fluorescence Microscopy

## 4.1 Introduction

Lasers are routinely used in fluorescence microscopy since it can be focused to a very tight and bright spot in the sample plane. However, a serious limitation of fluorescence microscopy has been the availability of different excitation wavelengths from spatially coherent or ultrashort light sources. There are a limited number of (tunable) laser excitation sources with few covering the visible spectrum, such that most instruments have had access only to a few excitation wavelengths. This placed a constraint on the design and selection, for example, of fluorescence probes for biological experiments, which currently need to be excitable by the standard laser excitation lines and this can limit the science that can be undertaken in many laboratories. The efficient transfer of laser energy to new wavelengths, especially extending into the green and blue, is therefore a highly attractive prospect. Such an excitation source might be useful in a multidimensional fluorescence approach, where many parameters are to be measured for each object / pixel. One of the most often used laser is Ti:Sapphire laser [218] that offers tunable radiation in a range from ~ 700 nm to 1.1 μm. However, to use it in fluorescence microscopy the radiation has to be frequency doubled and different wavelengths cannot be used at the same time.

This Chapter demonstrates that supercontinuum generated in microstructured optical fibres (MOF) can be successfully used in fluorescence microscopy. The first application of supercontinuum generated in microstructured optical fibre to fluorescence microscopy was demonstrated in 2004 by recording some confocal fluorescence intensity images [219]. This was followed by our group publication where pulsed and tunable nature of supercontinuum was exploited to record fluorescence lifetime images and excitation spectrum, respectively [220]. Later it was shown that supercontinuum has temporally short enough pulses to excite two-photon fluorescence and therefore to record two-photon excited fluorescence images [221, 222]. Tapered standard fibre was also demonstrated to be suitable for supercontinuum generation (430-1300 nm) and application to 3-D confocal and multiphoton microscopy [223]. Whole spectrum fluorescence detection with supercontinuum was demonstrated to enable simultaneous excitation of different fluorophores [224]. Chromatic confocal microscopy was established by focusing an entire bandwidth of generated supercontinuum on the sample [225]. This probed different depths of the sample because of chromatic aberration present in the employed objective. Supercontinuum is also successfully used in coherent Anti-stokes Raman spectroscopy microscopy [226-228]. A commercial confocal has been recently released by Leica incorporating supercontinuum source that allows continuous tuning of the excitation wavelength in the range of 470 to 670 nm with 1 nm increment (Leica TCS SP5 X).

In this Chapter the application of supercontinuum-based source to various types of fluorescence lifetime imaging microscopes is demonstrated. Those include Nipkow spinning disc, Wide-field and Hyperspectral microscopes. Supercontinuum generation in microstructured optical fibre was first demonstrated in 2000 [229]. However the underlying mechanisms of supercontinuum generation have only recently been elucidated. Supercontinuum generation in optical fibres is not new, having been demonstrated already in 1970's – long before the advent of the microstructured optical fibre in 1996 [230]. Moreover the various nonlinear processes that are involved in supercontinuum generation in microstructured optical fibre like self-phase modulation (SPM), four wave mixing (FWM) and Raman scattering were well known. Such nonlinear processes are stronger in microstructured optical fibre than in standard fibre due to the unique microstructure that permits reduced effective core sizes and optimised dispersion profiles that can enhance nonlinear processes. Therefore, when an ultrashort laser pulse travels down the microstructured optical fibre, it experiences enhanced spectral broadening – supercontinuum generation due to many nonlinear processes involved at the same time. It is difficult to isolate individual processes involved in supercontinuum generation and therefore it is not easy to interpret such spectra. Thus was not straightforward to understand which parameters of laser pulses and those of microstructured optical fibre are most important and how they affect supercontinuum generation. Initially supercontinuum generation was optimised empirically but now various algorithms have been developed to model supercontinuum generation in order to predict conditions for optimal spectral broadening [231]. One reason that made supercontinuum generation with microstructured optical fibres so widespread was the availability of tunable Ti:Sapphire laser technology that could provide radiation matching the anomalous dispersion region of microstructured optical fibres used. This enabled widespread research and application of the supercontinuum generation [16, 17, 232].

This Chapter introduces supercontinuum generation in microstructured optical fibres. It covers the various nonlinear processes involved in supercontinuum generation and describes the properties of microstructured optical fibre and laser parameters that control supercontinuum generation. At the end of this Chapter various experimental setups are presented that were used to generate supercontinuum for application to various fluorescence microscopes.

## 4.2     Introduction to supercontinuum generation

Continuum generation is a phenomenon of continuous new wavelength generation through nonlinear interaction of light with its medium, e.g. when, an intense pulse travels in a nonlinear material and experiences spectral broadening. Supercontinuum generation refers to an extreme spectral broadening when very intense pulses or unusually highly nonlinear material is used [233, 234].

### 4.2.1    Supercontinuum generation

#### Bulk media

Supercontinuum generation was first observed in 1970 in bulk material by Alfano where radiation spanning from 400 nm upwards was generated in bulk material from 780 nm input ultrashort *nJ* laser pulses [235, 236]. Soon after the demonstration in bulk material, a similar phenomenon was observed in fibres [237] but to a lesser extent since there was a limited choice of appropriate lasers at the time and the fibres used did not have the right dispersion and nonlinear properties. However, supercontinuum generation in fibres is preferred over bulk material because a spectrally smoother output is generated due to absence of filamentation that occurs in bulk or liquid material [238]. The supercontinuum generated in bulk material normally also has complex spatio-temporal natures, complicating their interpretation.

Supercontinuum generation in fibres is easier to analyse since it happens in one spatial dimension – provided that fibre supports only a single transverse mode. Because single mode operation provides a clean and spatially coherent supercontinuum output, it is important in numerous applications like microscopy and tomography.

### *Standard optical fibres*

Spectral broadening in standard fibres is facilitated by extreme light confinement in the core running along the entire length of a single mode fibre that is typically long compared to the spot length of focused light. The core therefore provides a long interaction length for nonlinear processes, even though the material (silica) has a low nonlinear coefficient [239]. The broadest supercontinuum is obtained when light is coupled in the anomalous spectral part of the dispersion curve. The supercontinuum then tends to form around the zero-dispersion wavelength (ZDW) of the fibre potentially extending to the ultraviolet on the one side and to the infrared on the other. However, the zero-dispersion wavelength of 1.3 μm for standard single mode optical fibre is too far in the infrared to allow efficient generation of the visible radiation. It is therefore desirable that the fibre is modified to change its dispersion curve so as to shift the zero-dispersion wavelength closer to the visible spectral region.

### *Tapered standard optical fibres*

One way of varying zero-dispersion wavelength is by changing the material or the refractive index of the core, as it has been performed in the so called 'dispersion shifted' fibres with the zero-dispersion wavelength at 1.5 μm. Alternatively the size of the core can be changed since this will change waveguide dispersion. A simple way of doing it is tapering a standard fibre by heating and stretching it. This induces a progressively decreasing diameter of the core along the fibre, which in turn makes the zero-dispersion wavelength vary along the fibre. The zero-dispersion wavelength can be decreased down to 500 nm if the core of the fibre is made very small (below 1 μm) [240]. The cross-section of such a microstructured optical fibre is similar to web-like structure where a core is suspended in the air with thin sheets of silica, like one shown in Figure 33 (b). Therefore tapering 'brings' supercontinuum generation closer to the visible range by shifting its zero-dispersion wavelength towards shorter wavelength. It also enhances nonlinear effects due to the reduced size of the core. The progressive change of zero-dispersion wavelength (and dispersion curve) along the fibre plays another trick to push the 'blue' part of supercontinuum further to shorter wavelengths. This happens due to the change of the phase matching curve for four wave mixing along the fibre which then allows an abundant number of frequencies to interact with each other to produce new frequencies including in 'blue' spectral region. This will be discussed later in the context of the ultraviolet-enhanced supercontinuum generation with tapered microstructured optical fibre and fibre laser pump source in Section 4.3.3. In fact, supercontinuum generation with tapered standard fibres was realised shortly after the advent of microstructured optical fibre [241, 242]. It demonstrated that similar supercontinuum could be generated without using the more elaborate structures of microstructured optical fibres. However, there are more degrees of freedom with microstructured optical fibres because it allows optimisation of more fibre parameters. They can now be easily manufactured with sophisticated control of the dispersion curve and light confinement in the core achieved by managing the spatial arrangement and the size of the holes in the cladding that are running along the fibre. This allows manufacturing of microstructured optical fibre with specific properties to be used with a particular laser and to obtain supercontinuum with desired properties.

### 4.2.2    *Microstructured optical fibres*

Microstructured optical fibres (MOF) are otherwise known as photonic crystal fibres (PCF) due to their similarity to the photonic crystals [243, 244]. In fact, fibres with photonic crystal structure were originally produced with the expectation that their guiding mechanism will rely on band gap properties, as is normally observed in photonic crystals. However, due to the technical difficulties, the first fibres made had a solid core rather than a hole in the middle of the fibre as originally intended [230]. This limited the guiding mechanism to total internal reflection as in standard fibres. However, unique and useful properties were observed shortly afterwards. Two years later, a true photonic band gap fibre was manufactured [245].

#### *Different types of microstructured optical fibres*

The variety of microstructured optical fibres is not limited to the fibres with solid core as discussed above. The other most common type of microstructured optical fibre, as mentioned above, is the *Hollow core MOF*, shown in Figure 33 (a). The Hollow core microstructured optical fibre guides the light because the photonic crystal structure of the fibre cladding disallows light of certain wavelengths to propagate through it because of photonic band-gap [245]. Therefore the light is trapped inside the hollow core and has to travel down the fibre. The most important feature of this kind of fibre is that due to low interaction of the light in the core with the material the fibre exhibits unusually low nonlinearity. The most obvious application is the delivery of the high peak power pulses with such a fibre [246].

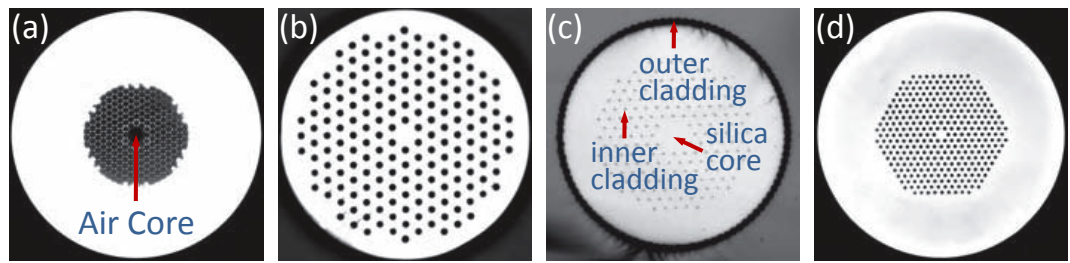

Figure 33. Scanning electron micrographs of different microstructured optical fibres. (a) Typical hollow-core photonic band gap fibre. (b) Endlessly single mode microstructured optical fibre with solid core. (c) Air-clad microstructured optical fibre. (d) Microstructured optical fibre with a small core possessing high nonlinearity and the zero-dispersion wavelength in the visible (suitable for supercontinuum generation). Figures obtained from *'Crystal Fibre'*.

The other type of microstructured optical fibre of particular interest is *Air-clad MOF* shown in Figure 33 (c). This is used in fibre laser oscillators to allow efficient interaction between the seed and the pump sources [247, 248].

### *Manufacturing*

Microstructured optical fibre is made from preforms consisting of stacks of many tubular rods. A stack is stretched to reach a diameter of micrometers to make the final fibre. The tubular rods in preform stacks can be arranged with many degrees of freedom, enabling control of the density, size and arrangement of the holes in the final microstructured optical fibre. This in turn has a great effect on the dispersion and nonlinear properties of the fibre that are crucial in supercontinuum generation.

### *Single mode operation*

Microstructured fibre can be made to support only a single mode and therefore could be a single mode fibre (SMF). The structure of the holes has a clear effect on the wave-guiding properties: the fibre can be made single mode over a broad spectral region if the right structure is manufactured. Single mode operation in microstructured optical fibre can be understood by considering the structure of the cladding. In the cladding with the comparatively big holes there is little room for the light to leak away between the holes and most of the light will undergo total internal reflection [243]. Only very high order modes will escape. At this configuration the fibre is multimode. If the size of the holes gets smaller there is more and more room between the holes for the light to escape until the holes are so small that all but single mode leaks out of the fibre [243]. In Figure 33 (b) an example of such microstructured optical fibre is shown. The fibre will be endlessly single mode for any wavelength if:

$$d / \Lambda < 0.45 \ [249]$$

eq. 27

Where $d$ – size of the hole and $\Lambda$ – pitch, distance between the holes. However the endlessly single mode fibre has a different structure from that optimum for supercontinuum generation (the structure for broad supercontinuum has small core as illustrated in Figure 33 (d)). Nevertheless the microstructured optical fibre for the efficient supercontinuum generation normally supports single mode operation over its entire spectral region of the supercontinuum since it is difficult to switch to multimode operation because of different wave-vectors between single and multimode operations.

### *Polarisation preservation*

The microstructured optical fibre can be made to maintain polarisation within the fibre if appropriate asymmetry is introduced in the core of the fibre, as first demonstrated in 2000 [250].

### *Nonlinearity*

The cladding is made of the same refractive index material as the core but holes running along the cladding modify the overall refractive index experienced by the light. This can effectively make the refractive index difference between the core and the cladding very high. Such a high refractive difference confines the light more strongly in the core (the effective area in the core might be reduced by 2 orders of magnitude compared to standard fibre), and therefore increases its intensity, which enables stronger nonlinear interactions between the light and the medium – leading to broader supercontinuum. The effective nonlinearity of a fibre is defined as:

$$\gamma = 2\pi n_2 / \lambda A_{eff}$$

eq. 28

where $A_{eff}$ is the effective mode area [233] of the fundamental guided mode and $n_2$ is the effective nonlinear refractive index of the material (for silica fibre $n_2 \approx 2.6 \times 10^{-20}$ m$^2$ / W). For telecommunication fibres $\gamma \sim 1$ (W × km)$^{-1}$ and for microstructured optical fibres $\gamma$ could potentially be around 100 (W × km)$^{-1}$. To parameterise the nonlinear properties of the fibre, a characteristic nonlinear length can be defined as:

$$L_{NL} = 1 / \gamma P_0 = \lambda A_{eff} / 2\pi n_2 P_0$$

eq. 29

where $P_0$ – Peak power of the laser pulses.

### *Dispersion*

Pulse spreading in fibre is determined by the second order derivative of refractive index with respect to the wavelength, which is referred to group velocity dispersion (GVD), $D$:

$$D = - \lambda / c \ (d^2 n / d\lambda^2)$$

eq. 30

Sometimes other notation is used to define group velocity dispersion, which is related to $D$:

$$\beta_2 = - (\lambda^2 / 2\pi c) \ D$$

eq. 31

To parameterise the dispersive properties of a fibre, a characteristic dispersion length is normally used:

$$L_D = T^2 / |\beta_2| = 2\pi c T^2 / (D \lambda^2)$$

eq. 32

where $T$ – pulse temporal width. The strong confinement due to the refractive index difference between the core and the cladding, together with the spatial arrangement of the holes in the cladding, has a strong effect on the waveguide dispersion. Since the overall fibre dispersion depends on material and waveguide dispersion, the structure of microstructured optical fibre influences the total dispersion curve. Importantly the dispersion in microstructured optical fibre can be modified so as to shift the zero-dispersion wavelength closer towards visible spectral region to enhance visible supercontinuum generation, as will be discussed below.

### 4.2.3    Nonlinear processes

The individual nonlinear processes that are involved in supercontinuum generation were indentified and understood a long time ago [239]. However, since they are happening at the same time it is not easy to separate their roles. Nevertheless, conditions exist in supercontinuum generation where a particular process can dominate.

#### Self-phase modulation (SPM)

Self-phase modulation is more significant for shorter pulses and is dominant in normal dispersion region of the fibre [251]. It is caused by materials experiencing high intensities that change their refractive index:

$$n = n_0 + \Delta n$$
eq. 33

where $\Delta n = n_2 I$; and $n_2$ – nonlinear refractive index.

The self-phase modulation produces an increase in spectral bandwidth approximated by:

$$\Delta \omega = (2\pi / \lambda) \, \omega_0 n_2 P_0 L / T_0$$
eq. 34

where $\omega_0$ – initial spectral bandwidth of the pulse; $n_2$ – nonlinear refractive index; $P_0$ – initial peak power; $L$ – effective length of the interaction. $T_0$ – initial pulse temporal width.

### *Four-wave mixing (FWM)*

Four-wave mixing process dominates for longer pulses. It generates sidebands spaced at equal frequency interval around the pump frequency. These are described as Stokes and anti-Stokes components. Four-wave mixing has to satisfy the phase matching and the conservation of energy equations:

$$2k_{pump} = k_{signal} + k_{idler} + 2\gamma P$$

eq. 35

and:

$$2\omega_{pump} = \omega_{signal} + \omega_{idler}$$

eq. 36

where $k$ are the wavevectors of the modes and $\omega$ – frequencies.

### *Raman scattering*

Raman scattering is an elastic scattering of photons by the molecules that occurs alongside self-phase modulation and four-wave mixing in supercontinuum generation. Rotational levels of the molecule are involved in the process during which red shifted photons are generated along with vibrational phonons. Two different Raman cases exist: in the normal dispersion region, two sidebands are generated, whereas in the anomalous dispersion region intrapulse Raman scattering causing solitons to continuously shift to longer wavelengths. A characteristic Stokes shift of 13 THz is observed in Raman scattered photons in silica material.

### 4.2.4    *Solitons in supercontinuum generation*

### *Fundamental soliton*

In case of $L_D = L_{NL}$, a soliton is formed from interplay of self-phase modulation and group velocity dispersion. Solitons can only form in the anomalous group velocity dispersion spectral region since negative dispersion is needed to counterbalance the effect of self-phase modulation; in the normal dispersion spectral region the pulse rapidly spreads out in time and this can be used to stretch pulses temporally. A soliton can be transmitted many kilometres in optical fibres without spreading in time and therefore it is useful, for example, in signal transmission (optical communication). While propagating in optical fibre, solitons experiences a continuous self-frequency shift towards the infrared because of intrapulse Raman scattering. The eventual soliton fate will depend on the pulse peak power and temporal pulse width. If there is sufficient peak power ($L_D >> L_{NL}$ case), a higher order soliton will form. The order $N$ of the soliton is defined through:

$$N^2 = L_D \, / \, L_{NL}$$

eq. 37

Theoretically, propagating higher order soliton should undergo periodic contraction and expansion in spectrum and temporal profile with a period defined as the characteristic soliton length:

$$z_{sol} = \pi \, L_D \, / \, 2$$

eq. 38

The pulse duration of higher order solitons is shortest when the soliton spectrum is widest. However, in practice, this periodic behaviour is easily disturbed by other non-linear processes leading to the pulse split-up, within its first soliton period, to fundamental solitons.

### *Soliton fission*

The perturbation splitting the higher order soliton could be higher order dispersion or Raman scattering. Which of the two dominates depends mostly on the input temporal width. For pulses with broader spectra (shorter temporal width), the dispersive splitting will dominate whereas Raman dominates for longer pulses. Figure 34 shows Raman induced third order soliton fission into fundamental solitons. We can identify individual solitons (three of them) in the spectral (a) and temporal (b) domain (Figure 34). The generated solitons differ in power and temporal width [252]. However, the picture above shows an ideal case of soliton fission that is not influenced by higher order dispersion. In the case of higher order dispersion, a soliton might transfer energy in the normal group velocity dispersion region to a dispersive wave that has a narrow band resonance and is group-delay matched to the soliton.

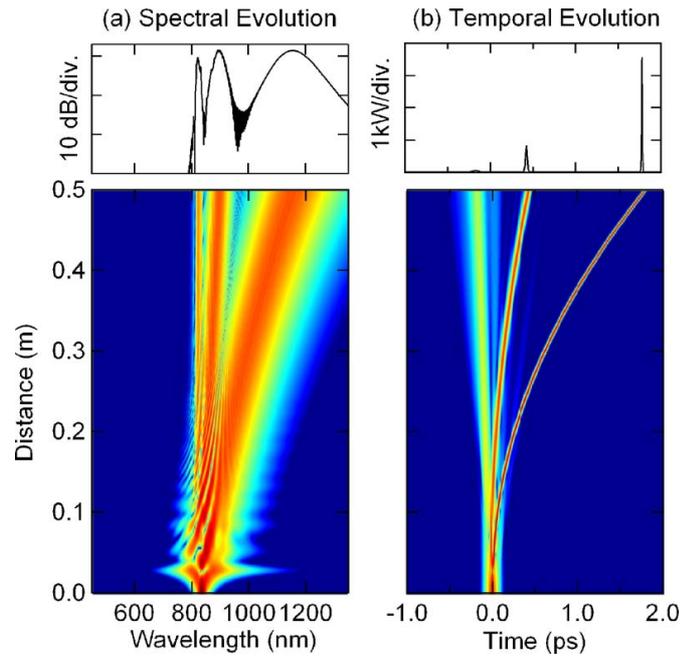

Figure 34. Results from numerical simulations showing spectral and temporal evolution for Raman induced fission of an incident N=3 soliton as described in the text. Top curves show the output profiles after 0.5 m propagation. Figures taken from [16].

As the solitons themselves are shifted to longer wavelengths, similar to the behaviour shown in Figure 34, the corresponding dispersive wave is 'blue' shifted. This will be further explained with Figure 37. Two other useful characteristic distances can be defined: a point at which soliton fission happens:

$$L_{fission} \sim L_D \, / \, N$$

eq. 39

A distance over which soliton features becomes apparent, which is called the separation length, $L_{sep}$:

$$L_{sep} \sim 5L_D$$

eq. 40

### 4.2.5 Supercontinuum generation as a function of laser parameters

A critical aspect in the supercontinuum generation is characteristics of the pump laser. As already discussed, the microstructured optical fibre has to be manufactured appropriate with properties to match the available pump lasers, e.g. an ultrashort pulse mode-locked Ti:Sapphire laser to generate a broad supercontinuum output. The most important laser parameters are wavelength, power and temporal pulse width (if the laser is operating in a pulsed mode). The supercontinuum generation can be differentiated into two clearly distinct operational regimes according to whether the pump wavelength falls into the normal or anomalous dispersion spectral region. Regarding temporal pulse width, two classes can also be distinguished: sub-picosecond pulses (femtosecond) or longer (picosecond, nanosecond, cw).

### *Pump wavelength*

The pump wavelength, or rather its position with respect to the zero-dispersion wavelength, will determine how broad a supercontinuum will form. This is due to the fact that a broad supercontinuum is formed only when pumping in the anomalous dispersion region, either through soliton fission (in sub-picosecond case) or modulation instability (for longer pulses).

### *a)  Pumping in anomalous dispersion region*

Simulations and experiments show that broader supercontinuum is generated when the pump wavelength is longer than the zero-dispersion wavelength. The sub-picosecond case is illustrated in Figure 35.

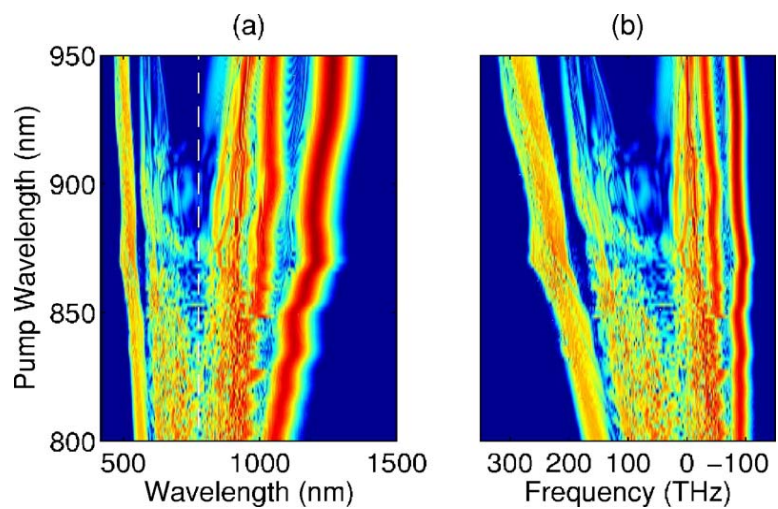

Figure 35. The variation in supercontinuum spectral density with pump wavelength (left axis) shown in density plot representation. The input pulse peak power is 10 kW and duration (FWHM) is 50 fs for all cases. The supercontinuum spectra are plotted as a function of (a) wavelength and (b) relative frequency. To facilitate comparison between these curves, the frequency axis in (b) is reversed. The dashed line in (a) shows the zero-dispersion wavelength. Figures taken from [16].

Figure 35 (b) shows the spectra with respect to frequency relative to the zero-dispersion wavelength in order to show that the blue part of supercontinuum is dependent on pump wavelength and shorter supercontinuum wavelengths are generated for longer pump wavelengths, whereas longer wavelength supercontinuum generation does not depend on the pump wavelength. This is unexpected since the soliton number decreases for longer pump wavelengths and therefore narrower supercontinuum should be generated. However, $L_{sep}$ (defined in eq. 40) scales linearly with $L_D$ whereas the soliton number scales with a square root dependence and $L_D$ is reduced for longer pump wavelengths. This leads to the soliton number being smaller for the longer wavelengths with distinct features appearing at shorter fibre lengths but the net effect is broader supercontinuum for longer pump wavelengths. This is not the case for longer pulses.

### *b)   Pumping in normal dispersion region*

If fibre is pumped in the normal dispersion spectral region then, for femtosecond pulses, self-phase modulation will dominate the broadening. If the pump wavelength is close enough to the zero-dispersion wavelength, the radiation can leak into the anomalous dispersion region and, provided there is enough power, can form higher order solitons that can split and generate supercontinuum extending to the blue. In case of longer pulses, the broadening mechanism could be four-wave mixing or Raman scattering, or both depending mainly on the pump wavelength relative to the zero-dispersion wavelength as explained in the next section.

### *Pulse temporal width*

The temporal pulse width dependence was partly discussed in the preceding section. In general, different nonlinear mechanisms dominate according to the pulse temporal length. For short pulses soliton fission / dispersive wave generation dominates whereas for longer pulses the main processes are – modulation instability / four-wave mixing.

### *a)   Pumping with femtosecond pulses*

In case of femtosecond pulse pumping, two different regimes can be considered discussed: normal and anomalous pumping. When pumping in the normal group velocity dispersion spectral region, self-phase modulation will dominate the broadening, and lead to temporal broadening that will limit any nonlinear interactions. Soliton dynamics are responsible for spectral broadening in the anomalous pumping regime where the supercontinuum exhibits broader spectra. The soliton splits at the point where its periodically cycling spectrum is broadest and drives the gain of the blue non-dispersive wave generation. Similarly to explanation of supercontinuum dependence on wavelength, the temporal dependence can also be explained through soliton number $N$ and soliton separation distance $L_{sep}$. $L_D$ is reduced for shorter pulses and therefore $N$ and $L_{sep}$ are also reduced. This leads to broader supercontinuum for longer pulses.

### b)   Pumping with picosecond or longer pulses

For picosecond or longer pulses, four-wave mixing / modulation instability dominates the spectral broadening and generally three different supercontinuum regimes exist [253]:

$$\lambda_{pump} << \lambda_0, \qquad \lambda_{pump} <\approx \lambda_0, \qquad \lambda_{pump} > \lambda_0$$

eq. 41

In case of $\lambda_{pump} > \lambda_0$, it is modulation instability that dominates the initial stages of supercontinuum generation. Modulation instability is a phenomenon where a cw radiation splits into the train of periodic pulses thus forming solitons [254]. In the spectral domain it is seen as appearance of two sidebands around the input wavelength. The sidebands, however, are initiated from noise and are not coherent therefore a supercontinuum generated this way is noise dominated and also not coherent. The process also occurs for longer pulses. The scale length, $L_{MI}$ characterising modulation instability can be written as:

$$L_{MI} \sim 4 / \gamma P_0$$

eq. 42

The condition $L_{MI} << L_{fission}$ marks the regime where modulation instability dominates over soliton fission since it develops too quickly for soliton fission to occur. Figure 36 shows a phase matching curves calculated from eq. 35 and eq. 36 at three different power levels for microstructured optical fibre (SC-5.0-1040, Crystal-Fibre) used in this thesis. As follows from eq. 35, the curves should change with power as is seen in the Figure 36.

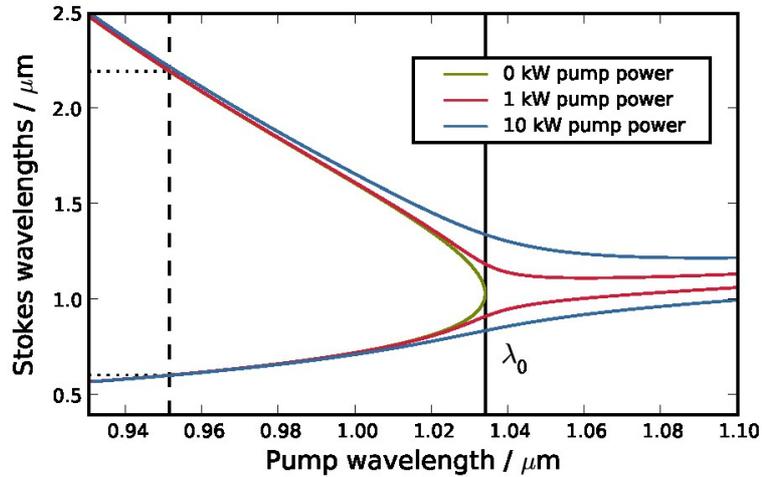

Figure 36. Phase matching curves for four-wave mixing in the fibre used in this thesis (SC-5.0-1040) as pumped by the picoseconds fibre laser for different pump power. Figure taken from [255].

When the difference between the pump wavelength and the zero-dispersion wavelength is small ($\lambda_{pump} \lesssim \lambda_0$) then the four-wave mixing dominates the supercontinuum generation. In this case, phase matching conditions will decide which new sidebands will be generated [256]. If the pulses are close to the zero-dispersion wavelength, then generated sidebands can be widely separated therefore this can be used as widely tunable parametric generator [257]. If the pump and the zero-dispersion wavelength difference is very large and the pump is in the normal dispersive region ($\lambda_{pump} << \lambda_0$), then the Raman scattering dominates since it follows from the phase-matching curve that the red part of the signal (Stokes line) should be suppressed because of the water absorption window, which in turn halt the competing four-wave mixing process.

### *Spectrograms*

As we can see from Figure 34, soliton dynamics can be very complex and sometimes not straightforward to interpret it from spectral or temporal profiles alone. However, if the spectral and temporal information is combined in a two-dimensional plot, additional insight into the supercontinuum evolution can be obtained. Figure 37 shows two spectrograms of supercontinuum generation in two different microstructured optical fibres that were used in this thesis (and pumped with two different lasers).

Visual examination of the plots leads to recognition of solitons as elliptical shaped objects (especially in Figure 37 (a)) with the dispersive waves below them. The spectrograms show that solitons in both pictures are Raman shifted, and that they are phase matched with dispersive waves. In addition the parabolic shape of the structure gives information about the dispersion.

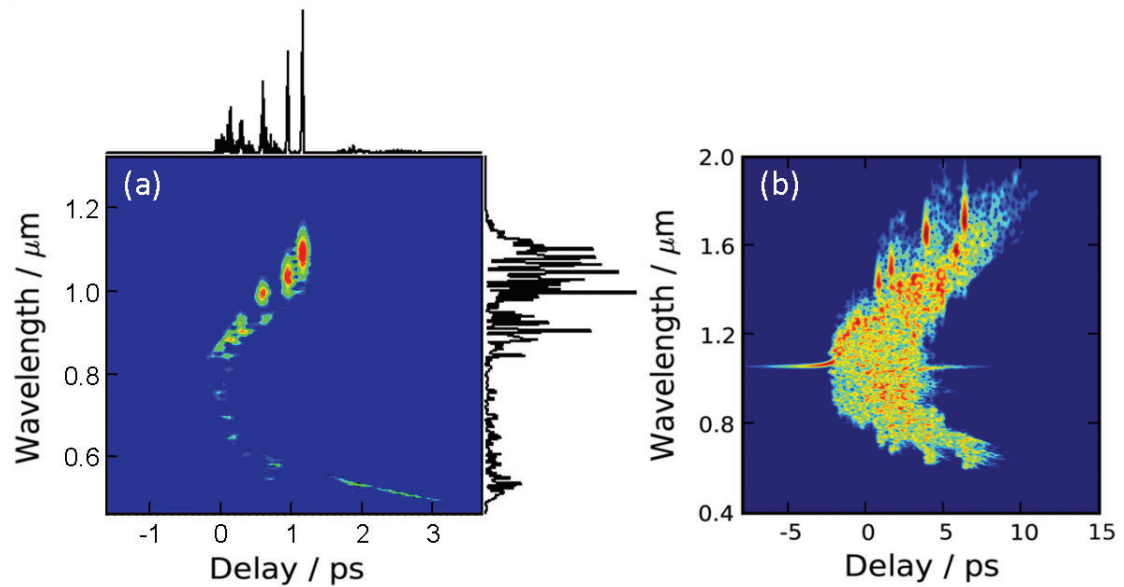

Figure 37. Simulated spectrograms of the supercontinuum generation initial stages when the governing mechanism is: (a) soliton dynamics and (b) modulation instability. (a) Supercontinuum in the beginning of the NL-740-2.0 fibre as pumped by femtosecond pulses from Ti:Sapphire laser operating at 830 nm. Figure taken from [258]. (b) Supercontinuum in the beginning of the SC-5.0-1040 fibre as pumped by the picoseconds pulses at 1.06 μm from the fibre laser. Figure taken from [255]. Both fibres were used in this thesis.

### *Supercontinuum coherence*

The coherence across a supercontinuum spectrum will depend on the laser and fibre parameters discussed above, but in general, shorter pulses favour more spectrally coherent output and pump wavelengths close to the zero-dispersion wavelength give worse coherence [259-261].

### *Different laser sources used*

Different laser sources can be used to pump microstructured optical fibre such as mode-locked Ti:Sapphire and fibre lasers, and Q-switched Nd:glass lasers. They can work in different temporal regimes: femtosecond, picoseconds and nanosecond respectively. Ti:Sapphire lasers were the first to be used for pumping the microstructured optical fibre due to their widespread availability at the time when the first microstructured optical fibres were manufactured and because they could provide pump wavelengths in the regions where soliton dynamics could be initiated in the those fibres [229]. Another solid state laser – Nd:YAG microchip nanosecond laser was also explore due to its low price and small size [262, 263]. Mode-locked fibre lasers have become sources of choice primarily because of much simpler handling and operation and also because of the higher average powers available [264]. It is worth noting that cw fibre lasers can also be used for supercontinuum generation with the output as high as 29 W [265]. Supercontinuum extending to 650 nm have been demonstrated with 50 W Ytterbium fibre laser [266, 267].

## 4.3    Application of supercontinuum generation to fluorescence microscopy

### 4.3.1    *Application of Ti:Sapphire laser pumped supercontinuum source*

Here the application of supercontinuum as an excitation source to the Nipkow disc microscopy is presented. Supercontinuum was generated in microstructured optical fibre pumped by Ti:Sapphire laser. The microscope was able of acquiring wide-field time-gated fluorescence lifetime images. The work presented here was published in Ref. [268]. The Author's input here was the characterisation of microstructured optical fibre.

<u>*Supercontinuum generation in microstructured optical fibre*</u>

The Ti:Sapphire laser is usually operated in femtosecond regime with the microstructured optical fibre being pumped in its anomalous dispersion region so that the supercontinuum generation is dominated by the soliton dynamics as discussed in Section 4.2.4. A short length of the microstructured optical fibre was used because the femtosecond pulses provided by the laser are comparatively spectrally broad and therefore leads to the short dispersion length, $L_D$, (eq. 32), which in turn enables soliton dynamics to evolve over the shorter length of the fibre (see eq. 39). A Ti:Sapphire (Spectra-Physics) laser, operating at 790 nm (76 MHz, 100 fs) was coupled into 75 cm of microstructured optical fibre with the zero-dispersion wavelength at 740 nm (NL-740-2.0, Crystal Fibre), as shown in Figure 38. An isolator consisting of the Faraday rotator and two polarisers prevented back reflections of the laser radiation from the tip of the fibre and other optical elements going back to the laser and compromising the mode-locking.

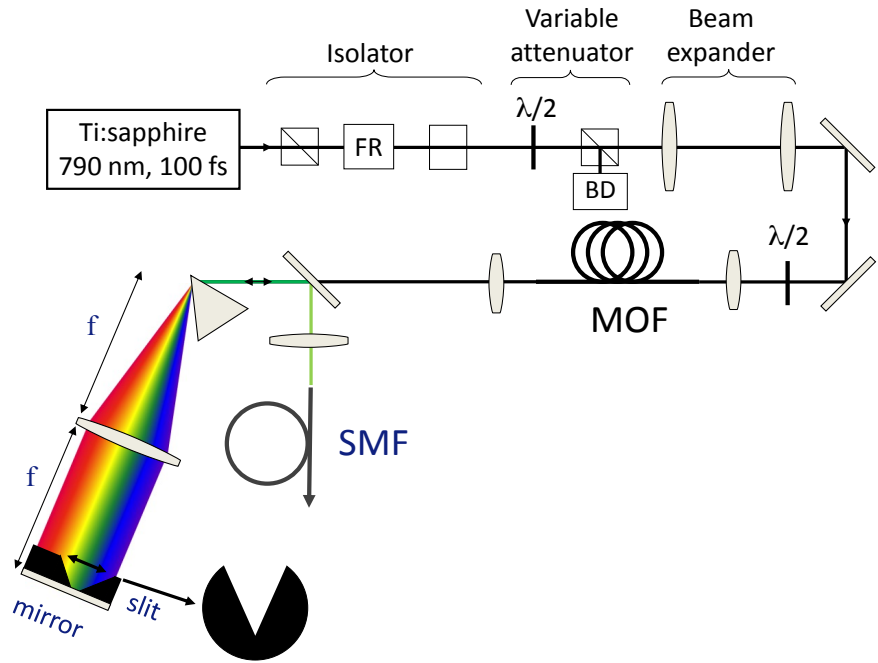

Figure 38. Supercontinuum generation setup with Ti:Sapphire laser and microstructured optical fibre (NL-740-2.0, Crystal Fibre).

Radiation from the supercontinuum was spectrally selected with a spectral selection kit consisting of a prism, a lens and an electronically tunable slit with a mirror behind it. The prism disperses supercontinuum, which is then collimated with the lens placed at the distance equal to its focal length from the prism. The collimated light is reflected by the mirror, in front of which is placed a V-shaped slit [220] to let through the light only of a particular spectral region. The light goes back through the lens and prism arrangement. The retro-reflecting mirror is slightly tilted so that the light can be picked off with a mirror in front of the prism and a tunable spatially coherent output beam is thus produced. Supercontinuum spectra were measured as a function of pump wavelength, as shown in Figure 39, illustrating how shorter wavelengths are generated if longer pump wavelengths are used, as discussed previously. The spectrogram shown in Figure 37 (a) illustrates that supercontinuum development expected in the beginning of the microstructured optical fibre that was used here. We can see an apparent structure of solitons showing that supercontinuum generation mechanism is dominated by soliton dynamics as expected. Figure 39 (b) and (c) shows pulse-to-pulse noise dependence versus pump wavelength. It shows that if the pump wavelength is detuned from the zero-dispersion wavelength towards the infrared, the supercontinuum generation becomes noisier because of modulation instability starting to dominate the initial stages of supercontinuum generation. The supercontinuum is least noisy when it is pumped in the normal dispersion region (710 nm). Similarly, the blue part of the supercontinuum is noisier because it is detuned furthest away from the pump wavelength.

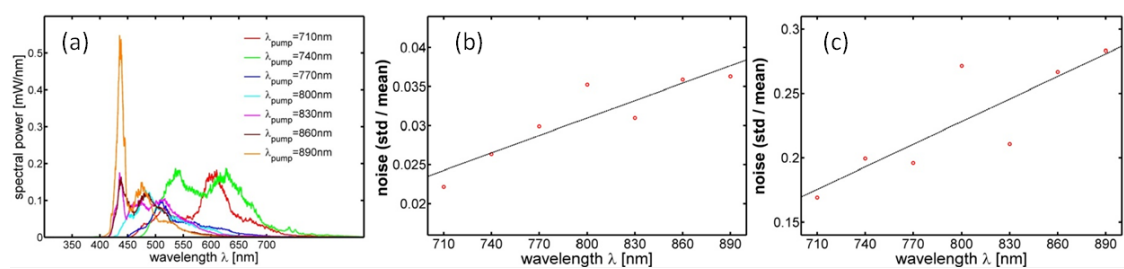

Figure 39. Characterisation of the microstructured optical fibre (NL-740-2.0). (a) Spectral power density spectrum as a function of pump wavelength (as provided by Ti:Sapphire laser). Pulse to pulse noise, as integrated over all supercontinuum spectrum (b) and as integrated in the region of 550 nm ±20 nm (c), as a function of the pump wavelength.

### *Nipkow disc fluorescence lifetime imaging microscopy setup*

Ti:Sapphire laser output (800 mW, 100 fs, 860 nm) was used to generate supercontinuum with an average power of 220 mW. Figure 40 shows the experimental setup of the microscope incorporating supercontinuum generation system (Figure 38) and the Nipkow disc FLIM system. The inverted Nipkow disk microscope was implemented on an Olympus IX71 frame. The supercontinuum was spectrally filtered (430-470 nm) and coupled into a single-mode fibre connected to the fibre input port of the Nipkow disk unit (Yokogawa Electrical Corporation, CSU10). Inside the unit, the excitation light was expanded to overfill a microlens array that focused the light through the array of pinholes, producing 20,000 point sources that were imaged onto the sample. Both lenslet and pinhole arrays disks were spun at 30 rotations per seconds and 12 complete images were swept out per rotation, providing wide-field images of up to 360 Hz. The resulting fluorescence was imaged back through the pinhole array and directed via a dichroic beam splitter located between the lenslet and the pinhole disks, to a wide-field detector – either a CCD camera for intensity imaging or a gated optical intensifier (GOI) for FLIM.

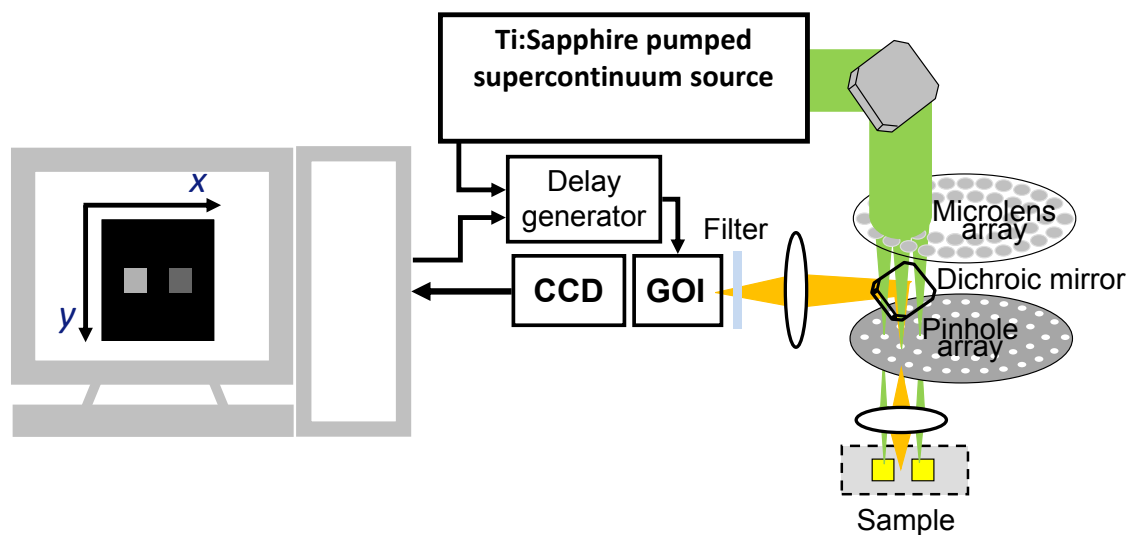

Figure 40. Nipkow disc microscope setup incorporating supercontinuum generation unit (Figure 38).

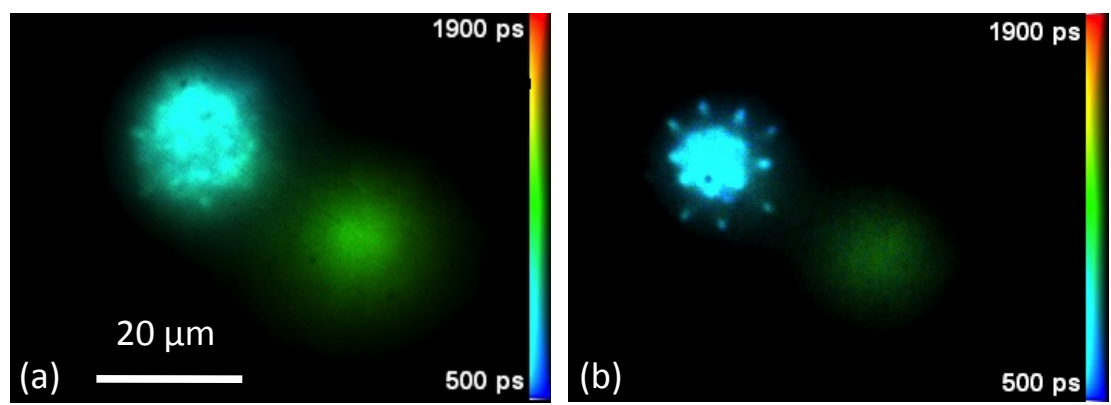

Figure 41. FLIM images of pollen grains as taken with the conventional (a) and sectioning (b) microscope employing Nipkow disc unit. Excitation: 450 / 40 nm, detection: > 490 nm. Acquisition time – 4 s.

Optical sectioning is important for FLIM since out of focus light not only degrades the spatial resolution but can also degrade the lifetime contrast. This is apparent in Figure 41, where FLIM images of stained pollen grains were acquired using the time-gated imaging described in Section 3.3.1. The image on the left shows 'conventional' unsectioned FLIM image and the one on the right – its Nipkow-sectioned counterpart. The sectioning image shows increased spatial information and therefore enhanced fluorescence lifetime contrast between the different pollen grains, due to the reduction of out-of-focus light contributions. This system was later modified by incorporating a more powerful excitation source to allow high speed optically sectioned FLIM [269].

### 4.3.2    Application of fibre laser pumped supercontinuum source

Fibre lasers were rapidly adopted as a pump sources following the initial Ti:Sapphire laser based setups for supercontinuum generation, owing to their advantages over solid state lasers. Here, it was used as an excitation source for wide-field FLIM microscopy as described below. Author's input was to build and characterise the supercontinuum generation setup.

#### *Fibre laser*

The fibre laser used here was developed at Imperial College London [264] and commercialised by IPG Photonics (Oxford, Massachusetts, USA). The laser (YLP-8-1060-PS) operates at 1.06 μm and delivers picosecond pulses with up to 8 W of average power (peak power up to 50 kW), at 51 MHz repetition rate.

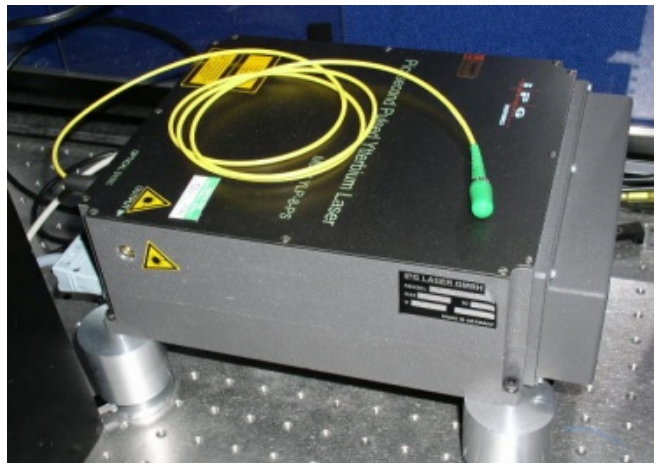

Figure 42. Photo of the fibre laser (YLP-8-1060-PS, IPG).

This fibre-based laser (Figure 42) cost around ~ £ 17 k, which is considerably cheaper than a femtosecond Ti:Sapphire laser and it is much smaller (25 cm × 15 cm × 10 cm) with a turnkey operation. A more detailed explanation of operation principle of the laser can be found in [264], but briefly the source is an all-fibre integrated, comprising a seed source and fibre amplifier, as shown in Figure 43. The seed source is fibre ring laser incorporating Ytterbium doped fibre amplifier, fibre isolator, and two fibre polarisation controllers (PC) with fibre polariser in between of them, as shown in the figure. The seed source generates a train of pulses through stable self-starting passively mode-locked operation. The mode-locking was based on nonlinear polarisation evolution [233] with the fibre polariser acting as the discriminating element. The radiation is amplified in Ytterbium doped fibre pumped by a diode laser operating at 960 nm. The pumping is delivered with multimode fibres through a fused fibre coupler. The fibre ring is of 3 m length, producing a pulse train of 51 MHz repetition rate. The pulsed radiation is coupled out of the seed source with another fused fibre coupler.

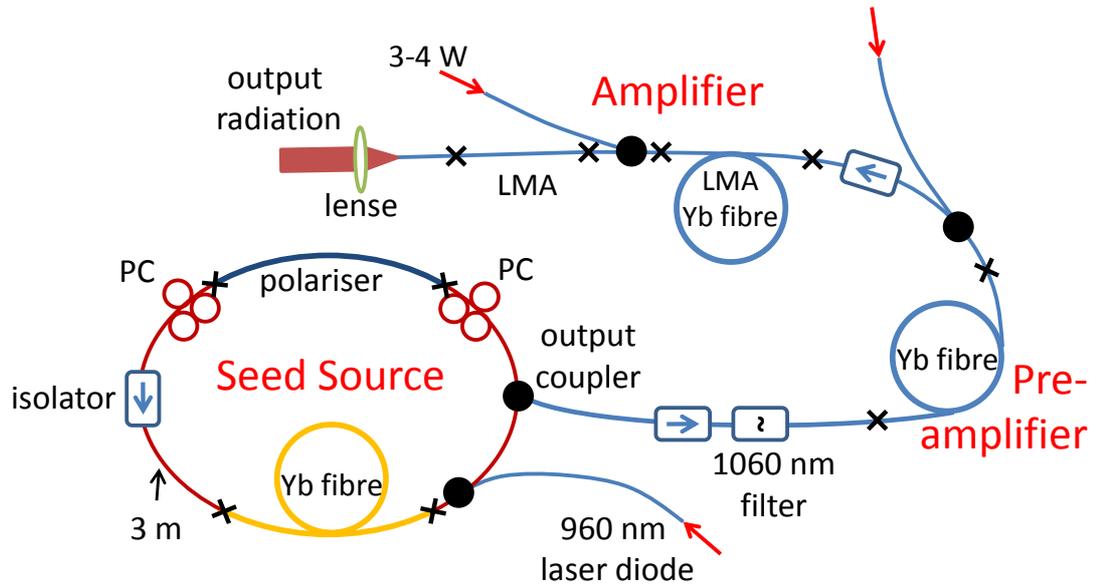

Figure 43. Schematics of the fibre laser: PC, polarisation controller; Yb, Ytterbium; LMA, large mode area.

The output is chirped, due to the normal dispersion of the fibre ring combined with self-phase modulation, therefore a filter at 1.06 µm is used to reduce the bandwidth and the temporal width of the pulses. The seed pulses are amplified in a two stage amplification process: first they are pre-amplified in 1 m Ytterbium doped fibre pumped by a single diode laser and then amplified further in 1.5 m of low non-linearity large mode area (LMA) Ytterbium doped fibre, where 3-4 W in total of diode laser pump power can be used to amplify the light. The pulse duration and spectral bandwidth increase during the amplification process due to self-phase modulation in the fibres and this spectral broadening is power dependent. The polarisation of light coming out of the fibre laser at high amplifier power is complex and is linear only at low power, therefore limiting its application where polarisation control is necessary. The spectral bandwidth increased to ~ 40 nm at the maximum power as shown in Figure 44. The power dependence of the spectral output is shown as a function of diode laser current drawn in the 2nd amplifier, which is not linear, as shown in Figure 44 (b).

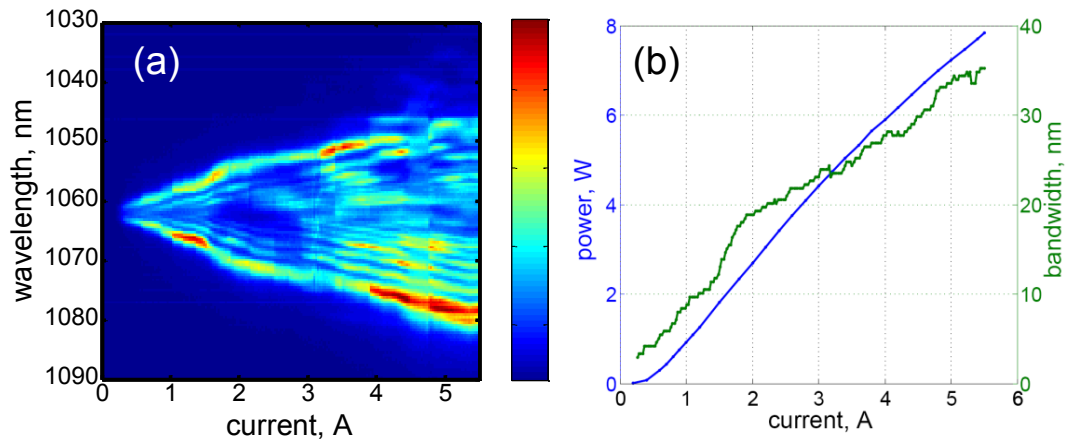

Figure 44. (a) Fibre laser output spectrum evolution with output power. (a) Normalised spectrum dependence on the laser current drawn in the amplifier. (b) Output power and spectrum bandwidth (full-width at 1 / exp of maximum) dependence on the current.

The broad spectrum at high power made it impossible to measure the temporal pulse length with second harmonic generation autocorrelation but, assuming dispersion of ~ 20 ps / nm × km, it was calculated to increase up to ~ 4 ps at maximum power [264]. The output beam diameter was determined to be of 2 mm (FWHM) from knife-edge scanning measurements across the beam.

### *Supercontinuum generation microstructured optical fibre*

Microstructured optical fibre (SC-5.0-1040, Blaze Photonics) with the zero-dispersion wavelength at 1.04 μm was used for supercontinuum generation with the Ytterbium fibre laser. 20 meters of the fibre was chosen, following work in Ref. [264], in order to generates supercontinuum with shortest wavelength. It was reported there that for the fibre lengths beyond 20 m the supercontinuum do not develop below 525 nm but rather is scattered out. The zero-dispersion wavelength was chosen to match closely the fibre laser emission wavelength since it was expected that the modulation instability should play here an important role and the broadest supercontinuum are formed when the pump is close to the zero-dispersion wavelength, as explained above. The fibre had these parameters: Λ = 3.36 μm and d / Λ = 0.47 as calculated from the scanning electron microscopy images of the microstructured optical fibre cross-sections, shown in Figure 45 (a). At the pump wavelength of 1.06 μm, the dispersion of the fibre was 5 ps / nm × km, as can be seen from dispersion curve in Figure 45 (b), and the nonlinear coefficient $\gamma = 13\ 1\ /\ (W \times km)$. The comparatively small dispersion and large nonlinearity of this fibre favours modulation instability rather than soliton fission, as follows from eq. 39 and eq. 42. The fibre had also a small attenuation ($\sim 2$ db / km) at the pump wavelength, as can be seen from Figure 45 (b).

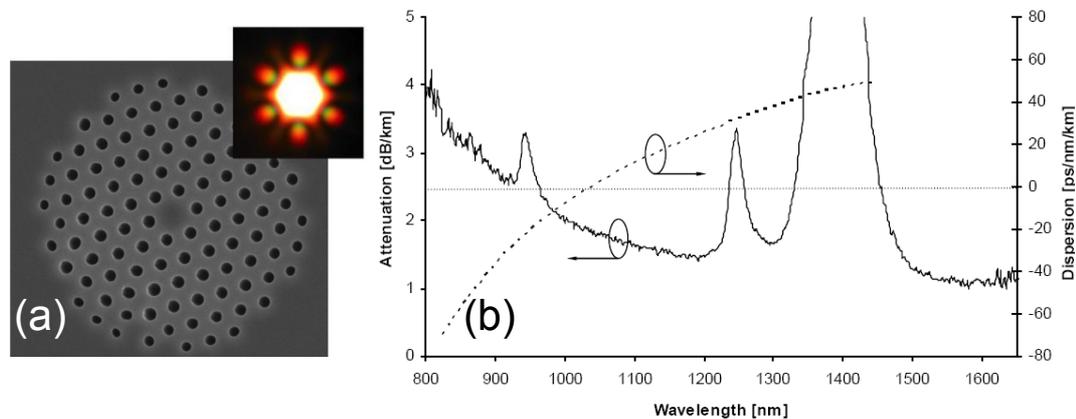

Figure 45. Properties of the microstructured optical fibre (SC-5.0-1040, Blaze Photonics). (a) Scanning electron microscope image (inset: near-field image). (b) Typical attenuation and dispersion spectra of the microstructured optical fibre. Figures adapted from *Blaze Photonics*.

If 1 kW of laser peak power is assumed, then $L_{MI} = 0.3$ m and $L_{fission} = 12$ m, therefore modulation instability will dominate and the extent of the supercontinuum generated will be largely governed by the phase matching curves for modulation instability and four-wave mixing. From the phase matching curve shown in Figure 36, it can be seen that at 1 kW two Stokes components at approximately 1 μm and 1.1 μm will be generated through four-wave mixing out of the pump wavelength at 1.06 μm. The anti-Stokes component (1 μm) will further generate two new Stokes components and so on. The limit to his process is the quenching of Stokes wavelengths by attenuation in the infrared. If a 2.2 μm limit is assumed (due to strong attenuation because of the water absorption at the air-silica interface) on the Stokes wavelength then the phase-matched anti-Stokes components will be at 0.6 μm as can be seen in Figure 36. The optical setup for supercontinuum generation is shown in Figure 46. The laser beam was aligned to be parallel to an optical axis and optical table by the two highly reflective mirrors. The half wave plates changed the light polarisation to be linear and to allow efficient coupling through the isolator. The beam splitters and Faraday rotator formed an isolator, which blocked any back reflected light from the reflecting optical components (especially from the microstructured optical fibre). The transmission of the wave plates and isolator optics was 77 %. The beam was focused by an aspheric anti-reflective coated lens (f = 5 mm, L1 in Figure 46) into the end of the microstructured optical fibre with coupling efficiency of 33 %.

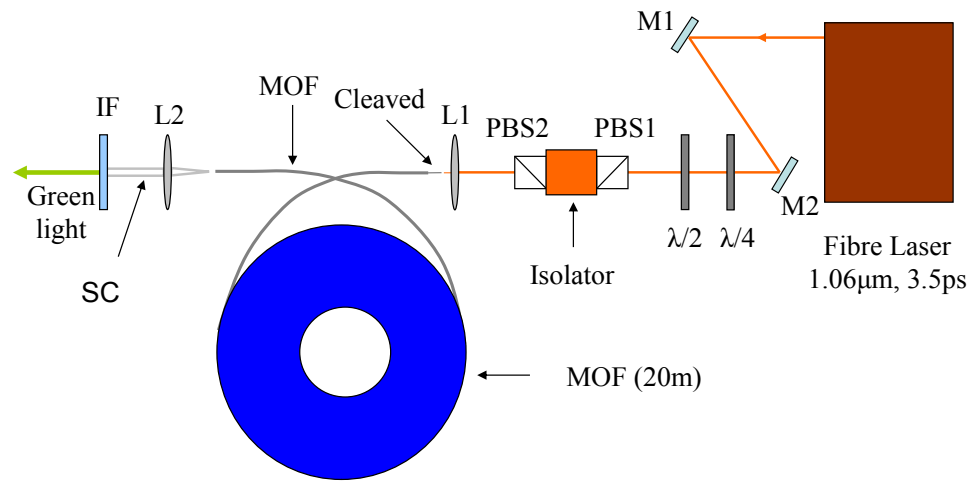

Figure 46. Supercontinuum generation experimental setup. M1-2, infrared coated dielectric mirrors. λ / 2, λ / 4 - quarter and half wave plates respectively. PBS1-2, polarizing beam splitters. L1, aspheric lens. L2, microscope objective (Nikon, M Plan x40, NA 0.65). MOF, microstructure optical fibre. IF, Interference filter – 550 nm / 50 nm.

The end of the fibre was stripped and the tip of the fibre was cleaved to make the tip perpendicular to the incoming beam. Cleaving also removed dust from the fibre since it cannot be cleaned by regular chemicals due to possible capillary effect along the holes of microstructured optical fibre. The end of the fibre was fixed on a 3-axis translation stage (not shown in the figure) to facilitate fine light coupling. The stage had piezo drivers on each translation axis that were electronically controllable through a 3-axis piezo controller that permitted alignment on tens of nanometres scale. The translation stage was aligned to be parallel to the laser beam by a pinhole-mirror device. As the emission power of the fibre laser was increased, by increasing the pump current, the spectrum (as measured with a fibre coupled spectrometer (Ocean Optics, ADC1000-USB))) of the beam coupled out of the microstructured optical fibre, started to broaden due to the non-linear processes, as shown in Figure 47 (a). The single mode fibre guiding maintained the spatial coherence of the laser light as was evident from looking at the near field output of the fibre. The spectrometer was not sensitive beyond 1100 nm; therefore spectra broadening cannot be seen in the infrared region. The output power increased with increased input power as shown in Figure 47 (b).

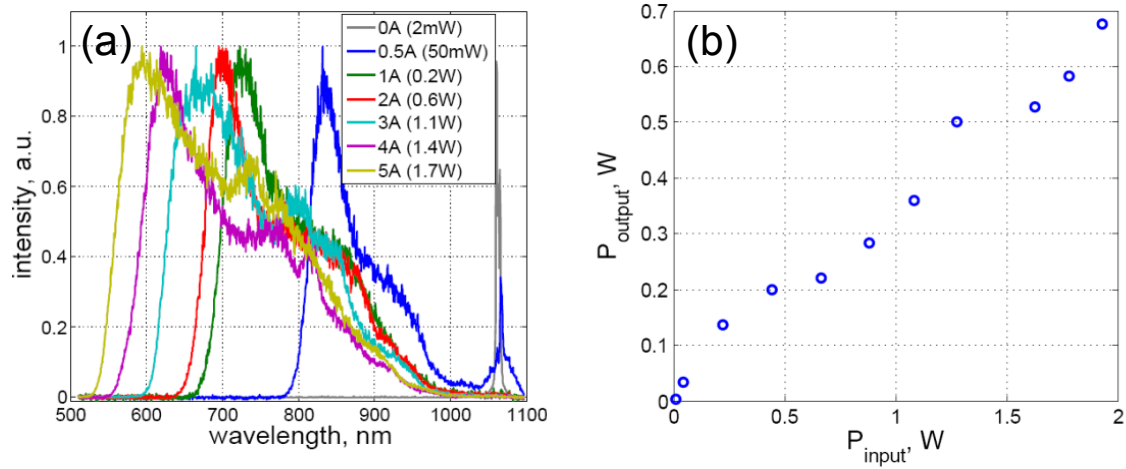

Figure 47. Supercontinuum generation with the fibre laser. (a) Supercontinuum spectra as a function of coupled-in power, $P_{input}$ (specified in brackets). The spectra are normalized to unit. (b) Generated supercontinuum power, $P_{out}$ as a function of input power, $P_{input}$.

Fluctuations in the output power and spectrum were observed at high pump powers, perhaps due to the thermal and other effects. In general, adjusting the position of the fibre on the stage through piezo drivers was normally sufficient to get back to the same output, but this was not always possible at the highest pump powers. The generated continuum was collimated with objective (L2 in Figure 46) and could then be filtered to get a particular wavelength range for use as an excitation source in a microscope. At the highest pump powers, blue light is observed leaking out in the first 40 cm of the fibre, as can be seen in Figure 48 (a). This is perhaps, because of second harmonic generation taking place [264] in the earlier stages of the nonlinear interaction in the microstructured optical fibre. However the fibre did not appear to efficiently support the guiding of blue light and a significant fraction therefore leaked out. The fact that supercontinuum reaches 525 nm, as seen in Figure 47, indicates that four-wave mixing, as dictated by the phase matching curves in Figure 36, is not the only nonlinear process taking place and other processes can participate in the supercontinuum generation. These could include cross phase modulation and soliton dynamics as shown in Figure 37 (a) where we can observe that modulation instability initiated spectral broadening evolve into a train of solitons. We can also observe that coupling between red solitons and the blue part of the supercontinuum through cross phase modulation or dispersive wave trapping [270] lead to new blue spectral component generation. As discussed earlier, the noise is expected to dominate the initial stage of this supercontinuum, as can also be seen in the spectrogram in Figure 37 (b) and therefore the generated supercontinuum will not be spectrally coherent. The fibre laser of 2 W average power was used to generate supercontinuum of 0.6 W average power (spectrum shown in Figure 48 (b)). Filtering with 550 / 20 nm filter reduced it to 5 mW.

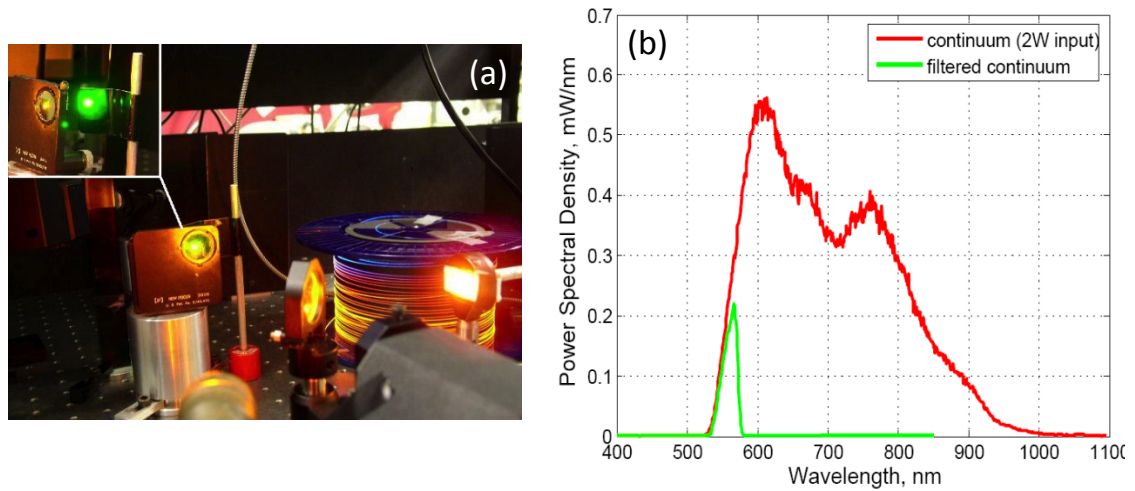

Figure 48. (a) Supercontinuum generation in microstructured optical fibre (inset: filtered supercontinuum). Blue light is not efficiently guided in the fibre and therefore leaks out (the top of the fibre reel). (b) Generated supercontinuum as pumped with 2 W of average power.

### *Wide-field fluorescence lifetime imaging microscopy setup*

The microscope experimental set up used here was similar to the one explained and used in Ref. [271].

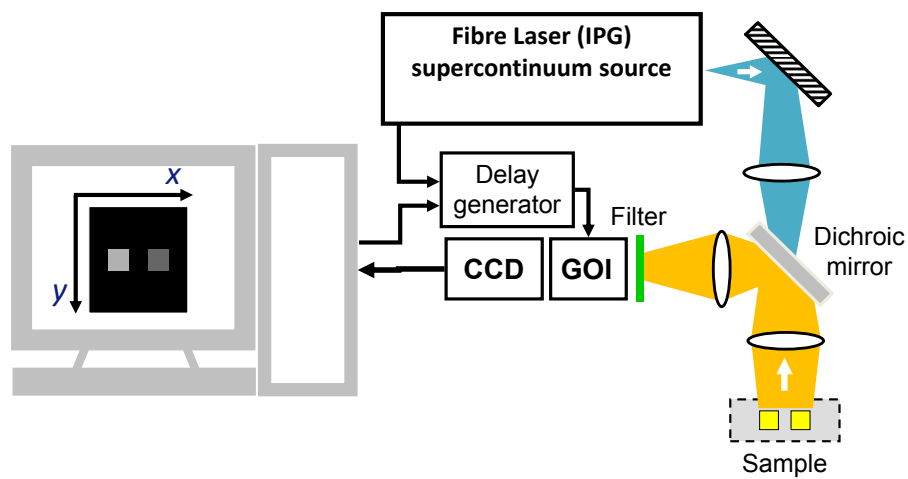

Figure 49. Wide-field time-gated FLIM setup comprising of supercontinuum source and inverted Nipkow disc microscope with the time-gated detection.

Filtered supercontinuum, after passing through a rotating diffuser to remove any coherent artefacts, was coupled into an inverted microscope (Olympus IX71) to provide Köhler illumination. An oil-immersion objective ( × 40, NA = 1.3) was used for the imaging. Sample fluorescence was recorded using gated optical intensifier (explained in Section 3.3.1) the output of which was coupled onto a CCD (ORCA-ER, Hamamatsu, Japan) via an optical relay, as shown in Figure 49. The gate-width of the gated optical intensifier was set to 800 ps and the time between gates was 400 ps, giving 15 gates that were then used to calculate the final lifetime image with home written LabView software by fitting single exponential decay. Figure 50 demonstrates that the system is capable of recording fluorescence lifetime images.

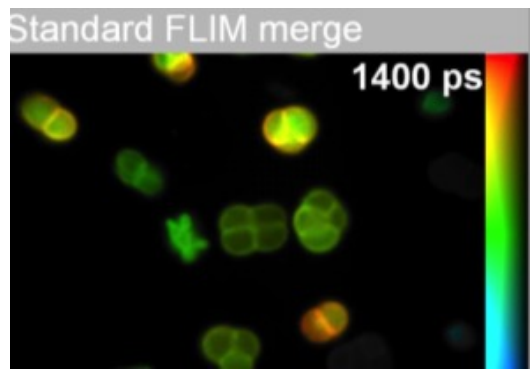

Figure 50. Fluorescence lifetime intensity-merged image of stained pollen grains as acquired with the wide-field time-gated FLIM microscopy setup. Image size – 100 μm across.

### 4.3.3   *Application of supercontinuum generated in tapered microstructured optical fibre*

In the previous section it was discussed that the blue part of the supercontinuum generated with picoseconds pulses was limited by the Stokes losses on the red edge of the spectrum [255], even though the supercontinuum extended further in the blue than four-wave mixing alone would allow. One can expect that optimising phase-matching curves for four-wave mixing could further push the 'blue' part of the supercontinuum towards the ultraviolet. Fibres with smaller cores and therefore with lower zero-dispersion wavelength, provide the phase matching conditions to allow blue anti-Stokes components to be created with Stoke components that are not quenched by the fibre losses. The shortest blue components can be generated in fibres with smallest core diameter and high d / Λ ratio. However, the core cannot became too small since the second zero-dispersion wavelength then comes too close to the visible region and starts limiting supercontinuum extending in the infrared and in turn limits supercontinuum extending to the shorter side of spectrum. A diameter of around 2 μm is found to be the optimum for the shortest wavelength generation [255]. However, the zero-dispersion wavelength of such a fibre is too far from Ytterbium fibre laser pump wavelength to allow efficient four-wave mixing to occur with anti-Stokes wavelengths in the visible spectrum. Instead the spectrum then broadens to the infrared through intra-pulse Raman scattering. The simple solution to that would be to use two stage supercontinuum generation – generating the supercontinuum in two microstructured optical fibres with the different zero-dispersion wavelength. The first microstructured optical fibre would generate supercontinuum from the 1.06 μm pump and this supercontinuum be used to pump the second microstructured optical fibre with a smaller core (and therefore lower zero-dispersion wavelength), which could have the right phase matching curve to generate short blue components. For example, it was demonstrated that the 732 nm spectral component can be generated with 35 % efficiency through four-wave mixing using 1.06 μm laser to pump normally dispersive fibre [253]. The output of 732 nm could further be used to pump anomalously dispersive fibre to produce components further into the blue through four-wave mixing [272]. Alternatively the first microstructured optical fibre could be an anomalously dispersive fibre that could generate Stokes components through modulation instability which would be further fed into another anomalously dispersive fibre to produce components further into blue through four-wave mixing [273]. The experiment can be carried out by splicing different fibres, as in above examples, or by tapering the properties of one fibre (i.e. by decreasing the core diameter along the fibre). A taper would allow continuous longitudinal change of dispersion and nonlinearity along the fibre compared to the abrupt change of dispersion in the case of two spliced fibres. Tapers can be produced by post processing the fibre or during its manufacturing process. In fact, tapered standard (non-microstructured) fibres can have the same dispersion and nonlinearity properties as the microstructured, leading to equally broad supercontinuum as demonstrated in 2000 [241], right after demonstration of supercontinuum with microstructured optical fibre [229]. However, tapered microstructured optical fibre have more degrees of control and are more robust since they preserve their outer cladding whereas in the case of standard fibre, core and cladding are merged into one and light is lead by the total internal reflection from the glass-air interface. The guidance is therefore very sensitive to the immediate surroundings of the fibre, which can be exploited to sense the environment [274]. Tapering microstructured optical fibre in the fibre drawing tower allows taper lengths of kilometres compared to centimetres in post-processed microstructured / standard fibres.

### *Tapered microstructured optical fibre*

The tapered microstructured optical fibre used here was developed at the University of Bath (Centre for Photonics and Photonic Materials). It was drawn from a 3 mm diameter preform by adjusting the outer diameter of the fibre within the drawing process and keeping the air-filling fraction in the cladding constant ($d / \Lambda$ = 0.7). The scanning electron microscopy of the beginning and the end of the fibre is shown in Figure 51.

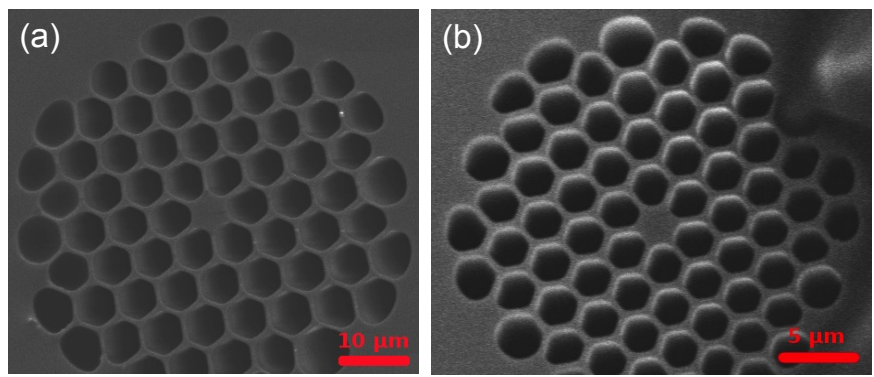

Figure 51. Scanning electron microscopy images of the tapered microstructured optical fibre ends. (a) input, (b) output. Figures taken from [255].

The outer diameter and the core size were proportionally decreasing, therefore it was possible to calculate the core size from the outer diameter. The core diameter changed with the taper length as shown in Figure 52 (UVT1 case). As can be seen from the graph, the core size varied from 6 μm in the beginning to 2.3 μm in the end, with the ratio of $d / \Lambda$ = 0.87 being constant over entire length. Changing core size changed the dispersion curve and therefore the zero-dispersion wavelength position. The new phase matching curves allowed shorter wavelength to be generated, as discussed previously. It was shown that the zero-dispersion wavelength and the shortest wavelength generated in supercontinuum varied as a function of taper length closely following the variation of pitch size, shown in **Figure 52** [255]. As discussed in the previous section, the blue extent of supercontinuum goes further than four-wave mixing alone would define because of the dispersive wave trapping. Moreover, the taper is expected to enhance that effect by preventing solitons spreading out in time while they are experiencing intrapulse-Raman scattering.

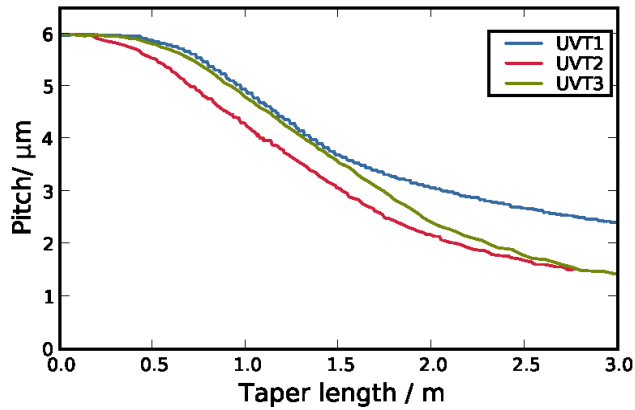

Figure 52. The size of the core of the microstructured optical fibre as a function of its length. Fibre labelled as UVT1 in the grah was used in this thesis. Core diameter is the same as pitch diameter. Figure taken from [255].

In the non-tapered case the intrapulse-Raman scattering would make solitons shift toward infrared wavelengths where the higher dispersion would spread them out in time. This would also reduce the soliton spectral bandwidth, which in turn eventually stops intra-pulse Raman scattering and halts the shift to the infrared and the generation of the dispersive waves further into the blue. By tapering the fibre, the dispersion is reduced continuously along the fibre, enabling solitons to preserve their spectral bandwidth [270]. This means that soliton trapping of dispersive waves is preserved for long fibre lengths. The initial dynamics of supercontinuum in tapered fibre is similar to that of non-tapered shown in Figure 37 (b).

*Supercontinuum generation in tapered microstructured optical fibre*

The setup used for shorter wavelength supercontinuum generation, as shown in Figure 53, was similar to the one described in the previous section, except Faraday isolator and microstructured optical fibre. The polarisation dependant isolator, used in the previous setup was changed here to the polarisation insensitive isolator (PII) (IO-2PI-1064-PBB, Optics for Research (OFR), Caldwell, New Jersey, USA), because it allowed a better power throughput of almost random fibre laser radiation polarisation at high power. Using the polarisation dependant isolator with a randomly polarised source limits transmission to 50 %. The fibre laser does not give a random polarisation but rather elliptically polarised light, which, if used with polarisation dependant isolator, would have transmittance of ~ 60 %.

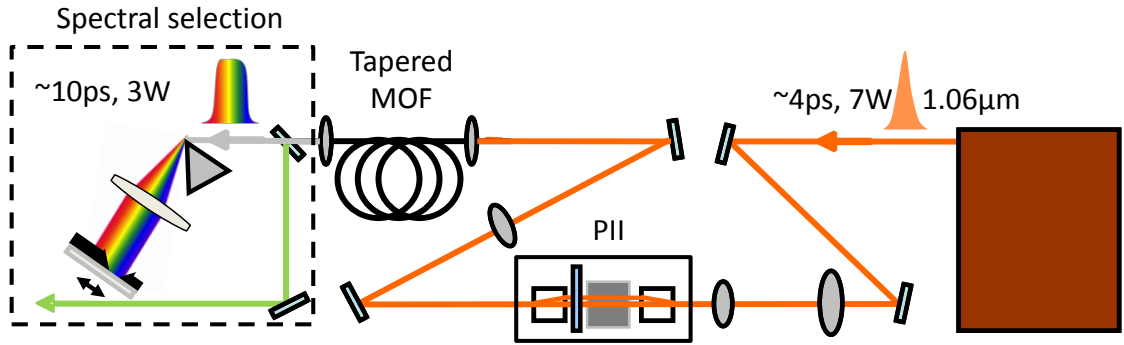

Figure 53. Supercontinuum generation setup consisting of the fibre laser and tapered microstructured optical fibre (MOF). PII – polarisation insensitive isolator. The spectral selection sub-setup is explained in Section 4.3.1.

Transmittance through polarisation insensitive isolator does not vary, regardless of the state of fibre laser polarisation. It works by splitting the incoming light in to two parallel beams with orthogonal polarisations with a walk-off crystal polariser and recombining them with another polariser of this type, after they have passed though the Faraday isolator. The light going backwards (reflected from the tip of fibre, for example), would not be recombined by the polariser and therefore dumped. A half wave plate between the polariser and the Faraday rotator helps to accurately adjust the orientation of two orthogonal polarisations in order to perfectly recombine the two beams.

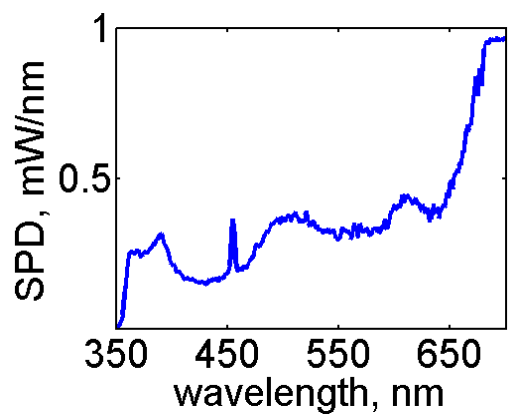

Figure 54. Spectral power density of the supercontinuum generated in the tapered microstructured optical fibre as pumped with the fibre laser.

Approximately 87 % (6.7 W) of the total fibre laser output (7.7 W) was incident on the tapered microstructured optical fibre (after the polarisation-insensitive isolator), $<\sim$ 70 % (4.7 W) of which was coupled into the fibre. A maximum of 3 W of supercontinuum radiation could be coupled out of the fibre (the other part is scattered along the fibre as can be seen in Figure 55), in the range of 0.35 µm - 2.0 µm, as shown in Figure 54 (ultraviolet-visible part of the spectrum). This radiation could be spectrally selected in one or, in principle, multiple channels using the tunable spectral selection setup, as shown in Figure 53.

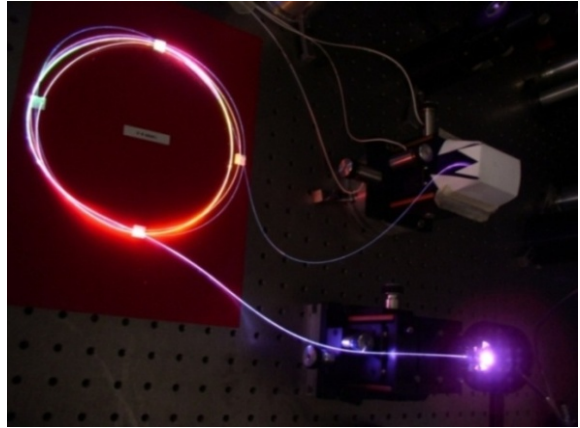

Figure 55. Picture of the supercontinuum being generated in the tapered microstructured optical fibre. Leaking bluish radiation can be seen in the last part of the fibre, in contrast to the leaking red radiation in the rest of the fibre. This indicates that the smaller core in the end of the fibre does not guide the blue part of the supercontinuum.

### *Hyperspectral fluorescence lifetime imaging microscopy setup*

In the work described below the author's input was to incorporate the developed tunable supercontinuum source, described in the Section above, into the hyperspectral FLIM microscope. Data acquisition, testing the performance of the system, was performed by Dylan Owen and others, and the results are summarised in Ref. [275]. Figure 56 shows schematics of the experimental setup incorporating the supercontinuum setup, shown in Figure 53, and a hyperspectral FLIM microscope that is explained in more detail in Ref. [276]. The supercontinuum output beam was coupled into the microscope using a cylindrical lens to form a line of excitation light on the sample and the resulting line of fluorescence was relayed to the input slit of an imaging spectrograph (ImSpector V8E, Specim, Oulu, Finland). The lens after the excitation slit was translatable to correct for chromatic aberration when tuning wavelengths.

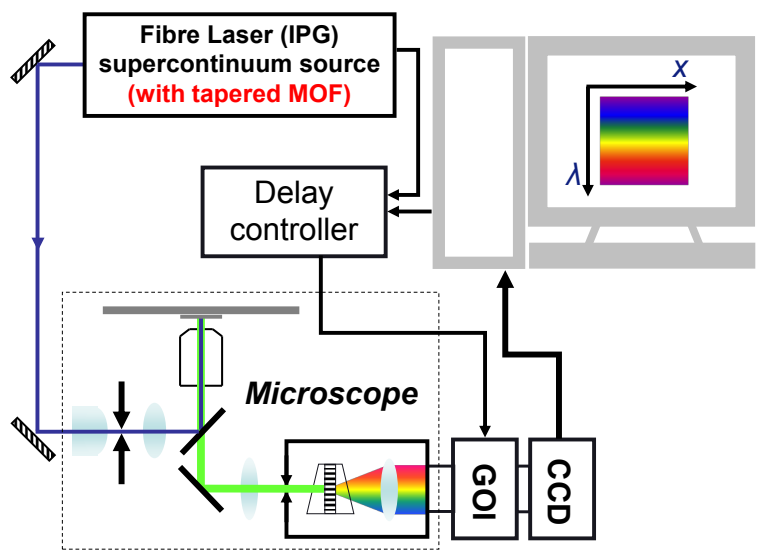

Figure 56. Schematic of excitation – resolved hyperspectral FLIM microscope.

Line illumination and slit detection rejected out of focus light thereby conferring semi-confocal optical sectioning. The emission was spectrally dispersed perpendicular to the slit and imaged onto a gated optical intensifier (the principles of which are explained in Section 3.3.1). This provided a series of time-gated spectrally-resolved line images that were read out by a cooled electron multiplying CCD camera (iXon DV887, Andor, Belfast, UK). The sample was stage-scanned in the direction perpendicular to the line illumination and image reconstruction and processing was undertaken using custom-written LabVIEW (National Instruments, Austin, TX) software. The acquired data was corrected for the spectral response of the spectrograph and high rate imager (HRI) using an absolute calibrated light source. This could be repeated for an arbitrary number of excitation wavelengths. Such acquired data provided rich information about fluorescing specimens since the fluorescence lifetime, together with excitation and emission spectra were recorded for each pixel in the image of the specimen.

### *Hyperspectral fluorescence lifetime imaging of tissue autofluorescence*

First, the potential of using the developed supercontinuum source (Figure 54) to excite molecules in the ultraviolet was explored. The power output around 360 nm can be used to excite the structural protein collagen. An unstained rat tail was used to check if the ultraviolet part of the supercontinuum (obtained by filtering with 360 nm filter with 10 nm bandwidth) was powerful enough to excite fluorescence. In total, 11 time gates were acquired with a width of 1ns each. The emission spectral resolution was set to 15 nm. The power at the sample was kept below 1 mW to limit photobleaching of the sample during the acquisition. Figure 57 shows fluorescence images as obtained by averaging fluorescence emission wavelength or lifetime, or both.

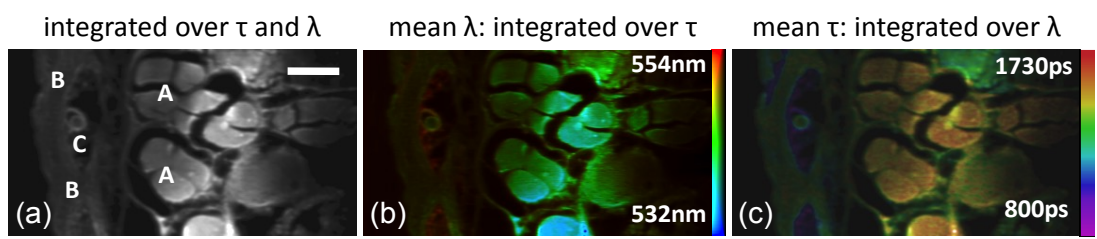

Figure 57. Fluorescence data from the unstained rat tail excited at 360 nm. Scale bar – 100 µm.

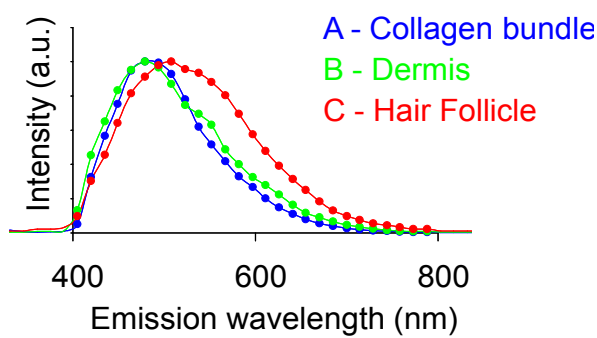

Figure 58. Emission spectra of the selected regions in the Figure 57.

Figure 58 shows the steady-state (time-integrated) emission spectral profiles of the labelled regions in Figure 57 (a). It is apparent that various tissue components have different fluorescence spectra, as is also apparent from the mean wavelength image in Figure 57 (b). In addition to that they differ in fluorescence lifetime, as can be seen in Figure 57 (c).

### *Excitation-emission-lifetime-resolved imaging of stained convallaria*

The developed supercontinuum source naturally offers itself to tuning because of the broad spectrum, therefore another sets of data were recorded for different excitation wavelengths by imaging a stained convallaria sample. A range of excitation wavelength bands from 430-580 nm (each of 10 nm width) were used with acquisition parameters set to 9 time gates of 1ns width and 15 nm spectral resolution. The acquired multidimensional fluorescence data stack can be manipulated in post-acquisition processing in many ways to optimize the contrast of different tissue compartments or fluorophores. Figure 59 shows data structure of this six dimensional fluorescence image stack containing three spatial dimensions, excitation, emission wavelength and lifetime.

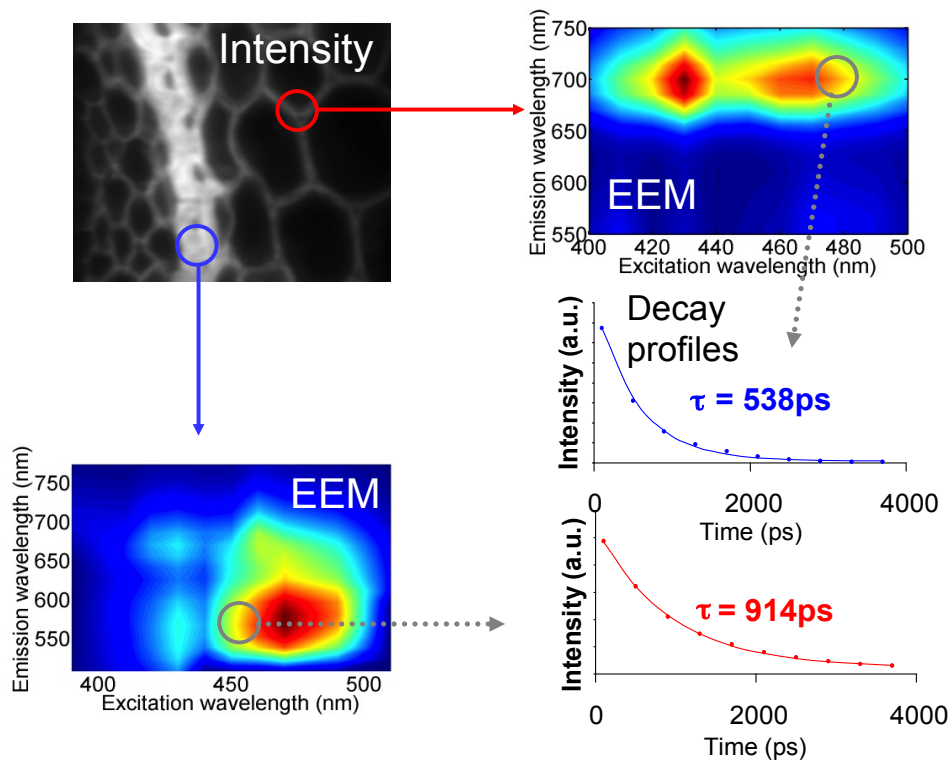

Figure 59. Excitation resolved hyperspectral data of the convallaria stained with three different dyes. Each pixel in the intensity image contains excitation-emission matrix (EEM). In turn each pixel in the matrix contains fluorescence decay.

For any pixel or region in the integrated fluorescence intensity image, it is possible to obtain a conventional excitation-emission matrix (EEM). From any region in these excitation-emission matrices, one can obtain the corresponding fluorescence decay profile. Figure 60 shows plot of mean fluorescence lifetime as a function of emission and excitation wavelength. This type of plot may prove useful in the separation of fluorophores whose spectra and lifetimes may strongly overlap. One can see that fluorescence lifetime is higher with lower excitation and emission wavelengths although is not clear why such a correlation is seen. It is common practice to use excitation-emission matrices to contrast fluorophores whose excitation and / or emission spectra otherwise overlap. Here it is demonstrated that fluorescence lifetime can provide further opportunities to enhance contrast. This can be achieved with inexpensive pulsed fibre laser-based supercontinuum generation in microstructured optical fibre that provides spectrally broad but yet powerful radiation.

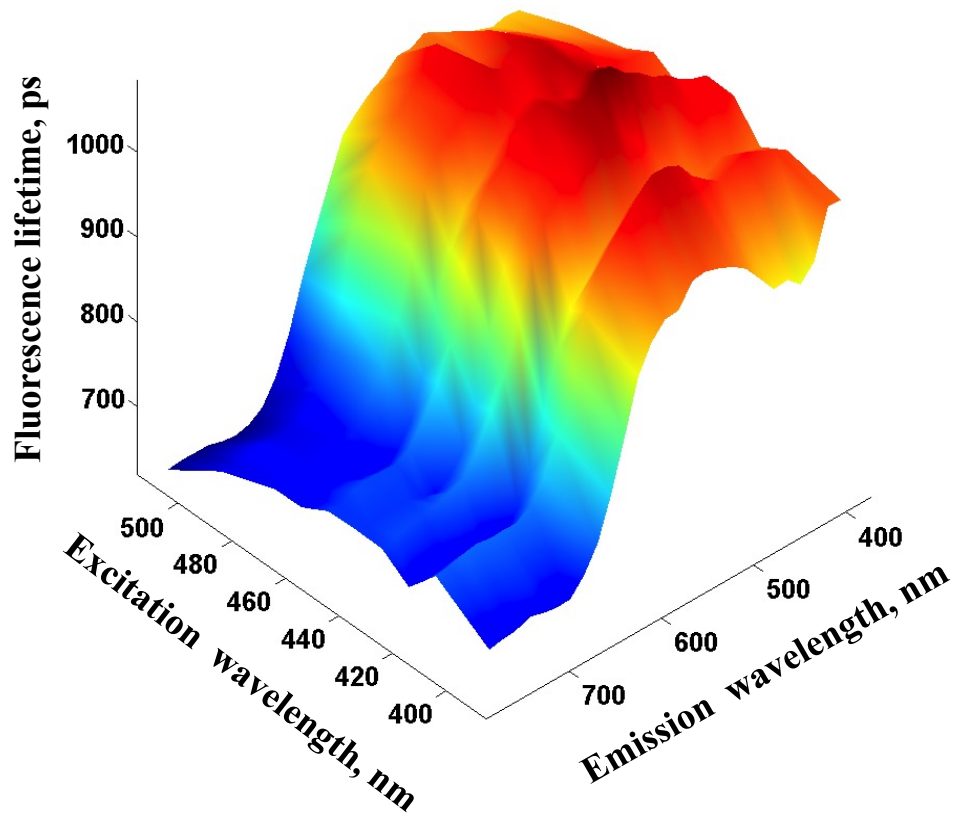

Figure 60. Plot of mean fluorescence lifetime as a function of emission and excitation wavelength in a single pixel. Colour is used to help contrasting different lifetime values.

## 4.4    Summary and Outlook

Broad supercontinuum can be generated in microstructured optical fibres if femtosecond or picosecond laser pulses are coupled in the anomalous dispersion region of the fibre, near to the zero-dispersion wavelength. In this case the dominating mechanism behind broad supercontinuum generation is soliton dynamics and the supercontinuum can cover the entire visible region and reach the ultraviolet if the zero-dispersion wavelength is close to the visible spectral region. Provided the laser pulses are powerful enough, higher order solitons are formed that, upon perturbation, split and generate dispersive waves across the zero-dispersion wavelength, in the visible part of spectrum. Nevertheless, supercontinuum generation is also possible with longer laser pulses – nanoseconds and even *cw* lasers have been demonstrated to generate broad supercontinuum. The most laser for supercontinuum generation at the moment seems to be picosecond Ytterbium fibre lasers that are bright, cheap, compact and easy to operate. On the other hand the best spectral coherence can be achieved with femtosecond pulses because longer pulses generate supercontinuum through the modulation instability (rather than soliton dynamics), which is more susceptible to the noise. The Ytterbium fibre lasers operate at 1.06 μm, which makes it difficult to generate supercontinuum covering the entire visible region, because the four-wave mixing process cannot generate anti-stokes components shorter then ~ 500 nm, due to the quenching of its Stokes component around 2.2 μm (as dictated by the phase-matching curve) by fibre losses. Tapered fibre that has continuously decreasing diameter of the core along the fibre along, which in turn changes the dispersion curve. This generates new phase-matching curves that allow supercontinuum to extend towards the ultraviolet. The changing dispersion curve along the fibre length decreases the dispersion experienced by solitons, which prevents them from spreading out as they experience intra-pulse Raman shift towards infra-red. Therefore solitons continue to interact with dispersive waves and pushes supercontinuum generation further to the ultraviolet.

In this thesis a variety of home built supercontinuum generation setups based on a range of microstructured optical fibres and laser supercontinuum sources were developed, as described in this Chapter. The first supercontinuum source was assembled by using a microstructured optical fibre with zero-dispersion wavelength at 740 nm and pumped by a femtosecond Ti:Sapphire laser. A fibre laser was then used to generate supercontinuum in microstructured fibre with the zero-dispersion wavelength at 1040 nm. Finally a tapered microstructured optical fibre was used, whose core changed from 6 to 2.3 μm along its length. This resulted in the ultraviolet-extended supercontinuum generation (down to 350 nm) through the soliton bandwidth preservation and enhanced four-wave mixing. It has been calculated [255] that for the efficient ultraviolet generation, the core size should change from 6 to 2 μm with d / Λ being close to unity. Therefore, supercontinuum with the deeper ultraviolet light could be generated if such a fibre were to be manufactured. Future improvements in the fibre laser technology should also enable generation of broader supercontinuum. The ultimate limit to the short wavelength edge is likely to be losses from Rayleigh scattering or two-photon absorption.

The application of the supercontinuum source as excitation for various FLIM microscopes was reported in this Chapter. It was applied to wide field, Nipkow disc and line scanning hyperspectral fluorescence lifetime imaging microscopes, with time resolved images being obtained using a gated optical intensifier in front of a CCD camera. Supercontinuum generation was carried out with various setups, of which the most promising seems to be supercontinuum generation pumped by a powerful picosecond fibre laser in conjunction with a tapered microstructured optical fibre, which allowed generation of UV-extended, but yet powerful supercontinuum. The supercontinuum was incorporated into a hyperspectral fluorescence lifetime imaging microscope that exploited virtually all its advantages including its broad spectrum, ultrashort pulse operation, brightness and single spatial mode. Data acquired with such a technique can give a wealth of information about fluorescing samples and enable different fluorescing species to be resolved by emission spectrum, excitation spectrum, fluorescence lifetime or combination of all of these. Supercontinuum sources should prove to be very useful in fluorescence microscopy and could replace the solid state and gas lasers currently used in many microscopes. There are now a few companies that commercially manufacture supercontinuum sources, including Fianium Ltd. and Koheras, which are already extensively used in fluorescence microscopy.

# 5. STED Microscopy: Control of the Point Spread Function

## 5.1 Introduction

In stimulated emission depletion (STED) microscopy a STED (depleting) beam has to be engineered so that it would efficiently narrow microscope's point spread function (PSF), formed by an excitation beam, through the process of fluorescence depletion. This Chapter reviews different ways of controlling wavefront of the STED beam so that it could form various relevant PSFs upon focusing with an objective. The Chapter then presents images of PSFs of the so called doughnut and the optical bottle beams as acquired with the STED microscopy setup (described in more details in Chapter 6). The wavefronts were engineered using blazed holograms (computed with a software written by Bosanta Boruah, Imperial College London) displayed on a spatial light modulator (SLM). The shapes of PSFs were optimised by correcting aberrations through control of various individual Zernike polynomials. This resulted in a near-ideal PSF shape.

## 5.2 Review of wavefront engineering in STED microscopy

### 5.2.1 Non-modified STED beams

In the first theoretical paper on STED microscopy in 1994, it was proposed that the lateral resolution of the scanning fluorescence microscope can be increased by simply quenching a part of the diffraction limited fluorescence spot by two spectrally red-shifted laser beams focused with the same objective but offset symmetrically with respect to the excitation beam [1]. Narrowing the PSF down to 35 nm in one lateral direction was then predicted. Such a scheme was finally realised in 1999 [114], where an overlap of the excitation and a single STED beam in the focal plane enabled resolution improvement, along the overlap direction by some 45 nm, compared to the confocal image. As was shown later, the same principle can be used to improve the axial resolution by exploiting the ear-shaped side lobe lateral structure of the PSF in the axial direction [277]. It was later demonstrated that by using a single STED beam offset along one lateral dimension, a resolution improvement by a factor of 2-3 was achieved simultaneously along the offset direction and along the optical axis. The STED beam in those experiments was displaced from the optical axis as well as from the excitation beam at an optimal distance in order to give the narrowest fluorescing spot but its wavefront was not modified in any way. A review on STED beam wavefront modification employing various techniques including phase plates (in various configurations) and spatial light modulator is given below.

### 5.2.2    *Wavefront modification with various phase plates*

The first wavefront engineering of the STED beam was performed in year 2000, by imprinting a circular $\pi$ phase distribution, shown in Figure 61 (a), on its wavefront [278]. Manufacturing a circular $\pi$ phase plate is relatively simple. A phase plate used for the phase modulation in this particular experiment was manufactured by depositing a layer of $MgF_2$ on a glass substrate. A cheap and easy way of producing such a phase plate has been recently reported, where a thin circular polymer film can be easily deposited on a glass slide [279]. The phase plate introduces a phase difference of $\pi$ radians between the central and outer parts of the beam, as illustrated in Figure 61 (b).

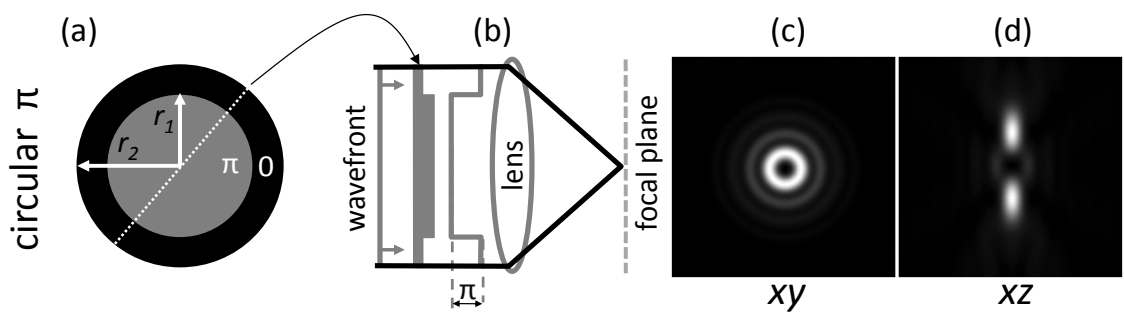

Figure 61. Properties of a circularly polarised optical bottle beam. A circular $\pi$ phase distribution (a) is introduced in the wavefront of a beam (b) that, when focused with a lens, results in a zero intensity in the focus because of the destructive interference. It has a doughnut-like PSF in the lateral plane (c) and the optical bottle beam shape in the axial plane (d). Figures (c) and (d) were taken from [280].

In order to introduce an exact π phase difference the layer has to be of $\lambda / (2(n-1))$, which for the given experimental conditions ($\lambda_{STED} \approx 765$ nm) was of ~ 1 μm. If such a modified wavefront is focused with a lens, it results in the destructive interference of the light, arriving from the two different parts of the phase plate. Since this holds true only for the light converging to the exact focus, the PSF features a dark region in the middle of a PSF surrounded by the light in all directions. The axial cross-section of the PSF features major lobes above and below the focal plane and minor lateral lobes in the focal plane, as shown in Figure 61 (d). The PSF in the literature is sometimes called the optical bottle beam and this name will be used throughout this thesis. The PSF has doughnut-like structure in lateral plane as demonstrated in Figure 61 (c). The optical bottle beam has enabled a simultaneous six-fold axial and a two-fold lateral resolution improvement in images of fluorescing beads and a three-fold axial improvement in images of live cells [278]. In fact, those were the first STED microscopy images of live cells. The axial resolution was more effectively improved than the lateral; therefore an initial fluorescing spot that had an anisotropic shape (due to the worse axial resolution compared to the lateral) resulted in a nearly isotropic spot after the depletion because of optical bottle beam shape of the STED beam. In order to achieve exactly zero intensity at the focus the energy going through the central and outer parts of the phase plate had to be equal. Thus, in case of uniform intensity illumination, the central and outer parts has to have the same area and therefore the radii of the phase feature should be chosen such that $r_1 / r_2 = 1 / (2)^{1/2} = 0.707$, where $r_1$ and $r_2$ are the radius of the inner and outer circles, as depicted in Figure 61 (a). Calculations show that the maximum intensity in the lateral lobes compared to the maximum intensity of the axial lobes is only 21 % [280]. This explains why the optical bottle beam improves resolution primarily in the axial direction. Also, its intensity in the focal plane has $r^4$ dependence, where $r$ is normalised optical co-ordinate, which means that the off-axis intensity (intensity around zero) is not very confined (compared to other PSFs), as will be explained in more details later (see Figure 66). To increase the lateral resolution more efficiently the use of two other phase plates, shown in Figure 62 (a & b), were later explored [281]. A semicircular π phase plate, shown in Figure 62 (a), shifts a half of the wavefront by π. If the polarisation of the beam is parallel to the phase dividing line, as shown in Figure 63 (a), the focused beam destructively interferes in the focal plane, and forms a valley-like cleft in the PSF, as shown in Figure 62 (c), which is oriented along the direction of the polarisation. In case of the polarisation being orthogonal to the phase dividing line, a strong *z* polarised component is created [282], as illustrated in Figure 63 (b). In fact, this simple PSF is the most effective to break the diffraction barrier albeit in one direction only [283].

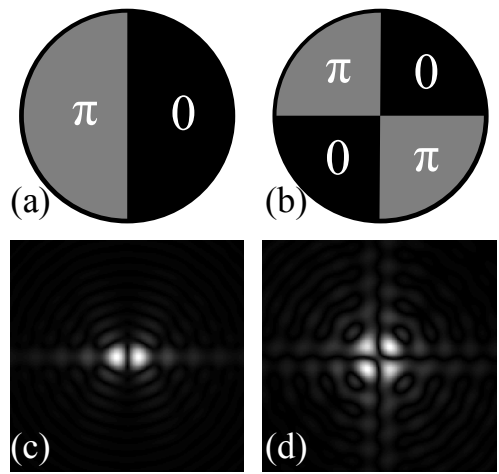

Figure 62. Phase plates (top) used to increase the lateral resolution in STED microscopy and their respective Fourier transforms (bottom). (a) Semicircular π phase plate. It can be used to improve resolution along one lateral dimension only. (b) Two quarters π phase plate. It is not useful in STED microscopy because it form too broad PSF. (c) Fourier transform of (a). (d) Fourier transform of (b).

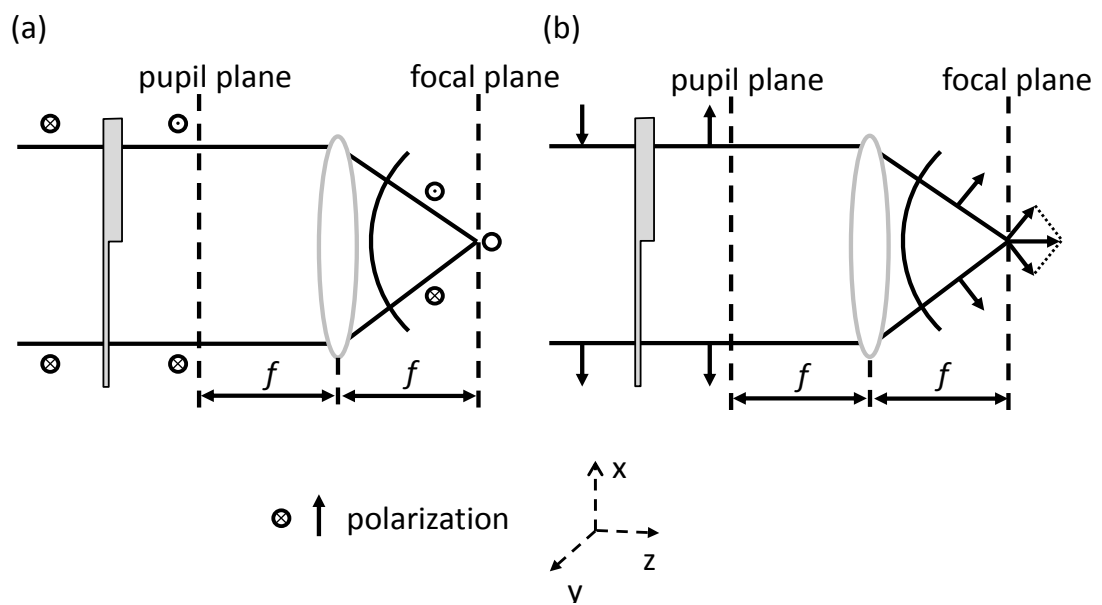

Figure 63. Focusing a beam that has half of its wavefront shifted by π with a semicircular π phase plate, shown in Figure 62 (a). The Electrical field orientation is depicted as parallel (a) and orthogonal (b) to the dividing line of the phase plate. The parallel orientation (a) does not result in the on-axis intensity in the focal plane, whereas orthogonal orientation (b) produces a strong *z* component.

The record resolution of 16 nm was achieved in year 2005, using this plate in combination of collinear orientation of the excitation and the STED beams as well as polarised detection at the same orientation [284]. Recently, the resolution of ~5-6 nm was achieved in a single direction using this phase plate in combination with very stable nitrogen vacancy colour centres in the bulk diamond [125]. Generally, the simultaneous resolution improvement in both directions is required for biological application. It was expected that the two-quarter phase plate, shown in Figure 62 (b), would produce two perpendicular valleys and consequently would improve the lateral resolution along two dimensions. However, it was found that such a phase plate cannot increase the resolution in both lateral directions simultaneously and with the same performance as the semicircular $\pi$ phase plate, because the former forms a broader PSF [284], as is evident in Figure 62 (d). A more successful approach to produce a doughnut-like PSF was demonstrated by using a Mach-Zehnder interferometer equipped with two polarising beamsplitters and two semicircular $\pi$ phase plates, as shown in Figure 64 [285].

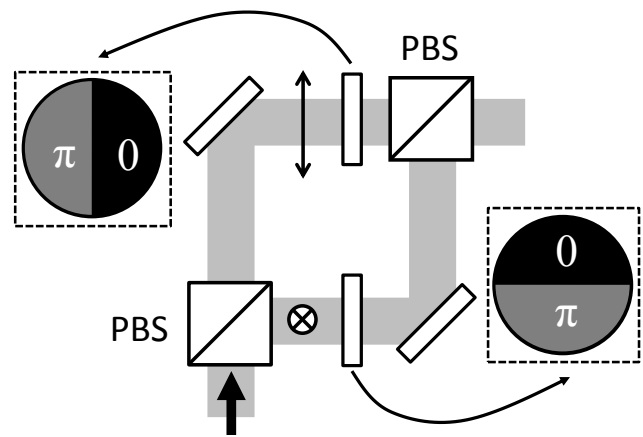

Figure 64. Mach-Zehnder interferometer to generate a doughnut-like PSF by using two semicircular phase plates that have their phase dividing lines oriented parallel to the direction of the polarisation. An incoming beam is split and later recombined with a pair of polarising beamsplitters (PBS). The phase plates are inserted in the beam path with phase diving line orientation as shown.

In case of the polarisation being parallel to the phase dividing line in each arm, a doughnut shaped PSF with a zero on-axis intensity is formed [281, 282]. Otherwise, when the orientation is perpendicular, a strong $z$ polarised component is created [282], which is of interest, for example, in $z$-polarised microscopy [286]. The two wavefronts can be overlapped either coherently or incoherently, which results in zero on-axis intensity for the both cases but with a slightly different properties off-axis. This is because the light from different arms can interfere between each other in the coherent case [282]. The most elegant way of creating a doughnut, however, is to use a spiral phase plate that imprints a helical phase distribution, shown in Figure 65 (a), on the beam's wavefront [287]. The distribution has an angle dependent phase that can be described as exp($il\,\varphi$), with $0 \le \varphi \le 2\pi$ being the azimuthal co-ordinate about the optic axis and $l$ − topological charge. For $l = -1$ the phase changes from 0 to $2\pi$ in the clockwise direction and for $l = +1$ − anticlockwise. A beam with topological charge is said to carry an orbital angular momentum, $lh$ [288], where $h$ is a Plank constant.

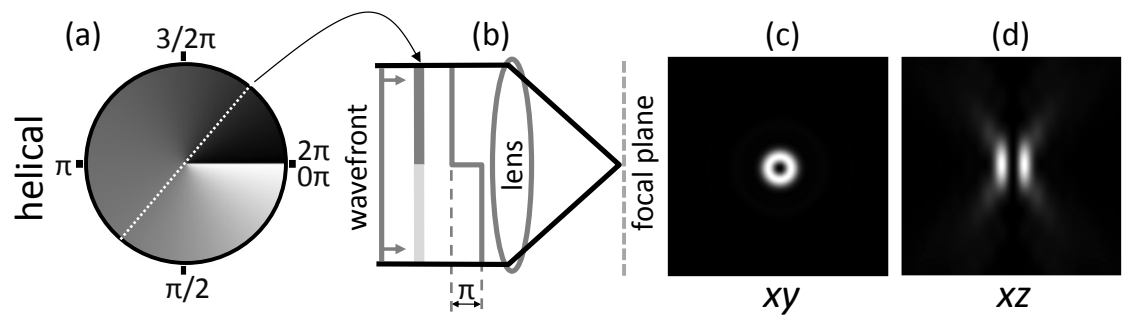

Figure 65. Properties of a doughnut beam. A helical phase distribution (a) is introduced in the wavefront of a beam (b) that, when focused with a lens, results in a zero intensity in the focus because of the destructive interference. It results in doughnut-like PSF in lateral plane (c) and in a cylinder type shape in the axial plane (d). Figures (c, d) are taken from [280].

If the light is circularly polarised it carries a spin angular momentum. In helical phase distribution a phase difference of $\pi$ radians is introduced for each pair of rays situated symmetrically around the optical axis, as illustrated in Figure 65 (b). When focused with a lens such a beam results in destructive interference along the optical axis and produces a cylinder-like axial PSF, as shown in Figure 65 (d). Laterally the beam forms a doughnut-like PSF, as shown in Figure 65 (c)Spiral (vortex) phase plates have recently become available commercially for ~ \$1,300 (RPC Photonics, Rochester, New York, USA). They are produced as a polymer replica on a glass substrate and are now routinely used in STED microscopy [289-295].

There are other various ways of producing doughnut PSF, for example, such as mode transformers using astigmatic lens systems [296], computer-generated sub-wavelength dielectric gratings [297] or polarisation controlled fibres [298] to name a few, but the possibility of using them in STED microscopy has not been explored. Figure 66 illustrates the fact that the intensity in the focal plane has $r^4$ dependence for the optical bottle beam and $r^2$ for the doughnut beam (also for the beam in Figure 62 (c)), where $r$ – normalised optical co-ordinate. Thus, the lateral resolution will be more efficiently improved with the doughnut shaped PSF [283].

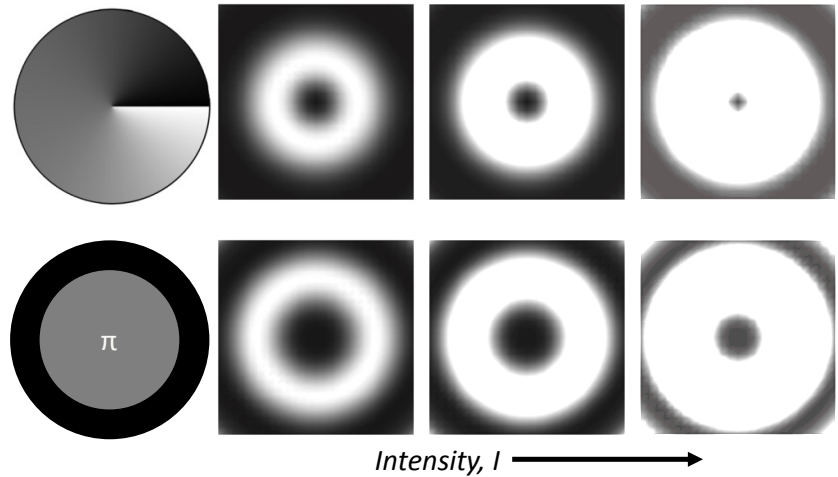

Figure 66. Qualitative illustration of intensity dependant focal distribution of the saturated intensity in the doughnut (top row) and the optical bottle beams (bottom row).

### 5.2.3    Focusing with high numerical aperture lens

In case of high numerical aperture (NA) focusing, polarisation effects start to play a noticeable part. For example, it was shown that focusing a linearly polarized doughnut beam, with a lens of NA = 1, a focal spot with the central intensity reaching 48.8 % of the maximum intensity, was created in the focal plane [299]. This effect is induced by the depolarisation [300] caused by the ray bending in the lens that couples, for example, the $x$ polarisation to the $y$ and $z$ polarisations. Figure 67 (c) shows an appearing $z$ component that illustrates the increased intensity in the direction orthogonal to the incident polarisation. Figure 67 (b) shows that in addition to that an elongation in the polarisation direction also occurs. However, it was shown that this non-zero on-axis intensity can be eliminated by circularly polarising the beam with an appropriate handedness [301] and it was experimentally tested with fluorescent beads [302]. In general, if the circular polarisation of the incoming beam creates an orbital angular momentum (circular polarisation handedness) parallel to the spin angular momentum of the photon, then zero on-axis intensity is produced that does not depend on the NA of the objective [303]. Simulations show that in such a case an axial component creates a doughnut shape intensity distribution as illustrated in Figure 67 (f). However, in the opposite case, when the spin and angular momentum are anti-parallel, a strong axial component is generated as shown in Figure 67 (i). It can be shown, that in the former case the time dependant electrical field in the pupil plane is exactly cancelled everywhere along the optical axis at any arbitrary time, as illustrated in Figure 68 (c), and for the latter case, the electrical field alternates between radial and azimuth orientations [302], as is illustrated in Figure 68 (b). Radial orientation of the electrical field creates a strong axial component [304], as can also be seen in geometrically representation in Figure 63 (b).

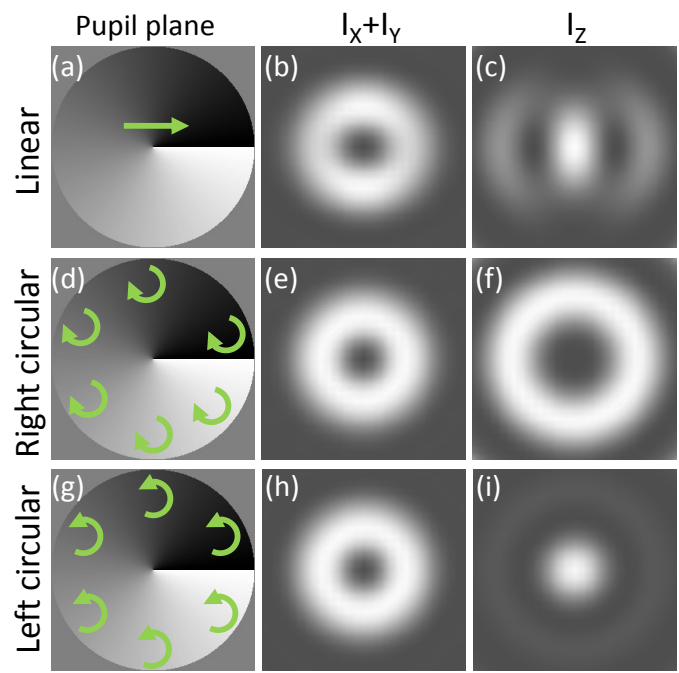

Figure 67. Calculated field intensity distributions. Beam with helical phase distribution (of topological charge, $l = + 1$) and varying polarisation is assumed. The modulus-squared of the lateral $(I_x + I_y)$ and axial $(I_z)$ components are shown for different polarisations. Linear polarisation (a) produces a doughnut-shaped intensity distribution laterally (b) with elongation along polarisation direction and complicated distribution axially (c) with non-zero on-axis intensity. Right circularly polarisation (d) produces a doughnut-shaped intensity distribution laterally (e) as well as axially (f). Therefore the overall distribution produces true zero on-axis intensity. Left circularly polarisation (g) produces a doughnut-shaped intensity distribution laterally (h) and non-zero on-axis intensity axially (i). Simulated using the Fourier transform form of the vectorial theory for semi aperture angle of $\alpha = 1.125$. Images are normalised to unit. Figures adapted from [280].

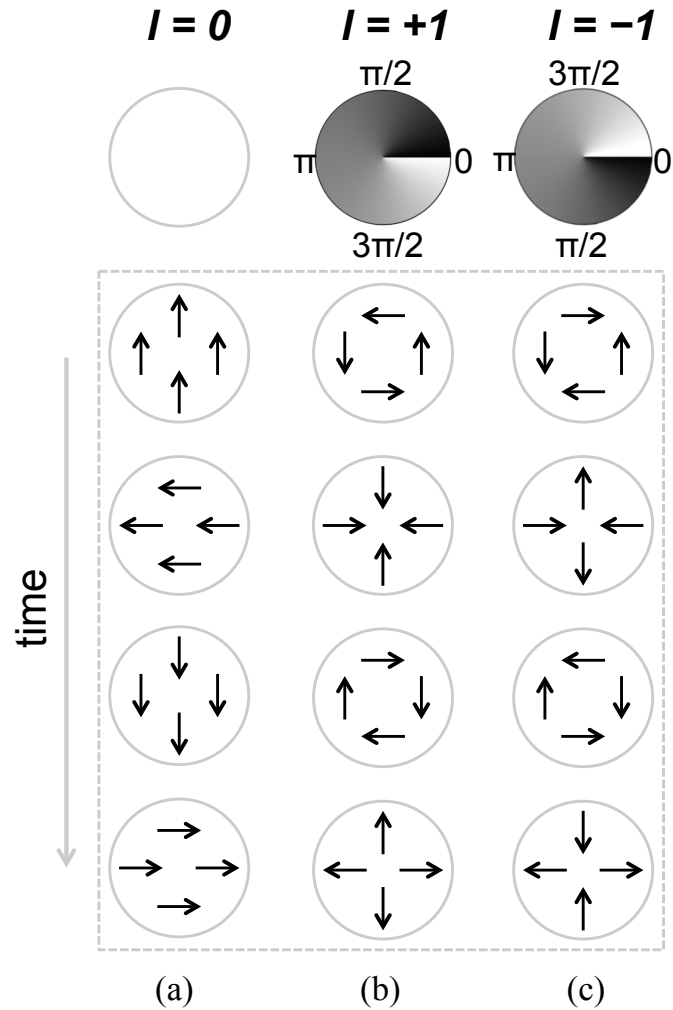

Figure 68. Electrical field vector evolution of a left circularly polarised light in the pupil plane. (a) Vector evolution of a beam with non-engineered wavefront (*l = 0*). (b) and (c) shows vector evolution with the helical phase distribution (with the topological charge of +1 and −1, respectively) imprinted on the wavefront. For the *l = +1* (b) the electrical field alternates between azimuthally and radially polarised light and therefore the focused light will have a strong axial component, as illustrated in Figure 63 (b). For the *l = −1* (c) the electrical field produces zero on-axis intensity since the electrical field cancels out each other along the axis, similar to Figure 63 (a).

### 5.2.4    Overlap of the doughnut and optical bottle beams

A doughnut beam and an optical bottle beam are the optimum choice for confining the fluorescence spot in lateral and axial directions, respectively [283, 305]. Ideally one wants both the axial and the later resolutions improved at the same time. The performance of the individual beams can be combined together with A Mach-Zehnder interferometer, as shown in Figure 69.

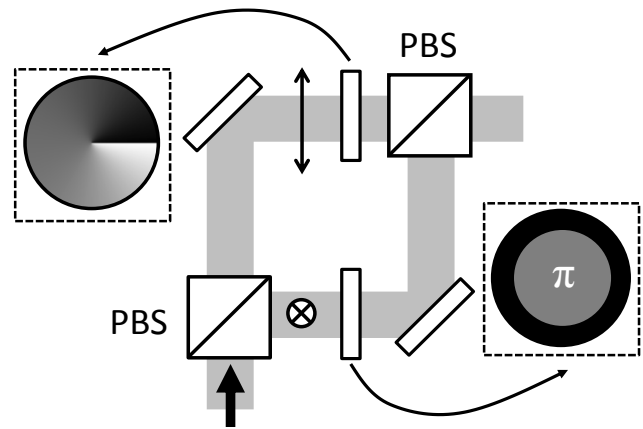

Figure 69. Mach-Zehnder interferometer combining the doughnut and optical bottle beams using a spiral and a circular π phase plates and incoherent overlap.

The incoherent overlap of the two beams was recently reported [294] resulting in the lateral and axial resolution of 43 nm and 125 nm, respectively. It is believed that the coherent overlap should lead to a more economic use of laser power due to the synergy effects. However, interference between the two beams can lead to the PSF distortion that might not be suitable for STED microscopy. If the two beams are selected to have a circular polarisation with opposite handedness, it is expected that such a combination would minimise interference between the two beams. Figure 70 shows simulated PSF of such overlap.

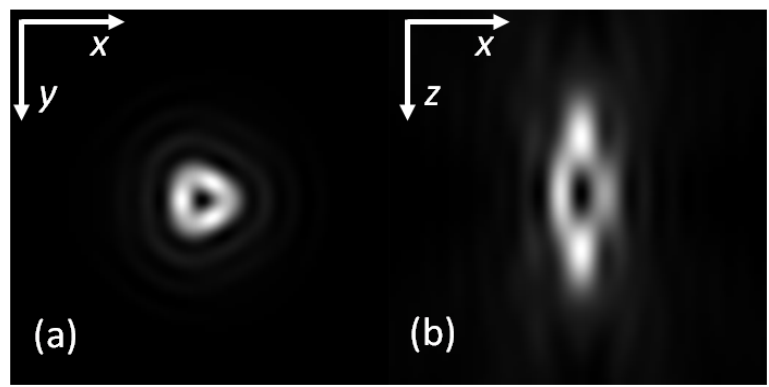

Figure 70. Coherent addition of doughnut and optical bottle beams with the right and left circular polarisations. Figures taken from [280].

The maximum intensity in the focal plane is 85 % of maximum intensity along the optic axis, which is a significant improvement from the 21 % in the case of the optical bottle beam shown in Figure 61. The intensity distribution in the lateral plane, shown in Figure 70 (a), also has a $r^2$ intensity dependence and a small amount of a 3-fold asymmetry due to the presence of a third order azimuthally dependent term [280].

However, the intensity variation around the PSF at the points of maximum intensity was about 20 % only. Simulations also showed that the focal plane polarisation of the overlapped beam can in general be defined as elliptical [280]. The beam can be generated for example, by using polarisation multiplexed holograms [306].

### 5.2.5    STED-4pi microscopy

As follows from the discussions above, the coherent [280] or incoherent [293] overlap of the doughnut and the optical bottle beams are not able to produce an isotropically fluorescent spot in stimulated emission depletion microscopy. Such a spot can be generated in a 4pi arrangement, where two pairs of counter propagating laser beams create an almost spherical depletion pattern. In principle, STED point spread function can be created by interfering two laser beams in the focal plane [117], since the created interference pattern has a sinusoidal-like periodic intensity distribution that features zero intensity valleys with the steep intensity rise around it. This principle is implemented in a related technique [130]. The narrowest valley that can be created with the two counter propagating beams is that of $\lambda / 4n$ (of FWHM), where n is refractive index. This can be carried out in the 4pi microscope configuration. Excitation and detection in this microscope can be performed through a single objective, whereas depletion through the opposing objectives. Such a kind of a microscope, called STED-4pi microscope [307], was able to achieve the axial resolution of 33 nm [66]. This, at the time of writing this thesis, was the best axial resolution demonstrated in the far-field optical microscope. However, contrary to the standard 4pi microscope, the relative phase of the counter propagating STED beams is adjusted to $\pi$ at the focal point to ensure the zero intensity in the focal spot through the destructive interference [308]. Nevertheless, the sidelobes problem, encountered in a standard 4pi microscope, is also present STED-4pi since the multiple minima in the STED beam create undepleted regions in the excitation fluorescing spot, which manifest itself as the side lobes in the resulting PSF. Similar to 4pi microscopy the side lobes can be as high as 50 % of the maximum intensity. The solution to this problem in STED-4pi microscope is wavefront engineering of the STED beam by, for example, a phase plate that leaves only a central minimum in the axial PSF of the STED beam [66, 307, 309]. Sidelobes could thus be reduced from 50 to 20 %, although theoretically the complete eradication is possible. The discrepancy between the theory and the experiment probably comes from the asymmetrical aberration present in the system, therefore, as stated therein the system would benefit from the active aberration correction to further reduce the side lobes [309]. Nevertheless, the system was demonstrated to be suitable for the immune-fluorescence imaging (imaging intracellular structure) with spatial resolution of 50 nm ($\lambda$ / 16) [310]. Recently the lateral resolution together with the axial resolution was improved with the STED-4pi microscopy by additionally introducing a counter propagating doughnut beams [311]. The beams created PSF with on-axis zero intensity in the focal plane and were incoherently added with the standard PSF that improves the axial resolution. To ensure the incoherent overlap the doughnut beams and the standard beams were provided by separate lasers. An additional advantage of the 4pi configuration was the counteraction of depolarisation caused by the high NA objective [300] ensuring a true 3 D zero. A nearly spherical fluorescing spot of 40-45 nm was thus created [311].

### 5.2.6    *Holographic wavefront control*

The phase of a wavefront can be modulated directly when a beam passes through or is reflected from an optical device, for example SLM. However, if a beam acquires phase modulation via the process of diffraction, then the phase of the beam is said to be controlled holographically [312]. In its simplest way a hologram can be recorded on a photographic film in the form of interference pattern (interferogram) between a plane wave and the beam one desires to produce. Once developed, the film can be illuminated by a plane wave to produce a first-order diffracted beam that has the desired properties. If two beams with plain wavefronts interfere, it gives a pattern with the straight fringes with the spacing between them determined by the intersection angle and the wavelength of the beams. However, if one of the beams has a helical wavefront then the interference result in a fork-like structure in the interferogram [313] as, for example, shown in Figure 71 (c). The interference pattern does not have to be experimentally recorded, but instead can be calculated [314] and printed on an optical device or displayed on SLM.

### 5.2.7 *Wavefront control with spatial light modulators*

As described above a spatial light modulator (SLM) is a device which is capable of creating almost arbitrary wavefronts [315, 316] with high spatial resolution. Various types of SLM exist including deformable mirror (DM) SLM, digital micro-mirror device (DMD) SLM and liquid crystal (LC) SLM. They can in general control polarisation, intensity and phase of the light. Commercial liquid crystal SLM are either optically (Hamamatsu) or electrically (Boulder, Holoeye) addressed. The easiest way to control wavefronts is to use a liquid crystal spatial light modulator that enables either direct [317] or holographic [318] phase modulation. A liquid crystal SLM has significant advantages over other optical devices owing to its reconfigurability and the ease with which phase discontinuities can be generated [319]. Continuous devices such as deformable mirrors [320], for example, cannot display helical phase distribution with high reproduction, because they are less able to implement a sudden phase change from 0 to $\pi$ compared to that of SLM. Liquid crystal SLM have been used to produce doughnut shaped beams in STED microscopy [116, 123, 321-329]. However, in the work reported therein the SLM (Hamamatsu) has been used in the direct phase modulation mode to display helical phase distribution. This means that the achieved accuracy of the helical wavefront imprinted on the beam was limited by absolute modulation afforded by the SLM e.g. number of phase levels that can be generated. In contrast the holographic approach has significant advantages, in that the achieved accuracy is a function of the shape of the displayed hologram fringes that depends only on the number of pixels on the SLM array. More importantly the holographic approach allows the separation of different diffraction orders including the zeroth diffraction order, which can increases on-axis zero intensity in generated PSF. In this thesis holographic wavefront control with a liquid crystal SLM, was used to generated doughnut and optical bottle beams and to correct for aberrations. Figure 71 shows computer generated phase distributions in a direct (a,b) and holographic mode (c,d). The software to compute the distribution was written by Bosanta Boruah (Imperial College London). Figure 71 (c) and (d) shows computer generated holograms displayed on the SLM that allowed imprinting an accurate reproduction of the helical and circular $\pi$ phase distributions (Figure 71 (a) and (b), respectively) to produce a doughnut and an optical bottle beams in the +1 diffraction order. Hologram shown in Figure 71 (c) is called *type I* and in Figure 71 (d) – *type II*, in this thesis.

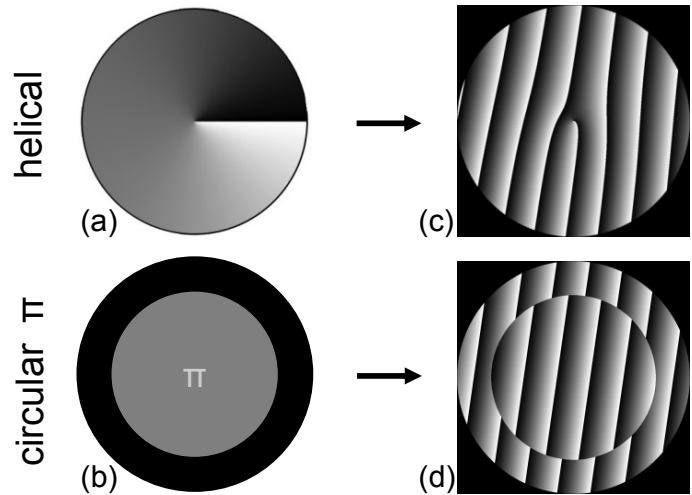

Figure 71. Computer generated patterns displayed on SLM. (a) Helical wavefront for generating a doughnut beam and (b) Circular π phase wavefront for generating an optical bottle beam in the direct phase modulation. (c) Hologram (*type I*) for generating a doughnut beam and (d) Hologram (*type II*) for generating an optical bottle beam in the +1 diffraction order.

### 5.2.8    *Aberration correction*

In a microscope, aberrations can be induced in the flat or spherical wavefront when it travels through the imaging path or the sample [330, 331]. As a result, aberrations distort the diffraction limited PSF and can therefore degrade the microscope's resolution and signal intensity. It is especially thought to be detrimental to nonlinear microscopies where the imaging process (two-photon absorption, SHG etc) is nonlinearly dependant on the focal intensity. In STED microscopy, resolution improvement is nonlinearly dependant on the focal intensity distribution of the STED beam [293], therefore even small aberrations can have enormous effect on the resolution. Indeed, it has been reported that aberrations are often a limiting factor in STED microscopy [301, 309]. The STED beam itself can be thought of as an aberrated beam with the helical aberration, for instance. If this aberration is removed it results in a vast resolution decrease. In STED-4pi microscope, for example, a certain aberration is induced in the STED PSF in order to remove sidelobes from the fluorescing spot [309]. In the work presented here, however, the purpose of the aberration correction in the STED microscopy context is to create a flat wavefront (aberration-free) upon which a helical or circular π phase distribution is imprinted (aberration correction and the phase distribution can be imprinted on one hologram together as will be shown later below). To appropriately describe aberrations, a mathematical model is needed. Figure 72 shows various Zernike circle polynomials that are often used to describe aberrations due to their convenient mathematical properties [29].

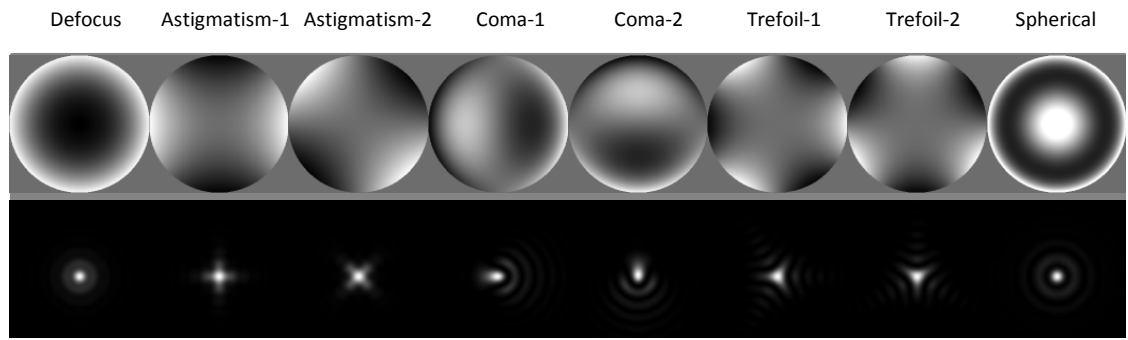

Figure 72. Zernike modes representing real aberrations. Phase distributions are shown in the top row and the corresponding PSFs in the bottom row. Figures taken from [280].

One important property of the Zernike functions is their orthogonality. In addition they are similar to traditional aberrations, such as astigmatism, coma or spherical aberrations. Usually only a small set of the polynomials are necessary to describe the overall aberration [332]. It is often possible to correct aberrations by pre-aberrating the beam wavefront so that it cancels out aberrations met in a microscope [333]. This requires knowledge of the aberrations present in the system, which can be measured with some sort of wavefront sensor [334], such as Shack–Hartmann sensor [335] that is popular in astronomy. However, it has limited application in microscopy due to its difficult implementation. Various indirect ways to measure aberrations has been developed that required minor microscope modifications. Some of those techniques relies on finding an optimum wavefront that would maximise the detected fluorescence signal [336] using various optimisation algorithms [337]. However these require long iterations and therefore might not be suitable for fast imaging. Moreover, it does not directly provide any information about the types of aberrations present in the wavefront. Another method, called modal wavefront sensor, has been developed that is able to measure aberrations in a form of Zernike modes [338] and to correct it in a few iterations [339]. Previously, it was demonstrated that it can correct aberrations for the two-photon [340] and confocal [341] microscopes and in this thesis it is used to correct for aberration in STED beam. To detect aberrations in the STED beam in a form of various Zernike mode coefficients and to subsequently correct it, the following implementation was used. The liquid crystal SLM, placed in the plane conjugate to the pupil plane, was employed to generate a binary hologram, like the one shown in Figure 73 (b). It splits the beam in to the two main diffraction orders, ±1, as shown in Figure 73 (f). The software used to generated holograms was Labview based and created by Bosanta Boruah (Imperial College London).

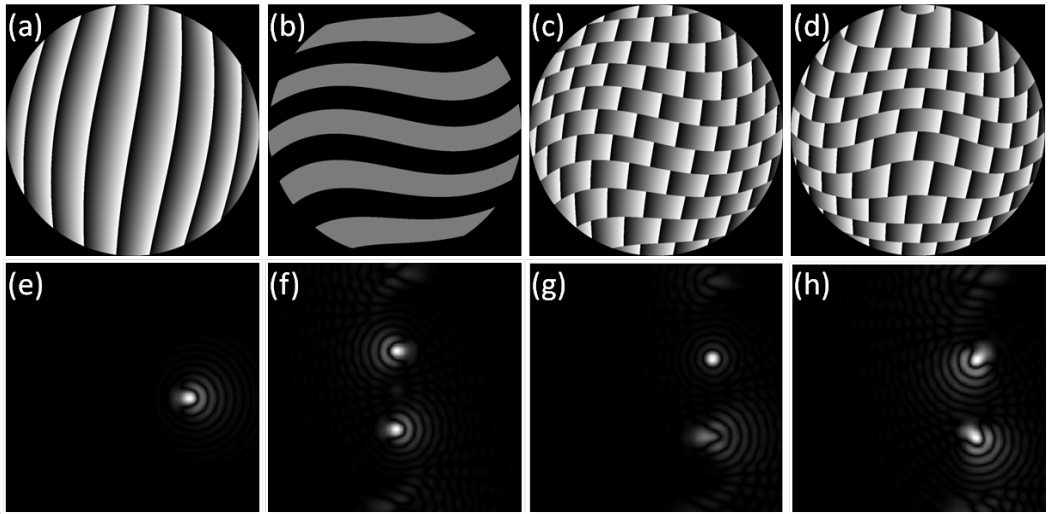

Figure 73. Computer generated holograms (top) and their respective Fourier transforms (bottom). (a) A hologram creating coma aberration in the +1 diffraction order. Its Fourier transform (e) shows PSF with introduced coma and spatial shift along the $x$ dimension. (b) Hologram generating the +1 and -1 diffraction orders with the negative and positive coma aberrations (of similar absolute amount as in (a)), respectively. Its Fourier transform (f) shows positive and negative coma PSF, shifted spatially along the $y$ dimension. (c) A multiplexed hologram, as obtained by combining (a) and (b) holograms into one. Its Fourier transform (g) shows that the two diffraction orders are not of the same intensities, which can be used to measure the amount of the introduced aberration by (b) hologram. (d) A multiplexed hologram (different from (c) hologram), which has an orthogonal coma aberration to that introduced with (b) hologram. (h) Fourier transform of (d) hologram.

The hologram, which also resembles a distorted grating, introduced positive and negative aberrations of a certain mode (coma in this example) in the +1 and −1 diffraction orders, respectively. If the input beam has a flat wavefront (no aberrations) then such a hologram produces two diffracted beams, which, upon focusing with the objective lens, creates two PSF of equal intensity in the focal plane, as is in Figure 73 (f), because both PSF suffer equally from the same amount (but of different sign) of aberrations. However, if the input beam has a positive aberration (generated, for example with the hologram in Figure 73 (a)) of the same type as that introduced by the distorted grating, then it increases the intensity in the focused −1 order and reduces it in the focused +1 order, as shown in Figure 73 (g). The difference of the intensities in the two spots can be used to measure the amount of aberration, in form of the specific Zernike mode that is displayed in the 'distorted grating' hologram. On the other hand, if the input beam has an aberration that is orthogonal to that superimposed by the hologram, then it affects both PSF by equally reducing their intensity, as illustrated in Figure 73 (h), and therefore the difference in the intensities between the two PSF equals to zero and no aberration is measured, which shows that the sensor measures only specific aberration. Once the wavefront of the aberrated beam is probed with a certain number of Zernike modes, and a degree of each presence is measured, the aberration correction can proceed. After the correction the procedure can be repeated till the correction converges [339]. Beam can be corrected by preaberrating its wavefront with another SLM or a separate hologram. Figure 73 (a) shows a hologram that could be used to correct the coma aberration in the +1 diffracted beam. However, the same SLM can be used to preaberrate the input beam and to detect the aberrations by using a multiplex hologram. Figure 73 (c) show such a hologram that preaberrates the beam with coma in the +1 diffraction order and splits it into two additional diffraction orders to check for any residual coma aberration. Sometimes it is not necessary to exactly measure the present aberrations since it might be easy to tell it by just looking at the PSF and correction can be subsequently applied. Doughnut beams, for example, can be used for that purpose due to their sensitivity to non-rotationally symmetric aberrations [342]. Figure 79 illustrates that the doughnut beam is more sensitive to aberrations than a Gaussian beam (uniform flat pupil). The on-axis intensity of the doughnut beam, however, is less sensitive to aberrations because the singularity in the helical phase distribution of the doughnut beam effectively acts as a diffraction grating with a very high spatial frequency so that it diffracts any light impinging on that part of the SLM (singularity) and thus from the optical axis. Therefore, as can also be seen in Figure 79, aberrations mostly affects off-axis intensity distribution. In this thesis both ways of detecting aberrations (with the detector or visually) were used. However, the visual inspection method is often found to be simpler and more convenient to implement.

## 5.3     Generating various PSF for STED microscopy

### 5.3.1     Gamma correction of spatial light modulator

Electrically controlled spatial light modulator (SLM) has nonlinear relation between the addressed video gray levels and the actual phase modulation [343]. This is due to the birefringence of the LC that has a nonlinear response to the applied voltage across it. The SLM used in this thesis (Figure 74) was a pure phase modulating, electrically addressed SLM (HEO 1080 P, Holoeye, Germany) based on parallel aligned liquid crystal on silicon (LCOS) technology [344].

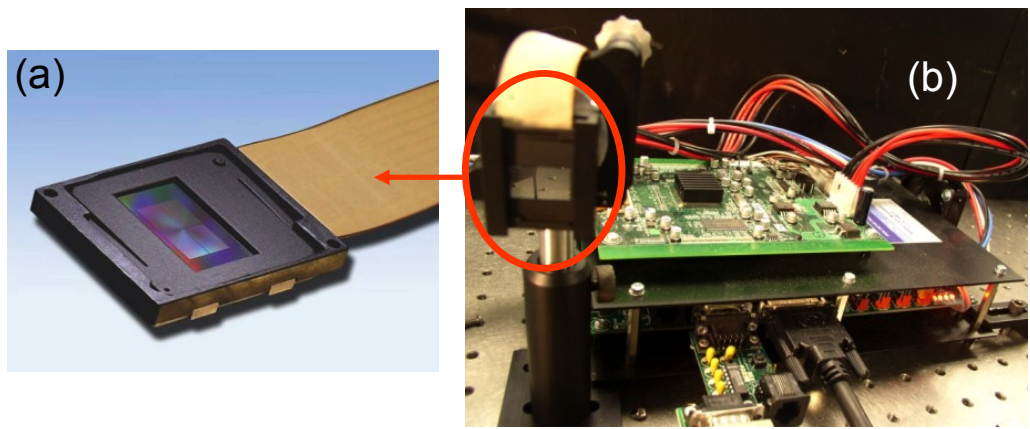

Figure 74. HDTV Phase Panel Developer Kit & HEO 1080 P, Holoeye. (a) SLM. (b) Interface drive electronics.

The SLM had 1920 × 1080 pixels in a 15.36 mm × 8.64 mm array that was controlled as an additional computer monitor via the digital video interface (DVI) on a personal computer. Other SLM properties are listed in Table 1.

**Table 1. Properties of the SLM (HEO 1080 P).**

| Pixel pitch | 8.0 μm |
|---|---|
| Phase Levels | 256 (8 bit) |
| Max. refresh frame rate | 60 Hz |
| Max. illumination | < 2 W / cm$^2$ |
| Depolarisation | <1 % |

In order to linearise the relation between the video gray levels and phase modulation a look-up table (LUT) had to be built that would convert video gray levels to the appropriate voltage levels, which would then result in the linear relation between the phase and the gray levels. To find the look-up table it was necessary to measure induced phase modulation as a function of addressed video gray levels. In the default hardware configuration (22 : 6 bitplane) 256 video gray levels were mapped to 1472 voltage levels. However, it was noticed that this configuration introduces fringes in recorded images because the liquid crystal molecules, addressed at 120 Hz, flicker due to their limited viscosity. The flicker could be effectively minimised by increasing the addressing frequency up to 300 Hz that can be achieved with the 5 : 5 bitplane configuration in this system.

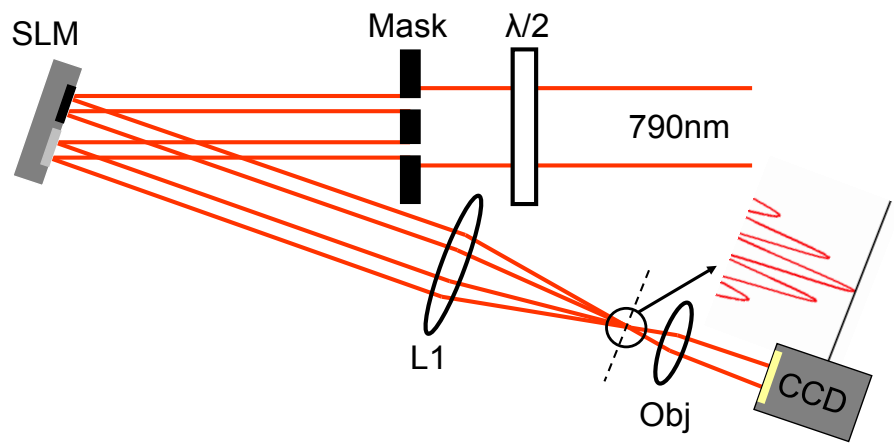

Figure 75. Setup to measure SLM phase modulation vs. addressed video gray level. Two beams are created with double hole mask and impinge on different parts of the SLM. Reflected beams are focused with a lens, L1 and the interference pattern in its focus is imaged onto a CCD camera with objective lens (x10), Obj. When video gray levels are changed in one of the parts from 0 to 255 the resulting interference pattern shifts spatially.

In this particular configuration only 192 voltage levels existed that were addressed by 256 video gray levels, therefore, it resulted in reduced phase resolution. The phase modulation with respect to the addressed video gray levels can be inferred from the interference pattern measurements of the two beams reflecting from the SLM at a small angle, as shown in Figure 75. When one part of the SLM, where one of the beams is impinging, is addressed with various video gray levels, it induces a phase shift of various degree that results in the interference pattern shift as shown in Figure 76. The spatial shift, $\Delta$ can be related to the phase shift, $\delta$ through the relation: $\delta = 2\pi \times (\Delta / \Gamma)$, where $\Gamma$ is the period of the interference pattern. Thus, if video gray levels from 0 to 255 are addressed subsequently, a series of the interference fringes can be acquired, as shown in Figure 77. Home written software in MatLab was used to evaluate the phase shift as a function of addressed video gray level in order to map the nonlinear relation. This is shown in Figure 77 (b), where it is also evident that the phase modulation of the SLM actually exceeds 2π. The relation was approximated by the polynomial curve in the region marked with the red line and was subsequently used to produce the gamma correction look-up table. The new gamma look-up table was written into the SLM hardware's memory that resulted in the linear response of the phase modulation to the addressed grey levels, as illustrated in Figure 77 (c).

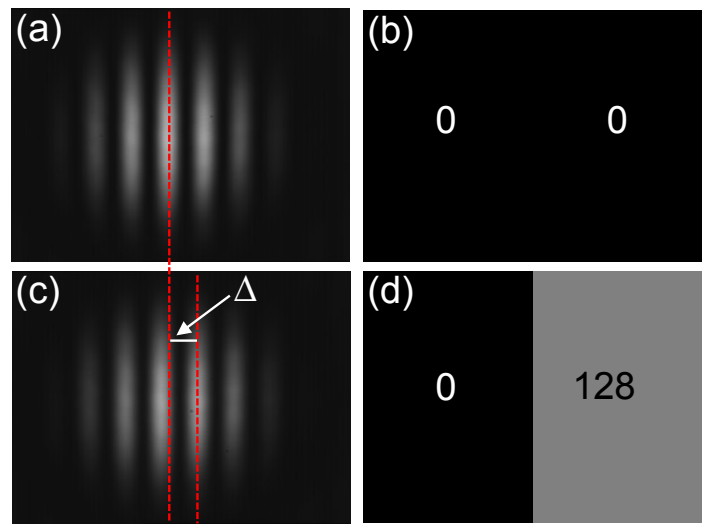

Figure 76. (a, c) interferograms and (b, d) the patterns displayed on the SLM. A spatial shift of Δ = 2 / 1.6 period of the interference pattern corresponds to a phase shift of δ = 1.6π, when one of the beams experiences a change in video grey level from 0 to 128.

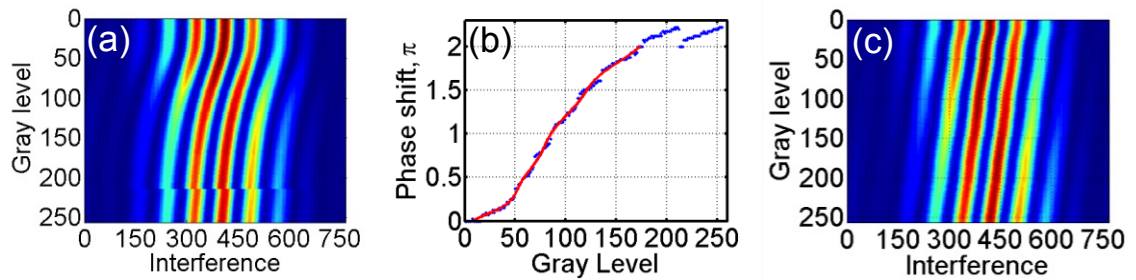

Figure 77. Linearisation of phase modulation with respect to addressed video gray levels. (a) interference fringes recorded when gray levels are displayed from 0 to 255 in one part of the SLM. (b) Phase modulation versus addressed gray levels extracted from (a). (c) The same as (a) but with the new gamma curve implemented, which was inferred from (b).

### 5.3.2    Generating beams of various shapes

Control of the STED beam point spread function was accomplished by placing the spatial light modulator in a plane conjugate to the pupil plane of the objective through the *4-f-like* lens system. Figure 78 shows the principle setup that can be used to generate various PSFs. Similar one was a part of the STED microscopy setup, which is shown in Figure 85. In practise the system had more relay lenses in order to comfortably control beam steering as explained in Section 6.2.4. Computer generated digital holograms of *type I* and *type II* (Figure 71) were displayed on the SLM that imprinted an accurate reproduction of the helical and circular $\pi$ phase distributions, respectively on the +1 diffracted beam, as depicted in Figure 78.

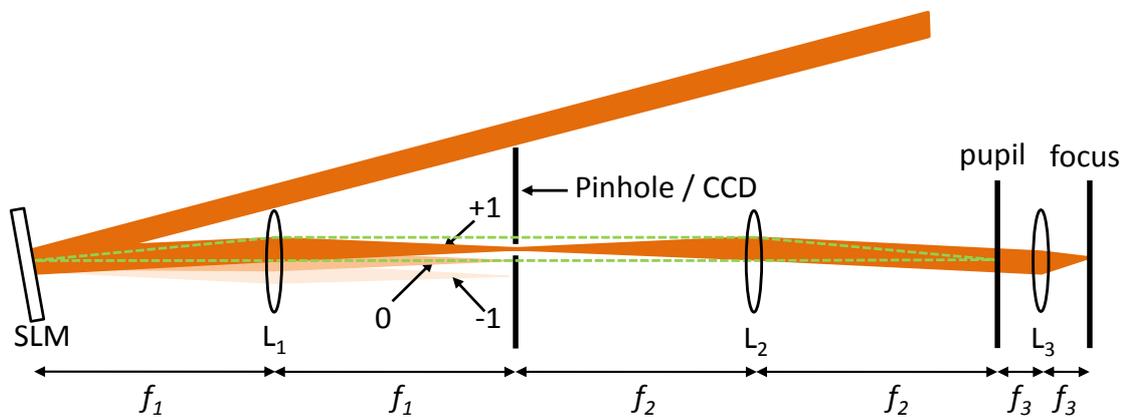

Figure 78. The principle scheme of controlling point spread function of the STED beam by means of spatial light modulator (SLM). Lens $L_1$ forms PSF in the focal plane (at the place of pinhole) that can be directly imaged using a CCD camera (acquired images shown in Figure 79). Alternatively, if a pinhole is used, the zeroth diffraction order (and others) can be blocked, and only the +1 diffraction beam let through. Lens $L_2$ collimates the beam and $L_3$ (objective) forms PSF in the focus. Green dashed curve depicts off-axis rays that images hologram displayed on the SLM onto the pupil plane.

In order to direct most of the energy to the +1 diffraction order, the generated holograms were blazed and gamma response of the SLM was carefully calibrated, as explained in Section 5.3.1. If a laser beam has a small diameter and lens ($L_1$) with a long enough focal length is used (low numerical aperture case) then it is easy to directly image PSF by placing a CCD camera in the focal plane (in the place of the pinhole in Figure 78). A lens ($L_1$ in Figure 78) focuses the diffracted beam from the SLM, which is equivalent to Fourier transform of the blazed hologram imprinted on the SLM, as illustrated in Figure 73. Figure 79 shows Gaussian and doughnut PSFs generated by using holograms that imprinted the same amount of various aberrations on the both beams. The picture also shows that the doughnut beam is more sensitive to aberrations than the Gaussian beam, as discussed in Section 5.2.8.

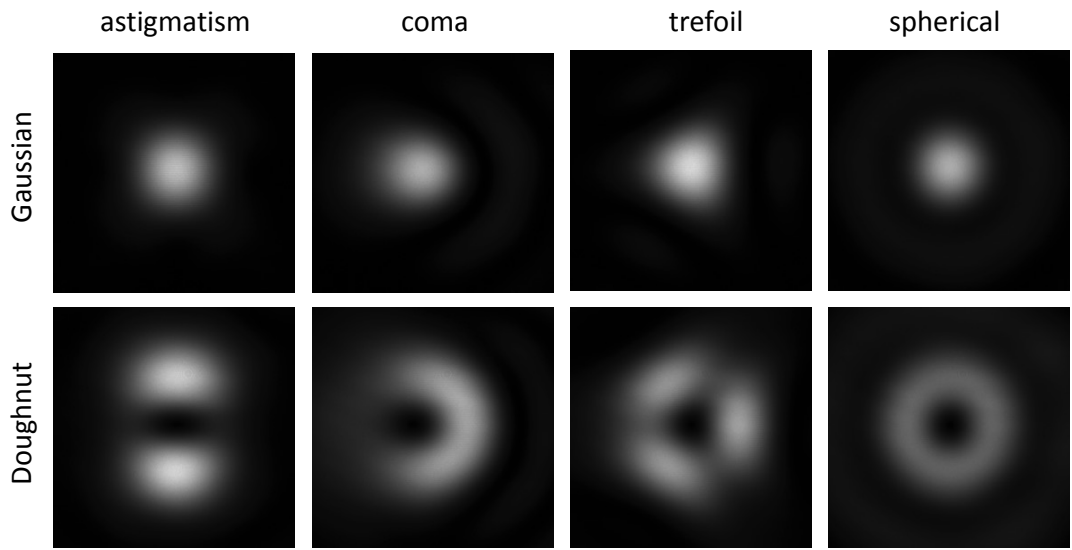

Figure 79. Comparison of the point spread functions of the Gaussian (top row) and doughnut (bottom row) beams subjected to the same amount of aberrations. Images were recorded with a CCD camera positioned in the focal plane of a lens with long focal length (see text for explanation). It is evident that the doughnut beam is more sensitive to aberrations.

### 5.3.3    Generating STED beams with high NA lens

In case of the high numerical aperture (NA) the PSF formed in the focal plane will depend on polarisation, as discussed in Section 5.2.3. Therefore, for generating PSF in confocal microscope with high NA objective, a quarter wave plate was put close to the pupil plane in the microscope body to create circularly polarised light. By rotating it to the appropriate angular position it was possible to create linear, left circular or right circular polarisations. A correct combination of circular polarisation handedness and wavefront topological charge of the helical distribution was chosen in order to produce near-zero on-axis intensity as explained in Section 5.2.6. The PSFs were imaged using the back-scattered light from 200 nm gold beads [345] and an open detector pinhole. Fine alignment of the STED beam within the microscope was possible by controlling the period and orientation of the blazed grating. Figure 80 (a) shows the PSF of the aberrated doughnut beam generated with a *type I* hologram. Aberrations in this system could be detected as discussed in Section 5.2.8, but simple visual inspection of the doughnut tells that some degree of coma and spherical aberration is present. Correction of it results in a near ideal shape doughnut with the centre intensity close to zero, as shown in Figure 80 (b). The spherical aberration is most evident in the axial image (not shown), however a halo around the doughnut in the lateral image can be seen, which is an indication of the aberration and therefore could be used to detect and correct it. It is important to produce an on-axis intensity as close to zero as possible. In order to measure this weak on-axis intensity, Fourier filtering was performed on the recorded images to remove the noise, as will be explained in more details in context of STED image processing, in Section 6.3.2.

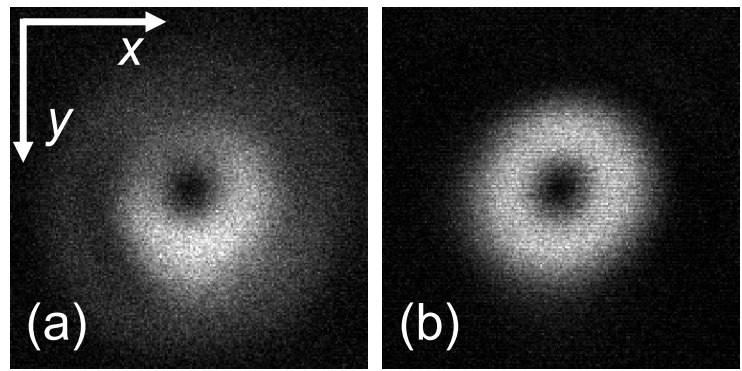

Figure 80. An example of aberrated (a) and aberration-corrected (b) doughnut recorded in the microscope.

Figure 81 shows that the doughnut has on-axis intensity less than 1 % of its maximum intensity. The on-axis intensity of the doughnut depends mainly on polarisation effects and aberrations, as discussed in Section 5.2.3. Figure 82 shows aberration-corrected images of the doughnut and the optical bottle beams.

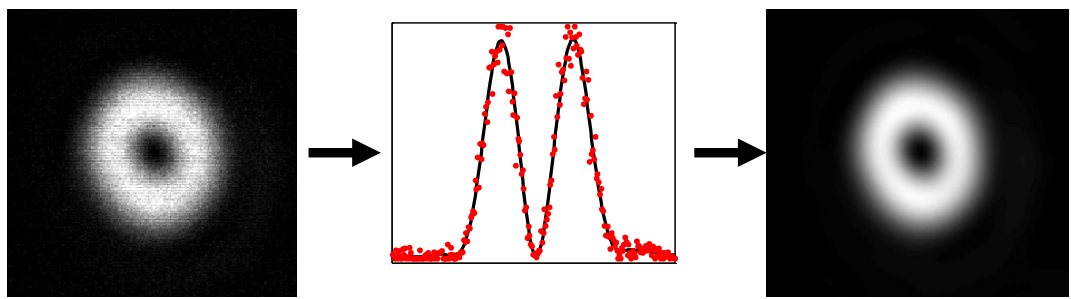

Figure 81. Fourier filtering of doughnut point spread function. It can be deduced that the on-axis intensity is less than 1 % of maximum intensity and FWHM of the doughnut hole is of 285 nm.

Instead of adjusting a quarter wave plate, in order to produce circular polarisation of different handedness, the sign of the topological charge was varied instead because it had the same effect on on-axis intensity, but was much easier to implement. As can be seen in Figure 82 (c), a strong *z* component is present in the axial image of the point spread function when a helical beam of +1 topological charge and left circularly polarisation is used. When the topological charge is changed to −1, a doughnut with almost zero on-axis intensity is created, as can be seen in Figure 82 (b). Different point spread functions were acquired by changing holograms displayed on the SLM and it can be in principle done almost instantaneously (60 Hz as afforded by the SLM). This demonstrates advantage of using SLM as an active wavefront changing device over fixed phase plates, where physical replacement is necessary in order to change the STED beam PSF.

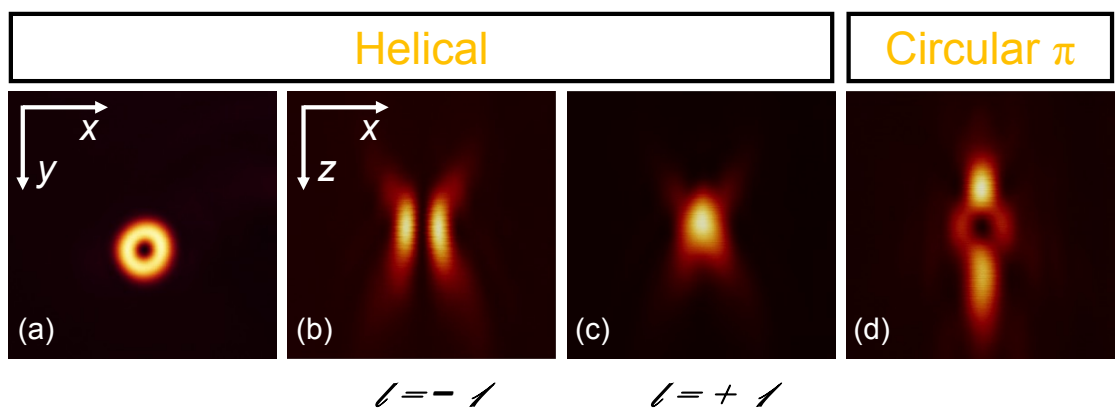

Figure 82. Aberration corrected point spread functions of doughnut (a, b and c) and optical bottle (d) beams in the microscope. Change of topological charge, $l$ results in different on-axis intensities for the doughnut beam (compare (b) and (c)). Images are 4.69 µm across. Measured by using a backscatter signal from 200 nm gold beads.

## 5.4    Summary and Outlook

Point spread function engineering of the STED beam is necessary in STED microscopy in order to achieve the optimal resolution improvement. Near ideal point spread functions with close to zero ($< \sim 1\ \%$) on-axis intensity of a doughnut and optical bottle beams were generated using helical and circular $\pi$ phase distributions with circularly polarised light and active aberration correction. The beams were generated employing two different holograms displayed on a phase only liquid crystal SLM that enabled to imprint an exact reproduction of a helical and circular $\pi$ phase distribution in the +1 diffraction order, to produce a doughnut and an optical bottle beams in the focal plane, respectively. The implementation was capable of switching between the two point spread functions that could be used to improve resolution either laterally or primarily axially. This prepared grounds for STED microscopy, which is presented in the next Chapter. In the future the experimental implementation of overlapped doughnut and optical bottle beams will be explored in more details by, for example, using either two different SLM [316] or one SLM with two different holograms imprinted on it [306, 346].

# 6.   STED Microscopy: Setup and Results

## 6.1    Introduction

The ultimate aim of the work presented in this Chapter is to develop a versatile STED microscope for biological FLIM-FRET applications, including studies of protein interactions localized on scales below the classical diffraction limit, such as microclusters of cell immune receptor activation [214], as explained in Section 3.4.3. To this end a STED microscope using a tunable supercontinuum excitation source [347] to provide relatively straightforward and low-cost spectral versatility in the excitation path has been demonstrated. Owing to the ultrashort pulse structure of the supercontinuum radiation the system was capable of fluorescence lifetime imaging (FLIM) through time-correlated-single-photon-counting (TCSPC) detection. The system also incorporated a programmable spatial light modulator (SLM) to facilitate convenient switching between different STED imaging modes and to compensate for aberrations in the depletion path. STED was implemented on an otherwise standard confocal scanning microscope for compatibility with conventional microscopy techniques and instrumentation. A review on STED microscopy is given along with explanation of various parts of the setup and results obtained with it. An introduction to STED microscopy was given in Section 2.6.1.

## 6.2    STED microscopy setup

This section presents the ongoing development of STED microscope. A FLIM microscope, described in Section 3.3.2, that was based on ultrashort pulsed laser, a laser scanning confocal microscopy and time correlated single photon counting (TCSPC) detection, was converted to STED operation. The modification essentially included addition of a microstructured optical fibre (MOF) for supercontinuum generation, a temporal pulse stretcher and a spatial beam shaping unit for STED beam shaping. The elements were built around the microscope as shown schematically in Figure 83 (the picture of the STED microscope is shown in Figure 84).

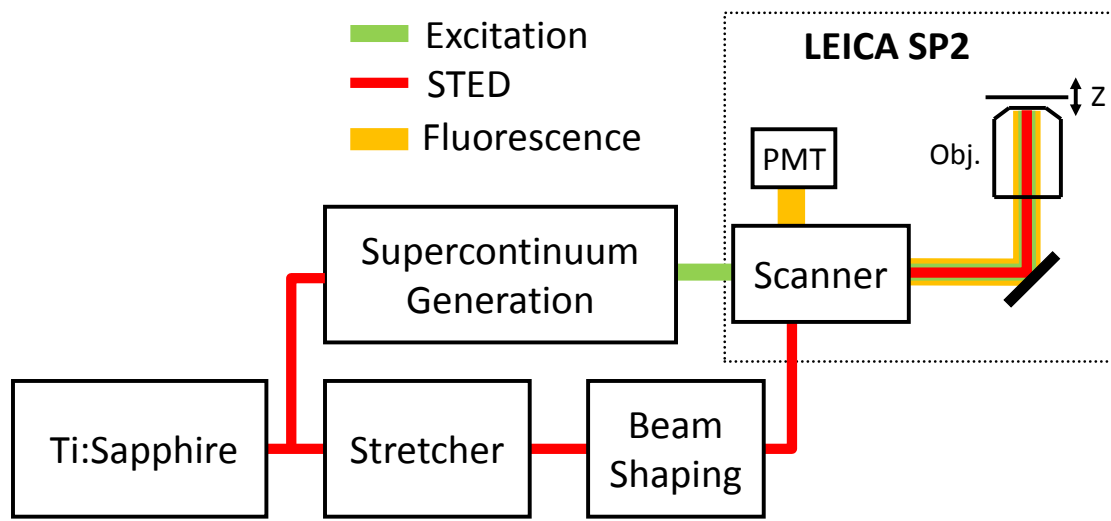

Figure 83. STED microscopy system schematics. Radiation from Ti:Sapphire is split into two beams – one for supercontinuum generation and the other for STED. Supercontinuum is filtered and STED beam is temporally and spatially shaped. Both beams are overlapped in the microscope.

Radiation from the mode-locked Ti:Sapphire laser (Tsunami, Spectra – Physics Inc), which was pumped by a 10 W argon-ion laser, was split into two beams, each of which was coupled in to an optical fibre. The pulses of the first beam, to be used as an excitation beam, were broadened spectrally in a microstructured optical fibre (through the supercontinuum generation, as explained in Chapter 4) and the pulses of the second beam, to be used as the STED beam, were broadened temporally in a single mode fibre (SMF) relying on the effects of group velocity dispersion and self phase modulation. The STED beam was then passed into the beam shaping unit consisting of a spatial light modulator for wavefront modification as described in Chapter 5.

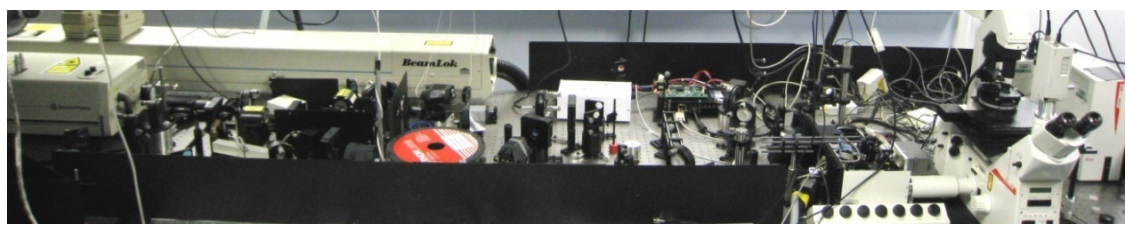

Figure 84. Picture of the STED microscopy system.

The wavefront engineered STED and spectrally filtered supercontinuum beams were then coupled into the commercial microscope (TCS SP2, Leica), where they were overlapped with the help of a 30:70 (Reflection : Transmission) beam-splitter (Figure 85) before being coupled to the *xy* galvanometer scanner. The collinear beams then went to the microscope objectives and were focused onto the sample. The setup is described in more details in the following sections.

### 6.2.1    *Beam path on the bench*

A detailed beam path is shown in Figure 85. The intensity of the laser operating at 780 nm ($\Delta\lambda$ = 6 nm, $\Delta\tau \sim$ 100 fs, 76 MHz) was controlled with a half wave plate in combination with a polariser. The ability to control the laser intensity was convenient for alignment purposes or whenever low laser intensities were necessary. A Faraday rotator (FR1 in Figure 85), followed by another polariser, oriented at 45 degree with respect to the first one, acted as an optical isolator.

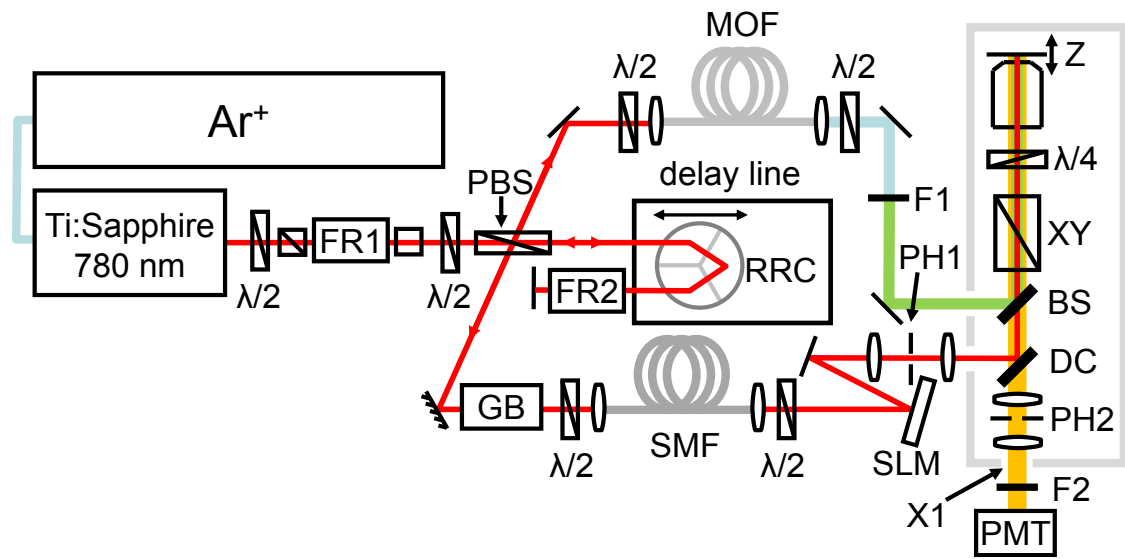

Figure 85. STED microscopy setup. FR, Faraday rotator; $\lambda$ / 2, achromatic half wave plate; $\lambda$ / 4, achromatic quarter wave plate; PBS, polarising beamsplitter; GB, SF57 Glass block; RRC, retroreflector corner-cube; MOF, microstructured optical fibre; SMF, single mode fibre; F1, bandpass filter (628 / 40 nm); F2, combination of short pass filter (750 nm) and bandpass filter (692 / 40 nm); SLM, Spatial Light Modulator; DC, (infrared port) dichroic; BS, 30 / 70 Reflection / Transmission beam splitter; XY, x-y galvanometer scanner; Z, galvanometer driven specimen stage; PMT, photomultiplier; PH1, pinhole; PH2, confocal pinhole. Note that the diagram is simplified (see the text).

The isolator protected the mode-locked laser from unwanted optical feedback coming from optical elements, including the two optical fibres (which gave the biggest contribution to the unwanted reflections). After the isolator, the beam was divided into the two parts with another polariser (PBS in Figure 85); here a Glan laser double escape window polariser (CPAD-8.0-670-1064, CVI) was used. A half-wave-plate ($\lambda / 2$ in Figure 85) in front of the polariser allowed the relative intensity to be adjusted between the two beams. One part of the beam was then used for supercontinuum generation and the other for the STED. The beam for supercontinuum generation first went into the delay line unit before being reflected by the polarising beam splitter and coupled in to the microstructured optical fibre. Both the excitation and STED beams consisted of trains of short pulses that were synchronized with each other on a picosecond timescale. The optomechanical delay line was introduced in order to control the delay between the excitation and STED pulses. It was chosen to vary the delay of the excitation pulses with respect to STED pulses rather than the other way round because the loss of the delay system was 40 % and therefore it was better to lose it from the less powerful beam – excitation beam. The delay line was introduced before the microstructured optical fibre to avoid any chromatic effects that might be induced by the delay optics over the broad spectral extent of the generated supercontinuum. It was important to make sure that movement introduced by the delay line unit did not change the position or orientation of the focused spot on the microstructured optical fibre, in order to make coupling efficiency into the fibre and, therefore, the intensity of the generated supercontinuum stable. This was achieved by using a combination of a retro-reflective cube (NT45-187, Edmund Optics), a Faraday rotator and a mirror. A retro-reflective cube reflected the beam back parallel to the direction of the beam impinging on it. However, the position of the beam might still 'walk' from side-to-side if the beam is not exactly parallel to the direction of the cube translation. Nevertheless, if the beam, after retro-reflecting from the cube, is sent back along the same path by simply reflecting it with a mirror then the beam going backwards would not change its orientation or position when the cube is translated. In order to separate the reflected beam from the initial beam with the polarising beam splitter, the polarisation of the backward beam has to be $90^{\circ}$ with respected to the initial beam, which is achieved by placing another Faraday rotator (FR2 in Figure 85) between the cube and the mirror. This makes the light going through the rotator in backward and then forward directions to change the polarisation by $90^{\circ}$. It was chosen to use the Faraday rotator rather than the less expensive quarter wave plate since it was found that the former produced the 90° angle more accurately. The beam, after being reflected-off the polarising beam splitter, went to the microstructured optical fibre (MOF in Figure 85) for supercontinuum generation.

### 6.2.2   Generating excitation beam

The first STED microscope reported in 1999, utilized a mode-locked Ti:Sapphire laser system and its second harmonic signal to provide STED and excitation beams, respectively [114]. This approach was limited to a few fluorophores that exhibited the necessary large Stokes shift. More flexible systems were later developed that employed optical parametric oscillators (OPO) to frequency up convert Ti:Sapphire radiation in order to produce tunable excitation beam [277, 278, 348], but these were complex and expensive. Subsequently simpler systems were demonstrated exploiting synchronised pulsed laser diodes for excitation and STED [349, 350]. In order to achieve sufficient STED intensity, either an incoherent overlap of two different diode lasers [349] or amplification ( × 8) of a single diode laser [350] were used. Pulsed diode lasers are now routinely used in STED microscopy to provide the excitation beam. They can be synchronised with the STED beam, which in turn can be provided by either a mode-locked Ti:Sapphire laser [289, 292, 295] or OPA system [322, 323, 325-329], or by both of them [324, 351]. Recently, continuous wave (*cw*) lasers were demonstrated to be a convenient and cheap alternative that significantly simplifies the laser instrumentation in STED microscopy [352, 353]. The disadvantage of *cw* lasers, however, is that they can only be used with dyes that have low or no triplet level build-up, in order to prevent fast photobleaching. The most obvious candidates for such imaging are colour centres in diamonds. TCSPC FLIM would also not be possible with *cw* lasers. In addition, the ultrashort lasers are better suited for supercontinuum generation. Nevertheless, powerful *cw* supercontinuum generation was recently demonstrated with the spectral extent to the visible spectral region [354], therefore, *cw* supercontinuum could potentially be used in STED microscopy as a cheap alternative.

Here, in this thesis it is demonstrated (later in this Chapter) that a broadly tunable supercontinuum source (similar to that reported in Ref. [220]) obtained by pumping a microstructured optical fibre with the ultrashort pulses from Ti:Sapphire laser can be used for STED microscopy. The generated supercontinuum can provide a spatially coherent train of excitation pulses that is inherently synchronized to the STED beam pulse train, as they both originate from the same laser oscillator. Following the publication of this work [347], another work demonstrated that supercontinuum originating from a fibre laser based source can be used to provide the excitation and STED beams simultaneously [291]. This significantly reduces the complexity and cost associated with the laser systems for STED microscopy but still preserves the spatial coherence and ultrashort pulses for both beams. The important result of this publication is that it shows that the STED beam obtained by slicing out a bandwidth of 20 nm from this source is sufficiently intense for STED microscopy, improving resolution by a factor of 8-9 beyond the diffraction limit. Potentially the resolution should be as good as ~ 20 nm. It is important to keep the bandwidth narrow in order to avoid introducing chromatic aberrations in the STED beam, which could potentially distort the PSF. The source, due to its broad spectrum, naturally lends itself to tunability of both the excitation and the STED wavelengths, as was also demonstrated by acquiring sub-resolution images at three different excitation and STED wavelength pairs in the same publication [291]. In this Chapter, the initial attempts to use similar fibre laser-based supercontinuum source to STED microscopy is presented. However, this approach was abandoned because of the too low pulse energy available from that particular model as explained below. Recently it was shown that Raman shift in standard fibres [355] can be employed to provide a range of new red shifted wavelengths that can be used as a STED beam [290].

In the STED setup in this thesis, Ti:Sapphire radiation was coupled into the 1 meter of polarisation maintaining microstructured optical fibre (MOF in Figure 85) with 1.8 μm core and two zero-dispersion wavelengths at 750 nm and 1260 nm (NL-PM-750, Crystal Fibre). A complete setup is shown in Figure 85. An achromatic infrared half wave-plate (ACWP-700-1000-10-2, CVI) was used before the microstructured optical fibre to rotate polarisation of the laser beam to match the fast axis of the fibre. Polarisation of the output supercontinuum was then controlled with another achromatic visible half wave plate (ACWP-400-700-10-2, CVI). The microstructured optical fibre used here had an asymmetry in the core of the fibre which induces birefringence and leads to a polarisation maintaining property of the fibre. Therefore, the supercontinuum coming out of the fibre is polarised with high polarisation extinction. Coupling efficiency of 40 % into the fibre was typically achieved. The microstructured optical fibre was a single mode fibre with a cut-off wavelength of < 650 nm. Some other properties of the MOF are listed in Table 2.

**Table 2 Optical properties (at 780 nm) of the microstructured optical fibre.**

| Birefringence | $3 \cdot 10^{-4}$ |
|---|---|
| Numerical aperture | 0.38 |
| Mode field diameter | 1.6 μm |
| Nonlinear coefficient | 95 (Wkm)$^{-1}$ |

The zero-dispersion wavelength shift from 1.28 μm (for bulk silica) to 0.75 μm (MOF) due to microstructured fibre design influence on the waveguide dispersion is illustrated in Figure 86 (a). In Figure 86 (b) the spectra of generated supercontinuum are shown as a function of coupled-in pump power at 780 nm. It was found that the supercontinuum output fluctuates less in power if the MOF is pumped with higher powers.

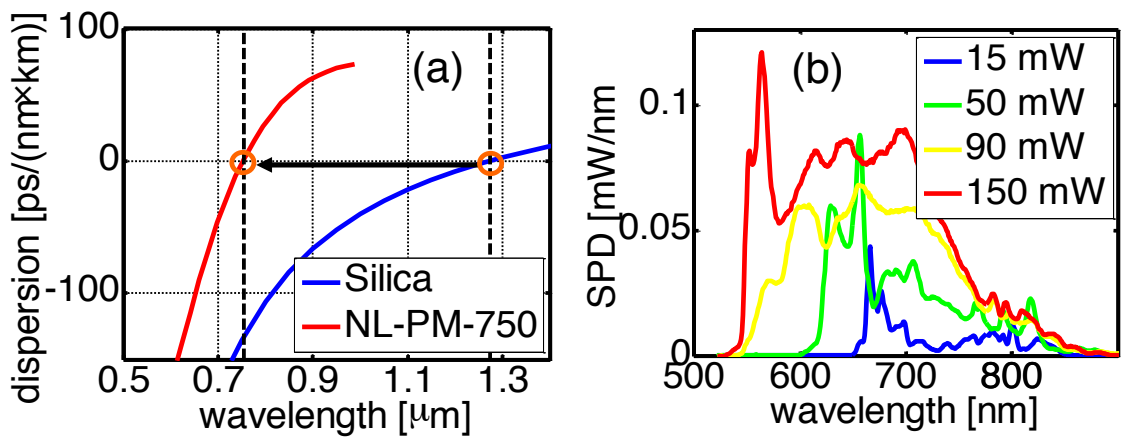

Figure 86. (a) Comparison of the dispersion curve of the NL-PM-750 fibre (data from Crystal – Fibre) and the bulk silica (calculated). (b) Supercontinuum generated in the fibre as a function of pumped power at 780 nm (as measured in front of the fibre).

### 6.2.3 Generating STED beam

STED pulses have to have high peak power in order to outperform spontaneous emission. High power might induce unwanted fluorescence through the two-photon excitation or re-excite molecules right after the molecules has been depleted [123] if the STED pulses are 'blue' enough to cause the excitation through the single photon absorption, as illustrated in Figure 87. Temporal broadening of STED pulses helps to reduce these effects as well as to minimise photobleaching [356].

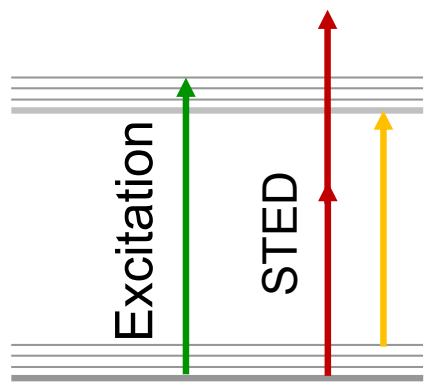

Figure 87. A molecule can be excited with the visible excitation pulse (green arrow) as well as with the red / infrared STED pulse. STED pulse can excite molecule through the two photon absorption (dark red arrow) or, if a more energetic single photon is used, through the single photon absorption (yellow arrow). The two-photon absorption usually happens with femtosecond STED pulses, whereas the single photon absorption can happen if the STED pulses are 'blue' enough. The pulses can be spectrally broadened towards blue if femtosecond pulses are stretched by means of optical fibre and resulting spectral broadening, induced through self-phase modulation by ultrashort pulses, is broad enough to reach the absorption band of the molecule.

A pair of gratings was used to stretch pulses temporally in the first series of STED experiments [66, 114, 278, 307, 310, 348]. One of the disadvantages of using gratings, however, is that they can degrade the quality of the beam if they are not used in the correct alignment. The beam produced by the Ti:Sapphire laser is known to often be of an elliptical shape, therefore, to improve the quality of the beam, and also to stretch its pulses temporally, a single mode fibre (SMF) can be used instead of gratings [322, 357-360]. Temporal stretching in single mode fibre is a consequence of the group velocity dispersion. However, if femtosecond pulses (with high average power) are focused to a fibre, a spectral broadening can occur, mostly due to self-phase modulation, as discussed in Chapter 4. This is good if we want to get longer STED pulses (normally the case) since self-phase modulation can further broaden the pulses, but this can be a problem because spectrally broader pulses might 'leak' into the detection channel on the 'red' side of the spectrum, or even excite fluorophores through a single-photon excitation on the 'blue' side of the spectrum, as shown in Figure 87. To reduce the self-phase modulation (and therefore spectral broadening), the peak power of pulses can be reduced by temporally pre-stretching them in a glass block, before coupling them into the single mode fibre. Sometimes a combination of a pair of gratings and then a single mode fibre is used, in order to adequately stretch pulses temporally and to produce a high quality beam [123, 324], but reflection losses at the gratings and coupling losses in the fibre, in such a system, results in inefficient use of power. Stretching of the STED beam temporally might not be necessary if the laser produces sufficiently long pulses. For example, a mode-locked krypton-ion laser [311], and Q-switched microchip laser [361] were successfully used as STED beams. As discussed earlier, the diode laser [349], *cw* laser [353] or supercontinuum source [291] can be used to produce STED beam, which do not require any stretching. Sometimes a single mode fibre is used anyway just to improve the quality of the beam.

In this thesis, a STED pulse stretcher consisting of 19 cm of bulk SF57 glass (SCHOTT) followed by 100 metres of polarisation-maintaining single mode fibre (FS-LS-4616, Thorlabs) was used. The fibre was a cheap alternative to other commercially available polarisation maintaining fibres. Polarisation control was implemented throughout using achromatic half wave-plates. Depending on the power coupled into the fibre, the pulses could be stretched temporally over a range from 60 ps (linear broadening due to group velocity dispersion in bulk silica) to > 300 ps (caused by the combination of the group velocity dispersion and power-dependent self-phase modulation in the normal dispersion region). The pulse temporal width depends on the input power due to the role of self-phase modulation. The induced nonlinear pulse broadening occurs in a first few centimetres of the fibre and the subsequent reduced peak power means that the pulses do not change much spectrally in the remaining ~ 100 meters. However, the pulses continue to be stretched temporally over the remaining length of the fibre due to the group velocity dispersion. The spectral broadening might be not as dramatic as in microstructured optical fibre, nevertheless it could, for example, reach ~ 80 nm broadening at the 10 μW / nm level, when pumped with 520 mW average power at 780 nm, as shown in Figure 88.

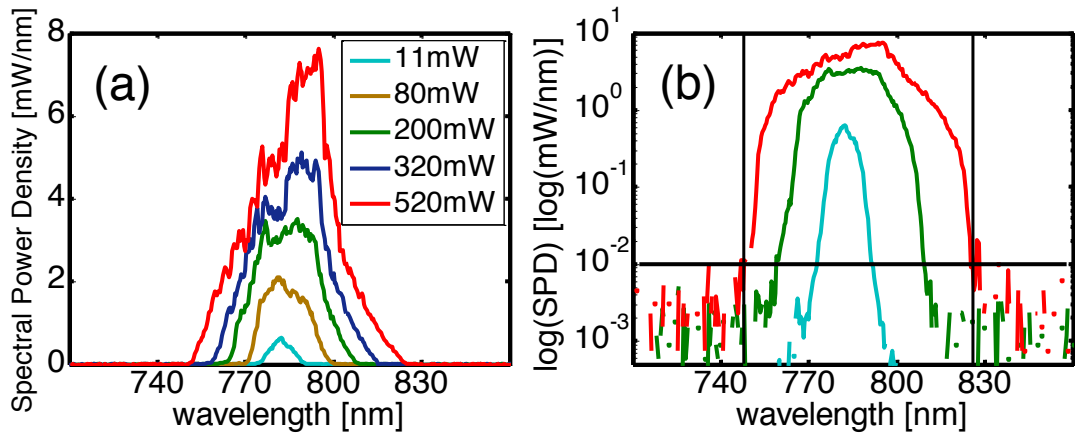

Figure 88. Spectral broadening of STED pulses in a the single mode fibre as a function of the average coupled-in power, with a glass block in front of it. (a) In spectral power density (SPD) units. (b) in log(SPD). It can be seen in (b) that the spectrum broadens from 745 nm to 825 nm (Δ80 nm) at the 10 μW / nm level when pulses of 520 nm average power are coupled into the fibre.

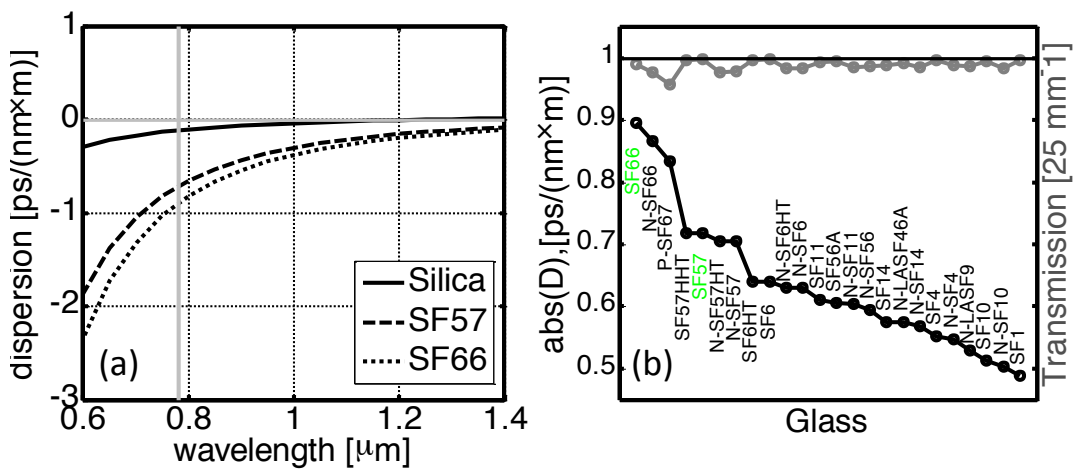

Figure 89. (a) Calculated dispersion of bulk Silica, SF57 and SF66 materials. A horizontal gray marker marks the dispersion values at 780 nm (Silica: −0.12 ps / (nm × m), SF57: −0.72 ps / (nm × m), SF66: −0.9 ps / (nm × m)). (b) Absolute dispersion values at 780 nm, calculated using Sellmeier constants (SCHOTT data), and transmission values (SCHOTT data) at 700 nm of various highly dispersive SCHOTT glasses.

This spectral broadening could cause problems in STED microscopy experiments because the 'blue' part of the STED pulses might be absorbed by the sample if it overlaps with the 'red' part of the fluorophore excitation spectrum. Since the STED pulses are powerful compared to the excitation beam, the fluorescence excited by the STED pulses alone might become comparable to the fluorescence excited with the supercontinuum. As discussed earlier, to limit self-phase modulation in the fibre, a glass block can be introduced to pre-stretch the pulses before they are focused on to the fibre core. Glasses can effectively stretch visible pulses but infrared pulses are stretched less effectively, as is evident looking at the dispersion curves for various materials, presented in Figure 89 (b). Therefore, a glass material had to be carefully selected in order to maximally stretch pulses at 780 nm. The dispersion values of different SCHOTT glasses were calculated at 780 nm and plotted in Figure 89 (b), using Sellmeier equation [29] and the Sellmeier constants obtained from the SCHOTT website. The glass was also chosen on the basis of its transmission values since it is important that the STED beam would not be too much attenuated. The multiplication of the two values was used as a figure of merit in selecting the glass material – the higher the factor the more suitable glass. The most chromatically dispersive glasses for infrared pulses (~ 780 nm) were found to be of the *SF* series with the *SF66* topping the list. However, since *SF66* was more expensive than *SF57*, the latter was chosen (both marked green in Figure 89 (b)). Commercially available blocks of the glass were up to a maximum of 19 cm in length (from SCHOTT). Calculations show that such a length of the *SF57* stretches pulses of 100 fs to ~ 0.7 ps. Nevertheless it reduces the pulse peak power and consequently the self-phase modulation in the fibre, which significantly reduces the spectral broadening, as will be discussed below. The reduced peak power also reduces the potential damage of the tip of the fibre. Nevertheless, same spectral broadening is still apparent, as can be seen in Figure 88, where all the data shown were acquired with the glass block in front of the single mode fibre. Measurements of the spectral and temporal broadening of the stretcher were performed as a function of input power to find out how the power coupled in to the fibre affected the spectral and temporal characteristics of the STED pulses. A spectrometer (Ocean Optics) was positioned after the single mode fibre to monitor the spectral broadening and a PMH-100 detector (B&H), working together with TCSPC (B&H) card, was used to measure the temporal profile. The detector had a typical IRF of ~ 183 ps and therefore it was not suitable to measure short pulses and measurements of longer pulses required deconvolution with the PMH-100 instrument response function to get the actual pulse shape. It is apparent from the Figure 90, that pulses broaden spectrally with high power and the central wavelength shifts towards the red (presumably because of Raman scattering). The shape of the spectrum changes as can be seen in Figure 91 (a). Pulses also broaden temporally with the high powers and, in addition, the centre of the pulses is delayed in time (Figure 90).

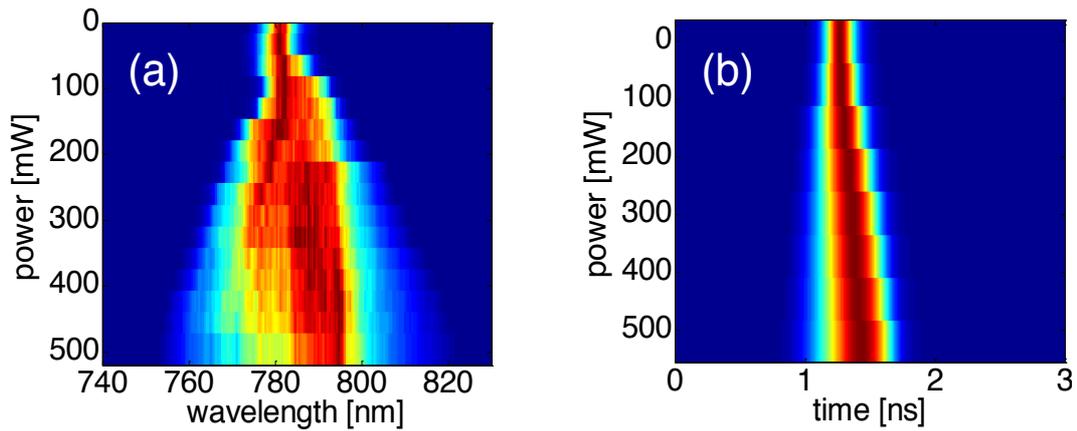

Figure 90. Spectral (a) and temporal (b) broadening of STED pulses versus power in 100 metres of a single mode fibre (+glass block in front). (a) Normalised spectral profiles *vs.* power as measured with Ocean Optics spectrometer. (b) Normalised temporal profiles *vs.* power as measured with B&H TCSPC card (IRF of ∼ 183 ps). A shift in time with power is observed because of the high nonlinear refractive index at high average powers. A slight spectral shift towards the infrared with power is observed most probably due to the Raman scattering.

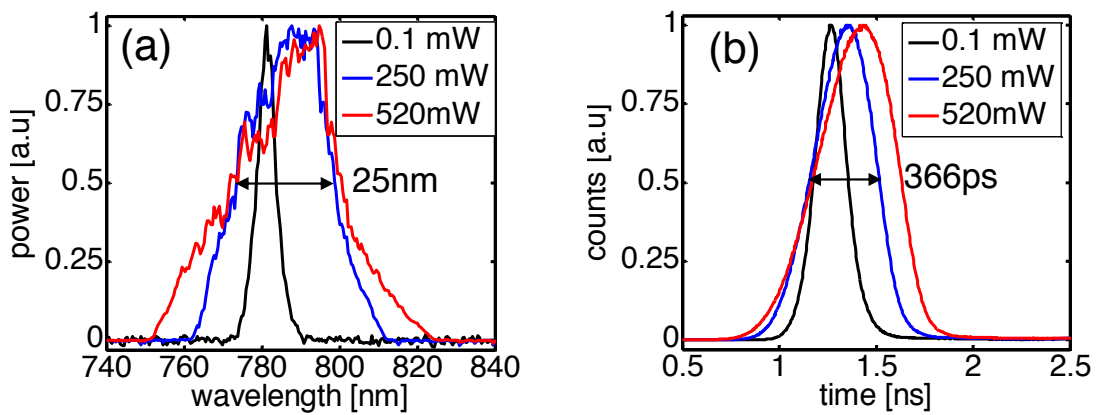

Figure 91. Spectral (a) and temporal (b) broadening at three different power levels: (black) 0.1 mW (5.2 nm, 183 ps); (blue) 250 mW (25 nm, 366 ps); (red) 520 mW (29 nm, 464 ps). All values in FWHM. Temporal values measured by B&H TCSPC card with IRF of ∼ 183 ps. The blue curves, with FWHM values as shown in the figure, were typically used in the STED experiments.

This can be due to the group velocity dispersion and increase in the refractive index, which is power dependant because of the nonlinear refractive index, $n_2$ (eq. 33). It was also observed that if the laser mode-locking is lost at any time during the measurement then it ruins the data acquisition, since the point of the reference is lost (if the laser is mode-locked again the pulse arrival time might be shifted and pulse temporal profile might change due to different spectral properties of the laser after the re-mode-locking). As an example of typical pulse broadening, 250 mW of average power coupled into the fibre broadened the pulses by 25 nm spectrally and to 300 ps temporally (all data in FWHM units and temporal measurements deconvolved with the IRF of 183 ps, from a measured FWHM of ~ 366 ps). This could also be validated by the following calculations: if the dispersion $D$ of silica is assumed to be of $-0.117$ ps / (nm × m) at 780 nm, then pulses with spectral width $\Delta S$ of 25 nm, in $L =$ 100 m of single mode silica fibre would roughly broaden by $\Delta\tau = D \times L \times \Delta S$ and therefore $\Delta\tau = 293$ ps, in approximate agreement with the experimental data. This estimate confirms that self-phase modulation happens at the beginning of the fibre and then pulses further broaden temporally because of the group velocity dispersion.

## 6.2.4    *Overlap of excitation and STED beams*

The precise alignment of the excitation and STED beams spatially (as well as temporally) is one of the most important factors in order to successfully perform STED microscopy measurements on a nanometre scale. To overlap the beams spatially, each beam was controlled with two steering mirrors located on the bench, which had a pupil and sample planes overlaid on them. These planes, conjugate to microscope's pupil and sample planes, were created outside the microscope using appropriately positioned relay lenses. The position of the focal spot at the sample plane was controlled by adjusting a mirror with the pupil plane overlaid on it. The equivalent effect was used to adjust the position of the beam on the pupil of the objective. This allowed nearly independent control of the position of the beam in both the sample and pupil planes. To relay the pupil and the sample planes required building a telescope with four lenses, as shown in Figure 92. The steering mirrors were placed in $S$ and $P''$ planes as shown in the figure. The telescope magnification was chosen such that it would match the working area of ~ 1 cm on the SLM to 5 mm on pupil plane of the objective. A conjugate pupil plane ($P'$) formed by the scan-lens in the scanner unit was found to be inside the scanner unit some 30 cm away from the exit of the infrared port and with the diameter of 2-3 mm. Therefore, the telescope was designed to magnify the image in $P'$ four times in order to match the area on the SLM. The position of $P'$ prevented building a real *4-f* system, since that would have require the first lens to have 30 cm focal length and even longer lengths for some of the following lenses in order to magnify the beam. This all would make a long telescope occupying considerable space on the optical bench. Because the telescope was not *4-f* system it did not allow completely independent control of pupil and sample planes. Consequently, when the beam is moved in the sample plane, it moves slightly on pupil planes, and vice versa. However, this did not affect the performance of the STED because the beams could nevertheless be overlapped on top of each other and along optical axis, and once aligned the beams would not need to be realigned. In the later configuration (not reported in this thesis) a *4-f* system was built by removing some of the original infrared port components in order to be able to place a short focal length lens (f = 10 cm) closer to the scanning mirror, where a pupil plane was overlaid.

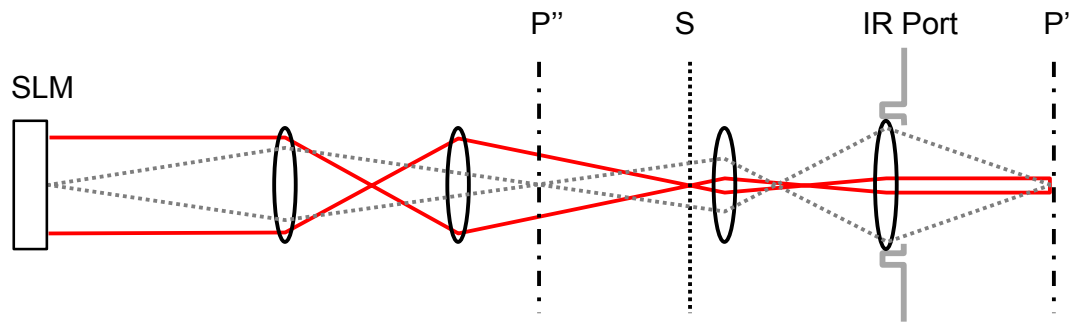

Figure 92. Lens system enabling nearly independent control of sample and pupil planes in the STED beam. Beam path (image formation) is marked in red. Illumination (pupil) is marked in dotted gray. Steering mirrors are placed in *S* and *P''* planes to allow control over pupil and sample planes respectively. IR – infrared. Drawn not to scale.

The excitation and STED beams were coupled into the confocal scanning unit (TCS SP2, Leica) through the ultraviolet and infrared ports using RT30 / 70 (reflection / transmission) intensity and infrared dichroic beam splitters, respectively, as shown in Figure 85. After passing through the galvanometric *xy* scanner, the overlapped beams were diverted to the inverted confocal microscope (DMIRE2, Leica) and both beams were focused onto the sample plane by the objective lens (NA = 1.4, oil immersion, HCX PL APO PH 3, x100, Leica). An infrared quarter wave plate (ACWP-700-1000-10-4, CVI) in the pupil plane of the microscope objective was used to create a circularly polarized STED beam. This arrangement minimized the effects of the excitation and STED anisotropy and, most importantly, ensured a true zero intensity in the centre of the STED beam PSF, as discussed in Chapter 5. The alignment of the excitation and STED beams was performed while continuously imaging gold beads [345]. The reflected (or rather backscattered) signal was used to build simultaneous images, as shown in Figure 93, of the single gold bead with the excitation and STED beams (by using appropriate filers in front of the two separate detectors). First, it was necessary to ensure that the excitation beam was on-axis by maximising the signal going through the confocal pinhole. Then, the STED beam was overlapped on the excitation beam with the help of the steering mirrors that independently controlled sample and pupil planes.

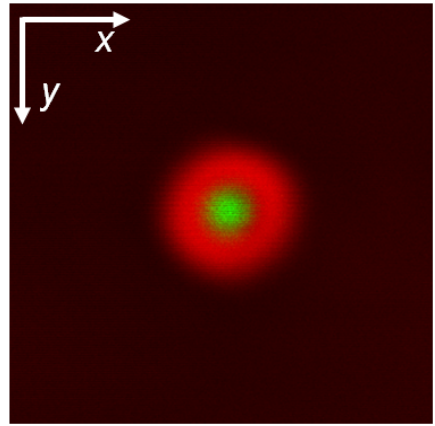

Figure 93. Overlaid images of the backscattered signal from gold beads of the excitation (green) and STED (red) beams. The image represent the corresponding point spread functions.

The excitation and STED beams were also overlapped in the *z* direction by moving one of the lenses in the telescope next to *P''* or *S* planes. The final adjustments axially were done with the spatial light modulator by adjusting the out-of-focus Zernike aberration as explained in Chapter 5. In the initial experiments, the alignment of the STED beam was carried out using the non-descanned detector (NDD) by inserting a 50 / 50 beam splitter cube in the microscope body so that the backscattered light would be reflected to the non-descanned detector. This detection was not optimal since images of STED PSF this way would be different from the ones where the signal would go all the way back (de-scanned) to the scanner and further to the dichroic and pinhole. The later represents the true PSF since it is the path that fluorescence would normally take and, therefore, it should be used to record the true PSF. This is also important in, for example, aberration corrections. However, the first approach was used initially because the de-scanned detectors inside the scanner unit box could not be used due to a multiphoton filter installed in the scanner in its original configuration. After removing this filter, the light could reach the *X1* (external) port, but not the internal detector, since the configuration of the spectral selection mechanism [362] was optimized to detect the visible rather than the infrared light. Therefore, most of the experiments were performed with the Leica external PMT (for intensity imaging) or with the B&H PMT (PMH-100) (for time resolved measurements) mounted on the *X1* port (see Figure 20 and Figure 85).

Temporal synchronization of excitation and STED pulses was performed with the delay line by shifting excitation pulses in time, as is explained in more details in Section 6.3.1. There, fluorescence signal rather than backscattered signal from the gold beads was monitored in order to determine the optimal separation between the two pulses. The pulses were adjusted in time so that the fluorescence depletion, induced by STED pulse, would maximally quench overall fluorescence. The TCSPC card, running in 'oscilloscope' mode, was used to perform repetitive measurements that were displayed like on an oscilloscope in real time to facilitate temporal adjustment. The fluorescent measurements are described below.

### 6.2.5   Fluorescence excitation and detection

For STED experiments a sample of 200 nm diameter "dark red" fluorescent beads (Molecular Probes) were used dispersed on a cover-slip and mounted in Mowiol mounting medium. Both excitation and STED beams filled the back aperture of the objective and the confocal pinhole size was set to the diameter equivalent to the radius of 1 Airy disk (see Figure 8 for explanation). The power incident on the microstructured optical fibre was chosen such that it would be just enough to generate a stable supercontinuum in order to save the rest of the infrared beam for STED.

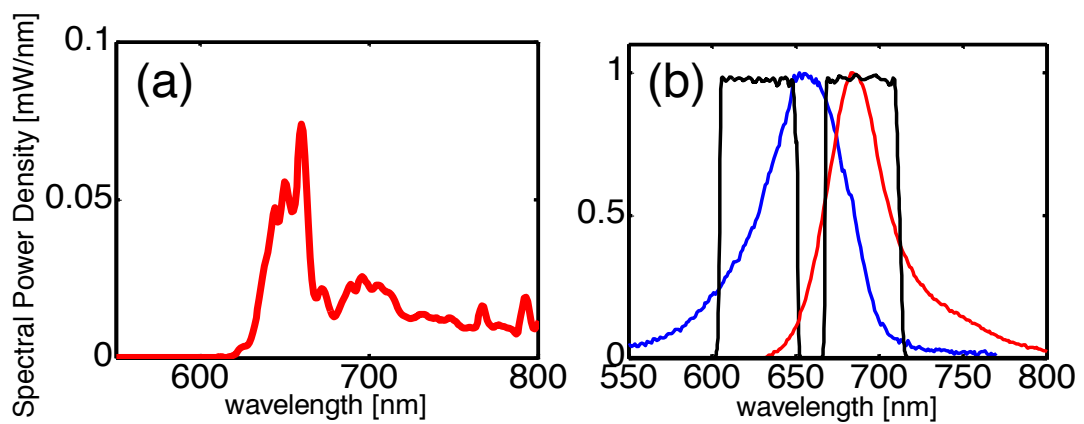

Figure 94. (a) Supercontinuum used in STED microscopy as generated with 30 mW of input power. (b) Excitation (band-pass 628 / 40 nm) and Emission (band-pass 692 / 40 nm) filters with excitation and emission spectra of 'Dark Red' (Molecular probes) dye.

A total average output power of 12 mW was generated from 30 mW average incident laser power. This resulted in the supercontinuum shown in Figure 94 (a). After spectral selection with a band-pass excitation filter (BrightLine HC 628 / 40 nm, Semrock), shown in Figure 94 (b), the excitation power was reduced to 1 mW. A neutral density filter was used to further attenuate the excitation light if necessary to prevent saturation of the excited state. Fluorescence, like in reflection measurements (in the Section 6.2.4) was also detected through the *X1* port using PMT (PMH-100) and TCSPC card (SPC830, Becker & Hickl) with the emission filter in front (BrightLine HC 692 / 40 nm, Semrock), shown in Figure 94 (b).

### 6.2.6    *Scanning of excitation and STED beams*

Once the beams were overlapped, they had to be scanned across the sample to build up a super-resolved image. Piezo scanning stages, similar to ones used in scanning near-field optical microscopy (discussed in Section 2.7.1), can be used to ensure the highest accuracy available that is required in super-resolution imaging. The stages with a clear aperture are preferred, because a sample can be firmly held from all sides. Piezo stages can guarantee the precise incremental movement and repeatability in 3 D since the scanning can be operated in a closed loop regime, where sensors are used to monitor the motion induced by piezos. A cheap 3 D scanning piezo stage (NanoBlock, MellesGriot, Irvine, CA), alternative to expensive piezo systems, has been routinely used in Stefan Hell's lab. Nevertheless, the stage scanning is inherently slow because a sample has to be moved. In the work presented here, the rapid beam scanning was used with galvano-mirrors as provided in the original configuration of the commercial microscope system. Moreover, if resonant galvanometer scanners are employed, then the scanning speeds of up to 16 kHz can be achieved. At the time when this work was carried out, the beam scanning approach was novel in STED microscopy, since the previous reported STED systems had used piezo-scanning stages. Beam scanning is, however, less accurate and may suffer from off-axis aberrations, especially when imaging on the edge of a field of view. The inaccuracy manifests itself as a jitter – each scanned line along $x$ may start at different $x$ coordinates on the sample. This comes from the hysteresis of the galvanometer scanner (the mirror would not came back to the same point), which is higher than in piezo based scanner systems. Nevertheless, a faster than video rate imaging (80 frames per second) was recently demonstrated by scanning along *x* with resonant scanner and along *y* with piezo [363]. Video rate STED imaging of a live neuron with 60 nm lateral resolution was thus demonstrated [295]. Here we used galvanometer scanner approach to scan in *xy* plane and a stage scanning galvono-mirror to scan in the *z* axis, as available in the Leica confocal microscope that was used here. The scanned area was chosen to be in the centre of the microscope's field of view in order to minimize the off-axis aberrations and, in particular, to minimise chromatic aberrations that would cause a beam shift between the excitation and STED beams. A separated experiment (data not shown), where the gold beads were imaged in the two separate spectral channels using the excitation and STED beams, showed that the beams are overlapped in the centre of field of view but not in the edges of it (field of view – 150 μm). Therefore, this demonstrated that there is a significant amount of lateral chromatic aberration present in the objective used.

## 6.3    STED experiments

### 6.3.1    *STED dynamics*

A spatially aligned and temporally synchronised train of excitation and STED pulses were used to study depletion as a function of the STED beam intensity and the arrival time of STED pulse with respect to excitation pulse. Figure 95 shows fluorescence decay curves that are depleted with STED pulses of various intensities ~ 2 ns after the excitation. The amount of fluorescence that is quenched is proportional to the STED beam intensity, *I* and it is known to vary exponentially (~ $exp(-I / I_s)$, where $I_s$ – saturated intensity) [293]. Unfortunately, the STED beam alone is also able to excite fluorescence, as discussed above and is evident in Figure 95. Fluorescence depletion can be observed when STED pulse arrives after the excitation pulse and there are still fluorophores in the excited state. When the STED pulse arrives before the excitation pulse, the fluorescence decay curve remains largely unchanged (as will be shown in Figure 97). However, to induce the maximum fluorescence quenching, the arrival time of the STED pulse has to be optimised with respect to the excitation pulse. Figure 96 (a) shows the shape of the fluorescence decay curve where the maximum fluorescence quenching has been achieved by optimising the arrival time of the STED pulse with respect to the excitation pulse.

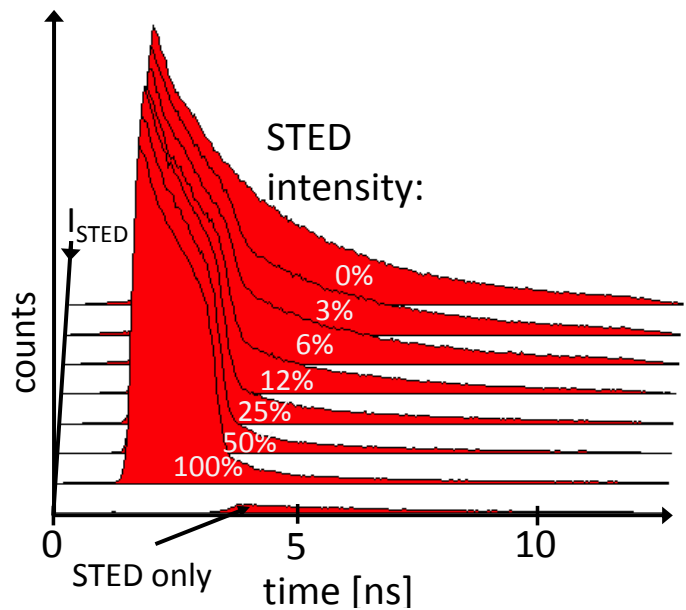

Figure 95. Fluorescence decay as a function of STED intensity when the arrival time of the excitation (EXC) and STED pulses remain fixed – *F(t, $t_{EXC}$=const., $t_{STED}$=const., $I_{STED}$)*. It can be seen that the fluorescence depletion varies nonlinearly with depletion intensity. Note the noticeable fluorescence decay induced by the STED-only excitation.

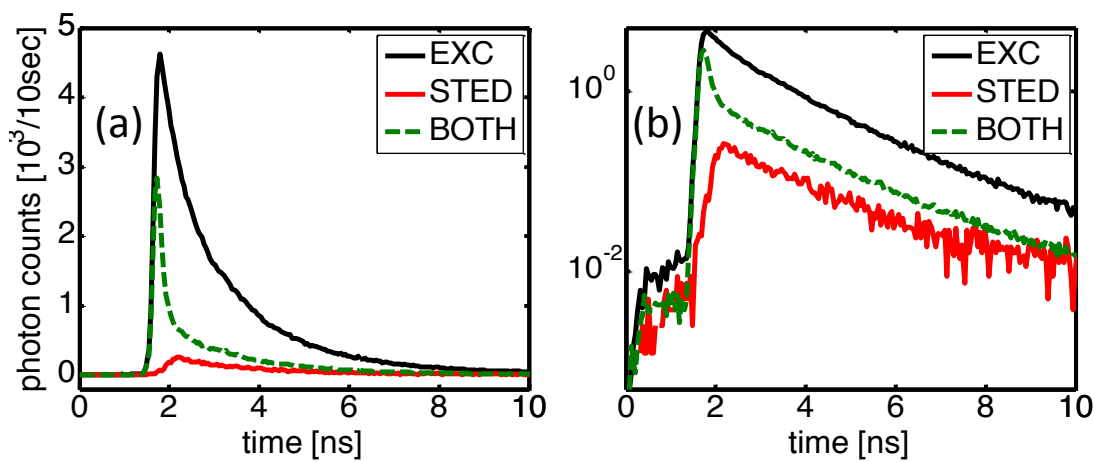

Figure 96. Fluorescence decay curves induced by the excitation pulses (EXC), STED pulses only and the combination of the excitation and STED (BOTH). (a) Linear and (b) log plots. 'EXC' and 'BOTH' fluorescence decays were collected in 10 s., whereas STED in 60 s.

However, under this condition, the beginning of the decay curve displays a peak that can be clearly seen in the log scale of the decay, shown in Figure 96 (b). The peak is not due to the leakage of the scattered excitation or the STED light since the decay curves of fluorescence, shown in Figure 96 (b), excited with the respective beams only do not exhibit such a peak. The peak, as will be shown later, conveys no 'super-resolved' information (the image reconstructed from these early photons was similar to the conventional confocal image, as shown in Figure 104) and therefore could be discarded in imaging experiments, with only later arriving photons used to build up an image. What is also obvious from the Figure 96 (b) is that the decay curves before and after the depletion stay parallel after the peak. This shows that, although the STED pulse affects the fluorescence decay profile, it can still give correct information about fluorescence lifetime of the molecule. That has important consequences for the work presented in this thesis, because it follows that fluorescence lifetime images can be recorded beyond the diffraction limit. The (unwanted) STED-only induced fluorescence decay curve, shown in Figure 96 (b), is less parallel to the other curves in the figure due to the noise present in the tail of the decay. In order to understand why the peak appears in the decay curve, a separate pump-probe type experiment was carried out by recording the fluorescence decay curves as a function of the time delay between the excitation and STED pulses.

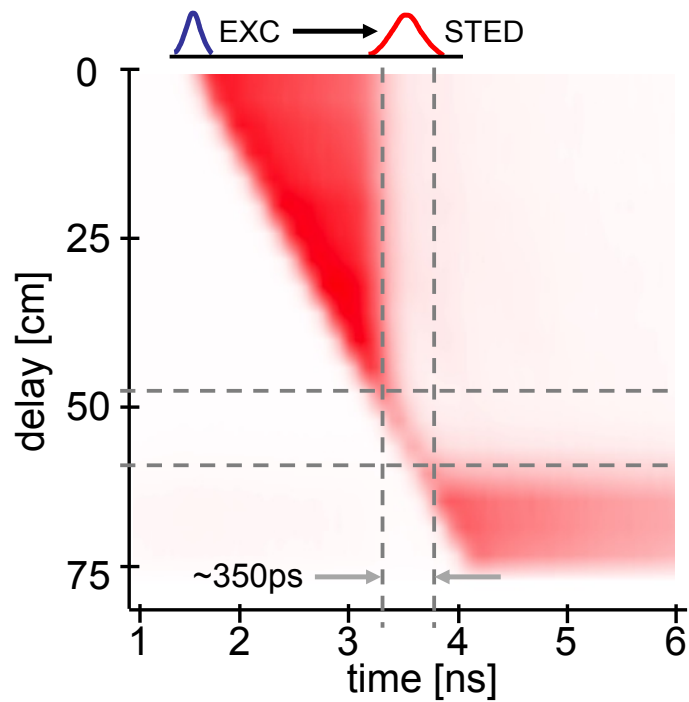

Figure 97. Fluorescence decay as a function of the excitation pulse arrival time with the respect to the STED pulse − $F(t, t_{EXC}, t_{STED}=const., I_{STED}=const.)$. Excitation pulse is shifted in time with the delay line. Dashed lines mark the region where the fluorescence decay curves exhibits a peak at its beginning. This region has the width of the STED pulse indicating that the peak appears when the excitation pulse scans over the STED pulse.

The curves were recorded in the *F(t,T)* mode of the B & H software (see Chapter 3 for more explanations), where the measurement were repeated in specified time intervals, and the results were written into subsequent memory block of the TCSPC module [364]. The excitation pulses rather than STED pulses were chosen to vary in time, by using a mechanical delay line, as explained in Section 6.2.1. Figure 97 shows the fluorescence decay change when the excitation pulse is shifted over ~ 75 cm with the delay line that corresponds to ~ 2.5 ns shift in the pulse arrival time. The STED pulses were of maximum power of ~ 500 mW that was available in this experiment. It follows from Figure 97 that the peak appears when excitation and STED beams physically overlap (any part of it). To understand what may cause the appearance of the initial peak in the fluorescence decay, and to find out how the fluorescence is excited with the STED beam alone, the temporal profiles of the excitation and STED pulses were recorded. The profiles were recorded in the same experiment as the decay curves depicted in Figure 96, but using a mirror instead of the fluorescing sample and replacing the filter in front of the detector with the stack of the neutral density filters. Figure 98 (a) shows temporal shapes of the excitation and STED pulses, which do not represent the actual temporal profiles because the instrument response function of 162 ps was not sufficiently short for measuring the picosecond pulses. Nevertheless, it was possible to accurately determine the pulse arrival time because the TCSPC card afforded superior temporal resolution. For example, Figure 98 (a) shows that, for the experimental conditions used to record the decay curves in Figure 96, the temporal separation between the centres of the excitation and STED pulses are of 280 ps. The measurements could also accurately pinpoint the changes in the STED pulse arrival time as a function of its power. The pulse shapes (STED-1 and STED-2) shown in Figure 98 (a) were recorded at the two different power levels (0.1 mW and 550 mW respectively) that resulted in a shift of the pulse centre by ~ 200 ps (not shown). Figure 98 (b) shows the temporal profile of the excitation pulse and the corresponding fluorescence decay. The graph also shows a function (red curve) of the integrated pulse with respect to the time. It represents the fluorescence build-up induced by the pulse. The function matches the left part of the decay curve confirming that the fluorescence decay is induced by the excitation pulse. In case of the STED pulse induced fluorescence the excitation could, in principle, happen through two-photon or single-photon processes. The two-photon-excitation usually takes place when the STED pulses are shorter than 1 ps, whereas the single-photon excitation happens when the STED pulses are spectrally broad, as explained earlier. Figure 98 (d) shows
the STED-alone induced fluorescence. In addition to a regular STED pulse, a squared
shape of it is plotted that represents the effective pulse profile, which would induce
fluorescence through the two-photon-excitation.

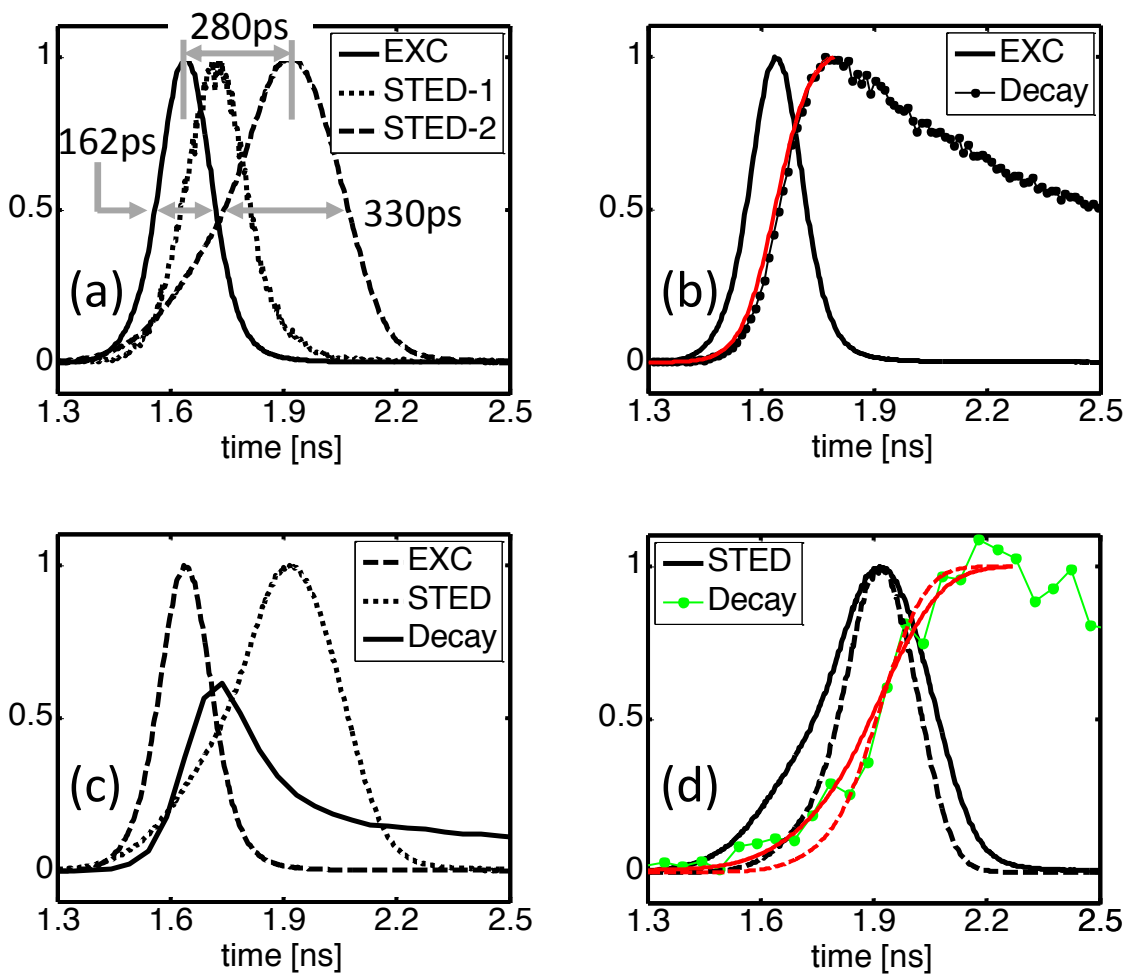

Figure 98. Temporal pulse profiles and fluorescence decay curves. (a) Excitation (EXC) and
STED pulses at two different powers coming out of the stretcher: STED-1 – 0.1 mW, STED-2 –
550 mW. (b) Excitation pulse and its induced fluorescence decay. (c) Fluorescence decay curve
(Decay) that is induced by the excitation (EXC) and STED pulses. (d) STED pulse, its induced
fluorescence decay and the squared STED pulse (dashed curve). Red curves in the graphs depict
a function of an integrated pulse with respect to the time. Dashed red curve is expected
fluorescence build up as induced by the STED pulse though two-photon excitation.

The red curves in the graph depict the fluorescence build-up induced with the STED
pulse through the single and two-photon excitation. The single-photon excitation
fluorescence build-up curve better matches the rising part of the induced fluorescence
temporal profile. Therefore, this suggests that the STED-alone excitation happens
through the single-photon excitation with the 'blue' part of the STED pulse. Figure

98 (c) shows the fluorescence decay curve resulting from the interplay between the excitation and STED pulses. The timing of the pulses is optimised to give a maximum fluorescence quenching as explained above. It was noticed that the temporal measurements of the pulses, presented in Figure 98, as performed with the detector mounted on the microscope's scanner unit, are different from the measurements performed on the optical bench (presented in Section 6.2.3). This discrepancy is illustrated in Figure 99 (a), where it shows that the FWHM of the STED pulse of 520 mW is 465 ps, as measured outside the microscope, and 330 ps − inside the microscope.

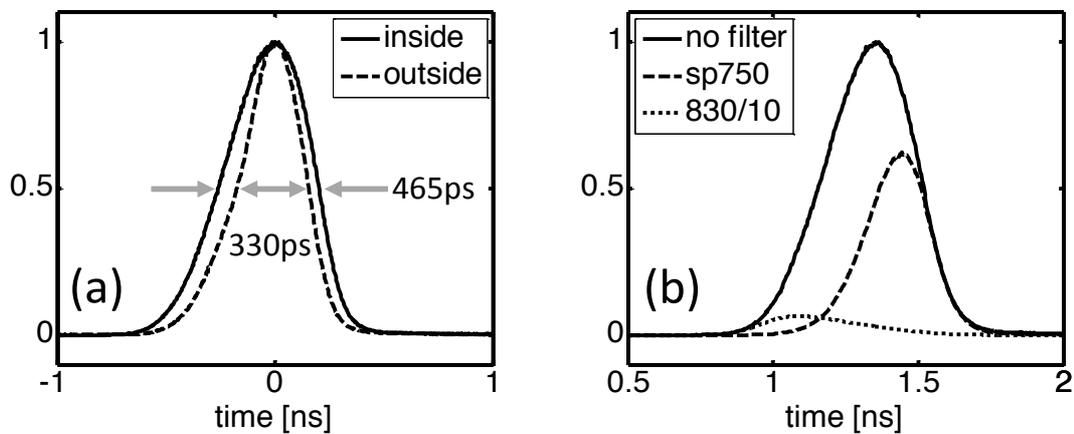

Figure 99. (a) Comparison of the STED pulse shapes measured inside and outside of the microscope. (b) Spectral characteristics of STED pulse showing that the 'red' part of the pulse (square dots) arrives before the 'blue' pulse (dashed). Measurements were done with the short-pass 750 nm and band-pass 830 / 10 nm filter, respectively in front of the detector. The profiles were arbitrary normalised.

The difference must come from the optical path inside the microscope and the scanner unit that the STED pulse takes − the optical components are probably suppressing some parts of the pulse. Figure 99 (b) shows that the STED pulse is spectrally up-chirped. It is to be noted that, if the STED pulse power is changed, the arrival time of the excitation pulses have to be readjusted accordingly, due to the STED pulse arrival time change as a function of power, as explained above. Since the trains of the excitation and STED pulses travel through the single mode and microstructured optical fibres, which have significantly different dispersion profiles, therefore any change in the pulse temporal or spectral profile before the fibres will lead to the considerable changes in the arrival times and the spectral contents of the pulses after the fibres. This can happen, for example, if the laser mode-locking is lost during the experiment, since the re-mode-locking will usually change the spectral properties and timing of the pulses produced by the Ti:Sapphire laser.

### 6.3.2    STED microscopy experiments

After spatial and temporal alignment of the excitation and STED pulses, STED microscopy experiments were performed on the fluorescent beads. The STED power was typically adjusted to provide 40 mW in the pupil plane. At this power level the spectral profile of the pulses broadened to ~ 30 nm in the single mode fibre (due to self-phase modulation), as shown in Figure 91. The STED beam was arranged to overfill the hologram on the SLM that resulted in an overall efficiency of 40 % for the doughnut beam generation, as explained in Chapter 5. Lateral dispersive effects introduced by the holograms (acting effectively as a grating) were minimized by limiting the number of fringes displayed to 30 in this particular experiment. The number could be further reduced to decrease the spectral dispersion in the +1 diffracted order, but in the configuration used here, this would have restricted the blocking of the zeroth diffraction order. The images were acquired over several frames with the lowest possible line scanning speed frequency of 200 Hz, as afforded by the scanner. Figure 100 shows confocal (a) and STED (b) images of 200 nm 'dark red' fluorescing beads (Molecular Probes), as acquired with Leica's PMT mounted on the *X1* port. The increase of resolution is evident.

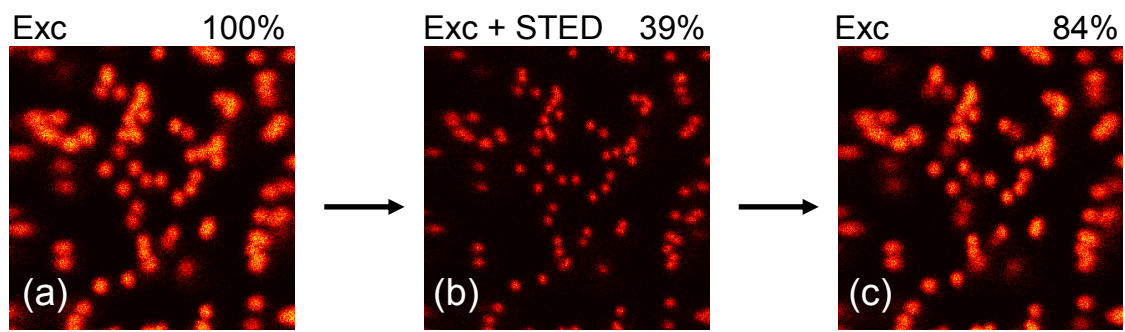

Figure 100. Confocal (a) and STED (b) images of 200 nm 'dark red' fluorescence beads, acquired over 5 s (2 Leica's frames, 512 × 512, 200 Hz line scan speed). STED beam reduces the total fluorescence by 39 %, but some of that is due the photobleaching since the next confocal image (c) displays fluorescence recovery of up to 84 % only. Excitation of 0.4 mW and STED of 40 mW at the pupil planes were used. All images are 18 µm across.

The STED image was acquired using a doughnut shaped STED beam. These images also demonstrate an apparent photobleaching since after 5 seconds of acquisition the overall fluorescence intensity is decreased to 84 % of the initial fluorescence signal. Fourier filtering was employed to reduce noise, since these images were oversampled. The signal-to-noise improvement factor, $G$, gained with the Fourier filtering, is equivalent to the ratio of the area of the Fourier plane (normally − unit), $S$ to that of the mask, $S_{mask}$:

$$G = S / S_{mask}$$

eq. 43

Figure 101 (b) shows that the mask (black colour) of unit radius of 0.12 is blocking out most of the noise and retaining a large part of the signal (yellow).

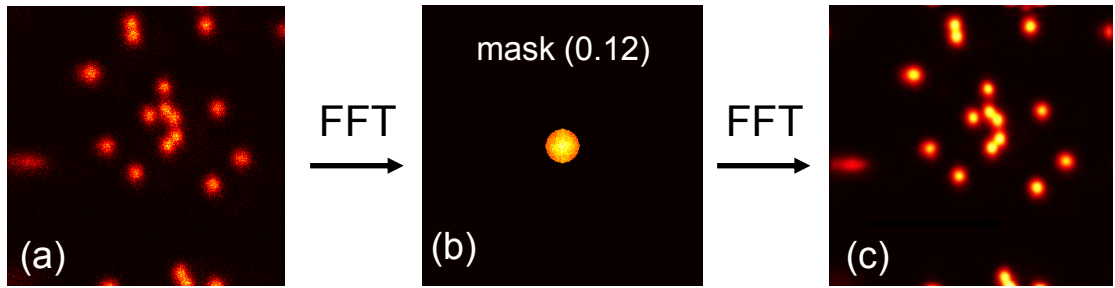

Figure 101. The Fourier filtering of the STED image (a) of 200 nm 'dark red' fluorescence beads that is oversampled. A mask (b) of 0.12 unit radius was used to produce the filtered image (c) with higher signal-to-noise ratio.

The signal-to-noise ratio improvement, $G$ can thus be calculated using eq. 43: $\pi \times 1^2 / (\pi \times 0.12)^2 = 69.4$. Such filtering also reduces image spatial resolution since it is equivalent of convolving the measured data with the Fourier transform of the mask i.e. an Airy disk (eq. 9), with the FWHM equal to pixel size divided by the unit radius: $9.15\ nm\ /\ 0.12 \approx 76\ nm$. To test the resolution, the zoomed-in images were recorded as shown in Figure 102. As can be seen from the Fourier filtered line profiles in Figure 102 (c), the bead FWHM is reduced in the STED image to 200 nm compared to a FWHM of 330 nm in the confocal image. The excitation power used here was sufficient to saturate the fluorophores and therefore the resulting excitation PSF was broadened. This should not limit the final resolution of the STED microscope because the STED beam should deplete the broadened excitation PSF and the resolution ultimately should depend mainly on the shape and the power of the STED beam.

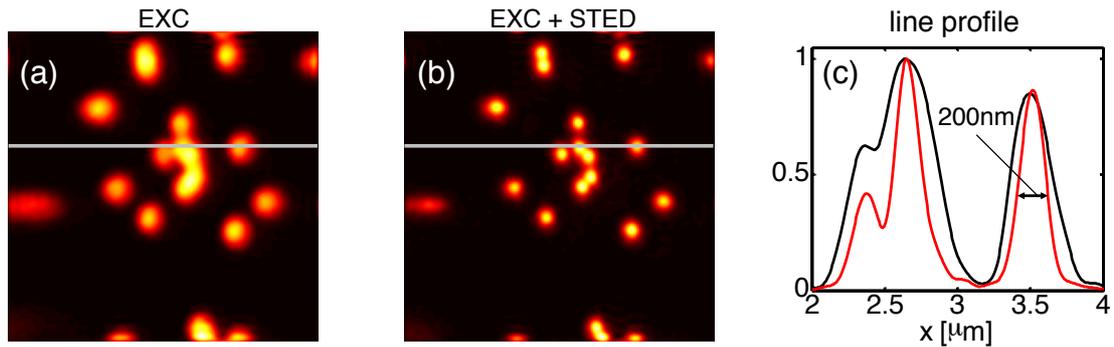

Figure 102. (a) Fluorescence intensity image and (b) Superresolved image of 200 nm 'dark red' fluorescence beads. (c) Line profile from (a) and (b) images. The images acquired with saturated excitation. Acquisition parameters: line rate - 200 Hz, image size – 512 x 512 – 9 μm across., Integration – 25 sec. (10 Leica's frames), Exc – 0.4 mW.

However, it was found in practise that saturated excitation worsened the final resolution in the STED image. Therefore, a new set of images were recorded with the decreased excitation (30 μW at the pupil plane). This resulted in further resolution increase. Figure 103 (a) shows a standard confocal image (*x-y*), whereas Figure 103 (b) and (c) – STED images acquired scanning overlapped excitation and STED beams of the doughnut and the optical beam shape respectively. The shapes of the beams where changed with the spatial light modulator applying *type I* and *type II* holograms, shown in Figure 71, respectively. Figure 103 (g) shows line profiles through the images, which show that STED improves the FWHM of a single bead from 273 nm to 150 nm by applying *type I* hologram and to 210 nm by applying *type II* holograms. It can clearly be seen that some unresolved pairs of beads in the standard confocal image become distinguishable in the STED image, for example, - the pair of beads pointed with the white arrow in Figure 103 (a) and (b), respectively. Some beads in Figure 103 (c) have become less bright due to the enhanced axial resolution achieved using the STED beam with the optical bottle beam shape. This is evident in the red curve in Figure 103 (g) (derived from the image in Figure 103 (c)), which shows that the bead on the left hand side has significantly lower fluorescence intensity as it now falls outside the focal plane imaged due to the improvement in the axial resolution. The axial resolution improvement is more evidently illustrated in the axial (*xz*) images, shown in Figure 103 (d) *to* (f). The FWHM of the bead, shown in Figure 103 (d), is reduced from 0.73 μm to 0.34 μm, by applying the *type II* hologram, as is apparent from line profiles shown in Figure 103 (h).

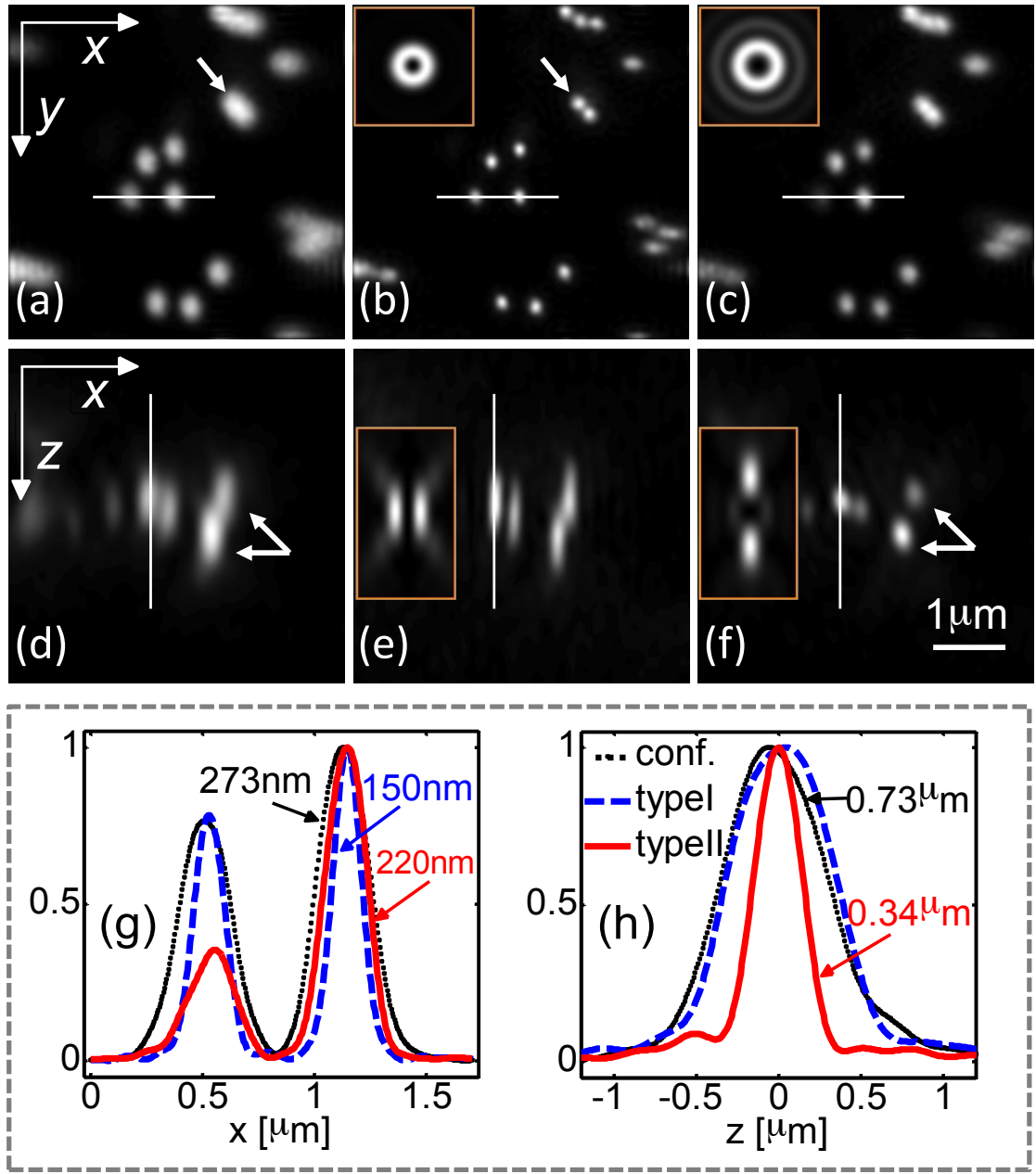

Figure 103. Comparison of confocal and STED images of 200 nm beads. First row (a-c) shows lateral images and second row (d-f) – axial images. The first column, (*a* and *d*), contains images acquired in the confocal mode and the rest of the images are acquired with the STED mode: the second column (*b* and *e*) – with the doughnut shaped STED beam and third column – with the STED beam shaped as an optical bottle beam. (g and h) shows normalised fluorescence intensity line profiles of lateral and axial images, respectively, with FWHM specified for corresponding curves. White arrows shows pair of beads that are unresolved in confocal mode but become resolved in the STED mode. Inset – the shape of the STED beam PSFs. Properties of the images: size – 1024 × 1024, acq. time – 40 sec, line scan speed – 200 Hz, pixel size – 9.19 × 9.18 – axial, 4.61 × 4.61 – lateral. Acquired during 8 Leica's frames.

The white arrows in the corresponding images show beads that are unresolved in the confocal mode but are resolved in the STED mode. The acquired images were over-sampled in all spatial dimensions and therefore Fourier filtering was employed to remove out-of-band noise by suppressing high Fourier frequencies, as explained previously. The results show that the resolution could be improved axially and laterally by using two different STED PSFs. In principle the super-resolved information contained in those two types of images can be combined into one image though some mathematical algorithms, like in Ref. [120].

### 6.3.3    STED-FLIM microscopy

In the experiments described above the time correlated single photon counting (TCSPC) card was configured to acquire images without time resolved information (analog-to-digital converter (ADC) resolution = 1). However, the offset of the time-to-amplitude converter (TAC), described in Section 3.2.3, was set such that only photons arriving after the peak (explained above) would be recorded, since the first photons carried no super-resolution information. Nevertheless, 64 temporal bins can be acquired for each pixel in the $512 \times 512$ image, as afforded by the B&H SPC-830 card. Thus, images of $512 \times 512$ were acquired in 60 s by collecting fluorescence signal over 20 scan frames with a line scan speed of 200 Hz and a dwell time of 60 µs (per FWHM of the excitation PSF). The image had to be acquired over 20 frames to collect enough photons.

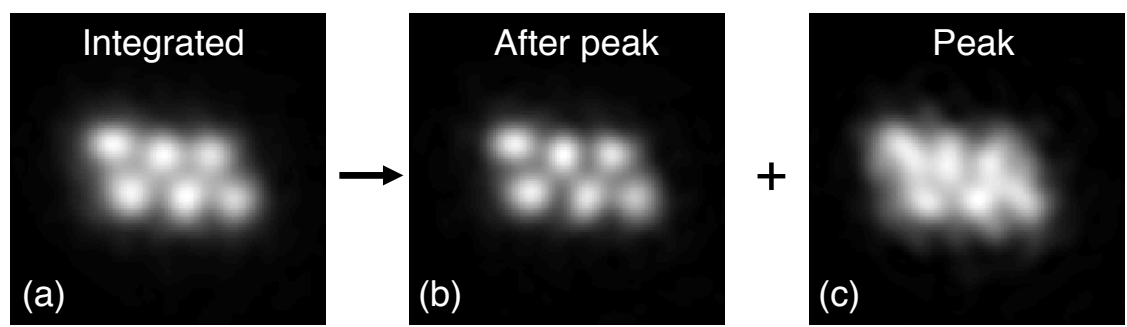

Figure 104. The effect of the peak-removal from the fluorescence decay curve in each pixel of the image. If the whole fluorescence decay curve is integrated (a), then the image of the beads shows less structure compared to the peak-removed image (b), and therefore carries less super-resolved information. (c) An image obtained from the peak-only part of fluorescence curve.

After image acquisition, the procedure was the following: Fourier filtering (as shown in Figure 101), rejection in time of the very first part of the decay (as shown in Figure 104), single exponential fitting to calculate fluorescence lifetime and merging with integrated intensity image (shown in Figure 106). The image formed from the peak alone, shown in Figure 104, shows less structure than the temporally integrated image and therefore can be removed as carrying less super-resolved information. Fourier filtering was used to remove out-of-band noise at high spatial frequencies as explained before. For the image mask of 0.15 unit radius used here, in the Fourier plane, the photon counts per pixel decay were effectively increased from 20 to 900. It also had an effect of convolving the images with Airy disk of 61 nm, as explained before. The Fourier filtering was performed on each time plane individually.

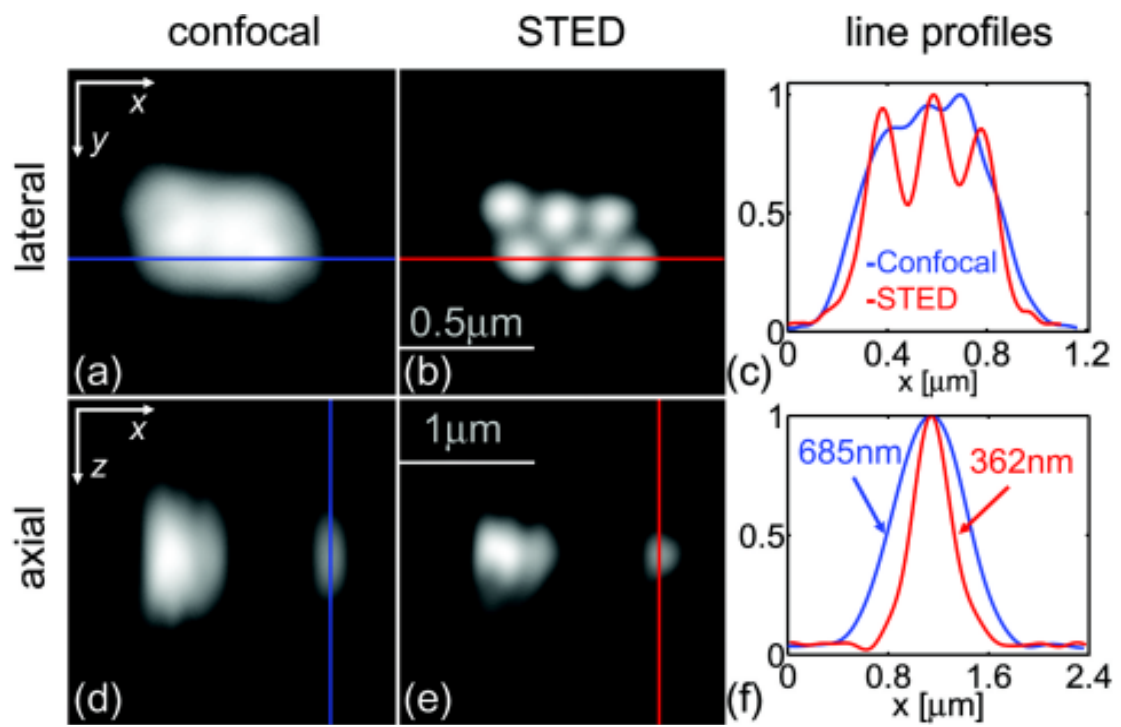

Figure 105. Lateral (a, b) and axial (d, e) fluorescence images of 200 nm beads with: (a, d) confocal acquisition, (b, e) acquisition in STED mode (b – with type I hologram, e – with type II hologram). (c) Normalized fluorescence intensity line profiles of lateral images and (f) axial images with FWHM specified for corresponding curves.

Figure 105 shows images with integrated fluorescence (integrated over the fluorescence decay time) recorded in confocal and STED modes (using the doughnut STED beam). From images and line profiles in Figure 105 it is clear that STED resolves densely packed beads beyond the resolution of standard confocal imaging. The corresponding axial resolution improvement is illustrated in images in Figure 105 (d, e) with STED beam of the optical bottle shape. Single exponential decays were fitted to acquire fluorescence decay images in order to calculate fluorescence lifetime image. A resulting fluorescence lifetime map is shown in Figure 106 (b). An exponential fit was used with a thresholding image mask (as can be seen in Figure 106 (b) as an abrupt change from black to coloured region) to save computational time and to avoid fitting areas with very few photons. It is standard practice in group's lab to merge (modulate) our FLIM maps with the integrated intensity images in order to give pixels with higher photon counts more weight and to retain spatial information. This results in composite images as shown in Figure 106 (c).

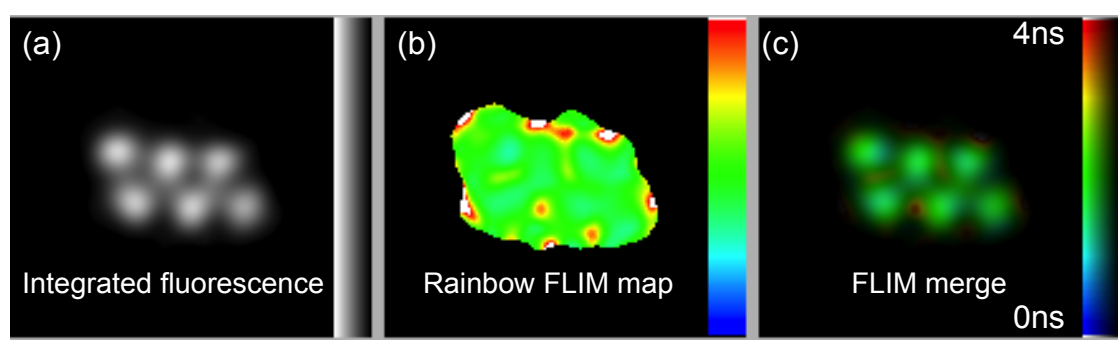

Figure 106. Merging FLIM with the intensity image. (a) Integrated fluorescence image. (b) A FLIM map as obtained by fitting a single exponential to each pixel of the image. (c) A merge of the (a) and (b) images.

Figure 107 compares fluorescence lifetime images acquired in confocal and STED modes. It can be clearly seen that the spatial resolution is improved in the STED mode, but at the same time the fluorescence lifetime stays the same. This more evident looking at integrated fluorescence decay curves in Figure 108. This indicates that STED can be combined with FLIM. The fluorescence decay curve (integrated over the beads) derived from the STED image remains parallel to the normal confocal curve, apart from the initial peak discussed earlier, as can be seen in the logarithmic plot shown in Figure 108 (c), indicating that the STED images returned the correct lifetime. Only later-arriving 'super-resolving' photons were used to calculate the STED and fluorescence lifetime images. The correspondence in lifetime can also be seen in Figure 108 (b) where the centres of the two lifetime histograms are the same. The STED histogram is broader, due to decreased signal in the STED image for which the fitting is less stable, giving rise to the more diverse lifetimes. This is evident in Figure 108 (a) where it is shown that the lifetime fitting slightly depends on the number of the counts and is more stable with higher count rates. Figure 108 (b) shows how many pixels in the image were fitted with specific lifetime.

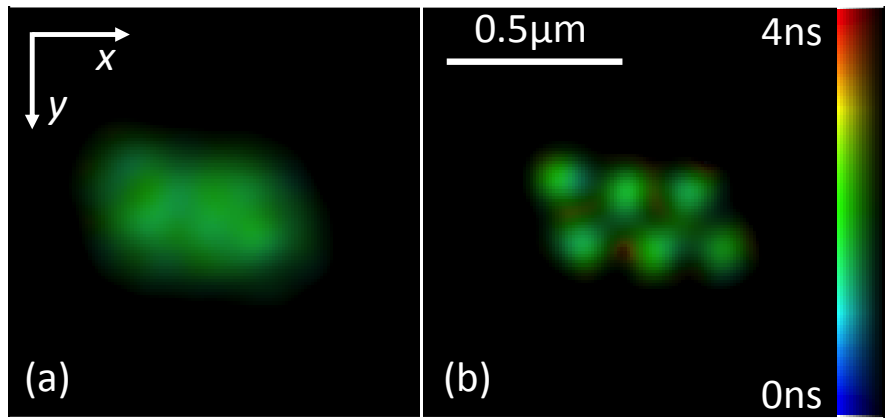

Figure 107. STED-FLIM. Intensity merged fluorescence lifetime images (x-y plane) obtained from the same data set as shown in Figure 105, recorded in (a) confocal mode and (b) STED mode with the doughnut shaped STED beam.

Both curves follow Poissonian distribution, the centres of which is almost in the same position and equals to ~1.7 ns. The STED curve has longer lifetime fitted because of a fever photons in the pixel.

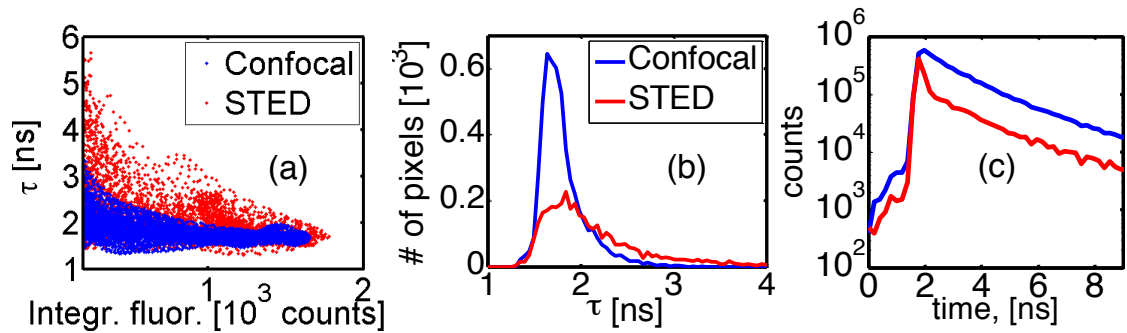

Figure 108. Lifetime distributions. (a) Fitted lifetime as a function of photon counts in the overall fluorescence decay curve in confocal (blues) and STED images (red). (b) Lifetime histogram in confocal (blues) and STED images (red). (c) *log* plot fluorescence decay (integrated over the beads) curves of confocal (blue) and STED (red) images, respectively.

### 6.3.4    STED with fibre laser-based supercontinuum source

In STED microscopy the spectral tunability of the excitation laser source is useful, as it is in any other fluorescence microscopy. However the situation in STED microscopy is more complicated since the tunability in STED beam is also desired along with tunability in the excitation. It is further complicated by the difference in the power levels required in the two beams. Supercontinuum stemming from a microstructured optical fibre being pumped by Ti:Sapphire (as described in this Chapter) was not powerful enough to deplete fluorescence and therefore a part of Ti:Sapphire radiation was used directly due to its higher power spectral density. However, with the new powerful fibre lasers it is now possible to obtain both excitation and STED beams from a single compact fibre laser-based supercontinuum source.

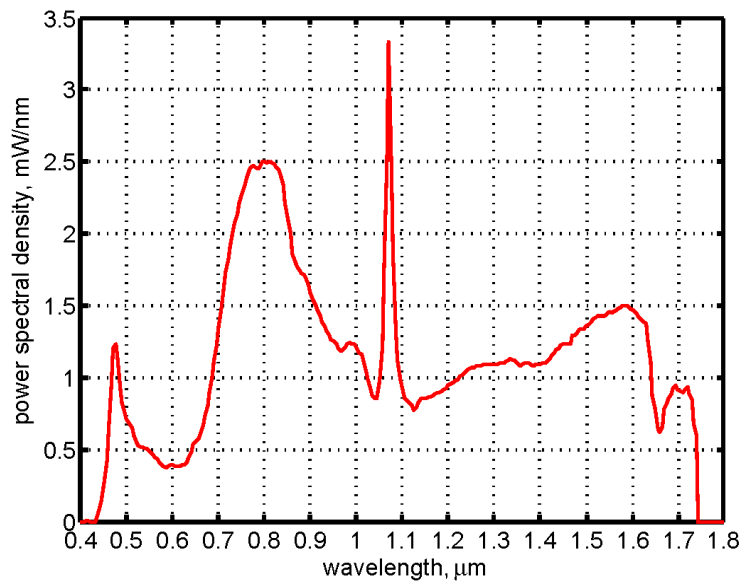

Figure 109. Emission spectrum of fibre laser-based supercontinuum source (SC450-2, Fianium Ltd, Southampton, UK).

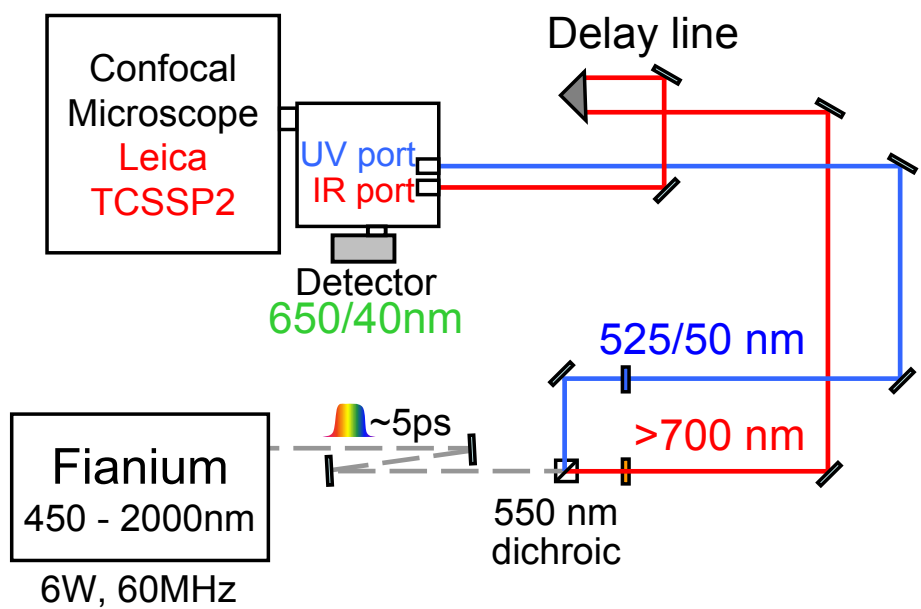

Figure 110. STED microscopy setup based on Supercontinuum Fibre Laser Source (Fianium Ltd).

Such sources can now provide high ($> mW / nm$) average powers with spectral coverage from the near infrared to the ultraviolet [365]. Therefore, for the second approach in this Chapter, a high power fibre laser-pumped supercontinuum source was used, which provided ~ 1.5 W average power in the visible spectrum (~ 450-800 nm) at 20 MHz, built by Fianium Ltd following our specifications. The output from this source (spectrum shown in Figure 109) was split using a 550 nm dichroic to provide the excitation and STED beams as shown in Figure 110. The achieved fluorescence depletion was of 70 % for Pyridin-2 molecules, as can be seen from fluorescence decay curves in Figure 111 (b). Therefore, it demonstrated, at the time when those experiments were carried out, that the fibre laser-based supercontinuum sources could be potentially employed in STED microscopy.

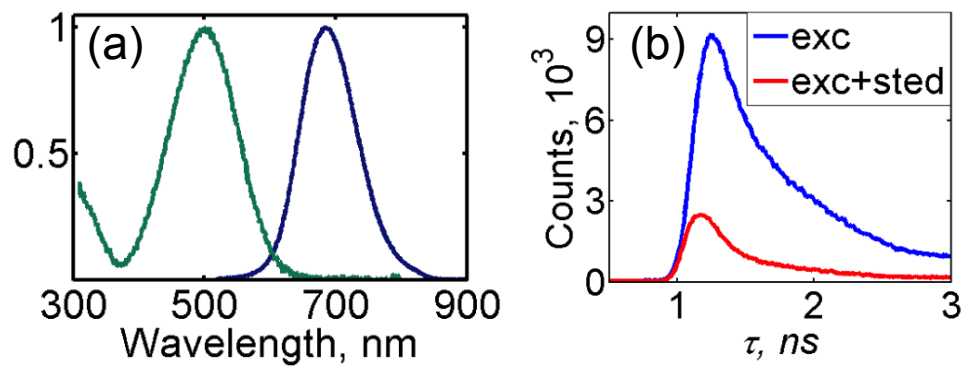

Figure 111. (a) Excitation and fluorescence spectra of Pyridin-2 (in ethanol). (b) Fluorescence decays curves of Pyridine-2.

However, the technology of those sources at the time did not allow to have enough power for STED beam and therefore Ti:Sapphire laser was employed instead, as described earlier in this Chapter. Recently, as mentioned previously in the Section 6.2.2, a new model of the fibre laser-based supercontinuum source (SC-450-PP-HE, Fianium Ltd, Southampton, UK) was successfully employed to improve resolution, as described in Ref. [291]. The source had a pulse picker installed before amplifier (see Figure 43 as a reference) that allowed changing repetition rate of the laser as well as generating higher pulse energy and therefore achieving better fluorescence depletion when used in STED microscopy.

## 6.4     Summary and Outlook

This Chapter demonstrated for the first time that the fluorescence lifetime images can be acquired at spatial resolutions below the diffraction limit. It was also the first demonstration of the fact that supercontinuum generated in a microstructured optical fibre can be used as a relatively cheap alternative tunable excitation source to some of the lasers currently used in STED microscopy. In addition, a control of the STED beam wavefront was demonstrated using a spatial light modulator that enabled convenient and fast switching between different imaging modes (to improve resolution laterally or axially). Furthermore, the holographic control of the STED beam with the spatial light modulator enabled to imprint the wavefront of the STED beam with an accurate reproduction of the phase distribution displayed on the spatial light modulator that allowed precise control of the point spread function and sensitive correction of aberrations present in the STED beam illumination path. Time resolved study revealed that there is a peak in the beginning of the decay curve, which appears when the maximum quenching of fluorescence is achieved. The peak seems to appear when there is a temporal and spatial overlap of the excitation and the STED beams, suggesting that it is due to some kind of nonlinear process. However, the peak can be rejected and only following part of fluorescence decay profile used to reconstruct the intensity image. It can be also used to derive fluorescence lifetime images beyond the diffraction limit by fitting single exponential curves to the recorded decay profiles. Thus STED-FLIM concept was introduced and is expected to be useful to report FRET with better resolution. Moreover STED microscopy was demonstrated here in a standard commercial confocal laser-scanning microscope, allowing the flexibility and convenience of sophisticated instrumentation developed for imaging biological samples. This is expected to be useful, for example, to look at protein interactions using FLIM-FRET in order to study nanoclusters of cell signalling or to study sub-resolution features such as "lipid rafts" with the better resolution. The stability of the galvanometric scanners in this particular commercial instrument unfortunately imposed an ultimate limit to the achievable resolution improvement, comparing in favourably to that reported elsewhere using piezo stage scanning. Therefore, more stable scanners should be employed in the future. It is likely that the fibre laser based supercontinuum source will be more commonly employed in the future, since it can access almost all wavelengths necessary for both excitation and STED of the dyes most commonly used in biological research. Recently, it has been shown there is enough spectral power density in the red and the infrared spectral region of a supercontinuum source to induce considerable STED without having to integrate over a large spectral bandwidth. The fibre supercontinuum sources are also cheap compared to the Ti:Sapphire lasers that are routinely used for nonlinear microscopy and FLIM.

# 7.   Conclusions and Outlook

This thesis concerned the development and application of fluorescence lifetime imaging (FLIM) microscopy, with particular focus on utilising supercontinuum generation in microstructured optical fibres (MOF) to provide tunable ultrafast excitation sources and the development of a super-resolving FLIM microscope exploiting the technique of stimulated emission depletion (STED).

Optical microscopy, and particularly fluorescence microscopy, is one of the most widespread imaging techniques in cell biology due to its potential for molecular (spectroscopic) contrast, high sensitivity, non-invasiveness, imaging speed and ability to image in 3 D. It is therefore often preferred over other higher resolution imaging modalities such as electron microscopy or scanning probe microscopy. Various optical microscope techniques have been developed to image with a better contrast, imaging speed or resolution, as reviewed in Chapter 2. To increase molecular contrast, various parameters, such as fluorescence excitation and emission spectra, polarisation and fluorescence lifetime, can be recorded in addition to the signal intensity and many different excitation / detection geometries can be used, such as dark field or confocal configurations, for instance. Furthermore, nonlinear processes can be exploited, including two-photon excitation, second harmonic and coherent anti-Stokes Raman scattering microscopies, with the latter two being able to provide information complementary to fluorescence. In order to increase imaging speed, various approaches can be used, including Nipkow disk and line scanning microscopy. The latter can be used to rapidly acquire spectrally resolved images if combined with a spectrograph and wide-field (2 D) detector. To improve the resolution of the optical microscopy beyond the diffraction-limited has been a major challenge for optical molecular imaging and has only recently been successfully addressed. Several techniques have been developed that can increase resolution beyond what is possible from conventional instruments, for example, by using a larger effective numerical aperture to illuminate and / or collect light and using spatially structured illumination techniques, such as in I$^5$M, 4pi or standing wave microscopies. Various sample nonlinearities can be also exploited, especially to increase resolution axially, since nonlinearities such as two-photon absorption or second harmonic generation occurs only in the focal plane. However, a theoretically unlimited spatial resolution can be achieved with the so-called reversible saturable optical (fluorescence) transition (RESOLFT) techniques that rely on the nonlinear behaviour of reversible saturable optical transitions. It includes STED microscopy and saturated structured illumination microscopy. STED microscopy is arguably the most promising super-resolution technique in that it is able to achieve highest resolution to date, including in 3 D. It can also provide video rate imaging speeds and can be successfully applied to live cell imaging. So far some fundamental biological questions were answered with this technique. Recently, it has been demonstrated that, in principle, diffraction unlimited resolution can also be achieved with stochastic optical reconstruction and photo-activated localisation microscopy techniques, which are based on the localisation of individual fluorescent molecules.

Fluorescence lifetime can be employed to contrast different fluorophores, even if they have similar fluorescence excitation / emission spectra. In addition, the measurement of fluorescence lifetime is insensitive to a fluorophore's concentration, the path length or attenuation of the detected photons and, therefore, fluorescence lifetime can be used to measure changes in the local fluorophore's environment. Thus, fluorescence lifetime imaging (FLIM) can provide a robust means to map out fluorophore interactions with other molecules if they change the local environment, e.g. due to excited-state dynamics as in the case of FRET. FLIM can be performed in a wide-field or scanning microscopes using, for example, a gated optical intensifier (GOI) and time correlated single photon counting (TCSPC), respectively. Chapter 3 demonstrated FLIM applications to biology where FLIM was either used to sense changes in the molecular environment or to report FRET. Specifically, green fluorescent protein (GFP)-labelled killer immunoglobulin-like receptor (KIR) and Cy3 labelled anti-phosphortyrosine specific molecular antibody pair was used to indicate the receptor phosphorylation at immune synapses. It revealed that the KIR and therefore the signalling was spatially restricted to the intercellular contact and occurred in clusters. These microclusters were of the order of the resolution limit of the fluorescence microscope and so it is hoped that super-resolved microscopy could resolve the microclusters' structure and provide additional insight into the KIR micro-organisation. FLIM was also investigated as a potential tool to study different actomyosin states in skeletal muscle fibres by monitoring fluorescence lifetime change of an ATP-analogue labelled with a coumarin-based fluorophore, which could bind to the actomyosin. It was also shown that FLIM can be used to report changes in the actomyosin complex during muscle contraction.

Excitation sources based on the supercontinuum generation in microstructured optical fibres have recently become popular for spectroscopy and microscopy applications, due to their broad spectral tunability, ultrashort pulse operation and high spatial coherence. The most promising pump source for practical supercontinuum generation seems to be a picosecond Ytterbium fibre laser since it is powerful, cheap, easy to operate and is also able generate broad supercontinuum as demonstrated in Chapter 4. However, such supercontinuum generation is typically limited to ~ 450 nm on the short wavelength side of the spectrum (when microstructured optical fibre with a fixed zero-dispersion wavelength near the pump wavelength is used) and therefore limits its application to fluorescence microscopy. However, if microstructured optical fibre is tapered, such that its core diameter (and therefore the nonlinear refractive index and the dispersion profile), change along its length, supercontinuum can extend to the ultraviolet. Chapter 4 presented various supercontinuum generation setups that were used as the excitation source for different FLIM microscopes, including wide field, Nipkow disc and line scanning hyperspectral microscopes. For the latter microscope, the supercontinuum extending to the ultraviolet was used that was generated by pumping tapered microstructured optical fibre with the fibre laser.

To maximally increase resolution in STED microscopy, the PSF of the STED beam has to be engineered in such a way that it would result in optimal final excitation PSF. For improving the lateral resolution, a 'doughnut' shaped PSF can be used. However, this does not increase the axial resolution. Axial super-resolution can be realised using the 'optical bottle' PSF but this does not provide the same improvement in lateral resolution as the doughnut. Therefore, these two PSF have to be used either simultaneously or one after another to gain super-resolved image information in 3 D. The most convenient way to generate these PSF is to use a phase plate that can imprint a particular phase distribution on the STED beam's wavefront. However, using a computer controlled spatial light modulator (SLM) allows the phase distribution to be changed on a fast time scale and therefore permits PSF control in real time, which can also be used for adaptive correction of phase aberrations, e.g. arising from the microscope optics. Chapter 5 presented a holographic STED beam PSF control approach that enabled a more precise control of the phase distribution because its accuracy was not limited by the number of gray levels afforded by the SLM, but by the accuracy with which fringes could be generated (which depends on the size of the SLM array). The holographic control also permits the zeroth diffraction order, which contributes to non-zero on-axis intensity, to be separated from the STED beam by diffracting them at different angles. Furthermore, by using a blazed grating-like structured hologram most of the energy can be concentrated in to the one diffraction order. Aberration correction can be performed by adjusting the contributions of individual Zernike modes, the proportion of which could be found by either visually inspecting the beam profile or by using a modal wavefront sensor. A phase only SLM was used that, after calibration, could display various computer generated holograms with 0 to $2\pi$ phase modulation capability allowing an exact phase distribution imprint. Using the SLM to produce the appropriate PSF, the temporally synchronised and spatially overlapped STED and the excitation beams, with the later being derived from a supercontinuum source, were scanned across the sample with galvanometric scanners to realise STED microscopy. Chapter 6 presented STED microscopy images that demonstrated resolution improvement beyond the diffraction limit in the axial and lateral planes. In addition, the temporal fluorescence decay profiles were measured using TCSPC to provide fluorescence lifetime images with resolution beyond the diffraction limit. This was the first demonstration of STED-FLIM, which is expected to be useful to record sub-diffraction FRET images. The laser scanning confocal FLIM microscope used for much of the work reported in this thesis was based on several commercial sub-systems consisting of the conventional confocal microscope, the TCSPC system and the ultrafast laser system. It was modified to STED microscope by introducing a supercontinuum generation setup and an SLM for wavefront engineering. Leica Microsystems GmbH now manufactures a microscope that integrates STED with confocal microscopy. The microscope developed in this thesis uniquely combines STED microscopy, FLIM and supercontinuum excitation, but it is expected that all these modalities will soon be available in a commercial instrument and will find wide application for biology and other fields of research.

In summary, this thesis describes some developments in FLIM instrumentation and applications to biological research including the extension of FLIM microscopy to multidimensional fluorescence imaging, with various applications of tunable ultrafast supercontinuum excitation sources, and the development of the first super-resolved FLIM microscope. It is hoped that this work will lead to new opportunities for studying cell signalling, tissue autofluorescence and other topics in the life sciences.

# Publications

### Appearance in science and technology magazines

1. D. J. Palmer, "Two methods marry for superresolution imaging," Laser Focus World **44**, 17-20 (2008)
   (Also in *BioOptics World* (March 2008), p. 16.)
2. P. Gwynne, "Supercontinuum Lasers begin to shine in Biomedicine," in *BioOptics World* (March 2008), p. 10.

### Conference proceedings

1. H. Manning, D. Owen, E. Auksorius, P. d. Beule, S. Oddos, C. Talbot, C. Dunsby, I. Munro, A. Magee, M. Neil, and P. French, "Applications of rapid time-gated hyperspectral FLIM: live cell imaging of membrane order and 6-D microscopy," in *Confocal, Multiphoton, and Nonlinear Microscopic Imaging III*, A. Periasamy, ed., Vol. SPIE Volume 6630 of Progress In Biomedical Optics And Imaging (Optical Society of America, 2007), paper 6630_44.
2. D. Grant, E. Auksorius, D. Schimpf, D. S. Elson, C. Dunsby, J. Requejo-Isidro, I. Munro, M. A. Neil, P. M. French, and P. Courtney, "Fluorescence Lifetime Imaging Microscopy Using a Tunable Continuum Source and a Nipkow Disk Confocal Microscope," in *Confocal, Multiphoton, and Nonlinear Microscopic Imaging II*, T. Wilson, ed., Vol. SPIE Volume 5860 of Progress In Biomedical Optics And Imaging (Optical Society of America, 2005), paper ThF8.

### Selected conference presentations

1. E. Auksorius, B. R. Boruah, P. M. P. Lanigan, G. Kennedy, C. Dunsby, M. A. A. Neil, and P. M. W. French, "FLIM beyond the diffraction limit using STED microscopy with a supercontinuum excitation source and holographic PSF control," in BiOS, Photonics West (San Jose, USA, 2009).
2. E. Auksorius, B. R. Boruah, C. Dunsby, P. M. P. Lanigan, G. Kennedy, M. A. A. Neil, and P. M. W. French, "STED microscopy with supercontinuum source, holographic PSF control and FLIM," in Focus on Microscopy (Osaka-Awaji, Japan, 2008).
   (Also mentioned in *Nature Photonics*: "View from...FOM 2008: Beyond the limit," Nat Photon **2**, 338-339 (2008))
3. E. Auksorius, B. R. Boruah, C. Dunsby, P. M. P. Lanigan, G. Kennedy, M. A. A. Neil, and P. M. W. French, "Stimulated emission depletion microscopy with a supercontinuum source and fluorescence lifetime imaging," in Photonex07 (Coventry, UK, 2007).
4. E. Auksorius, D. M. Owen, H. B. Manning, P. De Beule, D. M. Grant, S. Kumar, P. M. P. Lanigan, C. B. Talbot, J. McGinty, C. W. Dunsby, M. A. A. Neil, and P. M. W. French, "Application of Tunable Continuum Sources to Fluorescence Imaging and Metrology," in BiOS, Photonics West (San Jose, USA, 2007).